\pdfoutput=1 
\documentclass[12pt,letterpaper]{article}

\usepackage{amssymb,amsmath}
\usepackage{graphicx}
\usepackage{color}

\newcommand{\bea}{\begin{eqnarray}}
\newcommand{\beq}{\begin{equation}}
\newcommand{\eea}{\end{eqnarray}}
\newcommand{\eeq}{\end{equation}}

\newcommand{\epsb}{ \overline{\epsilon} }

\newcommand{\epsR}{ {\epsilon_R} }

\newcommand{\tgb}{ t_\beta }

\newcommand{\lsim}{\stackrel{<}{_\sim}}

\newcommand{\be}{\begin{equation}}
\newcommand{\ee}{\end{equation}}

\newcommand{\ba}{\begin{array}}
\newcommand{\ea}{\end{array}}

\newcommand{\vcb}{|V_{cb}|}

\newcommand{\vus}{|V_{us}|}

\newcommand{\nn}{\nonumber}

\newcommand{\bd}{\begin{displaymath}}
\newcommand{\ed}{\end{displaymath}}

\newcommand{\bi}{\begin{itemize}}
\newcommand{\ei}{\end{itemize}}

\newcommand{\heff}{\mathcal{H}_{\rm eff}^{\Delta F = 2}}
\newcommand{\ov}{\overline}

\begin{document}


\begin{titlepage}


\begin{flushright}
{TUM-HEP-727/09}
\\
{MPP-2009-133}
\end{flushright}
\vskip1.2cm


\begin{center}
{\Large \bf \boldmath 
Anatomy and Phenomenology of \vspace{0.3em}\\ FCNC and CPV Effects in SUSY Theories
}

\vskip1.1cm

{\bf
Wolfgang~Altmannshofer$^{1}$,
Andrzej~J.~Buras$^{1,2}$,
Stefania~Gori$^{1,3}$, \vspace{0.2em}\\
Paride~Paradisi$^{1}$
and
David~M.~Straub$^{1}$
}

\vskip0.5cm

$^1$ {\em Physik-Department, Technische Universit\"at M\"unchen, D-85748 Garching, Germany}\\
\vskip0.3cm
$^2$ {\em TUM Institute for Advanced Study, Technische Universit\"at M\"unchen, D-80333 M\"unchen, Germany}\\
\vskip0.3cm
$^3$ {\em Max-Planck-Institut f\"ur Physik (Werner-Heisenberg-Institut), D-80805 M\"unchen, Germany}

\vskip1.5cm
{\large\bf Abstract\\[10pt]}

\parbox[t]{0.9\textwidth}{

We perform an extensive study of FCNC and CP Violation within Supersymmetric (SUSY) theories with
particular emphasis put on processes governed by $b\to s$ transitions and of their correlations 
with processes governed by $b\to d$ transitions, $s\to d$ transitions, $D^0-\bar D^0$ oscillations,
lepton flavour violating decays, electric dipole moments and $(g-2)_{\mu}$.
We first perform a comprehensive model-independent analysis of $\Delta F=2$ observables and we 
emphasize the usefulness of the $R_b-\gamma$ plane in exhibiting transparently various tensions 
in the present UT analyses.
Secondly, we consider a number of SUSY models: the general MSSM, a flavour blind MSSM, the MSSM
with Minimal Flavour Violation as well as SUSY flavour models based on abelian and non-abelian 
flavour symmetries that show representative flavour structures in the soft SUSY breaking terms.
We show how the characteristic patterns of correlations among the considered flavour observables
allow to distinguish between these different SUSY scenarios. Of particular importance are the
correlations between the CP asymmetry $S_{\psi\phi}$ and $B_s\rightarrow\mu^+\mu^-$, between
the anomalies in $S_{\phi K_S}$ and $S_{\psi\phi}$, between $S_{\phi K_S}$ and $d_e$, between
$S_{\psi\phi}$ and $(g-2)_{\mu}$ and also those involving lepton flavour violating decays.
In our analysis, the presence of right-handed currents and of the double Higgs penguin 
contributions to $B_s$ mixing plays a very important role.
We propose a ``DNA-Flavour Test'' of NP models including Supersymmetry, the Littlest Higgs model
with T-parity and the Randall-Sundrum model with custodial protection, with the aim of showing a
tool to distinguish between these NP scenarios, once additional data on flavour changing processes
become available. As a byproduct, we present the SM prediction for
$\text{BR}(B^+\to\tau^+\nu)=(0.80\pm 0.12)\times 10^{-4}$ that follows solely from an analytical formula for this branching ratio in terms of $\Delta M _{s,d}$ and $S_{\psi K_S}$ asymmetry and which does not involve $V_{ub}$ and $F_B$ uncertainties.
}

\end{center}
\end{titlepage}

\renewcommand{\thefootnote}{\arabic{footnote}}
\renewcommand{\theequation}{\arabic{section}.\arabic{equation}}

\newpage
\setcounter{tocdepth}{2}
\tableofcontents
\newpage

\section{Introduction and Motivation}

The Standard Model (SM) of elementary particles has been very successfully tested at the loop level both in flavour-conserving electroweak (EW) physics at the LEP and the SLC and also in low-energy flavour physics.

On the other hand, it is a common belief that the SM has to be regarded as an effective field theory, valid
up to some still undetermined cut-off scale $\Lambda$ above the EW scale. Theoretical arguments based on a
natural solution of the hierarchy problem suggest that $\Lambda$ should not exceed a few TeV.

Besides the direct search for New Physics (NP) at the TeV scale (the so-called {\em high-energy frontier})
that will be performed at the upcoming LHC, a complementary and equally important tool to shed light on
NP is provided by high-precision low-energy experiments (the so-called {\em high-intensity frontier}).
In particular, the latter allows to indirectly probe very short distances even beyond those accessible by
direct detection.

In the last years, the two $B$ factories have established that the measured $B_d$ flavour- and CP-violating processes are well described by the SM theory up to an accuracy of the $(10-20)\%$ level \cite{Barberio:2008fa,Buchalla:2008jp}. Unfortunately, irreducible hadronic uncertainties and the overall good agreement of flavour data with the SM predictions still prevent any conclusive evidence for NP effects in the quark sector.

This immediately implies a tension between the solution of the hierarchy problem that requires NP at the 
TeV scale and the explanation of the Flavour Physics data in which this NP did not show up convincingly. 
An elegant way to avoid this tension is provided by the Minimal Flavour Violation (MFV) hypothesis~\cite{Buras:2000dm,Buras:2003jf,D'Ambrosio:2002ex,Chivukula:1987py,Hall:1990ac}, where flavour violation, even beyond the SM, is still entirely described by the CKM matrix. As a result, it turns out
that a low energy NP scale at the level of few TeV is still fully compatible with the flavour data within
this minimalistic scenario \cite{Bona:2007vi,Hurth:2008jc}.

In this context, the question we intend to address in this work is whether it is still possible (and to which extent) to expect NP phenomena to appear in the $B_s$ system where the SM has not been experimentally tested
at the same accuracy level as in the $B_d$ system. In particular, it is well known that $b\to s$ transitions represent a special ground where to perform efficient tests of NP scenarios \cite{Moroi:2000tk,Chang:2002mq,Harnik:2002vs,Ciuchini:2002uv,Foster:2005wb,Blanke:2008zb,Blanke:2008yr,Blanke:2006sb,Barger:2009qs}. Indeed, CP violation in $b\to s$ transitions is predicted to be very small in the SM, thus, any experimental evidence for sizable CP violating effects in the $B_s$ system would unambiguously point towards a NP evidence. Recent messages from the CDF and D0 experiments \cite{Aaltonen:2007he,:2008fj} seem to indicate that this indeed could be the case \cite{Bona:2008jn,Lenz:2006hd}.

On the theoretical side, there exist many well motivated NP scenarios predicting large effects especially
in $b\to s$ transitions. Among them are supersymmetric (SUSY) flavour models based on abelian \cite{Nir:1993mx,Ellis:2000js,Joshipura:2000sn,Maekawa:2001uk,Kakizaki:2002hs,Babu:2002tx,Jack:2003pb,Dreiner:2003yr,Berezhiani:1996nu,Choi:1996se,Froggatt:1998xc,Shafi:1999rm,Leontaris:1999wf,Agashe:2003rj,Dudas:1995eq} and non-abelian \cite{Berezhiani:2000cg,Roberts:2001zy,Chen:2003zv,Pomarol:1995xc,Hall:1995es,Carone:1996nd,Barbieri:1996ww,Dermisek:1999vy,Blazek:1999hz,Barbieri:1995uv,Barbieri:1997tu,King:2001uz,King:2003rf,Ross:2004qn,Antusch:2007re} flavour symmetries naturally leading to large NP contributions in $b \to s$ processes while maintaining,
at the same time, the NP contributions occurring in $s \to d$ and (sometimes) $b \to d$ transitions, under control. Moreover, also Grand Unified Theories (GUTs) represent a suitable ground where large NP effects
in $b\to s$ transitions can be generated \cite{Moroi:2000tk,Chang:2002mq,Ciuchini:2007ha,Hisano:2008df,Ko:2008zu,Dutta:2009iy,Trine:2009ns}.
In fact, GUTs link leptonic and hadronic sources of flavour and CP violation and the observed large
atmospheric neutrino mixing is transmitted to a large flavour violation in $b\leftrightarrow s$
transitions \cite{Moroi:2000tk,Chang:2002mq}.

In the present work, we focus on the NP predictions for the CP violating and CP conserving $b\to s$
transitions within SUSY models and their correlations with other observables, measured in coming years
in dedicated flavour precision experiments in $K$, $B$, $D$ and charged lepton decays. In particular,
when we deal with specific models, the source of the flavour and CP violation for $b \to s$ transitions
will simultaneously generate not only $\Delta B=2$ and $\Delta B=1$ processes that will turn out to be 
correlated, but will also have impact on observables outside the $B$ meson system.
The major aim of the present work is twofold:
\begin{itemize}
\item[\bf i)] to quantify the NP room left for $b\to s$ transitions compatible with all the available
experimental data on $\Delta F=2$ and $\Delta F=1$ processes,
\item[\bf ii)] to outline strategies to disentangle different NP scenarios by means of a correlated
analysis of low energy observables, including also $K$ and $D$ systems as well as lepton flavour
violation, electric dipole moments (EDMs) and the anomalous magnetic moment of the muon ($(g-2)_{\mu}$).
\end{itemize}

Since many analyses along this subject appeared in the literature~\cite{Buchalla:2008jp,Harnik:2002vs,Ciuchini:2002uv,Ball:2003se,Foster:2005kb,Goto:2007ee,Dutta:2009iy,Gabbiani:1996hi}, we want to emphasize here the novelties of our study. In particular,

\begin{itemize}

\item We consider a very complete set of $\Delta F=2$, $\Delta F=1$ and $\Delta F=0$ processes, i.e. the
EDMs and the $(g-2)_\mu$. To best of our knowledge, the current analysis represents the most complete
analysis present in the literature in this subject as for i) the number of processes considered and
ii) the inclusion of all relevant SUSY contributions. We believe that the steps i) and ii) are extremely
important in order to try to understand the pattern of deviations from the SM predictions that we could
obtain by means of future experimental measurements.
\item We perform an updated analysis of the bounds on the flavour violating terms in the SUSY soft sector,
the so-called Mass Insertions (MIs), in the general MSSM, in the light of all the available experimental
data in flavour physics.

\item In addition to the general MSSM, we study several well motivated and predictive SUSY frameworks,
such as the MSSM with MFV, a flavour blind MSSM (FBMSSM) and in particular SUSY flavour models based
on abelian and non-abelian flavour symmetries and compare their predictions with those found in the 
Littlest Higgs Model with T-parity (LHT) and a Randall-Sundrum (RS) model with custodial protection.

\item We outline a comprehensive set of strategies to disentangle among different NP scenarios by
means of their predicted patterns of deviations from the SM expectations in many low energy processes.
At the same time, we exploit the synergy and interplay of low energy flavour data with the high energy
data, obtainable at the LHC, to unveil the kind of NP model that will emerge, if any.

\end{itemize}

Our paper is organized as follows. In sec.~\ref{sec:model_independent}, we present a model independent 
analysis of $B_d$, $K^0$ and $B_s$ mixing observables, analyzing in particular the existing tensions
between the data and the SM in the $R_b-\gamma$ plane. We also investigate what room for NP is still
present in the $B_d$ and $B_s$ systems.

In sec.~\ref{sec:DF012}, we list the formulae for $\Delta F=0,1,2$  processes in a general MSSM that
are most relevant for our purposes.

In sec.~\ref{sec:SUSY_breaking}, we briefly review basic features of SUSY models with abelian and non-abelian flavour symmetries as well as models respecting the MFV principle. We also address the
question of the stability of the squark mass matrix textures, predicted by the flavour models, under
RGE effects from the GUT scale to the EW scale.

As our paper involves many observables calculated in several supersymmetric models, we outline in sec.~\ref{sec:strategy} our strategy for the numerical analysis that consists of five steps. The first step
is a model independent analysis within the general MSSM framework resulting in the allowed ranges of the
MIs. The next two steps deal with three specific supersymmetric flavour models in which right-handed
currents play an important role: the abelian model by Agashe and Carone (AC)~\cite{Agashe:2003rj} based
on a $U(1)$ flavour symmetry and the non-abelian models by Ross, Velasco-Sevilla and Vives (RVV)~\cite{Ross:2004qn} (or, more precisely, a specific example of the RVV model, i.e. the RVV2
model~\cite{Calibbi:2009ja}) and by Antusch, King and Malinsky (AKM)~\cite{Antusch:2007re}
based on the flavour symmetry $SU(3)$. These three models are then compared in the last two steps
with a flavour model with left-handed (CKM-like) currents only \cite{Hall:1995es,Nir:2002ah}, with
a flavour blind MSSM (FBMSSM) \cite{Baek:1998yn,Baek:1999qy,Bartl:2001wc,Ellis:2007kb,Altmannshofer:2008hc} and with the MFV MSSM with additional CP phases~\cite{Colangelo:2008qp,Mercolli:2009ns,Paradisi:2009ey}.
The numerical results of our strategy are presented systematically in sec.~\ref{sec:numerics}.
The correlations between different observables, offering very powerful means to distinguish between various models, play an important role in this presentation.

In sec.~\ref{sec:comparison}, we review very briefly the patterns of flavour violation in the Littlest Higgs Model with T-Parity (LHT) \cite{Blanke:2006eb,Blanke:2006sb,Blanke:2007db,Blanke:2007wr,Goto:2008fj,delAguila:2008zu,Blanke:2009am} and a Randall-Sundrum (RS) model with custodial protection \cite{Blanke:2008zb,Blanke:2008yr} and compare them with the ones identified in supersymmetric models in sec.~\ref{sec:numerics}. The main result of this section is a table summarizing and comparing the sensitivity of various observables
to NP effects present in each model.
This table can be considered as a ``DNA-Flavour Test'' of the considered extensions of the SM.

In sec.~\ref{sec:flavour_vs_lhc}, we exploit the complementarity and the synergy between flavour and
LHC data in shedding light on the NP scenario that is at work.

Finally, in sec.~\ref{sec:summary}, the main results of our paper are summarized and an outlook for
coming years is given. A compendium of one loop functions can be found in appendix~\ref{sec:appendix} and in appendix~\ref{sec:appendixRosiek} some details on the convention of the MSSM parameters used throughout our paper are shown.

\section{\boldmath Model-independent Analysis of $\Delta F=2$ Observables} \label{sec:model_independent}
\setcounter{equation}{0}

\subsection[The $B_d$ System]{\boldmath The $B_d$ System} \label{sec:Bd_system}
%
The present unitarity triangle (UT) analyses are dominated by $\Delta F=2$ processes. We begin by reviewing
the status of the UT trying to outline transparently possible hints of NP and related tests to falsify or
to confirm them.

We consider the following two sets of fundamental parameters related to the CKM matrix and to the unitarity triangle:
\begin{equation}\label{eq:UT_parameter_1}
\vus \equiv \lambda ~,~~~ \vcb ~,~~~ R_b ~,~~~ \gamma~,
\end{equation}
\begin{equation}\label{eq:UT_parameter_2}
\vus \equiv \lambda ~,~~~ \vcb ~,~~~ R_t ~,~~~ \beta~.
\end{equation}
Here,
\begin{eqnarray} \label{eq:Rb}
R_b &\equiv& \frac{| V_{ud}^{}V^*_{ub}|}{| V_{cd}^{}V^*_{cb}|}
= \sqrt{\bar\varrho^2 +\bar\eta^2}
= \left(1-\frac{\lambda^2}{2}\right)\frac{1}{\lambda}
\left| \frac{V_{ub}}{V_{cb}} \right| ~,
\\ \label{eq:Rt}
R_t &\equiv& \frac{| V_{td}^{}V^*_{tb}|}{| V_{cd}^{}V^*_{cb}|}
= \sqrt{(1-\bar\varrho)^2 +\bar\eta^2}
= \frac{1}{\lambda} \left| \frac{V_{td}}{V_{cb}} \right| ~,
\end{eqnarray}
are the length of the sides of the UT opposite to the angles $\beta$ and $\gamma$, respectively. 
The latter are defined as follows
\begin{equation} \label{eq:beta_gamma}
V_{td} = |V_{td}| e^{- i \beta} ~,~~~ V_{ub} =  |V_{ub}| e^{- i \gamma}~,
\end{equation}
and one has
\begin{equation} \label{eq:sin2beta_tangamma}
\sin2\beta = \frac{2\bar\eta (1-\bar\varrho)}{\bar\eta^2 + (1-\bar\varrho)^2} ~,~~~ \tan\gamma = \frac{\bar\eta}{\bar\varrho}~.
\end{equation}

The parameter set in (\ref{eq:UT_parameter_1}) can be obtained entirely from tree level decays,
hence, it should be unaffected by any significant NP pollution (see however~\cite{Crivellin:2009sd}).
The corresponding UT is known as the reference unitarity triangle (RUT)~\cite{Goto:1995hj}.
In contrast, $R_t$ and $\beta$ in the parameter set in (\ref{eq:UT_parameter_2}) can only be extracted from loop-induced FCNC processes and hence are potentially sensitive to NP effects. Consequently, the corresponding UT, the universal unitarity triangle (UUT)~\cite{Buras:2000dm} of models with constrained minimal flavour violation (CMFV)~\cite{Buras:2003jf,Blanke:2006ig}, could differ from the RUT signaling NP effects beyond not only the SM but also beyond CMFV models. Thus a comparative UT analysis performed by means of these two independent sets of parameters may represent a powerful tool to unveil NP effects.
The dictionary between these two sets of variables is given by
\begin{equation} \label{eq:Rb_gamma}
R_b=\sqrt{1+R_t^2-2 R_t\cos\beta} ~,~~~
\cot\gamma=\frac{1-R_t\cos\beta}{R_t\sin\beta}~,
\end{equation}
\begin{equation} \label{eq:Rt_beta}
R_t=\sqrt{1+R_b^2-2 R_b\cos\gamma} ~,~~~
\cot\beta=\frac{1-R_b\cos\gamma}{R_b\sin\gamma}~.
\end{equation}

Assuming no NP, the parameters $R_t$ and $\beta$ can be related directly to observables
\begin{equation} \label{eq:Rt_sin2beta_SM}
R_t = \xi \frac{1}{\lambda} \sqrt{\frac{m_{B_s}}{m_{B_d}}} \sqrt{\frac{\Delta M_d}{\Delta M_s}} ~,~~~ \sin2\beta = S_{\psi K_S}~,
\end{equation}
where $\Delta M_d$ and $\Delta M_s$ are the mass differences in the neutral $B_d$ and $B_s$ systems,
$S_{\psi K_S}$ represents the mixing-induced CP asymmetry in the decay $B_d \to \psi K_S$
and the value of the non-perturbative parameter $\xi$ is given in
tab.~\ref{tab:utinputs}.
In the presence of NP however, these relations are modified and one finds
\begin{equation} \label{eq:Rt_sin2beta_NP}
R_t = \xi \frac{1}{\lambda} \sqrt{\frac{m_{B_s}}{m_{B_d}}} \sqrt{\frac{\Delta M_d}{\Delta M_s}} \sqrt{\frac{C_{B_s}}{C_{B_d}}} ~,~~~  \sin(2 \beta + 2 \phi_{B_d}) = S_{\psi K_S} ~,
\end{equation}
where $\Delta M_q = \Delta M_q^{\rm SM} C_{B_q}$ and $\phi_{B_d}$ is a NP phase in $B_d$ 
mixing defined analogously to~(\ref{eq:M12s}) below.

\begin{figure}[t]
\includegraphics[width=1.0\textwidth]{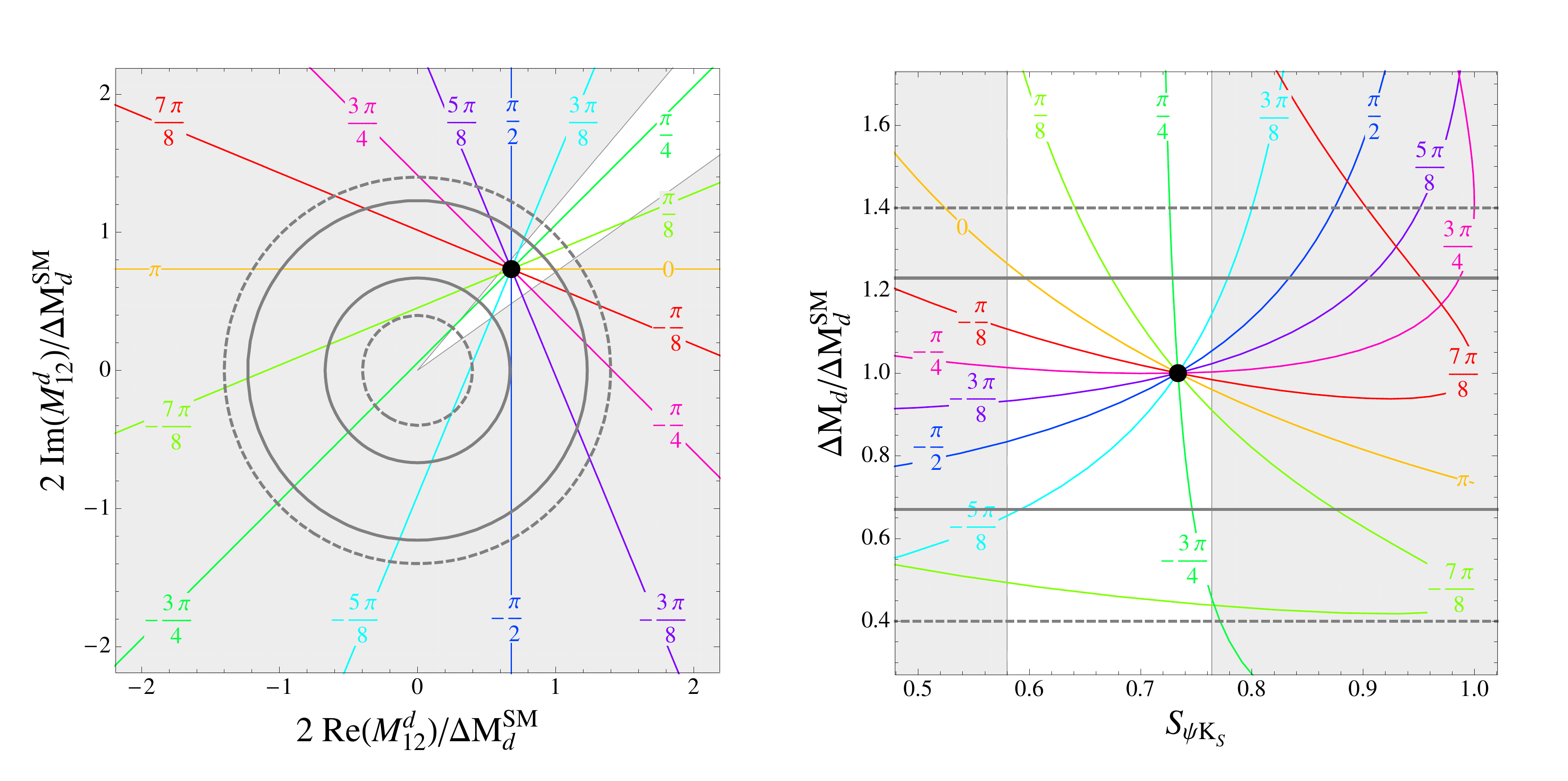}
\caption{\small
Left: The complex $M_{12}^d$ plane. The thick solid circles show the constraint on the $B_d$ mixing amplitude from the measurement of $\Delta M_d / \Delta M_s$ assuming no NP in $\Delta M_s$, while the thick dashed circles correspond to the constraint coming from $\Delta M_d$ alone. The thin lines correspond to NP contributions to $M_{12}^d$ with the indicated fixed phases $\theta_d$, defined analogously to (\ref{eq:M12s}). The white region corresponds to the allowed region of $S_{\psi K_S}$ (see tab. \ref{tab:DF2exp}) and the SM value corresponds to the black point. Right: Same as in the left panel, but now in the $S_{\psi K_S}$-$\Delta M_d / \Delta M_d^{\rm SM}$ plane.
}
\label{fig:M12d}
\end{figure}

The experimentally measured values of $\Delta M_{d}$ and $S_{\psi K_S}$ can be found in tab.~\ref{tab:DF2exp}
along with their SM predictions. While in $\Delta M_{d}$, there is still room for a NP contribution at the 25\% level, scenarios with large new CP violating phases are strongly constrained by the bound on $S_{\psi Ks}$. This is also illustrated by fig.~\ref{fig:M12d}, showing the experimental constraints in the complex $M_{12}^d$ and the $S_{\psi K_S}$-$\Delta M_d/\Delta M_d^{\rm SM}$ plane, respectively.

\begin{table}[t]
\addtolength{\arraycolsep}{3pt}
\renewcommand{\arraystretch}{1.3}
\centering
\begin{tabular}{|c|c|c|c|}
\hline
observable & experiment & SM prediction & exp./SM \\
\hline\hline
$\Delta M_K$ & $(5.292 \pm 0.009) \times 10^{-3}$ ps$^{-1}$~\cite{Amsler:2008zzb} && \\
$|\epsilon_K|$ & $(2.229\pm 0.010) \times 10^{-3}$~\cite{Amsler:2008zzb} & $(1.91 \pm 0.30) \times 10^{-3}$ & $1.17 \pm 0.18$ \\
\hline
$\Delta M_d$ & $(0.507\pm 0.005)$ ps$^{-1}$~\cite{Barberio:2008fa} & $(0.51\pm 0.13)$ ps$^{-1}$ & $0.99 \pm 0.25$ \\
$S_{\psi K_S}$ & $0.672\pm 0.023$~\cite{Barberio:2008fa} & $0.734\pm 0.038$ & $0.92 \pm 0.06$ \\
\hline
$\Delta M_s$ & $(17.77\pm 0.12)$ ps$^{-1}$ ~\cite{Abulencia:2006ze} & $(18.3\pm 5.1)$ ps$^{-1}$ & $0.97 \pm 0.27$ \\
$\Delta M_d/\Delta M_s$ & $(2.85 \pm 0.03) \times 10^{-2}$ & $(2.85 \pm 0.38) \times 10^{-2}$ & $1.00 \pm 0.13$\\
\hline
\end{tabular}
\caption{\small
Experimental values and SM predictions for $\Delta F=2$ observables. The SM predictions are obtained using
CKM parameters from the NP UTfit~\cite{Bona:2005eu}. The last column shows the ratio of the measured value
and the SM prediction, signaling the room left for NP effects in the corresponding observable.
We do not give a SM prediction for $\Delta M_K$ because of unknown long distance contributions.}
\label{tab:DF2exp}
\end{table}

\subsection[The $K$ System]{\boldmath The $K$ System} \label{sec:K_system}

In the $K$ system the central role at present is played by $\epsilon_K$, that represents another 
crucial ingredient of any UT analysis.

The CP-violating parameter $\epsilon_K$ can be written in the SM as follows
\begin{equation} \label{eq:epsK_SM}
|\epsilon_K|^{\rm SM}=\kappa_\epsilon C_\epsilon\hat B_K |V_{cb}|^2 |V_{us}|^2
\left(\frac{1}{2} |V_{cb}|^2 R_t^2 \sin 2\beta \eta_{tt}S_0(x_t)+
R_t\sin\beta(\eta_{ct}S_0(x_c,x_t)-\eta_{cc}x_c)\right),
\end{equation}
where $C_\epsilon$ is a numerical constant 
\begin{equation}
C_\epsilon = \frac{G_F^2 M_W^2 F_K^2 m_{K^0}}{6 \sqrt{2} \pi^2 \Delta M_K} \simeq 3.655 \times 10^4
\end{equation}
and the SM loop function $S_0$ depends on $x_i=m_i^2(m_i)/M_W^2$ (where $m_i(m_i)$ is the quark mass
$m_i$ computed at the scale $m_i$ in the $\overline{\textnormal{MS}}$ scheme) and can be found in
appendix~\ref{sec:appendix}. The factors $\eta_{tt}$, $\eta_{ct}$ and $\eta_{cc}$ are QCD corrections known at the NLO level~\cite{Buras:1990fn,Herrlich:1993yv,Herrlich:1995hh,Herrlich:1996vf}, $\hat B_K$ is a non-perturbative parameter and $\kappa_\epsilon$ is explained below.

While $\epsilon_K$ has been dormant for some time due to a large error in the relevant nonperturbative parameter $\hat B_K$ and CKM parameter uncertainties, the improved value of $\hat B_K$, the improved determinations of the elements of the CKM matrix and in particular the inclusion of additional corrections to $\epsilon_K$~\cite{Buras:2008nn} that were neglected in the past enhanced the role of this CP-violating parameter in the search for NP.

Indeed it has been recently stressed \cite{Buras:2008nn} that the SM prediction of $\epsilon_K$ implied by
the measured value of $\sin2\beta$ may be too small to agree with experiment. The main reasons for
this are on the one hand a decreased value of $\hat B_K=0.724\pm 0.008\pm 0.028$~\cite{Aubin:2009jh} (see also \cite{Antonio:2007pb}), lower by 5--10\% with respect to the values used in existing UT fits \cite{Bona:2005eu,Charles:2004jd}, and on the other hand the decreased value of $\epsilon_K$ in the SM arising from a multiplicative factor, estimated as $\kappa_\epsilon=0.92\pm 0.02$~\cite{Buras:2008nn,Buras:2009pj}. 

Given that $\epsilon_K \propto\hat B_K \kappa_\epsilon$, the total suppression of $\epsilon_K$
compared to the commonly used formulae is typically of order 15\%.
Using directly (\ref{eq:Rt_sin2beta_SM}) together with~(\ref{eq:epsK_SM}), one finds then~\cite{Buras:2009pj}\footnote{Using instead 
the values of $\bar\varrho$ and $\bar\eta$ from tab.~\ref{tab:utinputs} one finds the SM prediction 
in tab.~\ref{tab:DF2exp}.}
\begin{equation} \label{eq:epsK_SMnumber}
|\epsilon_K|^{\rm SM} = (1.78 \pm 0.25) \times 10^{-3}~,
\end{equation}
to be compared with the experimental measurement \cite{Amsler:2008zzb}
\begin{equation} \label{eq:epsK_exp}
|\epsilon_K|^{\rm exp} = (2.229 \pm 0.010) \times 10^{-3}~.
\end{equation}
The 15\% error in~(\ref{eq:epsK_SMnumber}) arises from the three main sources of uncertainty that
are still $\hat B_K$, $|V_{cb}|^4$ and $R_t^2$.

On general grounds, the agreement between (\ref{eq:epsK_SMnumber}) and (\ref{eq:epsK_exp}) improves for higher values of $\hat B_K$, $R_t$ or $|V_{cb}|$ and also the correlation between $\epsilon_K$ and $\sin 2 \beta$ within the SM is highly sensitive to these parameters. Consequently improved determinations of all these parameters is very desirable in order to find out whether NP is at work in $S_{\psi K_S}$ or in 
$\epsilon_K$ or both. Some ideas can be found in~\cite{Lunghi:2008aa,Buras:2008nn,Buras:2009pj,Lunghi:2009sm}. We will now have a closer look at possible tensions in the UT analysis in a somewhat different manner than done in the existing literature.

\subsection[A New Look at Various Tensions: $R_b$-$\gamma$ Plane]{\boldmath A New Look at Various Tensions: $R_b$-$\gamma$ Plane}
\label{sec:UT_tensions}

\begin{figure}[t]
\includegraphics[width=1.\textwidth]{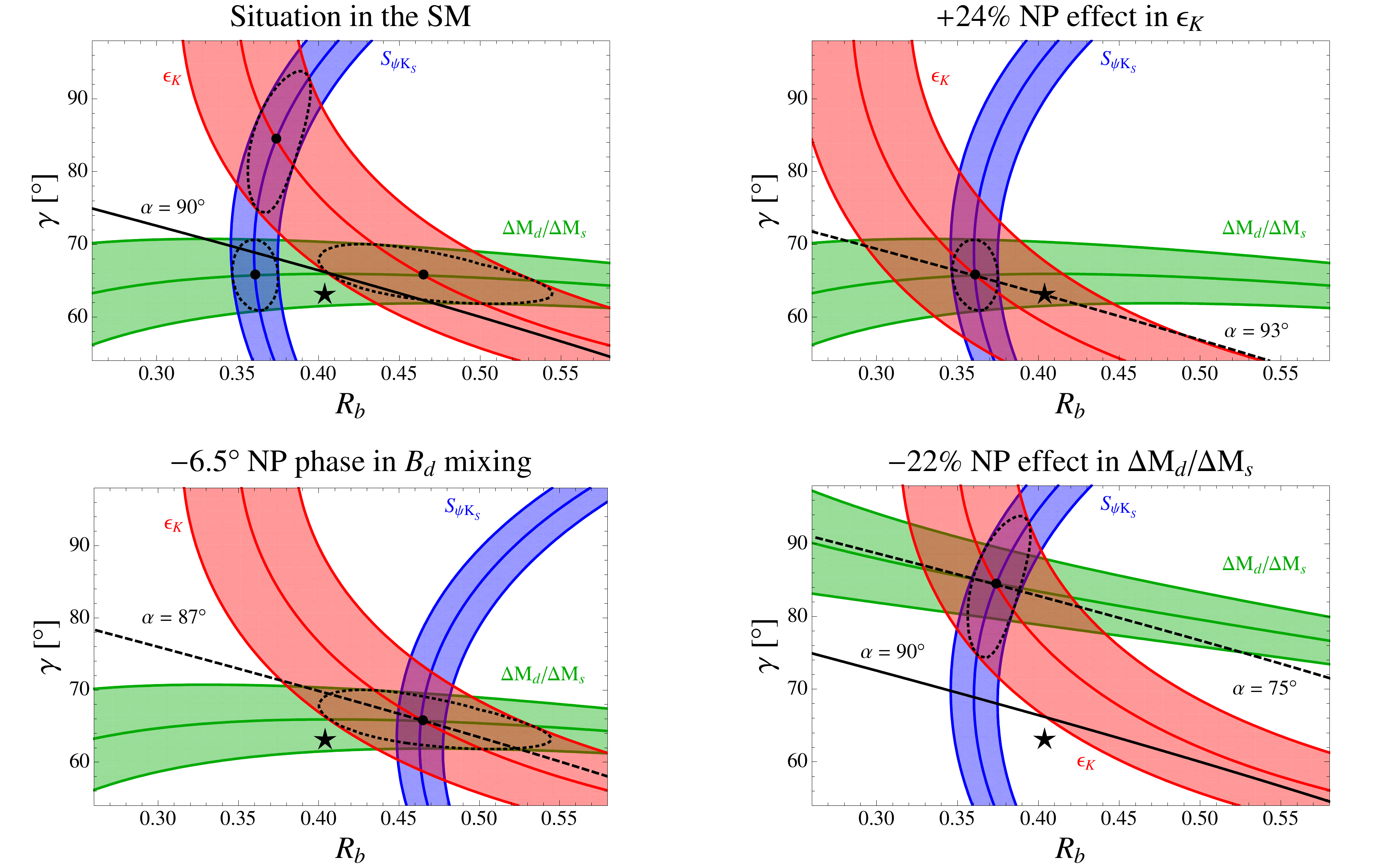}
\caption{\small
The $R_b-\gamma$ plane assuming: {\bf i)} $\sin 2\beta$, $R_t$ and $\epsilon_K$ not affected by NP effects (upper left), {\bf ii)} $\sin 2\beta$ and $R_t$ NP free while $\epsilon_K$ affected by a positive NP effect at the level of $+24\%$ compared to the SM contribution (upper right), {\bf iii)} $\epsilon_K$ and $R_t$ NP free while $\sin 2\beta$ affected by a NP phase in $B_d$ mixing of $-6.5^\circ$ (lower left), {\bf iv)} $\epsilon_K$ and $\sin 2\beta$ NP free while $\Delta M_d / \Delta M_s$ affected by a negative NP effect at the level of $-22\%$ compared to the SM contribution (lower right).
The black star indicates the values for $R_b$ and $\gamma$ obtained in the NP UT fit of \cite{Bona:2005eu}.}
\label{fig:UTfit}
\end{figure}

While the above tension can be fully analyzed by means of the standard UT analysis, we find it more transparent to use the $R_b-\gamma$ plane~\cite{Buras:2002yj,Altmannshofer:2007cs} for this purpose.
In fig.~\ref{fig:UTfit}, in the upper left plot the {\it blue} ({\it green}) region corresponds to the 1$\sigma$ allowed range for $\sin2\beta$ ($R_t$) as calculated by means of~(\ref{eq:Rt_sin2beta_SM}). The {\it red} region corresponds to $|\epsilon_K|^{\rm SM}$ as given in (\ref{eq:epsK_SM}). Finally the solid black line corresponds to $\alpha=90^\circ$ that is close to the one favoured by UT fits and the determination from $B\to\rho\rho$ \cite{Bartsch:2008ps}.

The numerical input parameter that we use to obtain this plot are collected in tabs.~\ref{tab:DF2exp} and~\ref{tab:utinputs}.
It is evident that there is a tension between various regions as there are three different values of $(R_b, \gamma)$, dependently which two constraints are simultaneously applied.

Possible solutions to this tension can be achieved by assuming NP contributions just in one of 
the three observables $\epsilon_K$, $\sin2\beta$, $R_t$:
\begin{itemize}
\item[\bf 1)] a positive NP effect in $\epsilon_K$ (at the level of $+ 24\%$ compared to its SM value)
while taking $\sin 2\beta$ and $\Delta M_d/\Delta M_s$ SM-like~\cite{Buras:2008nn}, as shown by the upper
right plot of fig.~\ref{fig:UTfit}.
The required effect in $\epsilon_K$ could be for instance achieved within models with CMFV by a positive shift in the function $S_0(x_t)$~\cite{Buras:2009pj} which, while not modifying $(\sin2\beta)_{\psi K_S}$ and $\Delta M_d / \Delta M_s$, would require the preferred values of $\sqrt{B_{d,s}} F_{B_{d,s}}$ to be by $\simeq 10\%$ lower than the present central values in order to fit $\Delta M_d$ and $\Delta M_s$ separately. Alternatively, new non-minimal sources of flavour violation relevant only for the $K$ system could solve the problem.
\item[\bf 2)] $\epsilon_K$ and $\Delta M_d/\Delta M_s$  NP free while $S_{\psi K_S}$ affected by a NP phase in $B_d$ mixing of $-6.5^\circ$ as indicated in (\ref{eq:Rt_sin2beta_NP}) and shown by the lower left plot of fig.~\ref{fig:UTfit}.
The predicted value for $\sin2\beta$ is now $\sin 2\beta= 0.823^{+0.088}_{-0.090}~$
\footnote{
The value we give here is slightly smaller than the one found in~\cite{Lunghi:2008aa,Buras:2008nn,Lunghi:2009sm} performing comprehensive fits of the UT.}.
This value is significantly larger than the measured $S_{\psi K_S}$ which allows to fit the
experimental value of $\epsilon_K$.
\item[\bf 3)] $\epsilon_K$ and $S_{\psi K_S}$ NP free while the determination of $R_t$ through
$\Delta M_d/\Delta M_s$ affected by NP as indicated in (\ref{eq:Rt_sin2beta_NP}) and shown by
the lower right plot of fig.~\ref{fig:UTfit}.
In that scenario one finds a very high value for $|V_{td}| \simeq 9.6 \times 10^{-3}$ that
corresponds to $\Delta M_d^{\rm SM} / \Delta M_s^{\rm SM} = (3.66^{+0.52}_{-0.53})\times 10^{-2}$,
much higher than the actual measurement. In order to agree exactly with the experimental central
value, one needs a NP contribution to $\Delta M_d / \Delta M_s$ at the level of $-22\%$.
Non-universal contributions suppressing $\Delta M_d$ ($C_{B_d} < 1$) and/or enhancing
$\Delta M_s$ ($C_{B_s} > 1$) could be responsible for this shift as is evident from (\ref{eq:Rt_sin2beta_NP}). The increased value of $R_t$ that compensates the negative effect
of NP in $\Delta M_d/\Delta M_s$ allows to fit the experimental value of $\epsilon_K$.
\end{itemize}

\begin{table}[t]
\addtolength{\arraycolsep}{4pt}
\renewcommand{\arraystretch}{1.5}
\centering
\begin{tabular}{|l|c|c|c|c|c|c|}
\hline
& $\bar\rho$ & $\bar\eta$ & $\alpha [^\circ]$ & $\gamma [^\circ]$ & $R_b$ & $|V_{ub}|\times 10^3$ \\
\hline\hline
{\bf 1)} & $0.148^{+0.029}_{-0.029}$ & $0.329^{+0.018}_{-0.017}$ & $93.2^{+5.0}_{-5.0}$ & $65.7^{+4.9}_{-4.9}$ & $0.361^{+0.014}_{-0.014}$ & $3.44^{+0.17}_{-0.16}$ \\
\hline
{\bf 2)} & $0.191^{+0.059}_{-0.049}$ & $0.424^{+0.063}_{-0.055}$ & $86.6^{+4.2}_{-3.8}$ & $65.7^{+4.3}_{-3.9}$ & $0.465^{+0.080}_{-0.065}$ & $4.44^{+0.77}_{-0.64}$ \\
\hline
{\bf 3)} & $0.036^{+0.064}_{-0.062}$ & $0.372^{+0.023}_{-0.019}$ & $74.5^{+10.0}_{-9.0}$ & $84.4^{+9.4}_{-10.2}$ & $0.374^{+0.021}_{-0.018}$ & $3.57^{+0.23}_{-0.19}$\\
\hline
UTfit & $0.177\pm0.044$ & $0.360\pm0.031$ & $92\pm7$ & $63\pm7$ & $0.404\pm0.025$ & $3.87\pm0.23$ \\
\hline
\end{tabular}
\caption{\small
Predictions of several CKM parameters in the three scenarios as discussed in the text. For comparison,
in the last line results are also shown from a global NP fit of the UT~\cite{Bona:2005eu}.}
\label{tab:UT_tensions}
\end{table}

Possibility 3) has not been discussed in~\cite{Lunghi:2008aa,Buras:2008nn,Lunghi:2009sm}. Interestingly, there are concrete and well motivated SUSY extensions of the SM compatible with scenario 3) as, for
instance, abelian flavour models (see the following sections) or SUSY GUTs with right-handed neutrinos.
In such cases, if the $b\to s$ transition contains, in addition to a large mixing angle, also a natural $\mathcal{O}(1)$ CPV phase, then solution 3) also implies a non standard value for $S_{\psi\phi}$.

From fig.~\ref{fig:UTfit} it is clear that each of the solutions corresponds to particular values of
$R_b$ and $\gamma$. In tab.~\ref{tab:UT_tensions} we show the values of the relevant CKM parameters 
corresponding to each case, where the values of the two variables characteristic for a given scenario
are assumed not to be affected by NP. We observe
\begin{itemize}
\item {\bf Solution 1)} corresponds to $\gamma \simeq 66^\circ$, $R_b \simeq 0.36$ and  $\alpha \simeq 93^\circ$ in accordance with the usual UT analysis.
\item {\bf Solution 2)} is characterized by a large value of $R_b \simeq 0.47$, that is significantly larger than its exclusive determinations but still compatible with the inclusive determinations. The angles $\gamma \simeq 66^\circ$ and $\alpha \simeq 87^\circ$ agree with the usual UT analysis.
\item {\bf Solution 3)} finally is characterized by a large value of $\gamma\simeq 84^\circ$ and $\alpha$
much below $90^\circ$. The latter fact could be problematic for this solution given the improving determinations of $\alpha$.
\end{itemize}

As seen in tab.~\ref{tab:UT_tensions}, these three NP scenarios characterized by black points in fig.~\ref{fig:UTfit} will be clearly distinguished from each other once the values of $\gamma$ and $R_b$ from tree level decays will be precisely known. Moreover, if the future measurements of $(R_b,\gamma)$ will select a point in the $R_b - \gamma$ plane that differs from the black points in fig.~\ref{fig:UTfit}, it is likely that NP will simultaneously enter $\epsilon_K$, $S_{\psi K_S}$ and $\Delta M_d / \Delta M_s$.

We also note that the easiest way to solve the tensions in question is to require a particular sign in the NP contribution to a given observable: positive shift in $|\epsilon_K|$, $\phi_{B_d}\le 0$ and negative shift in $\Delta M_d/\Delta M_s$, with the latter implying increased values of $R_t$ and $\gamma$.

On the other hand, a negative NP contribution to $\epsilon_K$ would make the tensions in the $R_b-\gamma$ plane more pronounced, requiring stronger shifts in $\phi_{B_d}$ and $R_t$ than in the examples given above. This specific pattern of the tension in the first plot in fig.~\ref{fig:UTfit} points towards certain NP scenarios and rules out specific regions of their parameter space.

\begin{table}[t]
\addtolength{\arraycolsep}{3pt}
\renewcommand{\arraystretch}{1.3}
\centering
\begin{tabular}{|l|l||l|l|}
\hline
parameter & value & parameter & value \\
\hline\hline
$\hat B_K$ & $0.724 \pm 0.008 \pm 0.028$~\cite{Aubin:2009jh} 
& $m_t(m_t)$ & $(163.5 \pm 1.7 ) \; {\rm GeV}$~\cite{:2009ec,Chetyrkin:2000yt}\\
$F_{Bs}$ & $(245 \pm 25)\, \rm{MeV}$~\cite{Lubicz:2008am} 
& $m_c(m_c)$ & $(1.279 \pm 0.013 ) \; {\rm GeV}$~\cite{Chetyrkin:2009fv}\\
$F_B$ & $(200 \pm 20)\, \rm{MeV}$~\cite{Lubicz:2008am}
& $\eta_{cc}$ & $1.44 \pm 0.35$~\cite{Herrlich:1993yv,Nierste:2009wg} \\
$F_K$ & $(156.1 \pm 0.8)\, \rm{MeV}~$\cite{Flavianet}
& $\eta_{tt}$ & $0.57 \pm 0.01$~\cite{Buras:1990fn}\\
$\hat B_{B_d}$ & $1.22 \pm 0.12$~\cite{Lubicz:2008am}
& $\eta_{ct}$ & $0.47 \pm 0.05$~\cite{Herrlich:1995hh,Herrlich:1996vf,Nierste:2009wg}\\
$\hat B_{B_s}$ & $1.22 \pm 0.12$~\cite{Lubicz:2008am}
& $\eta_B$ & $0.55 \pm 0.01$~\cite{Buras:1990fn,Urban:1997gw}\\
$F_{Bs} \sqrt{\hat B_{Bs}}$ & $(270 \pm 30)\, \rm{MeV}$~\cite{Lubicz:2008am}
& $\lambda$ & $0.2258 \pm 0.0014$~\cite{Bona:2007vi}\\
$F_B  \sqrt{\hat B_{Bd}}$ & $(225 \pm 25)\, \rm{MeV}$~\cite{Lubicz:2008am}
& $A$ & $0.808 \pm 0.014 $~\cite{Bona:2007vi} \\
$\xi$ & $1.21 \pm 0.04 $~\cite{Lubicz:2008am}
& $\bar\varrho$ & $0.177 \pm 0.044$~\cite{Bona:2007vi} \\
$V_{cb}$ & $ (41.2 \pm 1.1) \times 10^{-3}$~\cite{Amsler:2008zzb}
& $\bar\eta$ & $0.360 \pm 0.031$~\cite{Bona:2007vi} \\
\hline
\end{tabular}
\caption{\small
Input parameters used in the numerical analysis.}
\label{tab:utinputs}
\end{table}

\subsection[The $B_s$ System]{\boldmath The $B_s$ System} \label{sec:Bs_system}

Since the $B_s$ system is central for our investigations, let us recall some known formulae. 
First the time-dependent mixing induced CP asymmetry
\begin{equation} \label{eq:A_CP}
A^{s}_\text{CP}(\psi\phi,t) 
\equiv \frac{\Gamma(\bar B_s(t) \to \psi\phi) - \Gamma(B_s(t) \to \psi\phi)}{\Gamma(\bar B_s(t) \to \psi\phi) + \Gamma(B_s(t) \to \psi\phi)}
\simeq S_{\psi\phi} \sin(\Delta M_s t)~,
\end{equation}
where the CP violation in the decay amplitude is set to zero. Next, the semileptonic asymmetry is 
given by
\begin{equation} \label{eq:A_SL}
A^s_\text{SL}
\equiv \frac{\Gamma(\bar B_s \to l^+ X) - \Gamma(B_s \to l^- X)}{\Gamma(\bar B_s \to l^+ X) + \Gamma(B_s \to l^- X)} 
= {\rm Im}\left(\frac{\Gamma_{12}^s}{M_{12}^s}\right)~,
\end{equation}
where $\Gamma_{12}^s$ represents the absorptive part of the $B_s$ mixing amplitude. The theoretical
prediction in the SM for the semileptonic asymmetry $A^s_\text{SL}$ improved thanks to improvements
in lattice studies of $\Delta B=2$ four-fermion operators~\cite{Becirevic:2001xt} and to the NLO 
perturbative calculations of the corresponding Wilson coefficients~\cite{Ciuchini:2003ww,Beneke:2003az}.

Both asymmetries are very small in the SM where they turn out to be proportional to $\sin 2|\beta_s|$ with $\beta_s\simeq-1^\circ$. The latter phase enters the CKM matrix element $V_{ts}$
\begin{equation} \label{eq:betas}
V_{ts} = - |V_{ts}| e^{- i \beta_s}~.
\end{equation}
As a consequence, both $A^s_\text{SL} $ and $S_{ \psi\phi}$ represent very promising grounds where
to look for NP effects.

In order to study NP effects in $A^s_\text{SL} $ and $S_{ \psi\phi}$, let us recall possible
parameterizations of the NP contributions entering the $\Delta F = 2$ mixing
amplitudes~\cite{Bona:2005eu,Ligeti:2006pm}
\begin{eqnarray} \label{eq:M12s}
M_{12}^s &=& \langle B_s| H_{\rm eff}^s |\bar B_s \rangle = (M_{12}^s)^{\rm SM}+ (M_{12}^s)^{\rm NP} = |(M_{12}^s)^{\rm SM}|e^{2 i \beta_s}+ |(M_{12}^s)^{\rm NP}|e^{i \theta_s}
\nonumber \\
&\equiv& C_{B_s}e^{2 i\phi_{B_s}}(M_{12}^s)^{\rm SM} 
~.
\end{eqnarray}
For the mass difference in the $B_s$ meson system, one then has
\begin{equation}
\Delta M_s = 2 |M_{12}^s| =  C_{B_s} \Delta M_s^{\rm SM} 
~.
\end{equation}
In the case of the time-dependent CP asymmetry one finds~\cite{Blanke:2006ig}
\begin{equation}
S_{\psi\phi}=-\sin\left[{\rm Arg}(M_{12}^s)\right] = \sin(2|\beta_s|-2\phi_{B_s})~,
\end{equation}
where we took the CP parity of the $\psi\phi$ final state equal to +1.

Concerning $A^s_\text{SL} $, we recall that, in the presence of NP, $A^s_\text{SL} $ is correlated
with $S_{\psi\phi}$ according to the following expression \cite{Ligeti:2006pm},
\begin{equation} \label{eq:A_SL_corr}
A^s_\text{SL} = -\left |{\rm Re} \left( \frac{\Gamma^s_{12}}{M^s_{12}}\right)^{\rm SM} \right|
\frac{1}{C_{B_s}}S_{\psi\phi}~,
\end{equation}
where we have neglected small contributions proportional to ${\rm Im}(\Gamma^s_{12}/M^s_{12})^{\rm SM}$
and used~\cite{Ciuchini:2003ww} 
\beq
|{\rm Re}(\Gamma^s_{12}/M^s_{12})^{\rm SM}| = (2.6 \pm 1.0) \times 10^{-3}\,.
\eeq
Note that even a rather small value of $S_{\psi \phi} \simeq 0.1$ would lead to an order of magnitude enhancement of $A^s_\text{SL} $ relative to its SM expectation, since $A^s_\text{SL} $ within the SM 
is predicted to be of order $10^{-5}$~\cite{Blanke:2006sb,Lenz:2006hd}.

An alternative formula for the $A^s_{\rm SL}-S_{\psi\phi}$ model-independent correlation, pointed out 
recently in~\cite{Grossman:2009mn}, uses only measurable quantities
\begin{equation} \label{eq:cornew}
A^s_{\rm SL}=-\frac{\Delta\Gamma_s}{\Delta M_s} \frac{S_{\psi\phi}}{\sqrt{1-S^2_{\psi\phi}}}\,,
\end{equation}
and can be used once the data on $\Delta \Gamma_s$ improves. In writing (\ref{eq:cornew}), we have 
assumed that $\Delta\Gamma_s> 0$, as obtained in the SM using lattice methods.

Concerning the experimental situation, HFAG gives the following value for the semileptonic 
asymmetry~\cite{Barberio:2008fa},
\begin{equation}
(A_{\rm SL}^s)_{\rm exp} = (-3.7 \pm 9.4) \times 10^{-3}~.
\end{equation}
In the case of $S_{\psi\phi}$, there have been several analyses of the data from CDF~\cite{Aaltonen:2007he,Tonelli:2008ey} and D0~\cite{:2008fj} on the $B_s\to\psi\phi$ decay. Taking into account additional constraints coming e.g. from the flavour-specific $B_s$ lifetime and the semileptonic asymmetry, tensions with the tiny SM prediction $S_{\psi\phi}^{\rm SM}=\sin2|\beta_s|\simeq 0.036$ at the level of $(2-3) \sigma$ are found~\cite{Bona:2008jn,Bona:2009tn,Tisserand:2009ja,Barberio:2008fa}. In the following we will use the results given by HFAG \cite{Barberio:2008fa}. For the $B_s$ mixing phase, they quote
\begin{equation}
|\beta_s| - \phi_{B_s} = 0.47^{+0.13}_{-0.21} \vee 1.09^{+0.21}_{-0.13}~,
\end{equation}
and give the following range at the $90\%$ C.L.
\begin{equation}
|\beta_s| - \phi_{B_s} \in [0.10 , 0.68] \cup [0.89 , 1.47]~.
\end{equation}
This corresponds to 
\begin{equation} \label{eq:Spsiphi_exp}
S_{\psi\phi} = 0.81^{+0.12}_{-0.32} ~~\textnormal{and}~~
S_{\psi\phi} \in [0.20 , 0.98]~,
\end{equation}
respectively.

\begin{figure}[t]
\includegraphics[width=1.0\textwidth]{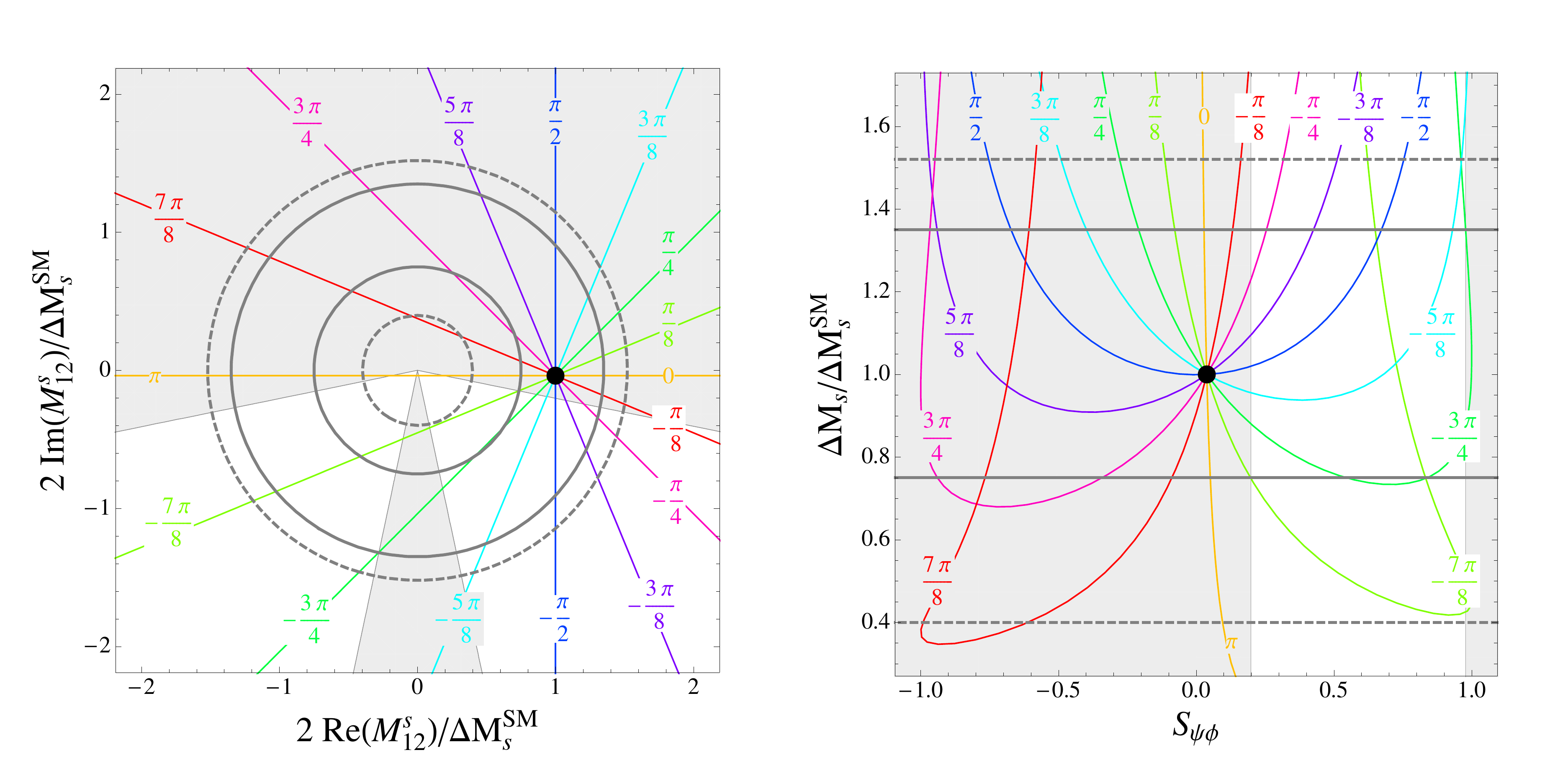}
\caption{\small
Left: The complex $M_{12}^s$ plane. The thick solid circles show the constraint on the $B_s$ mixing amplitude from the measurement of $\Delta M_d/\Delta M_s$ assuming no NP in $\Delta M_d$, while the thick dashed circles correspond to the constraint coming from $\Delta M_s$ alone. The thin lines correspond to NP contributions to $M_{12}^s$ with the indicated fixed phases $\theta_s$, see (\ref{eq:M12s}). The white regions correspond to the $90\%$ C.L. range for $S_{\psi\phi}$ in (\ref{eq:Spsiphi_exp}). The SM value corresponds to the black point. 
Right: Same as in the left panel, but now in the $S_{\psi\phi}$-$\Delta M_s/\Delta M_s^{\rm SM}$ plane.
}
\label{fig:M12s}
\end{figure}

In fig.~\ref{fig:M12s}, we show the experimental constraints in the complex $M_{12}^s$ plane and the $S_{\psi\phi}$-$\Delta M_s/\Delta M_s^{\rm SM}$ plane, analogous to fig.~\ref{fig:M12d} in the $B_d$ 
case. Interestingly enough, they show that the scenario with maximum CP violation, in which the phase
of $(M_{12}^s)_{\rm NP}$ is $-\pi/2$, is perfectly allowed and it would imply $S_{\psi\phi}\leq 0.7$ 
after imposing the constraint on $R_{\Delta M}=(\Delta M_d/\Delta M_s)/(\Delta M_d^\text{SM}/\Delta M_s^\text{SM})$ given in tab.~\ref{tab:DF2exp}. If we do not fix the
phase of $(M_{12}^s)_{\rm NP}$ then all values for $S_{\psi\phi}$ in the interval $[-1,1]$ are obviously 
possible, still being consistent with the constraint on $R_{\Delta M}$.

%

\section{\boldmath $\Delta F=0,1,2$ Transitions in a General MSSM} \label{sec:DF012}
\setcounter{equation}{0}

In this section, we discuss NP effects arising in general SUSY scenarios both in flavour violating
and flavour conserving processes. In the former case, we consider $\Delta F=2$ and $\Delta F=1$
transitions both in the $K$ and in the $B$ systems, in the latter, $\Delta F=0$ transitions such 
as the electron and the neutron EDMs $d_{e,n}$ are discussed.

Concerning the $\Delta F=2$ transitions, we present the SUSY contributions to the $B_{s,d}$ and $K^0$
mixing amplitudes that enter the predictions of CP conserving quantities such as the mass differences
$\Delta M_{s,d}$ and $\Delta M_K$ and CP violating ones such as the time-dependent CP asymmetries
$S_{\psi K_{S}}$, $S_{\psi\phi}$ and $\epsilon_K$. We also investigate the constraints coming from
$D^0-\bar D^0$ mixing.

Regarding $\Delta F = 1$ transitions, we give expressions for the CP asymmetries in $b\to s\gamma$, $B\to\phi(\eta^{\prime})K_{S}$, the rare decays $B_{s,d}\to\mu^+\mu^-$ and $B^+\to\tau^+\nu$ and
discuss the three T-odd CP asymmetries in $B\to K^{*}\mu^+\mu^-$.

In a general MSSM framework, there are various NP contributions to the FCNC processes that we consider.
In particular, in the case of flavour changing processes in the down quark sector, one has contributions
arising from one  loop diagrams involving charged Higgs bosons and the top quark, charginos and up
squarks, neutralinos and down squarks and also gluinos and down squarks.

Gluino loops typically give the dominant contributions if flavour off-diagonal entries in the soft SUSY
breaking terms are present. Such off-diagonal entries can be conveniently parameterized in terms of Mass
Insertions. Throughout the paper, we will adopt the following convention for the MIs
\begin{equation}
{\mathcal M}_{D}^2 = {\rm diag}(\tilde m^2) + \tilde m^2 \delta_d~,~~ {\mathcal M}_{U}^2 = {\rm diag}(\tilde m^2)
+ \tilde m^2 \delta_u~,
\label{eq:delta-def}
\end{equation}
where ${\mathcal M}_{Q}^2$ are the full $6\times6$ squark mass matrices in the super CKM (SCKM)
basis in the convention of~\cite{Rosiek:1995kg} (see also appendix~\ref{sec:appendixRosiek} for
the convention used) and $\tilde m^2$ is an average squark mass. The MIs are then further 
decomposed according to the ``chirality'' of the squarks
\begin{equation}
\delta_q = \left(
\begin{array}{cc}
\delta_q^{LL} & \delta_q^{LR} \\
\delta_q^{RL} & \delta_q^{RR} 
\end{array}
\right)~.
\end{equation}
Our convention for the trilinear couplings $A_u$ and $A_d$ follows also~\cite{Rosiek:1995kg} and
is such that
\begin{equation}
\tilde m^2 (\delta_u^{LR})_{33} = - m_t \left( A_t + \mu^* / t_\beta\right) ~,~~ \tilde m^2 (\delta_d^{LR})_{33} = - m_b \left( A_b+ \mu^* t_\beta\right)~,
\label{eq:Abt-def}
\end{equation}

with $m_t A_t=\frac{v_2}{\sqrt 2}\left(A_u\right)_{33}$ and $m_b A_b=\frac{v_1}{\sqrt 2}\left(A_d\right)_{33}$.

In the remainder of this section, we will report expressions in the Mass Insertion Approximation (MIA) for
the SUSY contributions entering in the processes we consider, since they are more suitable to understand
the physical results~\footnote{All the expressions we quote are relative to the MIA assuming a scenario
with degenerate sfermion families. The phenomenological implications in flavour physics arising from a
hierarchical sfermion scenario have been recently addressed in~\cite{Giudice:2008uk}.}.

In our numerical analysis instead, we work with mass eigenstates and we do not make use of the MIA as the
latter cannot be applied when the flavour violating mixing angles are $\mathcal O(1)$, as it is predicted
by many flavour models.

\subsection[$\Delta F = 2$ Processes]{\boldmath $\Delta F = 2$ Processes} \label{sec:DF2}

We begin our presentation with $\Delta F = 2$ transitions induced in the MSSM. To this end, let us briefly
recall, as already stated for the $B_s$ system in (\ref{eq:M12s}), that the meson-antimeson oscillations
are described by the mixing amplitudes $M^{(M)}_{12}$ $\equiv$ $\langle M |\heff|\ov M \rangle$, with
$M$ = $K^0$, $B_{d,s}$. Within the MSSM, the effective Hamiltonian has the form
\begin{equation} \label{eq:DF2_Heff}
\heff = \sum_{i=1}^5 C_i Q_i + \sum_{i=1}^3 \tilde C_i \tilde {Q}_i ~+{\rm h.c.}~,
\end{equation}
with the operators $Q_i$ given, in the case of $B_s$ mixing\footnote{Analogous formulae hold for the $B_d$
and $K^0$ systems, with the appropriate replacements of the quarks involved in the transition.}, by
\begin{eqnarray} \label{eq:DF2_ops}
Q_1 & = & 
(\bar s^{\alpha} \gamma_{\mu}P_{L}\, b^{\alpha}) \, (\ov s^{\beta} \gamma^\mu P_{L}\, b^{\beta})~,\nn \\
Q_2 & = & 
(\bar s^{\alpha} P_{L}\, b^{\alpha})\, (\ov s^{\beta} P_{L}\, b^{\beta})~,\nn \\
Q_3 & = & 
(\bar s^{\alpha} P_{L}\, b^{\beta})\, (\ov s^{\beta} P_{L}\, b^{\alpha})~,\nn\\
Q_4 & = & 
(\bar s^{\alpha} P_{L}\, b^{\alpha})\, (\ov s^{\beta} P_{R}\, b^{\beta})~,\nn \\
Q_5 & = & 
(\bar s^{\alpha} P_{L}\, b^{\beta})\, (\ov s^{\beta} P_{R}\, b^{\alpha})~,
\end{eqnarray}
where $P_{R,L}=\frac{1}{2}(1\pm\gamma_5)$ and $\alpha,\beta$ are colour indices. The operators 
$\tilde{Q}_{1,2,3}$ are obtained from $Q_{1,2,3}$ by the replacement $L \leftrightarrow R$.
In the SM only the operator $Q_1$ is generated, because of the $(V-A)$ structure of the SM
charged currents. On the other hand, within the MSSM, all operators typically arise.

In a MFV MSSM, the NP effects to the mixing amplitudes are quite small both in the low $\tan\beta$ regime~\cite{Altmannshofer:2007cs} and also at large $\tan\beta$~\cite{Buras:2002vd,Buras:2002wq}.

In the former case, the largest contributions to the amplitude $ M^{(M)}_{12}$ arise from the chargino
and charged Higgs effects by means of $C_{1}$ and $\tilde{C}_{3}$. All the other Wilson coefficients (WCs)
involve couplings that are highly suppressed by light quark Yukawa couplings. Moreover, it turns out that
the chargino box contributions to $\tilde{C}_{3}$ are the only non-negligible contributions sensitive to
flavour diagonal phases. In particular
\begin{equation} \label{eq:C1_cha_MFV}
C_1^{\tilde \chi^\pm}
\simeq
-\frac{g^4_2}{16\pi^2}
(V_{tb}V_{ts}^{\star})^2
\bigg[ \frac{m^4_t}{8 M^4_W} \frac{1}{\tilde m^2} f_1(x_\mu)
\bigg]~ + \mathcal{O}\left( \frac{v^2}{\tilde m^2} \right),
\end{equation}
\begin{equation} \label{eq:C3_cha_MFV}
\tilde{C}_3^{\tilde \chi^\pm}
\simeq
-\frac{g^4_2}{16\pi^2}
(V_{tb}V_{ts}^{\star})^2
\frac{m^2_b t^2_\beta}{(1+\epsilon t_\beta)^2}
\bigg[
\frac{m^4_t}{8 M^4_W}\frac{\mu^2A^2_t}{\tilde m^8}f_3(x_\mu)
\bigg]~,
\end{equation}
where $x_{\mu}=|\mu|^2/\tilde{m}^{2}$, $t_\beta = \tan\beta$, the loop functions are such that $f_1(1)=-1/12$, $f_3(1)=1/20$ and $\epsilon$ is the well known resummation factor arising from non-holomorphic ($t_{\beta}$ enhanced) threshold corrections~\cite{Hall:1993gn,Carena:1994bv,Isidori:2001fv,Buras:2002vd,Hofer:2009xb,Crivellin:2009ar}.
The dominant gluino contributions read
\begin{equation}
\epsilon \simeq \frac{2 \alpha_s}{3\pi} \frac{\mu M_{\tilde g}}{\tilde m^2} f(x_g)~,
\end{equation}
with $x_g=M_{\tilde g}^2/ \tilde m^2$ and the loop function $f$ given in the appendix.
One has $f(1) = 1/2$ such that $\epsilon\simeq\alpha_s/3\pi$ for a degenerate SUSY spectrum.

As anticipated, (\ref{eq:C1_cha_MFV}) clearly shows that $C_1$ is not sensitive to
flavour conserving phases, while $\tilde C_3$ can be complex for complex $\mu A_t$.
In~(\ref{eq:C3_cha_MFV}) we have omitted contributions sensitive to the phases of
the combinations $\mu^2 M_2^2$ and $\mu^2 A_t M_2$, as they vanish in the limit of
degenerate (left-handed) squark generations as a result of the super-GIM mechanism.

Moreover, we note that $\epsilon_K$ can receive sizable effects only through $C_1$, as the
contributions to $\epsilon_K$ from $\tilde{C}_{3}$ are suppressed by $m_s^2/m_b^2$ and thus
safely negligible. We find that the Wilson coefficient $C_1$ in~(\ref{eq:C1_cha_MFV}) has the
same sign as the SM contribution, thus we conclude that for flavour diagonal soft terms one
has $|\epsilon_K| > |\epsilon_K^{\rm SM}|$.

Within a MFV framework at large $\tan\beta$, there are additional contributions to
$\Delta F=2$ transitions stemming from the neutral Higgs sector~\cite{Buras:2001mb,Buras:2002wq,Buras:2002vd,Dedes:2002er,Gorbahn:2009pp,Hofer:2009xb}.
However, at least for $\mu >0$, these contributions turn out to be highly constrained by
the experimental limits on ${\rm BR}(B_s\to\mu^+\mu^-)$, hence, they can be safely neglected.

In the following, we therefore focus on the effects that arise in the presence of non-MFV structures
in the soft SUSY breaking sector. In such a setup, the leading contributions to the $K^0$, $B_d$ and
$B_s$ mixing amplitudes come from gluino boxes\footnote{In our numerical analysis we include the full set of contributions that can be found e.g. in~\cite{Altmannshofer:2007cs}.}. The relevant Wilson coefficients in the MIA read~\cite{Gabbiani:1996hi}\footnote{The NLO Wilson coefficients
in the MIA are given in~\cite{Ciuchini:2006dw} and in~\cite{Virto:2009wm,Crivellin:2009ar} in the mass eigenstate basis.}

\begin{eqnarray}
\label{eq:MIA2_C1}
C_1^{\tilde g}
&\simeq&-\frac{\alpha_s^2}{\tilde m^2}\left[(\delta_d^{LL})_{32}\right]^2 g_1^{(1)}(x_g)~, \\
\label{eq:MIA2_C1tilde}
\tilde C_1^{\tilde g}
&\simeq&-\frac{\alpha_s^2}{\tilde m^2}\left[(\delta_d^{RR})_{32}\right]^2 g_1^{(1)}(x_g)~, \\
\label{eq:MIA2_C4}
C_4^{\tilde g}
&\simeq&-\frac{\alpha_s^2}{\tilde m^2}\left[(\delta_d^{LL})_{32}(\delta_d^{RR})_{32}\right]g_4^{(1)}(x_g)~, \\
\label{eq:MIA2_C5}
C_5^{\tilde g}
&\simeq& -\frac{\alpha_s^2}{\tilde m^2}\left[(\delta_d^{LL})_{32}(\delta_d^{RR})_{32}\right] g_5^{(1)}(x_g)~,
\end{eqnarray}
where $x_g=M_{\tilde g}^2/ \tilde m^2$ and the analytic expressions for the loop functions $g_1^{(1)}$, $g_4^{(1)}$ and $g_5^{(1)}$ can be found in appendix~\ref{sec:appendix}. For the limiting case of degenerate masses we 
find $g_1^{(1)}(1)=-1/216$, $g_4^{(1)}(1)=23/180$ and $g_5^{(1)}(1)=-7/540$.

In the above expressions, we omitted the contributions arising from the $LR$ and $RL$ MI because, as we
will see in sec.~\ref{sec:Dipol_Wilson}, they are tightly constrained by BR$(b\rightarrow s\gamma)$.

The same argument does not apply to the $s\to d$ and $b\to d$ transitions as there is
no BR$(B\to X_s\gamma)$ analog here~\footnote{although, as a future perspective, $\rm{LR}$ and
$\rm{RL}$ MIs relative to the $b\to d$ transition will be probed by BR$(B\rightarrow \rho\gamma)$.}.
However, in the concrete flavour models we are dealing with, it turns out that the contributions
from $(\delta_d^{LL})_{ij}(\delta_d^{RR})_{ij}$ are always dominant compared to those from
$(\delta_d^{LR})_{ij}$ and $(\delta_d^{RL})_{ij}$, hence, the above expressions are accurate enough 
for all the $\Delta F = 2$ transitions~\footnote{Clearly the contributions of the MIs $(\delta_d^{LR})_{ij}$ and $(\delta_d^{RL})_{ij}$ are taken into account in our model independent analysis of sec.~\ref{subsec:bounds}.} .

An important observation comes directly from the Wilson coefficients (\ref{eq:MIA2_C1})--(\ref{eq:MIA2_C5}):
we expect large NP contributions in the mixing amplitudes from models that predict both nonzero $(\delta_d^{LL})_{32}$ and $(\delta_d^{RR})_{32}$ MIs, since in this way the operators $Q_4$ and $Q_5$,
that are strongly enhanced through QCD renormalization group effects~\cite{Ciuchini:1997bw,Buras:2000if},
have non-vanishing Wilson coefficients. We note in addition that especially the loop function $g_4^{(1)}$
entering the Wilson coefficient $C_4^{\tilde g}$ is roughly a factor 30 larger than the one entering
$C_1^{\tilde g}$.

Expressions (\ref{eq:MIA2_C1})-(\ref{eq:MIA2_C5}) are valid for the $B_s$ mixing amplitude, while the corresponding expressions for $B_d$ and $K^0$ mixing can be obtained by replacing the indices $(32)$ 
with $(31)$ or $(21)$, respectively.
Similarly, the corresponding expressions for $D^0-\bar D^0$ mixing are easily obtained from those
relative to $K^0-\bar K^0$ making the replacement $(\delta_d)_{21}\to(\delta_u)_{21}$.

In the case of $K^0$ mixing, it is important for our analysis to consider also the additional contribution 
coming from effective $(2 \to 1)$ mass insertions generated by a double flavour flip $(2\to 3)\times(3\to 1)$.
Results for the Wilson coefficients with one effective $(2 \to 1)$ transition are obtained in the third order
of the MIA. We find
\begin{eqnarray}
C_1^{\tilde g} &\simeq& -\frac{\alpha_s^2}{\tilde m^2} \left[ (\delta_d^{LL})_{21} (\delta_d^{LL})_{23} (\delta_d^{LL})_{31} \right] g_1^{(2)}(x_g)~, \\
\tilde C_1^{\tilde g} &\simeq& -\frac{\alpha_s^2}{\tilde m^2} \left[ (\delta_d^{RR})_{21} (\delta_d^{RR})_{23} (\delta_d^{RR})_{31} \right] g_1^{(2)}(x_g)~, \\
C_4^{\tilde g} &\simeq& -\frac{\alpha_s^2}{\tilde m^2} \frac{1}{2} \left[ (\delta_d^{LL})_{21} (\delta_d^{RR})_{23}  (\delta_d^{RR})_{31} + (\delta_d^{LL})_{23} (\delta_d^{LL})_{31} (\delta_d^{RR})_{21} \right] g_4^{(2)}(x_g)~, \\
C_5^{\tilde g} &\simeq& -\frac{\alpha_s^2}{\tilde m^2} \frac{1}{2} \left[ (\delta_d^{LL})_{21} (\delta_d^{RR})_{23}  (\delta_d^{RR})_{31} + (\delta_d^{LL})_{23} (\delta_d^{LL})_{31} (\delta_d^{RR})_{21} \right] g_5^{(2)}(x_g)~.
\end{eqnarray}
The loop functions are again given in appendix~\ref{sec:appendix} and for degenerate masses we find $g_1^{(2)}(1)=1/360$, $g_4^{(2)}(1)=-1/6$ and $g_5^{(2)}(1)=1/90$.

Finally we also have to consider the case where the $(2 \to 1)$ flavour transition is entirely generated by $(2 \to 3)$ and $(3 \to 1)$ transitions. We find
\begin{eqnarray}
C_1^{\tilde g} &\simeq& -\frac{\alpha_s^2}{\tilde m^2} \left[ (\delta_d^{LL})_{23} (\delta_d^{LL})_{31} \right]^2 g_1^{(3)}(x_g)~, \\
\tilde C_1^{\tilde g} &\simeq& -\frac{\alpha_s^2}{\tilde m^2} \left[ (\delta_d^{RR})_{23} (\delta_d^{RR})_{31} \right]^2 g_1^{(3)}(x_g)~, \\
C_4^{\tilde g} &\simeq& -\frac{\alpha_s^2}{\tilde m^2} \left[ (\delta_d^{LL})_{23} (\delta_d^{LL})_{31} (\delta_d^{RR})_{23} (\delta_d^{RR})_{31} \right] g_4^{(3)}(x_g)~, \\
C_5^{\tilde g} &\simeq& -\frac{\alpha_s^2}{\tilde m^2} \left[ (\delta_d^{LL})_{23} (\delta_d^{LL})_{31} (\delta_d^{RR})_{23} (\delta_d^{RR})_{31} \right] g_5^{(3)}(x_g)~,
\end{eqnarray}
with $g_1^{(3)}(1)=-1/3780$, $g_4^{(3)}(1)=37/630$ and $g_5^{(3)}(1)=-1/378$. The analytic expressions for 
these loop functions can again be found in appendix~\ref{sec:appendix}.

\begin{figure}[t]
\centering
\includegraphics[width=0.95\textwidth]{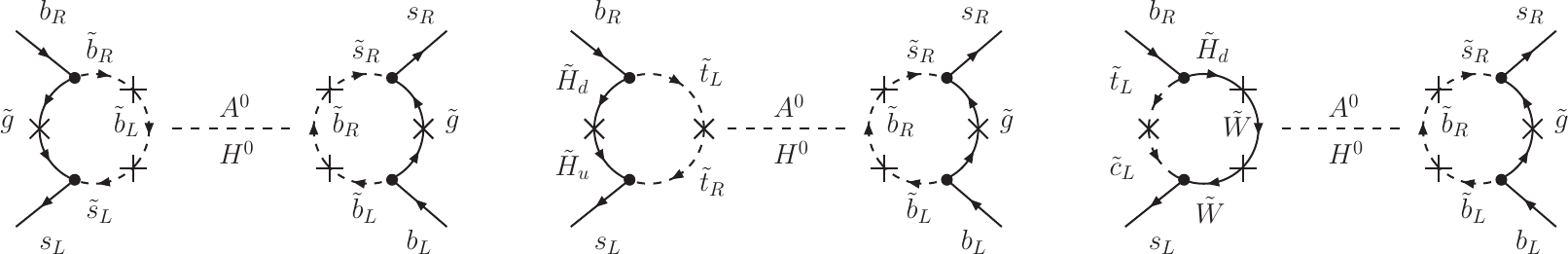}
\caption{\small
Feynman diagrams for the dominant Higgs mediated contributions to $\Delta B = 2$ transitions.
The leading contribution proportional to $\tan^4\beta$ comes from the self-energy corrections 
in diagrams where the Higgs propagators are attached to the external quark legs.
}
\label{diagrams_DeltaFeq2}
\end{figure}

In the large $\tan\beta$ regime, the gluino box contributions are not the dominant ones anymore but they have
to compete with double Higgs penguin contributions, see the Feynman diagrams of fig.~\ref{diagrams_DeltaFeq2},
that are enhanced by $\tan^4\beta$ as in the MFV case~\cite{Buras:2001mb,Buras:2002wq,Buras:2002vd}.
Taking into account both gluino and chargino loops and generalizing the formulae of~\cite{Buras:2001mb,Buras:2002wq,Buras:2002vd} to non-MFV contributions we find in the case of the $B_s$ system\footnote{The Wilson coefficient for $B_d$ mixing can be obtained by replacing $(32)$ with $(31)$ and $V_{ts}$ by $V_{td}$.}
\begin{eqnarray}
(C_4^H)_B
&\simeq&
-\frac{\alpha_s^2 \alpha_2}{4 \pi}\frac{m_b^2}{2 M_W^2}
\frac{t^{4}_{\beta}}{(1+\epsilon t_{\beta})^{4}}
\frac{|\mu|^2 M_{\tilde g}^2}{M_A^2 \tilde m^4}(\delta^{LL}_d)_{32}(\delta^{RR}_d)_{32}
\left[h_1(x_g)\right]^2+
\nonumber \\
&+&
\frac{\alpha_2^2 \alpha_s}{4\pi}\frac{m_b^2}{2 M_W^2}
\frac{t^{4}_{\beta}}{(1+\epsilon t_{\beta})^{4}}
\frac{|\mu|^2}{M_A^2\tilde m^2}\times\left[\frac{m_t^2}{M_W^2}\frac{A_t M_{\tilde g}}{\tilde m^2}
h_1(x_g) h_3(x_\mu) (\delta^{RR}_d)_{32} V_{tb} V_{ts}^*+ \right.
\nonumber \\
&&
+\left.\frac{M_2 M_{\tilde g}}{\tilde m^2}
(\delta^{LL}_u)_{32}(\delta^{RR}_d)_{32}
h_1(x_g) h_4(x_2,x_\mu) \right]~,
\label{eq:C4HB}
\end{eqnarray}
with the mass ratios $x_\mu = |\mu|^2/ \tilde m^2$, $x_2 = |M_2|^2/ \tilde m^2$ and the loop functions,
reported in appendix~\ref{sec:appendix}, satisfying $h_1(1) = 4/9$, $h_3(1) = -1/4$ and $h_4(1,1) = 1/6$. We stress
that, due to the presence of the RR MIs, these contributions are neither suppressed by $m_s/m_b$ nor
by other suppression factors, as it happens in contrast in the MFV framework \cite{Buras:2001mb,Buras:2002wq,Buras:2002vd,Gorbahn:2009pp}.

In the case of $K^0$ mixing, the most relevant effect from the neutral Higgses arises only at the fourth
order in the MI expansion and we find the following expression
\begin{equation}
(C_4^H)_K \simeq
-\frac{\alpha_s^2 \alpha_2}{4 \pi}
\frac{m_b^2}{2 M_W^2}
\frac{t^{4}_{\beta}}{(1+\epsilon t_{\beta})^{4}}
\frac{|\mu|^2 M_{\tilde g}^2}{M_A^2 \tilde m^4}
(\delta^{LL}_d)_{23}(\delta^{LL}_d)_{31}(\delta^{RR}_d)_{23}(\delta^{RR}_d)_{31}
h_2(x_g)^2~,
\end{equation}
with $h_2$ given in appendix~\ref{sec:appendix} and $h_2(1) = -2/9$.

In contrast to the case with gluino box contributions, now there are no analogous Higgs mediated
contributions for $D^0-\bar D^0$ mixing; in fact, the non-holomorphic ($\tan\beta$ enhanced)
threshold corrections lead to important FCNC couplings only among down-type fermions with the Higges.

\subsection[$\Delta F = 1$ Processes]{\boldmath $\Delta F = 1$ Processes} \label{sec:DF1}

\subsubsection{Wilson Coefficients of the Dipole Operator} \label{sec:Dipol_Wilson}

The part of the $\Delta F=1$ effective Hamiltonian relevant for the $b\to s\gamma$ transition that 
is most sensitive to NP effects reads
\begin{equation} \label{eq:H_eff_bsg}
\mathcal{H}_{\rm eff} \supset -\frac{4 G_F}{\sqrt{2}}V_{tb}V_{ts}^* \left(C_7 O_7 + C_8 O_8 + \tilde C_7 \tilde O_7 + \tilde C_8 \tilde O_8 \right)~,
\end{equation}
with the magnetic and chromomagnetic operators
\begin{equation}
O_7 = \frac{e}{16 \pi^2} m_b (\bar s \sigma^{\mu\nu} P_R b) F_{\mu\nu} ~,~~ 
O_8 = \frac{g_s}{16 \pi^2} m_b (\bar s \sigma^{\mu\nu} T^A P_R b) G^A_{\mu\nu}~.
\end{equation}
Here, $\sigma^{\mu\nu}=\frac{i}{2}[\gamma^\mu,\gamma^\nu]$ and $T^A$ are the $SU(3)_c$ generators. The operators $\tilde{O_i}$ are obtained by the corresponding operators $O_i$ by means of the replacement $L \leftrightarrow R$.

The dominant SUSY contributions to $C_{7,8}$ arise from the one-loop charged Higgs, chargino and gluino amplitudes $C^{\rm NP}_{7,8}=C^{H^\pm}_{7,8}+C^{\tilde{\chi}^\pm}_{7,8}+C^{\tilde{g}}_{7,8}$. 

The charged Higgs contribution reads
\begin{equation} \label{eq:C7_H}
C_{7,8}^{H^\pm} \simeq \left(\frac{1-\epsilon t_\beta}{1+\epsilon t_\beta} \right) \frac{1}{2} h_{7,8}(y_t)~,
\end{equation}
where $y_t=m_t^2 / M_{H^\pm}^2$, $\epsilon\sim 10^{-2}$ for
a degenerate SUSY spectrum and the loop functions can be found in appendix~\ref{sec:appendix}.

The corresponding expressions for the chargino contributions to $C_7$ and $C_8$ read
\begin{eqnarray} \label{eq:C7_cha}
\frac{4 G_F}{\sqrt{2}}~C_{7,8}^{\tilde \chi^\pm}
\simeq
\frac{g_2^2}{\tilde m^2}
\left[
\frac{(\delta_u^{LL})_{32}}{V_{tb}V_{ts}^*}
\frac{\mu M_2}{\tilde m^2}
f_{7,8}^{(1)}(x_2,x_\mu)+
\frac{m_t^2}{M_W^2}\frac{A_t \mu}{\tilde m^2}
f_{7,8}^{(2)}(x_\mu)
\right]\frac{t_{\beta}}{(1+\epsilon t_{\beta})}~.
\end{eqnarray}
The relevant loop functions are again defined in appendix~\ref{sec:appendix}.

Within a MFV SUSY framework, charged Higgs effects unambiguously increase the $b\to s\gamma$
branching ratio relative to the SM expectation, while the chargino ones (mostly from the Higgsinos)
can have either sign depending mainly on the sign and phase of $\mu A_{t}$.
Since in SUGRA inspired models the sign of $A_t$ is set (in almost the entire SUSY parameter space)
by the large (CP-conserving) RGE induced effects driven by the $SU(3)$ interactions, the final MFV
chargino effects will depend on the sign of the $\mu$ term that we assume to be real and positive
which is preferred by the $(g-2)_{\mu}$ constraint (see also sec.~\ref{sec:g-2}).
Then, the MFV contributions from the charged Higgs bosons and Higgsinos interfere destructively in
BR$(b\to s\gamma)$.

At the LO, both the charged Higgs and the chargino contributions to $\tilde C_{7,8}$ are suppressed
by $m_s/m_b$. However, this $m_s/m_b$ suppression can be avoided in non-MFV scenarios at the NLO
in the presence of RH currents by means of threshold corrections to the Yukawa interactions that
are induced by $\tan\beta$-enhanced non-holomorphic effects~\cite{Hisano:2008hn}. For simplicity,
we do not present these last contributions here, although they are systematically included in our
numerical analysis.

In the presence of non-CKM flavour structures, gluino mediated FCNC contributions also arise and the
corresponding Wilson coefficients governing the $b\to s\gamma$ transition read~\footnote{Gluino
contributions can also arise in the framework of the general MFV ansatz~\cite{D'Ambrosio:2002ex},
see e.g.~\cite{Wick:2008sz,Dudley:2008dp,Carena:2008ue}.}
\begin{eqnarray} \label{eq:C7_g}
\frac{4 G_F}{\sqrt{2}}~C_{7,8}^{\tilde g}
\simeq
\frac{g_s^2}{\tilde m^2}
\left[
\frac{M_{\tilde g}}{m_b}\frac{(\delta_d^{RL})_{32}}{V_{tb}V_{ts}^*}
g_{7,8}^{(1)}(x_g) +
\frac{M_{\tilde g}\mu}{\tilde m^2}\frac{t_{\beta}}{(1+\epsilon t_{\beta})}
\frac{(\delta_d^{LL})_{32}}{V_{tb}V_{ts}^*}
g_{7,8}^{(2)}(x_g)
\right]~,
\end{eqnarray}
\begin{eqnarray} \label{eq:tildeC7_g}
\frac{4 G_F}{\sqrt{2}}~\tilde C_{7,8}^{\tilde g}
\simeq
\frac{g_s^2}{\tilde m^2}
\left[
\frac{M_{\tilde g}}{m_b}
\frac{(\delta_d^{LR})_{32}}{V_{tb}V_{ts}^*}g_{7,8}^{(2)}(x_g)+
\frac{M_{\tilde g}\mu^*}{\tilde m^2} \frac{t_{\beta}}{(1+\epsilon t_{\beta})}
\frac{(\delta_d^{RR})_{32}}{V_{tb}V_{ts}^*}g_{7,8}^{(3)}(x_g)
\right]~.
\end{eqnarray}
The functions $g^{(i)}_{7,8}$ $(i=1,2)$ are given in appendix~\ref{sec:appendix}.
As is evident from~(\ref{eq:C7_g}) and~(\ref{eq:tildeC7_g}), $b \to s \gamma$ puts very strong constraints
on the helicity flipping mass insertions $(\delta_d^{RL})_{32}$ and $(\delta_d^{LR})_{32}$ because the corresponding terms are  chirally enhanced by a factor $M_{\tilde g}/m_b$.

In fact, once the constraint from ${\rm BR}(b\to s\gamma)$ is imposed, these helicity flipping MIs cannot 
generate large effects in $\Delta B=2$ observables, in particular in $S_{\psi\phi}$, anymore.

\subsubsection[Direct CP Asymmetry in $b\to s\gamma$]{\boldmath Direct CP Asymmetry in $b\to s\gamma$} \label{sec:ACP_bsg}

A very sensitive observable to NP CP violating effects is represented by the direct CP asymmetry in $b\to s\gamma$, i.e. $A_{\rm CP}(b\to s\gamma)$~\cite{Soares:1991te}. If NP effects dominate over the tiny SM contribution $A^\text{SM}_{\rm CP}(b\to s\gamma)\simeq - 0.5\%$, the following 
expression for $A_{\rm CP}(b\to s\gamma)$ holds~\cite{Kagan:1998bh,Kagan:1998ym}
\begin{multline} \label{eq:acp_bsg}
A_\text{CP}(b\to s\gamma) \equiv \frac{\Gamma(B \to X_{\bar{s}} \gamma) - \Gamma(\overline{B} \to X_s \gamma)}{\Gamma(B \to X_{\bar{s}}\gamma) + \Gamma(\overline{B} \to X_s\gamma)}\simeq \\
\simeq - \frac{1}{|C_7|^2} \left( 1.23~{\rm Im}[C_2 C_7^*] - 9.52~{\rm Im}[C_8 C_7^*] + 
0.10 ~{\rm Im}[C_2 C_8^*] \right)-0.5~~~({\rm in}~\%)~,
\end{multline}
where we assumed a cut for the photon energy at $E_{\gamma} \simeq 1.8$ GeV (see~\cite{Kagan:1998bh,Kagan:1998ym} for details). In~(\ref{eq:acp_bsg}), the Wilson coefficients $C_i$
are evaluated at the scale $m_b$ and they refer to the sum of SM plus NP contributions, i.e. $C_i=C^{\rm SM}_i+C^{\rm NP}_i$ (with $C^{\rm SM}_i$ real).

In order to take into account effects from the Wilson coefficients $\tilde C_i$ related to the operators $\tilde{O}_i$, (\ref{eq:acp_bsg}) has to be modified according to $C_i C_j\to  C_i C_j + \tilde C_i \tilde C_j$. Within a MFV scenario, $\tilde C_i$ are completely negligible, being suppressed by a factor of
$m_s/m_b$ compared to the corresponding Wilson coefficients $C_i$. This statement is no longer valid in the presence of NP in the right-handed currents, as is the case in some of the scenarios we are going to discuss.

Still, as $\tilde C_2$ is negligibly small and the phases of $\tilde C_7$ and $\tilde C_8$ are the same
to a very good approximation in all the frameworks that we consider, $A_{\rm CP}(b\to s\gamma)$ receives NP contributions only through the imaginary parts of $C_{7,8}^{\rm NP}$.

\subsubsection[Time-Dependent CP Asymmetries in $B_d\to\phi(\eta^{\prime}) K_S$]{\boldmath Time-Dependent CP Asymmetries in $B_d\to\phi(\eta^{\prime}) K_S$}

The time-dependent CP asymmetries in the decays of neutral $B$ mesons into final CP eigenstates $f$ 
can be written as
\begin{equation}
{\cal A}_f(t)=S_f\sin(\Delta M t)-C_f\cos(\Delta M t)~.
\end{equation}
Within the SM, it is predicted with good accuracy that the $|S_f|$ and $C_f$ parameters are universal for
all the transitions $\bar b\to\bar q^\prime q^\prime\bar s$ ($q^\prime=c,s,d,u$). In particular, the SM predicts that $-\eta_f S_f\simeq\sin2\beta$ and $C_f\simeq0$ where $\eta_f=\pm1$ is the CP eigenvalue of
the final state $f$. NP effects can contribute to\footnote{We assume that the asymmetry in the tree level transition $\bar b\to\bar cc\bar s$ is not significantly affected by NP.}
\begin{itemize}
\item[i)] the $B_d$ mixing amplitude \cite{Amsler:2008zzb};
\item[ii)] the decay amplitudes $\bar b\to\bar qq\bar s$ ($q=s,d,u$)~\cite{Amsler:2008zzb,Grossman:1996ke}.
\end{itemize}
In case i), the NP contribution shifts all $S_f$'s from $\sin2\beta$ in a universal way while the $C_f$'s
will still vanish. In case ii), the various $S_f$'s and also the $C_f$'s are, in general, not the same as 
in the SM.

The CP asymmetries $S_f$ and $C_f$ in $B_d\to f$ decays are calculated as follows. One defines a complex quantity $\lambda_f$,
\begin{equation}
\lambda_f=e^{-2i(\beta + \phi_{B_d})}(\overline{A}_f/A_f)~,
\end{equation}
where $\phi_{B_d}$ is the NP phase of the $B_d$ mixing amplitude, $M_{12}^d$, and $A_f$ ($\overline{A}_f$)
is the decay amplitude for $B_d(\overline{B_d})\to f$. $A_f$ and $\overline{A}_f$ can be calculated from
the effective Hamiltonian relevant for $\Delta B=1$ decays~\cite{Buchalla:1995vs}, in the following way
\begin{equation}
A_f=\langle f|{\cal H}_{\rm eff}|B_d\rangle ~,~~
\overline{A}_f=\langle f|{\cal H}_{\rm eff}|\overline{B_d}\rangle~,
\end{equation}
where the Wilson coefficients of the effective Hamiltonian depend on the electroweak theory while the
matrix elements $\langle f|O_i|B_d (\overline{B_d}) \rangle$ can be evaluated, for instance, by means
of QCD factorization~\cite{Buchalla:2005us}. We then have
\begin{equation}
S_f=\frac{2{\rm Im}(\lambda_f)}{1+|\lambda_f|^2} ~,~~
C_f=\frac{1-|\lambda_f|^2}{1+|\lambda_f|^2}~.
\end{equation}
The SM contribution to the decay amplitudes, related to $\bar b\to\bar q^\prime q^\prime\bar s$ transitions, can always be written as a sum of two terms, $A_f^{\rm SM}=A_f^c+A_f^u$, with $A_f^c\propto V_{cb}^*V_{cs}$
and $A_f^u\propto V_{ub}^*V_{us}$. Defining the ratio $a_f^u\equiv e^{-i\gamma}(A_f^u/A_f^c)$, we have
\begin{equation} \label{eq:def_a_fu}
A_f^{\rm SM}=A_f^c\left(1+a_f^u e^{i\gamma}\right)~,
\end{equation}
where the $a_f^u$ parameters have been evaluated in the QCD factorization approach at the leading order
and to zeroth order in $\Lambda/m_b$ in~\cite{Buchalla:2005us}. Within the SM, it turns out that 
$S_{\phi K_S}\simeq S_{\eta^{\prime}K_S} \simeq S_{\psi K_S} \simeq \sin2\beta$ with precise predictions
given in tabs.~\ref{tab:DF2exp} and~\ref{tab:observables}. The $a_f^u$ term provides only a negligible contribution to $B_d\to\psi K_S$, thus $\lambda_{\psi K_S}^{\rm SM}=-e^{-2i\beta}$. Also for charmless
modes, the effects induced by $a_f^u$ are small (at the percent level), being proportional to $|(V_{ub}V_{us}^{*})/ (V_{cb}V_{cs}^{*})|$.

The modification of $A_f$ from the SM expression~(\ref{eq:def_a_fu}) due to NP contributions can always 
be written as follows~\footnote{We thank Dominik Scherer for pointing out the correct expression for $A_f$~\cite{Hofer:2009xb}.}
\begin{equation} \label{eq:def_b_fu}
A_f=A_f^c\left[1+a_f^ue^{i\gamma}+\sum_i
\left(b_{fi}^c+b_{fi}^ue^{i\gamma}\right)\left(C_i^{\rm NP *}(M_W)+\zeta \tilde C_i^{\rm NP *}(M_W)\right)\right]~,
\end{equation}
where $C_i^{\rm NP}(M_W)$ and $\tilde C_i^{\rm NP}(M_W)$ are the NP contributions to the Wilson coefficients evaluated at the scale $M_W$, the parameters $b_{fi}^u$ and $b_{fi}^c$ calculated in \cite{Buchalla:2005us}
and $\zeta=\pm 1$ depending on the parity of the final state; for instance $\zeta=1$ for $\phi K_S$ and $\zeta=-1$ for $\eta^\prime K_S$~\cite{Gabrielli:2004yi}.

The generalization of this formalism to $B_s$ decays is straightforward.

\subsubsection[CP Asymmetries in $B\to K^*\mu^+\mu^-$]{\boldmath CP Asymmetries in $B\to K^*\mu^+\mu^-$}

The rare decay $B\to K^*(\to K\pi)\mu^+\mu^-$ represents a very promising channel to look for NP
in the $B$ system, since its angular decay distribution is sensitive to the polarization of the $K^*$ and gives access to many observables probing NP effects. Furthermore, for the neutral $B^0$ decay, the CP parity of the initial state can be unambiguously determined by measuring the charges of the kaon and pion in the final state. This ``self-tagging'' property allows a very clean access to CP-violating observables.

In this paper, we focus on those CP-violating observables with relatively small dependence on hadronic quantities and large sensitivity to NP, following closely the analysis of~\cite{Altmannshofer:2008dz}; other recent analyses can be found in~\cite{Bobeth:2008ij,Egede:2008uy}. In particular
\begin{itemize}
\item all the observables are evaluated in the dilepton mass range $1\,{\rm GeV}^2<q^2< 6\,{\rm GeV}^2$;
\item power-suppressed corrections are accounted for by means of the full QCD form factors in the naively factorized amplitude, and the ``soft'' form factors $\xi_{\perp,\parallel}$ in the QCD factorization corrections;
\item we focus on those observables which only depend on the ratios of form factors;
\item we assume NP effects in $C_{7,9,10,S,P}$ and $\tilde C_{7,9,10,S,P}$ but not in the other $C_i$.
\end{itemize}
In our analysis, we consider the T-odd CP asymmetries $A_7$, $A_8$ and $A_9$ which are not suppressed 
by small strong phases and can thus be sizable in the presence of new sources of CP violation (CPV)~\cite{Bobeth:2008ij,Altmannshofer:2008dz}. We use the conventions of~\cite{Altmannshofer:2008dz} 
to which we refer for the full expressions of the asymmetries.

In tab.~\ref{tab:Obs_vs_WC}, we just recall the main sensitivities of $A_7$, $A_8$ and $A_9$ to 
the relevant WCs of the $\Delta F=1$ effective Hamiltonian. Notice that $\langle A_9 \rangle$ is 
only sensitive to $\tilde C_7$, $\tilde C_9$ and $\tilde C_{10}$ hence it represents, in principle,
a golden channel to probe {\it right-handed} currents. However, as we will discuss in sec.~\ref{sec:numerics}, within the SUSY flavour models we are dealing with, the attained values for
$\langle A_9 \rangle$ seem to be far below the expected future experimental resolutions. This also
implies that any potential signal of NP in $\langle A_9 \rangle$ would disfavour the flavour models
under study pointing towards different scenarios.

We would also like to emphasize that $\langle A_9 \rangle$ is peculiar from an experimental point of
view since, in contrast to $\langle A_7 \rangle$ and $\langle A_8 \rangle$, it can be obtained from
a one-dimensional angular distribution and should thus be accessible at current $B$ factories.
In the conventions of~\cite{Altmannshofer:2008dz}, the CP-averaged differential decay distribution in
the angle $\phi$ reads
\begin{equation}
\frac{1}{\Gamma+\bar\Gamma}
\frac{d(\Gamma+\bar\Gamma)}{d\phi} = \frac{1}{2\pi} \left[
\langle S_2^c \rangle +
\langle S_3 \rangle \cos(2\phi) +
\langle A_9 \rangle \sin(2\phi)
\right] ~,
\end{equation}
where $\langle S_2^c \rangle=-F_L$ has already been measured by BaBar~\cite{:2008ju} and Belle~\cite{:2009zv} 
and $\langle S_3 \rangle$ is a sensitive probe of new CP conserving physics in right-handed
currents \cite{Lunghi:2006hc,Egede:2008uy,Altmannshofer:2008dz}.

\begin{table}[t]
\addtolength{\arraycolsep}{3pt}
\renewcommand{\arraystretch}{1.3}
\centering
\begin{tabular}{|l|l|l|}
\hline
process & CP asymmetry & sensitivity to the WCs \\
\hline\hline
$B\to K^*\mu^+\mu^-$ & $A_7$ & $C_7$, $\tilde C_7$, $C_{10}$, $\tilde C_{10}$\\
\hline
$B\to K^*\mu^+\mu^-$ & $A_8$ & $C_7$, $\tilde C_7$, $C_9$, $\tilde C_9$, $\tilde C_{10}$\\
\hline
$B\to K^*\mu^+\mu^-$ & $A_9$ & $\tilde C_7$, $\tilde C_9$, $\tilde C_{10}$\\
\hline
$b\to s\gamma$ & $A_\text{CP}(b\to s\gamma)$ & $C_7$, $C_8$\\
\hline
$B_d\to\phi K_S$ & $S_{\phi K_S}$ & $C_8$, $\tilde C_8$\\
\hline
$B_d\to\eta^\prime K_S$ & $S_{\eta^\prime K_S}$ & $C_8$, $\tilde C_8$\\
\hline
\end{tabular}
\caption{\small
CP asymmetries of various $\Delta F=1$ channels and the Wilson coefficients they are most sensitive to.
}
\label{tab:Obs_vs_WC}
\end{table}

\subsubsection[$B_s\rightarrow \mu^+\mu^-$]{\boldmath $B_s\rightarrow \mu^+\mu^-$}

The SM prediction for the branching ratio of the decay $B_s\to\mu^+\mu^-$ is (using the results in~\cite{Blanke:2006ig} with updated input parameters)
\begin{equation}
{\rm BR}(B_s \to \mu^+ \mu^-)_{\rm SM} = (3.60\pm 0.37)\times 10^{-9}~.
\end{equation}
This value should be compared to the present 95\% C.L. upper bound from
CDF~\cite{:2007kv} 
\begin{equation}
{\rm BR}(B_s \to \mu^+ \mu^-)_{\rm exp}   \lesssim 5.8 \times 10^{-8}~,
\end{equation}
that still leaves a large room for NP contributions~\footnote{An unofficial Tevatron combination
with D0 data \cite{D0-5344-CONF} yields an upper bound of $4.5\times 10^{-8}$ \cite{Punzi-EPS09}.}.
In particular, the MSSM with large $\tan\beta$ allows, in a natural way, large departures of 
${\rm BR}(B_s \to \mu^+ \mu^-)$ from its SM expectation~\cite{Hamzaoui:1998nu,Choudhury:1998ze,Babu:1999hn}.

The most relevant NP effects in the MSSM are encoded in the effective Hamiltonian
\begin{equation}
\mathcal{H}_{\rm eff} = - C_S Q_S - C_P Q_P - \tilde C_S \tilde Q_S - \tilde C_P \tilde Q_P~,
\end{equation}
with the scalar and pseudoscalar operators
\begin{equation}
Q_S = m_b \left(\bar s P_R b \right) \left(\bar \ell \ell \right) ~,~~ Q_P = m_b \left(\bar s P_R b \right) \left(\bar \ell \gamma_5 \ell \right)~,
\end{equation}
as well as the corresponding operators $\tilde Q_S$ and $\tilde Q_P$ that are obtained by the exchange
$L \leftrightarrow R$. For the corresponding Wilson coefficients in the MSSM, one has to a very good approximation
\begin{equation}
C_P \simeq -C_S ~,~~ \tilde C_P \simeq \tilde C_S~,
\end{equation}

\noindent with 

\begin{eqnarray}
C_S &=& -\frac{\alpha_2^2}{M^2_A} \frac{m_\ell}{4 M_W^2}
\frac{t^{3}_{\beta}}{(1+\epsilon t_{\beta})^{2}(1+\epsilon_{\ell}t_{\beta})}
\left[ \frac{m_t^2}{M_W^2}\frac{A_t\mu}{\tilde m^2} V_{tb}V_{ts}^* h_3(x_\mu)+\frac{M_2 \mu}{\tilde m^2} (\delta_u^{LL})_{32}h_4(x_2,x_\mu) \right]
\nonumber\\
&& + \frac{\alpha_2 \alpha_s}{M^2_A} \frac{m_\ell}{4 M_W^2} 
\frac{t^{3}_{\beta}}{(1+\epsilon t_{\beta})^{2}(1+\epsilon_{\ell}t_{\beta})}
\frac{M_{\tilde g}\mu}{\tilde m^2} (\delta_d^{LL})_{32} h_1(x_g)~,
\label{eq:CS}
\end{eqnarray}
\begin{eqnarray}
\tilde C_S &=& \frac{\alpha_2 \alpha_s}{M^2_A} \frac{m_\ell}{4 M_W^2} 
\frac{t^{3}_{\beta}}{(1+\epsilon t_{\beta})^{2}(1+\epsilon_{\ell}t_{\beta})}
\frac{M_{\tilde g}\mu^*}{\tilde m^2} (\delta_d^{RR})_{32} h_1(x_g)~.
\label{eq:CStilde}
\end{eqnarray}
The loop functions $h_1$, $h_3$ and $h_4$ already appeared in the discussion of the double Higgs
penguin contributions to $B_s$ mixing. This hints to correlations between NP effects in $B_s$
mixing and $B_s\to\mu^+\mu^-$ identified within the MSSM with MFV at large $\tan\beta$ in \cite{Buras:2002wq,Buras:2002vd}.

\begin{figure}[t]
\centering
\includegraphics[width=0.95\textwidth]{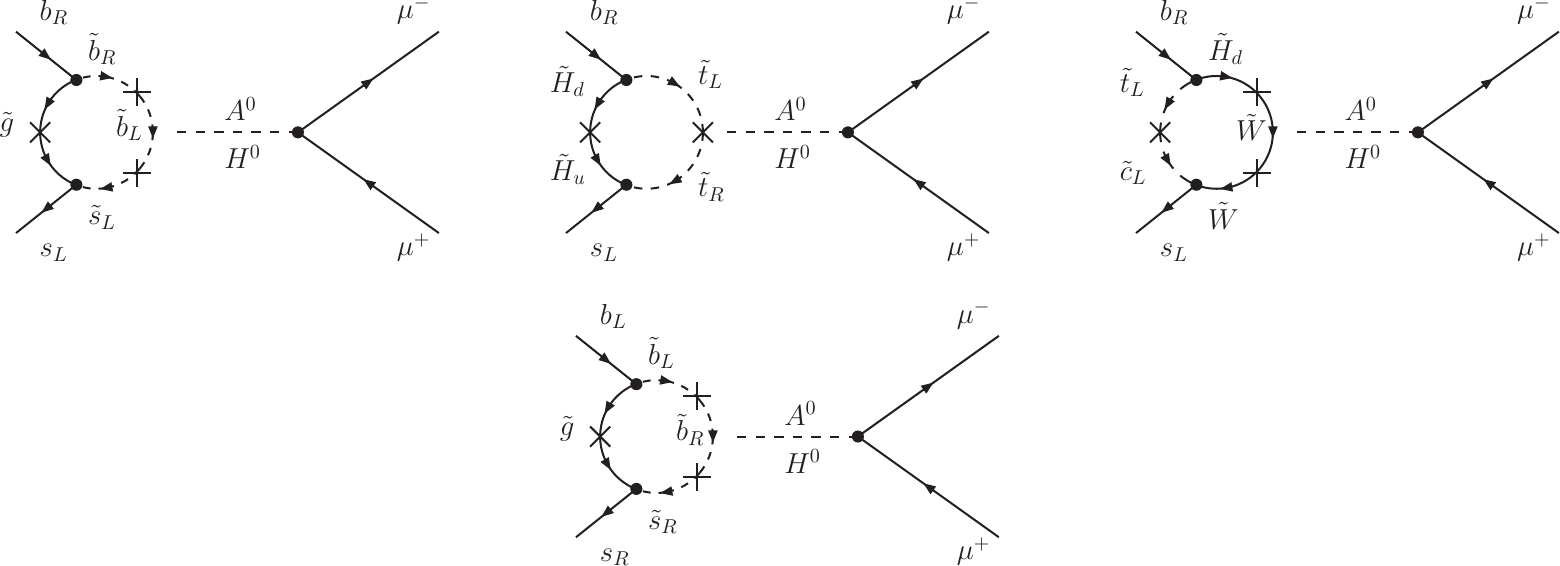}
\caption{\small
Feynman diagrams for the dominant Higgs mediated contributions to $B_s\to\mu^+\mu^-$.
The leading contribution to the decay amplitude proportional to $\tan^3\beta$ comes
from the self-energy corrections in diagrams where the Higgs propagators are attached
to the external quark legs. The diagrams in the first row correspond to (\ref{eq:CS}),
the one in the second row to (\ref{eq:CStilde}).
}
\label{diagrams_Bsmumu}
\end{figure}

In fig.~\ref{diagrams_Bsmumu}, we show the relevant SUSY Feynman diagrams for $B_s\to\mu^+\mu^-$
in the presence of new flavour structures.

The branching ratio for $B_s\to\mu^+\mu^-$ can be expressed in the following way
\begin{equation}
{\rm BR}(B_s \to \mu^+ \mu^-) = \frac{\tau_{B_s} F_{B_s}^2 m_{B_s}^3}{32 \pi} \sqrt{1-4 \frac{m_\mu^2}{m_{B_s}^2}} \left( |B|^2 \left( 1-4 \frac{m_\mu^2}{m_{B_s}^2} \right)+ |A|^2 \right)~,
\end{equation}
where the expressions $A$ and $B$ are given by the two linear combinations of the Wilson coefficients
\beq
A = 2 \frac{m_{\mu}}{m_{B_s}} C_{10}^{\rm SM} + m_{B_s}\left( C_P -\tilde C_P \right)~, 
\quad
B = m_{B_s}\left( C_S - \tilde C_S \right)~,
\eeq
where we assumed $C_{10}$ free of NP, which is approximately true in all the scenarios that
we consider. At leading order, the SM value for the Wilson coefficient $C_{10}$ is given by
\begin{equation}
C_{10}^{\rm SM} = \frac{g_2^2}{16 \pi^2} \frac{4 G_F}{\sqrt{2}} V_{tb} V_{ts}^{*} Y_0(x_t)~,
\end{equation}
and the loop function $Y_0$ can be found in appendix~\ref{sec:appendix}. The NLO QCD corrections to $Y_0$
have been calculated in \cite{Buchalla:1998ba} and found very small when $m_t(m_t)$ has been
used, which we do also in our analysis.

\subsubsection[$B^+ \to \tau^+ \nu$]{\boldmath $B^+ \to \tau^+ \nu$}

The SM expression for the branching ratio of the tree-level decay $B^+ \to \tau^+ \nu$ is given by
\begin{equation} \label{eq:Btaunu}
{\rm BR}(B^+ \to \tau^+ \nu)_{\rm SM} = \frac{G_F^2 m_{B^+} m_\tau^2}{8\pi} \left(1-\frac{m_\tau^2}{m^2_{B^+}} \right)^2 F_{B^+}^2 |V_{ub}|^2 \tau_{B^+}~.
\end{equation}
Its numerical value suffers from sizable parametrical uncertainties induced by $F_{B^+}$ and $V_{ub}$.

On the theoretical side, the $B\to\tau\nu$ process is one of the cleanest probes of the large $\tan\beta$ 
scenario due to its enhanced sensitivity to tree-level charged-Higgs exchange~\cite{Hou:1992sy,Akeroyd:2003zr,Isidori:2006pk}. In particular, a scalar charged current induced
by NP theories with extended Higgs sectors, leads to the following modification of the branching ratio
\begin{equation}
R_{B\tau\nu}= \frac{{\rm BR}(B^+ \to \tau^+ \nu)}{{\rm BR}(B^+ \to \tau^+ \nu)_{\rm SM}}= \bigg[1-\frac{m^2_{B^+}}{M^2_{H^+}}
\frac{t_\beta^2}{(1+\epsilon t_\beta)(1+\epsilon_\ell t_\beta)}
\bigg]^2~,
\label{Plnu}
\end{equation}
\noindent where we have included the $\tan\beta$ enhanced non-holomorphic corrections for the quark and 
lepton Yukawas. In the limit of degenerate SUSY particles, it turns out that $\epsilon\simeq\alpha_s/3\pi$~\cite{Buras:2002vd} and $\epsilon_{\ell}\simeq-3\alpha_2/16\pi$~\cite{Hisano:2008hn}.

Concerning the experimental situation, the HFAG collaboration~\cite{Barberio:2008fa} quotes
\begin{equation} \label{eq:Btaunu_HFAG}
{\rm BR}(B^+ \to \tau^+ \nu)_{\rm exp} = (1.43 \pm 0.37) \times 10^{-4}~,
\end{equation}
which is based on results by BaBar \cite{Aubert:2007xj} and Belle \cite{Ikado:2006un,:2008ch}.
Including additional preliminary results from BaBar \cite{:2008gx}, one finds the following new
World Average \cite{Tisserand:2009ja}
\begin{equation} \label{eq:Btaunu_exp}
{\rm BR}(B^+ \to \tau^+ \nu)_{\rm exp} = (1.73 \pm 0.35) \times 10^{-4}~,
\end{equation}
that is considerably higher than (\ref{eq:Btaunu_HFAG}).

Using the value for $|V_{ub}|$ quoted by the PDG~\cite{Amsler:2008zzb}, $|V_{ub}|=(3.95\pm
0.35)\times 10^{-3}$ and $F_{B^+}\simeq F_B$ given in tab.~\ref{tab:utinputs}, we find
the SM branching ratio given in tab.~\ref{tab:observables}, corresponding to
\begin{equation}
\label{eq:Btaunu_allowed}
\left(R_{B\tau\nu}\right)_\text{exp} = 1.57 \pm 0.53\,.
\end{equation}
In view of the parametric uncertainties induced in (\ref{eq:Btaunu}) by $F_{B^+}$ and $V_{ub}$,
in order to find the SM prediction for this branching ratio one can also use $\Delta M_d$ to find
\begin{eqnarray} \label{eq:Btaunu_DMd}
{\rm BR}(B^+ \to \tau^+ \nu)_{\rm SM} &=& \frac{3 \pi}{4
    \, \eta_B \, S_0(x_t) \, \hat B_{B_d}} \frac{m_\tau^2}{M_W^2} \left(1 -
  \frac{m_\tau^2}{m_{B^+}^2} \right)^2 \left\vert \frac{V_{ub}}{V_{td}}
  \right\vert^2 \tau_{B^+}~\Delta M_d ~.
\end{eqnarray}
Here $\Delta M_d$ is supposed to be taken from experiment and $\left\vert V_{ub}/V_{td}\right\vert^2$
is found using the formulae (\ref{eq:Rb}), (\ref{eq:Rt}) and (\ref{eq:Rb_gamma})
\begin{equation}
\left\vert \frac{V_{ub}}{V_{td}}\right\vert^2 =
\left( \frac{1}{1-\lambda^2/2} \right)^2
~\frac{1+R_t^2-2 R_t\cos\beta}{R^2_t} ~,
\end{equation}
with $R_t$ and $\beta$ determined by means of (\ref{eq:Rt_sin2beta_SM}).
In writing (\ref{eq:Btaunu_DMd}),
we used $F_B \simeq F_{B^+}$ and $m_{B_d}\simeq m_{B^+}$. We then find
\begin{equation}\label{eq:BtaunuSM1}
{\rm BR}(B^+ \to \tau^+ \nu)_{\rm SM}= (0.80 \pm 0.12)\times 10^{-4}
\end{equation}
that is by roughly a factor of two below the data in (\ref{eq:Btaunu_exp}). This result
agrees well with a recent result presented by the UTfit collaboration \cite{Bona:2009cj}.

It should be noted that the value of $\vert V_{ub}\vert$ used effectively in this procedure
turns out to be $3.50\times 10^{-3}$, which is close to $|V_{ub}|=(3.38\pm 0.36)\times 10^{-3}$ 
obtained from exclusive decays \cite{Antonelli:2009ws}.
On the other hand, it is significantly lower than the one quoted by the PDG and obtained from
tree level decays.

For $|V_{ub}|$ even higher than the PDG value, the experimental value for ${\rm BR}(B^+\to\tau^+\nu)$
can be reproduced, simultaneously making the appearance of a new phase in $B_d^0-\bar B_d^0$ likely
as discussed in sec.~\ref{sec:UT_tensions}. In fact, the solution 2) to the UT tensions presented
there would imply the central value ${\rm BR}(B^+ \to \tau^+ \nu)=1.32\times 10^{-4}$ in the SM and
be fully compatible with the experimental data, assuming no NP contributions to the latter decay.

This discussion highlights the importance of accurate determinations of $|V_{ub}|$ and of
${\rm BR}(B^+\to\tau^+\nu)$ in the future in order to be able to decide whether NP is at
work here or not. In particular, for low $|V_{ub}|$ values (exclusive determination) there
is a clear tension between the SM prediction for ${\rm BR}(B^+\to\tau^+\nu)$ and the data
and an even larger tension exists in the presence of charged Higgs contributions.

In our numerical analysis of NP contributions to various observables, we will use, to be
conservative, (\ref{eq:Btaunu_allowed}) as the constraint coming from this decay.

One could contemplate whether the charged Higgs correction in (\ref{Plnu}) could be so large that
$R_{B\tau\nu}$ becomes larger than unity improving the agreement of theory with the data. Such a
possibility, that would necessarily imply a light charged Higgs and large $\tan\beta$ values,
seems to be quite unlikely, even if not excluded yet~\footnote{For a detailed analysis on the
viability of NP scenarios with a heavy-light extended Higgs sector, we refer the reader to Ref.~\cite{Barenboim:2007sk}.}, in view of the constraints from other observables~\cite{Antonelli:2008jg}.

This discussion is at first sight independent of the values of the weak decay constants that cancel
in the ratio (\ref{eq:Btaunu_DMd}). Yet, the increase of the weak decay constants as suggested by recent non-quenched lattice calculations would certainly bring the SM value for ${\rm BR}(B^+\to\tau^+\nu)$
to agree better with the data. In (\ref{eq:Btaunu_DMd}), this effect is seen through the decreased
value of $|V_{td}|$ in order to agree with the experimental value of $\Delta M_d$ when the value of
the relevant weak decay constant is increased.

Another potentially interesting channel where to look for the presence of scalar charged currents
is represented by the purely leptonic Kaon decays. In particular, the NP effect on
$R_{K\mu\nu}=\Gamma^{\rm SUSY}(K\to\mu\nu)/\Gamma^{\rm SM}(K\to\mu\nu)$ is obtained from~(\ref{Plnu})
with the replacement $m_B^2\to m_K^2$~\cite{Isidori:2006pk}. Although the charged Higgs contributions
are now suppressed by a factor $m_K^2/m_B^2\simeq 1/100$, this is well compensated by the excellent 
experimental resolution~\cite{Flavianet} and the good theoretical control.
The best strategy to fully exploit the NP sensitivity of $K_{\ell2}$ systems is to consider the ratio $R^{'}=R_{K\mu\nu}/R_{\pi\mu\nu}$~\cite{Isidori:2006pk,Flavianet} instead of $R_{K\mu\nu}$. In fact,
while $R^{'}$ and $R_{K\mu\nu}$ have the same NP content (as $R_{\pi\mu\nu}$ is not sizably
affected by NP), $R^{'}$ depends on $(F_K/F_{\pi})^2$ instead of $F_K^2$ with $F_K/F_{\pi}$ determined
more precisely than $F_K$ by unquenched calculations in lattice QCD. However, given that the resolution 
on $F_K/F_{\pi}$ is at the \% level, the same level of NP sensitivity of $K\to\ell\nu$, we prefer to
not include the constraints from $K\to\ell\nu$ in the present analysis.
The above argument for $K\to \ell\nu$ does not apply to $B^+\to\tau^+\nu$. In fact, even if the hadronic uncertainties related to $F_B$ and $V_{ub}$ are much larger that those for $F_K/F_{\pi}$ and $V_{us}$,
they cannot hide in any way the huge NP effects that can affect ${\rm BR}(B^+\to\tau^+\nu)$.

\subsubsection[$K\to\pi\nu\bar{\nu}$ and $b\to s\nu\bar\nu$]{\boldmath $K\to\pi\nu\bar{\nu}$ and $b\to s\nu\bar\nu$}

Within the MSSM with $R$-parity conservation, sizable non-standard contributions to $K\to\pi\nu\overline{\nu}$ decays can be generated only if the soft-breaking terms have a non-MFV structure. The leading amplitudes giving rise to large effects are induced by: i) chargino/up-squark loops~\cite{Nir:1997tf,Buras:1997ij,Colangelo:1998pm,Buras:1999da} and ii) charged Higgs/top quark loops~\cite{Isidori:2006jh}. In the first case, large effects are generated if the trilinear couplings of 
the up-squarks have a non-MFV structure. In the second case, deviations from the SM are induced by non-MFV 
terms in the right-right down sector, provided $\tan\beta$ is large (30 to 50).

In the case of $b\to s\nu\bar\nu$ transitions like $B\to K\nu\bar\nu$, $B\to K^*\nu\bar\nu$ or  $B\to X_s\nu\bar\nu$, the second case above is prevented by the constraint on BR($B_s\to\mu^+\mu^-$), while chargino/up-squark loops with non-MFV trilinear couplings in the up-squark sector can also generate
sizable effects \cite{Yamada:2007me,Altmannshofer:2009ma}.

However, since the SUSY models we consider in sec.~\ref{sec:numerics} feature neither sizeable
off-diagonal entries in the trilinear couplings nor simultaneously large enough $(\delta_d^{RR})_{13}$
and $(\delta_d^{RR})_{23}$ mass insertions, both $K\to\pi\nu\bar{\nu}$ and $b\to s\nu\bar\nu$ decays
turn out to be SM-like.

\subsubsection[$\ell_i\to\ell_j\gamma$]{\boldmath $\ell_i\to\ell_j\gamma$}

Within SUSY models, LFV effects relevant to charged leptons originate from any misalignment between
fermion and sfermion mass eigenstates. Once non-vanishing LFV entries in the slepton mass matrices
are generated, irrespective of the underlying mechanism accounting for them, LFV rare decays like $\ell_i\to\ell_j\gamma$ are naturally induced by one-loop diagrams with the exchange of gauginos 
and sleptons. The present and projected bounds on these processes are summarized in 
tab.~\ref{tab:lfvtable}~\footnote{The 2008 data from MEG are already close
($\rm{BR}(\mu\to e\,\gamma)<3\times~10^{-11}$~\cite{Adam:2009ci}) to the present upper bound
from MEGA so that the 2009 data should be able to provide a new improved bound.}.

\begin{table}[t]
\addtolength{\arraycolsep}{3pt}
\renewcommand{\arraystretch}{1.3}
\centering
\begin{tabular}{|l|l|l|l|}
\hline
Process & Present Bounds & Expected Future Bounds  & Future Experiments\\
\hline\hline
BR($\mu \to e\,\gamma$) & $1.2~ \times~ 10^{-11}$ & $\mathcal{O}(10^{-13} - 10^{-14})$ & MEG, PSI\\
BR($\mu \to e\,e\,e$) & $1.1~ \times~ 10^{-12}$ & $\mathcal{O}(10^{-13} - 10^{-14})$ & ?\\
BR($\mu \to e$ in Nuclei (Ti)) & $1.1~ \times~ 10^{-12}$ & $\mathcal{O}(10^{-18})$ & J-PARC\\
BR($\tau \to e\,\gamma$) & $1.1~ \times~ 10^{-7}$ & $\mathcal{O}(10^{-8}) $ &SuperB\\
BR($\tau \to e\,e\,e$) & $2.7~ \times~ 10^{-7}$ & $\mathcal{O}(10^{-8}) $ &SuperB\\
BR($\tau \to e\,\mu\,\mu$) & $2.~ \times~ 10^{-7}$ & $\mathcal{O}(10^{-8}) $ &SuperB\\
BR($\tau \to \mu\,\gamma$) & $6.8~ \times~ 10^{-8}$ & $\mathcal{O}(10^{-8}) $ &SuperB\\
BR($\tau \to \mu\, \mu\, \mu$) & $2~ \times~ 10^{-7}$ & $\mathcal{O}(10^{-8}) $ &LHCb\\
BR($\tau \to \mu\, e\,e$) & $2.4~ \times~ 10^{-7}$ & $\mathcal{O}(10^{-8}) $ &SuperB\\
\hline
\end{tabular}
\caption{\small
Present~\cite{Amsler:2008zzb} and upcoming experimental limits on various leptonic processes at 90\% C.L.}
\label{tab:lfvtable}
\end{table}

The decay $\ell_i\to\ell_j\gamma$ is described by the dipole operator and the corresponding
amplitude reads
\bea
T=m_{\ell_i}\epsilon^{\lambda}\overline{u}_j(p-q)[iq^\nu\sigma_{\lambda\nu}
(A_{L}P_{L}+A_{R}P_{R})]u_i(p)\,,
\eea
where $p$ and $q$ are momenta of the leptons $\ell_k$ and of the photon respectively
and $A_{L,R}$ are the two possible amplitudes entering the process. The lepton mass 
factor $m_{\ell_i}$ is associated to the chirality flip present in this transition.
The branching ratio of $\ell_{i}\rightarrow \ell_{j}\gamma$ can be written as
\bea
\frac{{\rm BR}(\ell_{i}\rightarrow  \ell_{j}\gamma)}{{\rm BR}(\ell_{i}\rightarrow 
\ell_{j}\nu_i\bar{\nu_j})} =\frac{48\pi^{3}\alpha}{G_{F}^{2}}(|A_L^{ij}|^2+|A_R^{ij}|^2)\,.
\nonumber
\eea
In the MI approximation it is found that~\cite{Paradisi:2005fk}
\begin{eqnarray} 
\label{MIamplL}
A^{ij}_L&\simeq&\frac{\alpha_2}{4\pi}
\frac{\left(\delta^{LL}_{\ell}\right)_{ij}}{m_{\tilde \ell}^{2}}t_{\beta}
~\bigg[
\frac{\mu M_{2}}{(M_{2}^2-\mu^2)}\bigg(f_{2n}(x_2,x_\mu)+f_{2c}(x_2,x_\mu)\bigg)
\nonumber\\
&+& \tan^2\theta_{W}\,
\mu M_{1}\bigg(\frac{f_{3n}(x_1)}{m_{\tilde \ell}^{2}}+
\frac{f_{2n}(x_1,x_\mu)}{(\mu^2 - M_{1}^2)}\bigg)
\bigg]
\nonumber\\
&+& \frac{\alpha_1}{4\pi}~\frac{\left(\delta^{RL}_{\ell}\right)_{ij}}{m_{\tilde \ell}^2}~
\left(\frac{M_1}{m_{\ell_i}}\right)~2~f_{2n}(x_1)~,
\end{eqnarray}
\begin{eqnarray}
\label{MIamplR}
A^{ij}_R
\simeq
\frac{\alpha_{1}}{4\pi}
\left[
\frac{\left(\delta^{RR}_{e}\right)_{ij}}{m_{\tilde \ell}^{2}}\mu M_{1}t_{\beta}
\left(\frac{f_{3n}(x_1)}{m_{\tilde \ell}^{2}}-\frac{2f_{2n}(x_1,x_{\mu})}{(\mu^2-M_{1}^2)}\right)
+2\frac{\left(\delta^{LR}_{e}\right)_{ij}}{m_{\tilde\ell}^{2}}~
\left(\frac{M_1}{m_{\ell_i}}\right)~f_{2n}(x_1)
\right]~,
\end{eqnarray}
where $\theta_W$ is the weak mixing angle, $m_{\tilde\ell}$ is an average slepton mass,
$x_{1,2}=M_{1,2}^2/m_{\tilde \ell}^2$, $x_\mu=\mu^2/m_{\tilde \ell}^2$ and $f_{i(c,n)}(x,y)=f_{i(c,n)}(x)-f_{i(c,n)}(y)$. The loop functions $f_i$ are given in appendix~\ref{sec:appendix}.

In the case of $\mu\to e\gamma$, one has to consider also the contributions arising from double
MIs $\delta_{23}\delta_{31}$ as they can compete with the single MI contributions when $\delta_{21}\approx\delta_{23}\delta_{31}$; in particular, this will turn out to be the case
for the flavour models we will analyze. A particularly important effect is provided by the
following amplitude
\beq
A^{21}_L
\simeq
\frac{\alpha_1}{4\pi}\left(\frac{m_{\tau}}{m_{\mu}}\right)
\frac{\mu M_{1}t_{\beta}}{m^{4}_{\tilde\ell}}f_{4n}(x_1)
(\delta^{RR}_{e})_{23}(\delta^{LL}_{\ell})_{31}
\label{mueg_2mi}
\eeq
because of the enhancement factor $m_{\tau}/m_{\mu}$. The analog expression for $A^{21}_R$
is obtained by $A^{21}_R = A^{21}_L(L\leftrightarrow R)$. In~(\ref{mueg_2mi}), the loop
function is such that $f_{4n}(x)=f_0^{(3)}(x)/2$ with $f_0^{(3)}(1)=-1/15$ and $f_0^{(3)}(x)$
defined in appendix~\ref{sec:appendix}.
Other contributions, generated by the double MIs $(\delta^{LL}_{\ell})_{23}(\delta^{LL}_{\ell})_{31}$
and $(\delta^{RR}_{e})_{23}(\delta^{RR}_{e})_{31}$ and not enhanced by $m_{\tau}/m_{\mu}$,
can be still relevant; however, for simplicity, we do not report them here although they are
systematically included in our numerical analysis. In~(\ref{MIamplL}),~(\ref{MIamplR}),~(\ref{mueg_2mi}),
as well as in the remainder of sec.~\ref{sec:DF012}, we assume the $\mu$ term, the trilinear couplings
and the gaugino masses to be real and the latter also positive.

In the illustrative case of a degenerate SUSY spectrum with a common mass $m_{\tilde\ell}$, we find that
\begin{eqnarray}
  A^{21}_L
  &\simeq&
  \frac{\alpha_2}{60\pi}\frac{t_{\beta}}{m^{2}_{\tilde\ell}}
  (\delta^{LL}_{\ell})_{21}
  +\frac{\alpha_1}{48\pi}
  \frac{(\delta^{RL}_{\ell})_{21}}{m^{2}_{\tilde\ell}}
  \left(\frac{m_{\tilde\ell}}{m_{\mu}}\right)
 -\frac{\alpha_1}{120\pi}\frac{m_\tau}{m_\mu}\frac{t_{\beta}}{m^{2}_{\tilde\ell}}
  (\delta^{RR}_{e})_{23}(\delta^{LL}_{\ell})_{31}\,,
  \\
  A^{32}_L
  &\simeq&
  \frac{\alpha_2}{60\pi}\frac{t_{\beta}}{m^{2}_{\tilde\ell}}
  (\delta^{LL}_{\ell})_{32}
  +\frac{\alpha_1}{48\pi}
  \frac{(\delta^{RL}_{\ell})_{32}}{m^{2}_{\tilde\ell}}
  \left(\frac{m_{\tilde\ell}}{m_{\tau}}\right)\,,
  \\
  A^{21}_R
  &\simeq&
  -\frac{\alpha_1}{4\pi}\frac{t_{\beta}}{m^{2}_{\tilde\ell}}
  \left[
  \frac{(\delta^{RR}_{e})_{21}}{60}
  +\frac{m_\tau}{m_\mu}
  \frac{(\delta^{LL}_{\ell})_{23}(\delta^{RR}_{e})_{31}}{30}
  \right]
  +\frac{\alpha_1}{48\pi}
  \frac{(\delta^{LR}_{e})_{21}}{m^{2}_{\tilde\ell}}
  \left(\frac{m_{\tilde\ell}}{m_{\mu}}\right)\,,
  \\
  A^{32}_R
  &\simeq&
  -\frac{\alpha_1}{4\pi}\frac{t_{\beta}}{m^{2}_{\tilde\ell}}
  \frac{(\delta^{RR}_{e})_{32}}{60}
  +\frac{\alpha_1}{48\pi}
  \frac{(\delta^{RL}_{\ell})_{32}}{m^{2}_{\tilde\ell}}
  \left(\frac{m_{\tilde\ell}}{m_{\tau}}\right)\,.
\label{MI_degenerate}
\end{eqnarray}
Besides $\ell_{i}\to\ell_{j}\gamma$, there are also other promising LFV channels, such as $\ell_i\to\ell_j\ell_k\ell_k$ and $\mu$-$e$ conversion in nuclei, that could be measured
with the upcoming experimental sensitivities. However, within SUSY models, these processes
are typically dominated by the dipole transition $\ell_{i}\to\ell_{j}\gamma^{*}$ leading 
to the unambiguous prediction,
\begin{eqnarray}
\frac{{\rm BR}(\ell_i\rightarrow \ell_j\ell_k\bar{\ell}_k)}{{\rm BR}(\ell_i\rightarrow \ell_j\bar{\nu_j}\nu_{i})}
&\simeq&
\frac{\alpha_{el}}{3\pi}
\bigg(\log\frac{m^2_{\ell_i}}{m^2_{\ell_k}}-3\bigg)
\frac{{\rm BR}(\ell_i\rightarrow \ell_j\gamma)}{{\rm BR}(\ell_i\rightarrow \ell_j\bar{\nu_j}\nu_{i})}~,
\nonumber\\
{\rm CR}(\mu\to e~\mbox{in N})
&\simeq&
\alpha_{\rm em} \times {\rm BR}(\mu\rightarrow e\gamma)~.
\label{eq:dipole}
\end{eqnarray}
Thus, an experimental confirmation of the above relations would be crucial to prove the dipole
nature of the LFV transitions. This would provide a powerful tool to discriminate between
different NP scenarios as, for instance, SUSY and LHT models, as the latter do not predict
a dipole dominance for $\ell_{i}\to\ell_{j}\ell_{k}\ell_{k}$~\cite{Blanke:2007db}.

Additional and sizable contributions to LFV decays may arise from the Higgs sector by means of the
effective LFV Yukawa interactions induced by non-holomorphic terms~\cite{Babu:2002et}; hence, in
general, the expectations of~(\ref{eq:dipole}) can be violated~\cite{Brignole:2003iv,Brignole:2004ah,Kitano:2003wn,Arganda:2004bz,Arganda:2008jj,Paradisi:2005tk,Paradisi:2006jp}.
However, these effects become relevant only if $\tan\beta = \mathcal{O}(40-50)$ and if the Higgs
masses are roughly one order of magnitude lighter then the slepton masses~\cite{Brignole:2003iv,Brignole:2004ah,Kitano:2003wn,Arganda:2004bz,Arganda:2008jj,Paradisi:2005tk,Paradisi:2006jp}. The last condition never occurs in our scenarios,
hence Higgs mediated LFV effects are safely negligible in our analysis.

\subsection[$\Delta F = 0$ Processes]{\boldmath $\Delta F = 0$ Processes} \label{sec:DF0}

\subsubsection{Electric Dipole Moments} \label{sec:EDMs}

As the SM predictions for the electric dipole moments are very far from the present experimental
resolutions, the EDMs represent very clean probes of NP effects. Given that the EDMs are CP-violating
but flavour conserving observables, they do not require in principle any source of flavour violation,
hence, we refer to them as $\Delta F=0$ processes. Yet, they can also be generated by two $\Delta F=1$ transitions, in which case one refers to {\it ``flavoured''} EDMs.

Indeed, within a MSSM framework with flavour violating soft terms, large and potentially visible effects
in the {\it ``flavoured''} EDMs are typically expected~\cite{Hisano:2007cz,Hisano:2008hn,Hisano:2006mj}.
In particular, when NP sources of flavour violation generating $b\to s$ transitions are assumed, the 
chromo-EDM (CEDM) and the EDM of the strange quark are unambiguously predicted. Unfortunately, an issue
which is still unclear at present is the impact of the strange quark CEDM and EDM on the EDMs of physical
systems like the neutron or heavy atoms like the Thallium or the Mercury. The main source of uncertainty
comes from the evaluation of the relevant hadronic matrix elements that should be ultimately evaluated by
means of lattice QCD techniques.
As a result, it is not possible at present to correlate or to constrain CPV processes in $B$-physics by
means of the EDMs of physical systems that are  induced by the strange quark (C)EDM. Therefore, in our
analysis, we only monitor the predictions for the strange quark (C)EDM generated by sizable CP violating
effects in the flavour observables.

If in the future there will be theoretical improvements enabling us to relate in a reliable way the strange
quark (C)EDM to physical quantities, we could make use of additional observables, i.e. the EDMs of several systems, to test the NP theory that is at work, provided some NP signals in CPV $B_s$ systems would appear.

In the following, we report the relevant expressions for the {\it ``flavoured''} EDMs including the dominant beyond-leading-order (BLO) effects~\cite{Hisano:2007cz,Hisano:2008hn,Hisano:2006mj}. In fact, as shown in~\cite{Hisano:2007cz,Hisano:2008hn,Hisano:2006mj} BLO effects dominate over the leading-order (LO) ones
in a large region of the parameter space, hence, their inclusion in the evaluation of the hadronic EDMs is essential.

Although our numerical results have been obtained including the full set of contributions, for simplicity,
we report the dominant contributions to the hadronic EDMs.

The dominant gluino/squark contribution to the down-quark (C)EDMs is
\beq
\left\{\frac{d_{d_i}}{e},~d^c_{d_i}\right\}_{\tilde g}=
-\frac{\alpha_{s}}{4\pi}\frac{m_{b}}{{\tilde m}^{2}}
\frac{M_{\tilde g}\mu}{{\tilde m}^{2}}\tgb
\frac{f^{d}_{\tilde g}(x_g)}{1+\epsilon\tgb}
{\rm Im}\left[(\delta^{LL}_{d})_{i3}(\delta^{RR}_{d})_{3i}\right]\,,
\label{Eq:edm_d_gluino}
\eeq
where the loop functions satisfy $f^{d}_{\tilde g}(1)=$ $\{4/135,~11/180\}$.

Similarly, the corresponding prediction for the up-quark (C)EDMs is
\beq
 \left\{ \frac{d_{u}}{e},~d^{c}_{u}\right\}_{\tilde g}=
 -\frac{\alpha_s}{4\pi}\frac{m_{u_k}}{\tilde{m}^2}
 \frac{M_{\tilde g} A_{u_k}}{\tilde{m}^2}
 f^{u}_{\tilde g}(x_g)
 {\rm Im}\left[(\delta_{u}^{LL})_{1k}(\delta_{u}^{RR})_{k1}\right]\,,
 \label{Eq:edm_u_gluino}
\eeq
where $k=2,3$, $A_{u_2,u_3}=A_{c,t}$ and $f^{u}_{\tilde g}(1)=$ $\{-8/135,~11/180\}$.

The first $H^\pm$ effects to the (C)EDMs appear at the
BLO~\cite{Hisano:2006mj,Hisano:2007cz,Hisano:2008hn} and they are well approximated by
\beq
 \left\{ \frac{d_{d_i}}{e},~d^c_{d_i} \right\}_{H^\pm}
 =-\frac{\alpha_2}{16\pi}
 \frac{m_b}{M_{H^\pm}^2}
 \frac{m_{t}^{2}}{M_{W}^{2}}
 \frac{(1-\epsilon t_{\beta})\epsR t_{\beta}}{3(1+\epsilon t_{\beta})^{2}}
 {\rm Im}\left[V^*_{3i}(\delta^{RR}_{d})_{3i}\right]
 f_{H^\pm}(y_t)\,,
\label{Eq:edmH}
\eeq
where $f_{H^\pm}(1) = \{-7/9,-2/3\}$.

Charginos contribute to the EDMs already at the LO but the corresponding (C)EDMs are suppressed
by the light quark masses $m_{d_i}$. At the BLO, a new effect proportional to $m_{b}$ is generated
by the charged-Higgsino/squark diagrams leading to~\cite{Hisano:2006mj,Hisano:2007cz,Hisano:2008hn}
\begin{eqnarray}
 \left\{ \frac{d_{d_i}}{e},~d^c_{d_i} \right\}_{{\tilde H}^\pm}
 \!\!\!\!\!&=&\!\!\!
 \frac{\alpha_2}{16\pi}
 \frac{m_b}{{\tilde m}^2}
 \frac{m_t^2}{M_W^2}
 \frac{A_t \mu}{{\tilde m}^2}
 \tgb
\frac{\epsR\tgb}{3(1+\epsilon \tgb)^2}
  {\rm Im}\left[V^*_{3i}(\delta^{RR}_{d})_{3i}\right]
 f_{{\tilde H}^\pm}(x_{\mu})\,,
\label{Eq:FEDM_Higgsino}
\end{eqnarray}
where $x_{\mu}=\mu^2/{\tilde m}^2$ and $f_{{\tilde H}^\pm}(1)=\{-5/18,-1/6\}$. The full expressions of the
loop functions $f_{\tilde g}(x_g)$, $f_{H^\pm}(x)$ and $f_{{\tilde H}^\pm}(x)$ are listed in appendix~\ref{sec:appendix}.
For equal SUSY masses and $\mu>0$, it turns out that $\epsilon_{R}\!=\!\epsilon/3\!=\!\alpha_s/9\pi$.

So far, we have presented the dominant contributions to the quark (C)EDMs assuming the
presence of right-handed currents, hence of RR MIs. Since in the present work we are also
interested in models with purely left-handed currents, it is useful to show also the dominant
(one loop induced) contributions to the (C)EDMs arising within this scenario. They read
\begin{eqnarray}
 \left\{\frac{d_{d_i}}{e},~d^c_{d_i}\right\}_{{\tilde H}^\pm} =
 \frac{\alpha_2}{16\pi}
 \frac{m_{d_i}}{\tilde{m}^2}
 \frac{m_t^2}{m_W^2}
 \frac{A_t \mu}{\tilde{m}^2}
 \frac{\tgb}{(1+\epsilon\tgb)}~
  {\rm Im}\left[V^*_{3i}(\delta_{d}^{LL})_{3i}\right]
  g_{{\tilde H}^\pm}(x_{\mu})\,,
 \label{Eq:LO_Higgsino}
\end{eqnarray}
where $g_{{\tilde H}^\pm}(1)=\{2/15,~1/10\}$, with its complete expressions given in appendix~\ref{sec:appendix}. Eq.~(\ref{Eq:LO_Higgsino}) shows that, in this case, the
(C)EDMs are suppressed by the external light quark masses, in contrast to the case
where also RR MIs are non-vanishing. The analogous expression for up-type quarks, 
namely $d^{(c)}_{u_i}$, is of order $d^{(c)}_{d}/d^{(c)}_{u}\sim [m_t^2/(m_b^2\tgb^{2})]\times[A_t\mu/\mu^2]\times\tgb$ and thus safely negligible.

Passing to the leptonic sector, the dominant contribution to the electron EDM arises
from the one-loop exchange of binos/sleptons, and the corresponding EDM is given as
\bea
\label{Eq:lEDM_LO}
\frac{d_{e}}{e}
\!\!=\!\!
-\frac{\alpha_1}{4\pi}\frac{M_1}{m^{2}_{\tilde\ell}}\!\!\!
&\bigg\{&
\!\!\!{\rm Im}
[(\delta^{LR}_{\ell})_{1k}(\delta^{RR}_{e})_{k1} + (\delta^{LL}_{\ell})_{1k}(\delta^{LR}_{\ell})_{k1}]
\,f_{3n}(x_1)
+{\rm Im}[(\delta^{LL}_{\ell})_{1k}(\delta^{LR}_{\ell})_{kl}(\delta^{RR}_{e})_{l1}
\nonumber\\
&+&
(\delta^{LR}_{\ell})_{1k}(\delta^{RR}_{e})_{kl}(\delta^{RR}_{e})_{l1}+
(\delta^{LL}_{\ell})_{1k}(\delta^{LL}_{\ell})_{kl}(\delta^{LR}_{\ell})_{l1}]
\,f_{4n}(x_1)
\bigg\}\,,
\eea
where $k,l=2,3$, $(\delta^{LR}_{\ell})_{33}= -m_{\tau}(A_{\tau}+\mu\tgb)/m^{2}_{\tilde\ell}$ and
$f_{3n}(1)=-1/12$.

One of the most peculiar features of the {\it flavoured} EDMs is that they might be proportional to
the heaviest fermionic masses $m_t$, $m_b$ and $m_{\tau}$ instead of the lightest ones, as it happens
in the case of flavour {\it blind} phases. This huge enhancement factor can bring the (C)EDMs close
to the current and future experimental sensitivities, providing a splendid opportunity to probe the
flavour structure of the MSSM.

The main obstacles to fully exploit the NP sensitivity of the EDMs is that experimentally, one measures
the EDMs of composite systems, as heavy atoms, molecules or the neutron EDM while the theoretical
predictions are relative to the EDMs of constituent particles, {\it i.e.} quarks and leptons, thus
a matching between quarks and leptons EDMs into physical EDMs is necessary and this induces sizable 
uncertainties related to QCD, nuclear and atomic interactions.

The quark (C)EDMs and lepton EDMs can be obtained starting from the effective CP-odd Lagrangian
\beq
{\cal L}_{\rm eff}
    =
   -
   \sum_{i=u,d,s,e,\mu} i \frac{d_f}{2} \bar{\psi}_i (F\cdot \sigma)\gamma_5 \psi_i
   -
   \sum_{i=u,d,s} i \frac{d^c_f}{2} g_s \bar{\psi}_i (G\cdot \sigma)\gamma_5 \psi_i
   +
   \sum_{i,j} C_{ij}\, (\bar{\psi}_i \psi_i)(\bar{\psi}_j i\gamma_5 \psi_j)
   + \ldots,
\label{Eq:CPodd}
\eeq
where the first and the second terms of~(\ref{Eq:CPodd}) are the fermion EDMs and CEDMs,
respectively, while the coefficients $C_{ij}$ are relative to the dimension-six CP-odd
four-Fermi interaction operators.

Among the various atomic and hadronic EDMs, a particularly important role is played by
the thallium EDM ($d_{\rm{Tl}}$) and the neutron EDM ($d_n$). They can be estimated
as~\cite{LiuKelly,MP1,MP2,Pospelov:1999ha,Pospelov:2000bw}
\beq
 d_{\rm Tl} = -585\, d_e - e\, 43\, {\rm GeV}\, C_S^{(0)}~,
\label{Eq:dTl}
\eeq
\beq
d_n = (1\pm 0.5)\Big[ 1.4\,(d_d-0.25\, d_u) + 1.1\, e\,
(d^c_d+0.5\,d^c_u)\Big]~,
\label{Eq:dn_odd}
\eeq
where $C_s^{(0)}$ is given by a combination of the coefficients $C_{ij}$~\cite{Pospelov:1999ha,Pospelov:2000bw}.
However, when $d_{\rm{Tl}}$ and $d_n$ are generated by flavour effects, as in our case, the contributions they receive from $C_s^{(0)}$ are always very suppressed hence, safely negligible~\cite{Hisano:2008hn}.

\subsubsection{The Anomalous Magnetic Moment of the Muon} \label{sec:g-2}

The possibility that the anomalous magnetic moment of the muon (we define $a_\mu=(g-2)_{\mu}/2$),
which has been measured very precisely in the last few years \cite{Bennett:2006fi}, provides a
first hint of physics beyond the SM has been widely discussed in the literature.
Despite substantial progress both on the experimental and on the theoretical sides, the situation
is not completely clear yet (see~\cite{Jegerlehner:2009ry} for an updated discussion).

Most recent analyses based on $e^+e^-$ data converge towards a $3\sigma$ discrepancy~\footnote{
The most recent $\tau$-based estimate of the muon magnetic anomaly is found to be 1.9 standard
deviations lower than the SM prediction~\cite{Davier:2009ag}, coming closer to the $e^+e^-$ value.}
in the $10^{-9}$ range~\cite{Jegerlehner:2009ry}:
\beq
 \Delta a_{\mu} =  a_{\mu}^{\rm exp} - a_{\mu}^{\rm SM}
\approx (3 \pm 1) \times 10^{-9}~.
\label{eq:amu_exp}
\eeq
The possibility that the present discrepancy may arise from errors in the determination of the
hadronic leading-order contribution to $\Delta a_{\mu}$ seems to be unlikely, as recently stressed
in~\cite{Passera:2008jk}.

The main SUSY contribution to $a^{\rm MSSM}_\mu$ is usually provided by the loop exchange of charginos
and sneutrinos~\cite{Moroi:1995yh}. The supersymmetric contributions to $a_\mu$ are correctly reproduced
by the following approximate expression
\bea
\label{MIamplLsec}
a^{\rm MSSM}_\mu&=&
\frac{\alpha_{2}}{4\pi}
\frac{m^{2}_{\mu}}{m_{\tilde \ell}^{2}}t_{\beta}
~\bigg[
\frac{\mu M_{2}}{(M_{2}^2\!-\!\mu^2)}
\bigg(\frac{1}{2}f_{2n}(x_2,x_{\mu})\!-\!f_{2c}(x_2,x_{\mu})\bigg)
\nonumber\\
&+& \tan^2\theta_{W}\,
\bigg(\frac{\mu M_{1}}{m_{\tilde \ell}^{2}}f_{3n}(x_1)+
\frac{1}{2}\frac{\mu M_{1}}{(M_{1}^2\!-\!\mu^2)}f_{2n}(x_1,x_{\mu})
\bigg)
\bigg]\,.
\eea
In the limit of degenerate SUSY masses one can easily find that
\beq
\frac{a^{\rm MSSM}_\mu}{ 1 \times 10^{-9}}
\approx 1.5\left(\frac{\tan\beta }{10} \right)
\left( \frac{300~\rm GeV}{m_{\tilde \ell}}\right)^2 \text{sgn}\,\mu\,.
\label{eq:g_2}
\eeq
The most relevant feature of~(\ref{eq:g_2}) is that the sign of $a^{\rm MSSM}_\mu$ is fixed by
the sign of the $\mu$ term (given $M_2>0$) so that the solution $\mu>0$ is strongly favoured.

\begin{table}[t]
\addtolength{\arraycolsep}{3pt}
\renewcommand{\arraystretch}{1.4}
\centering
\begin{tabular}{|l|l|l|l|}
\hline
observable & SM prediction & exp. current & exp. future \\
\hline\hline
$S_{\psi\phi}$ & $\simeq 0.036$~\cite{Amsler:2008zzb} & $0.81^{+0.12}_{-0.32}$ \cite{Barberio:2008fa} &
$\simeq 0.02$ \cite{Blouw:2007px} \\
$S_{\phi K_S}$ & $\sin2\beta+0.02 \pm 0.01$~\cite{Buchalla:2008jp} & $0.44 \pm 0.17$~\cite{Barberio:2008fa} 
& $(2-3)\%$~\cite{Bona:2007qt} \\
$S_{\eta^{\prime} K_S}$ & $\sin2\beta+0.01 \pm 0.01$~\cite{Buchalla:2008jp} & $0.59 \pm 0.07$~\cite{Barberio:2008fa} & $(1-2)\%$~\cite{Bona:2007qt} \\
\hline
$A_\text{CP}(b\to s\gamma)$ & $\left(-0.44 ^{+0.14}_{-0.24}\right)\%$~\cite{Hurth:2003dk}& 
$\left(-0.4 \pm 3.6\right)\%$~\cite{Barberio:2008fa} & $(0.4-0.5)\%$~\cite{Bona:2007qt} \\
\hline
$\langle A_7 \rangle$ & $ (3.4^{+0.4}_{-0.5})10^{-3}$~\cite{Altmannshofer:2008dz} &  & \\
$\langle A_8 \rangle$ & $ (-2.6^{+0.4}_{-0.3})10^{-3}$~\cite{Altmannshofer:2008dz} &  & \\
$\langle A_9 \rangle$ & $ (0.1^{+0.1}_{-0.1})10^{-3}$~\cite{Altmannshofer:2008dz} &  & \\
\hline
$|d_e|~~(e\,$cm) & $\simeq 10^{-38}$~\cite{Pospelov:2005pr}& $< 1.6 \times 10^{-27}$ \cite{Regan:2002ta} &
$\simeq 10^{-31}$~\cite{Pospelov:2005pr} \\
$|d_n|~~(e\,$cm) & $\simeq 10^{-32}$~\cite{Pospelov:2005pr} & $< 2.9 \times 10^{-26}$ \cite{Baker:2006ts} &
$\simeq 10^{-28}$~\cite{Pospelov:2005pr} \\
\hline
BR$(B_s\to\mu^+\mu^-)$ & $ (3.60 \pm 0.37) 10^{-9}$ & $<5.8\times 10^{-8}$~\cite{:2007kv} &
$\simeq 10^{-9}$~\cite{Altarelli:2009ec} \\
BR$(B_d\to\mu^+\mu^-)$ & $(1.08 \pm 0.11)10^{-10}$ & $<1.8\times 10^{-8}$~\cite{:2007kv} &
 \\
BR$(B\to X_s\gamma)$ & $(3.15 \pm 0.23)10^{-4}$~\cite{Misiak:2006zs} & $(3.52\pm 0.25)10^{-4}$ ~\cite{Barberio:2008fa} & \\
BR$(B\to X_s \ell^+\ell^-)$ & $(1.59 \pm 0.11)10^{-6}$~\cite{Huber:2005ig} & $(1.59\pm 0.49)10^{-6}$ ~\cite{Aubert:2004it,Iwasaki:2005sy} & \\
BR$(B\to \tau\nu)$ & $(1.10 \pm 0.29)10^{-4}$  & $(1.73 \pm 0.35)10^{-4}$~\cite{Tisserand:2009ja}& \\
\hline
\end{tabular}
\caption{\small
SM predictions and current/expected experimental sensitivities for the observables most relevant
for our analysis. The branching ratio of $B\to X_s \ell^+\ell^-$ refers to the low dilepton
invariant mass region, $q^2_{\ell^+\ell^-} \in [1,6] \,\text{GeV}^2$. For the SM prediction of
BR$(B\to \tau\nu)$, see also (\ref{eq:BtaunuSM1}): BR$(B\to \tau\nu)=(0.80 \pm 0.12)\times 10^{-4}$.}
\label{tab:observables}
\end{table}

\subsection[Correlations between $\Delta F=0$ and $\Delta F=1$ Processes in the Leptonic Sector]
{Correlations between \boldmath $\Delta F=0$ and $\Delta F=1$ Processes in the Leptonic Sector}

In the following, we discuss the implications of a potential evidence or improved upper bound
of ${\rm BR}(\mu\to e\gamma)$ at the expected sensitivities of MEG, namely at the level of
${\rm BR}(\mu\to e\gamma)\gtrsim 10^{-13}$~\cite{Maki:2008zz}.
In particular, we will exploit the correlations among ${\rm BR}(\mu\to e\gamma)$, the leptonic
electric dipole moments (EDMs) and the SUSY contributions to $(g-2)_{\mu}$~\cite{Hisano:2009ae}.
Finally, we discuss the prospects for the observation of LFV signals in $\tau$ decays~\cite{Hisano:2009ae}. The
corresponding numerical analysis in a concrete model will be performed in sec.~\ref{sec:numerics}.

\subsubsection{\boldmath $(g-2)_\mu$ vs. ${\rm BR}(\ell_i\to\ell_j\gamma)$}
\label{sec:gm2_LFV}

An observation of $\mu\to e\gamma$ would provide an unambiguous evidence of NP but, unfortunately,
not a direct test of the LFV source, as ${\rm BR}(\mu\to e\gamma)$ depends also on other SUSY
parameters like the particle masses and $\tan\beta$. While the latter parameters should be
ultimately determined at the LHC/linear collider experiments, it would be desirable to access
them by exploiting the NP sensitivity of additional low energy observables. In particular, since
both $(g-2)_{\mu}$ and ${\rm BR}(\ell_i\to \ell_j\gamma)$ are governed by dipole transitions,
the SUSY contributions to these observables are well correlated and their combined analysis
provides a powerful tool to get access to the related LFV source.

For a natural choice of the SUSY parameters, $t_{\beta}=10$ and a degenerate SUSY spectrum at
$\tilde{m}=300\,{\rm GeV}$, it turns out that $\Delta a^{\rm SUSY}_{\mu}\simeq 1.5 \times 10^{-9}$
and the current observed anomaly can be easily explained. Assuming a degenerate SUSY spectrum,
it is straightforward to find the correlation between $\Delta a^{\rm SUSY}_{\mu}$ and the 
branching ratios of $\ell_i\to\ell_j\gamma$
\begin{eqnarray}
{\rm BR}(\mu\to e\gamma)&\approx&
2\times 10^{-12}
\left[\frac{\Delta a^{\rm SUSY}_{\mu}}{ 3 \times 10^{-9}}\right]^{2}
\bigg|\frac{(\delta^{LL}_{\ell})_{21}}{10^{-4}}\bigg|^2\,,\nonumber\\
{\rm BR}(\tau\to\mu\gamma)&\approx&
8\times 10^{-8}
\left[\frac{\Delta a^{\rm SUSY}_{\mu}}{ 3 \times 10^{-9}}\right]^{2}
\bigg|\frac{(\delta^{LL}_{\ell})_{32}}{10^{-2}}\bigg|^2\,,
\label{lfvgm2}
\end{eqnarray}
where we have assumed that the MIs $(\delta^{LL}_{\ell})_{ij}$ provide the dominant contributions to
${\rm BR}(\ell_i\to \ell_j\gamma)$.

Eq.~(\ref{lfvgm2}) tell us that, as long as the $(g-2)_{\mu}$ anomaly finds an explanation in
SUSY theories, ${\rm BR}(\ell_i\to \ell_j\gamma)$ are predicted once we specify the LFV sources.

We emphasize that the extraordinary experimental sensitivities of the MEG experiment looking
for $\mu\to e\gamma$ offer a unique chance to obtain the first evidence for NP in low-energy
flavour processes. Should this be the case, several leptonic observables related to $\mu\to e\gamma$
are also likely to show NP signals, i.e. the $(g-2)_{\mu}$, the electron EDM $d_e$ but also other 
LFV processes like $\mu\to eee$ and $\mu-e$ conversion in nuclei.

In order to get a more concrete idea of where we stand, in tab.~\ref{lfvtable2}, we report
the bounds on the MIs $(\delta^{LL}_{\ell})_{ij}$ arising from the current experimental bounds
on ${\rm BR}(\ell_i\to\ell_j\gamma)$ imposing $\Delta a^{\rm SUSY}_{\mu}=3\times 10^{-9}$,
corresponding to the central value of the $(g-2)_{\mu}$ anomaly. Moreover, in the last column
of tab.~\ref{lfvtable2}, we also show the expectations for the MIs $(\delta^{LL}_{\ell})_{ij}$
within a {\it non-abelian} $SU(3)$ model that we will analyze in detail in following sections:
the RVV2 model~\cite{Ross:2004qn,Calibbi:2008qt,Calibbi:2009ja}.
\begin{table}[t]
\addtolength{\arraycolsep}{3pt}
\renewcommand{\arraystretch}{1.4}
\begin{center}
\begin{tabular}{|l|l|c|c|}
\hline
Observable & Exp. bound on  $(\delta^{LL}_{\ell})_{ij}$ & $(\delta^{LL}_{\ell})_{ij}$ in RVV \\
\hline
 ${\rm BR}(\mu\to e\gamma)$ & $|(\delta^{LL}_{\ell})_{21}|< 3\times 10^{-4}$ & $\sim (0.3-1)\times10^{-4}$ \\
\hline
 ${\rm BR}(\tau\to e\gamma)$ & $|(\delta^{LL}_{\ell})_{31}|< 6\times 10^{-2}$ & $\sim (2-6)\times10^{-3}$\\
\hline
 ${\rm BR}(\tau\to\mu\gamma)$ & $|(\delta^{LL}_{\ell})_{32}|< 4\times 10^{-2}$ & $\sim (0.3-1)\times10^{-1}$\\
\hline
\end{tabular}
\end{center}
\caption{
Bounds on the effective LFV couplings $(\delta^{LL}_{\ell})_{ij}$ from the current experimental bounds
on the radiative LFV decays of $\tau$ and $\mu$ leptons (see tab.~\ref{tab:lfvtable}) by setting
$\Delta a^{\rm SUSY}_{\mu}=3 \times 10^{-9}$. The expectations for the $(\delta^{LL}_{\ell})_{ij}$'s
within the RVV2 model~\cite{Ross:2004qn,Calibbi:2008qt,Calibbi:2009ja} are reported in the last column.
The bound on $(\delta^{LL}_{\ell})_{21}$ scales as $[{\rm BR}(\mu\to e\gamma)_{exp}/1.2\times 10^{-11}]^{1/2}$. The scaling properties for the other flavour transitions are obtained analogously.}
\label{lfvtable2}
\end{table}
Interestingly enough, the expected experimental reaches of MEG (for $\mu\to e\gamma$)
and of a SuperB factory (for $\tau\to \mu\gamma$) would most likely probe the RVV model,
provided we assume the explanation of the $(g-2)_{\mu}$ anomaly in terms of SUSY effects.

\subsubsection{\boldmath Leptonic EDMs vs. ${\rm BR}(\ell_i\to\ell_j\gamma)$}

Within a SUSY framework, CP-violating sources are naturally induced by the soft SUSY breaking
terms through {\it i)} flavour blind $F$-terms~\cite{Pospelov:2005pr,Ellis:2008zy} and {\it ii)}
flavour dependent terms ~\cite{Hisano:2007cz}. It seems quite likely that the two categories {\it i)}
and {\it ii)} of CP violation are controlled by different physical mechanisms, thus, they may 
be distinguished and discussed independently.

In the case {\it i)}, the corresponding CP-violating phases generally lead to large electron and
neutron EDMs as they arise already at the one-loop level. For example, when $t_{\beta}=10$ and
$m_{\tilde{\ell}}=300\,{\rm GeV}$ it turns out that
\begin{equation}
d_e\sim 6\times 10^{-25}(\sin\theta_\mu + 10^{-2}\sin\theta_A)\,e\,{\rm cm}\,,
\end{equation}
while in the case {\it ii)} the leptonic EDMs, induced by {\it flavour dependent} phases (flavoured EDMs),
read
\beq
d_{e}\sim 10^{-22}\times{\rm Im}
((\delta^{RR}_{e})_{13}(\delta^{LL}_{\ell})_{31})\,e\,{\rm cm}\,.
\eeq
One of the most peculiar features disentangling the EDMs as induced by {\it flavour blind} or
{\it flavour dependent} phases regards their ratios. In particular,
\bea
\frac{d_e}{d_{\mu}}\!\!&=&\!\!\frac{m_e}{m_{\mu}}\quad\quad\quad\quad\quad\quad\quad\quad\quad\quad\quad\quad
{\it flavour\, blind}\,{\rm phases},
\nonumber\\
\frac{d_e}{d_{\mu}}\!\!&=&\!\!\
\frac{\Sigma_{k=2,3}{\rm Im}((\delta^{RR}_{e})_{1k}(\delta^{LL}_{\ell})_{k1})}
{{\rm Im}((\delta^{RR}_{e})_{23}(\delta^{LL}_{\ell})_{32})}\quad\quad {\it flavour\, dependent}\, 
{\rm phases}\,.
\eea
In the case of {\it flavour blind} phases, the current bound $d_{e}<1.7\times 10^{-27}\,e\,$cm \cite{Amsler:2008zzb} implies that $d_{\mu}\lesssim 3.5 \times 10^{-25}\,e\,$cm. On the contrary,
in the presence of {\it flavour dependent} phases, the leptonic EDMs typically violate the naive scaling
and values for $d_{\mu}> 2 \times 10^{-25}\,e\,$cm are still allowed.

Moreover, when the EDMs are generated by {\it flavour blind} phases, they are typically unrelated to
flavour violating transitions~\footnote{A relevant exception is represented by the FBMSSM where it has
been shown that there exist correlations among CP and flavour violating transitions in the $B$-meson
systems and the EDMs~\cite{Altmannshofer:2008hc}.}.
On the contrary, the {\it flavoured} EDMs are closely related to LFV processes as they both arise from
LFV effects and their correlated study would help to reconstruct the flavour structure responsible for
LFV transitions.

We recall that the predictions for the leptonic EDMs within a pure SUSY see-saw model are highly
suppressed~\cite{Ellis:2001yza,Masina:2003wt,Farzan:2004qu}, at a level well below any future
(realistic) experimental resolution.

After imposing the current experimental bound on ${\rm BR}(\mu\to e\gamma)$, it turns out that
$d_e \!\lsim \! 10^{-34}~e$cm, irrespective of the details of the heavy/light neutrino sectors.
On the contrary, when the see-saw mechanism is embedded in a SUSY GUT scheme, as $SU(5)_\text{RN}$,
$d_e$ may naturally saturate its current experimental upper bound. As we will show in sec.\ref{sec:numerics}, also the RVV model naturally predicts large (observable) values for $d_e$.

Hence, any experimental evidence for the leptonic EDMs at the upcoming experiments would point towards an underlying theory with either new sources of {\it flavour blind} phases or new CP and flavour violating
structures beyond those predicted by a MFV scenario. In fact, as shown by~(\ref{Eq:lEDM_LO}), a crucial 
ingredient to generate non-vanishing EDMs is the presence of right-handed mixing angles and hence of
right-handed currents. The latter are unavoidably generated in SUSY GUT scenarios as $SU(5)_\text{RN}$
or $SU(10)$ through the CKM matrix and also in a broad class of abelian and non-abelian flavour models.

A simultaneous evidence for the electronic EDM $d_e$ and of $\mu\to e\gamma$ at the MEG experiment,
could most likely suggest the presence of {\it flavoured} CP violating phases.

Noteworthily enough, the synergy of low-energy experiments, as the leptonic EDMs and LFV processes
(like $\mu\to e\gamma$), that are in principle unrelated, provides an important tool to unveil the
anatomy of the soft SUSY sector.

\section{Soft SUSY Breaking and FCNC Phenomena} \label{sec:SUSY_breaking}
\setcounter{equation}{0}

\subsection{Preliminaries}

The still elusive explanation of the pattern of SM fermion masses and mixing
angles constitutes one of the main issues of the SM: the so-called {\it ``SM flavour problem''}.
In addition, if nature is supersymmetric, the {\it ``flavour problem''} acquires a new aspect:
whenever fermions  and corresponding sfermions have mass matrices which are not diagonalized by
the same rotation, new flavour mixings occur at the gaugino-fermion-sfermion vertices, generally
leading to unacceptably large contributions to FCNC and/or CP-violating observables, of which
$K^0$ mixing, $\epsilon_K$ and $\mu\to e\gamma$ are the most problematic.

The most popular protection mechanisms to suppress such unwanted contributions are
\begin{itemize}
\item {\em Decoupling}. The sfermion mass scale is taken to be very high. Still, such a scenario may be
probed at the LHC through non-decoupling effects such as the super-oblique parameters~\cite{Cheng:1997sq}.
\item {\em Degeneracy}. The sfermion masses are degenerate to a large extent, leading to a
strong GIM suppression. Such degeneracy could naturally arise from models with gauge-mediated
supersymmetry breaking (GMSB) -- or with some other flavour-blind mechanism of SUSY breaking mediation --
if the mediation scale is low.\footnote{In GMSB models with a high messenger scale, gravity-mediated contributions cannot be neglected. The phenomenology of such gauge-gravity ``hybrid'' models is discussed in~\cite{Feng:2007ke,Hiller:2008sv}.}
\item {\em Alignment}. The quark and squark mass matrices are aligned, so that the flavour-changing gaugino-sfermion-fermion couplings are suppressed~\cite{Nir:1993mx}.
\item {\em MFV}. Flavour violation is assumed to be entirely described by the CKM
matrix even in theories beyond the SM~\cite{D'Ambrosio:2002ex}.
\end{itemize}

We recall here that the MFV symmetry principle in itself does not forbid the presence of
{\it flavour blind} CP violating sources~\cite{Baek:1998yn,Baek:1999qy,Bartl:2001wc,Ellis:2007kb,Colangelo:2008qp,Altmannshofer:2008hc,Mercolli:2009ns,Kagan:2009bn,Feldmann:2009dc},
hence a MFV MSSM suffers, in general, from the same SUSY CP problem as the ordinary MSSM.
Either an extra assumption or a mechanism accounting for a natural suppression of these CPV
phases is desirable.
The authors of~\cite{D'Ambrosio:2002ex} proposed the extreme situation where the SM Yukawa couplings
are the only source of CPV. In contrast, in~\cite{Paradisi:2009ey}, such a strong assumption has been
relaxed and the following generalized MFV ansatz has been proposed: the SUSY breaking mechanism is
{\it flavour blind} and CP conserving and the breaking of CP only arises through the MFV compatible
terms breaking the {\it flavour blindness}. That is, CP is preserved by the sector responsible for
SUSY breaking, while it is broken in the flavour sector.
While the generalized MFV ansatz still accounts for a natural solution of the SUSY CP problem, it also
leads to peculiar and testable predictions in low energy CP violating processes~\cite{Paradisi:2009ey}.

As discussed in the previous section, in the general SUSY framework it is useful to parameterize non-MFV interactions  in terms of the MIs $(\delta_{d,u}^{AB})_{ij}$ with $(A,B)=(L,R)$ and $(i,j=1,2,3)$ on
which the present data on FCNC processes put quite severe constraints~\cite{Gabbiani:1996hi,Ciuchini:2007ha,Paradisi:2005fk}.
Large departures from SM expectations are obviously still allowed in such a model-independent approach.

While a model-independent analysis gives a global picture of still allowed deviations from the SM, its weakness lies in the fact that the suppression of FCNC processes is not achieved by some symmetry principle but basically by fine tuning the MIs so that existing experimental bounds are kept under control. Moreover, such an approach does not address the question of the hierarchies of the quark mixings
in the CKM matrix nor the hierarchies of the quark masses. Analogous comments apply to the lepton sector.

Much more ambitious in this respect are supersymmetric models containing flavour symmetries that relate the structure of fermion and sfermion mass matrices. Such symmetries, while being at the origin of the pattern of fermion masses and mixings, can at the same time provide the sufficient suppression of FCNC and CP-violating phenomena by means of the {\it degeneracy} or {\it alignment} protection mechanisms discussed above. Moreover, as we will see in the context of our numerical analysis, SUSY flavour models imply certain
characteristic patterns of flavour and CP violation that can be confirmed or falsified with the upcoming experimental sensitivities.

Supersymmetric models with flavour symmetries have been considered extensively in the literature.
They can be naturally divided into two broad classes depending on whether they are based on abelian or
{\it non-abelian} flavour symmetries. They can be considered as generalizations of the Froggatt-Nielsen
mechanism~\cite{Froggatt:1978nt}: the flavour symmetry is spontaneously broken by the vacuum expectation
value of one or more ``flavon'' fields $\Phi_i$ and the hierarchical patterns in the fermion mass
matrices can then be generated by means of the suppression factors $\left(\langle\Phi_i\rangle/M\right)^n$, where $M$ is the scale of the breaking of the flavour symmetry and the power $n$ depends on the group
charges of the fermions involved in the Yukawa couplings, generating the mass terms.

\subsection{Abelian Flavour Models}

There is a rich literature on models based on abelian flavour symmetries
~\cite{Nir:1993mx,Ellis:2000js,Joshipura:2000sn,Maekawa:2001uk,Kakizaki:2002hs,Babu:2002tx,Jack:2003pb,
Dreiner:2003yr,Berezhiani:1996nu,Choi:1996se,Froggatt:1998xc,Shafi:1999rm,Leontaris:1999wf,Agashe:2003rj,Dudas:1995eq}.
While the simplest case, where a single $U(1)_F$ group is employed, is disfavoured since
it leads to unacceptably large contributions to FCNC processes as $\epsilon_K$ and $\Delta
M_K$~\cite{Nir:1993mx,Dudas:1995eq}, more successful models are realized through the abelian
flavour group $U(1)_{F1}\times U(1)_{F2}$~\cite{Nir:1993mx}.
In this last case, the tight FCNC constraints are naturally accounted for thanks to a precise
alignment of the down-quark and down-squark mass matrices~\cite{Nir:1993mx}.

On the other hand, the most prominent signature of this class of models are typically
large effects in $D^0-\bar D^0$ mixing.
In fact, abelian flavour symmetries do not impose any restriction on the mass splittings
between squarks of different generations therefore they are expected to be non-degenerate
with natural order one mass splittings.

In particular, a mass splitting between the first two generations of left-handed squarks
unavoidably implies a $(1 \leftrightarrow 2)$ flavour transition in the up-squark sector
of order $(\delta^{LL}_{u})_{21}\sim\lambda$.

This can be easily understood by recalling that the $SU(2)_L$ gauge symmetry relates the
left-left blocks of up and down squark matrices, i.e. $(M^{2}_{u})^{LL}$ and $(M^{2}_{d})^{LL}$
respectively, in such a way that $(M^{2}_{u})^{LL}= V^* (M^{2}_{d})^{LL} V^{\rm T}$.
In turn, the expansion of this relation at the first order in $\lambda$ implies that
\begin{equation}
(M_u^{2})^{LL}_{21} =
\left[V^*(M_d^{2})^{LL}V^{\rm T}\right]_{21}
\simeq
(M_d^{2})^{LL}_{21}+\lambda
\left(\tilde m_2^2 - \tilde m_1^2 \right)~.
\end{equation}
Thus, even for $(M_d^{2})^{LL}_{21} = 0$, which is approximately satisfied in alignment models,
there are irreducible flavour violating terms in the up squark sector driven by the CKM as long
as the left-handed squarks are splitted in mass. This is opposite to the case of non abelian
flavour symmetries which we discuss now.

\subsection{Non-abelian Flavour Models}

In contrast to abelian models, where there is a lot of freedom in fixing the charges of the SM 
fermions under the flavour symmetry, non-abelian models~\cite{Berezhiani:2000cg,Roberts:2001zy,Chen:2003zv,Pomarol:1995xc,Hall:1995es,Carone:1996nd,Barbieri:1996ww,Dermisek:1999vy,Blazek:1999hz,Barbieri:1995uv,Barbieri:1997tu,King:2001uz,King:2003rf,Ross:2004qn,Antusch:2007re}
are quite predictive for fermion mass matrices once the pattern of symmetry breaking is specified.

Moreover, non-abelian flavour models predict very small NP contributions for $(1\leftrightarrow 2)$
flavour transitions, since, if the non-abelian symmetry was an exact symmetry, then at least the first
two generations of squarks are degenerate, which would lead to a vanishing NP contribution to $K^0-\bar K^0$ and $D^0-\bar D^0$ mixings. However, as we will find in sec.~\ref{sec:nonabeliannumerics}, the violation of the symmetry is strong enough to produce large effects in the observable $\epsilon_K$.

There are many candidates for the flavour symmetry group $G_F$, each having distinct symmetry breaking
patters. In general, $G_F$ must be contained in the full global $U(3)^5$ symmetry group of the SM in
the limit of vanishing Yukawa couplings. In particular, a lot of attention is received by models with
a $U(2)$ symmetry~\cite{Barbieri:1995uv,Barbieri:1997tu} motivated by the large top mass, and also models
with $SU(3)$ symmetry~\cite{King:2001uz,King:2003rf} which are additionally able to naturally predict
an almost maximal atmospheric neutrino mixing angle $\theta_{23}\approx 45^\circ$ and to suggest a near
maximal solar mixing angle $\theta_{12}\approx 30^\circ$.

\subsection{Running Effects in Flavour Models}\label{sec:running}

Since the flavour models discussed in the previous subsections predict the pattern of off-diagonal elements
of the squark mass matrices at the GUT scale while for the calculation of physical observables, their values
at low energies are relevant, we now discuss the renormalization group (RG) evolution of these elements.
In particular, crucial questions are whether these off-diagonal elements are strongly reduced or enhanced 
in the running, whether they mix with each other and whether they are generated if they vanish at the initial scale. In short, the question is whether the textures of the squark mass matrices predicted by the flavour models are RG stable. We will disregard off-diagonalities in the trilinear couplings, i.e. LR and RL MIs,
as well as the slepton sector in the following discussion.

Studies of these running effects in the context of the MSSM with Minimal Flavour Violation have been
performed in \cite{Paradisi:2008qh,Colangelo:2008qp}.

A close inspection of the relevant RG equations (RGEs)~\cite{Martin:1993zk} shows that for ${\bf m}_U^2$ and ${\bf m}_D^2$, i.e. the RR MIs~\footnote{The trilinear coupling matrices $h_{u,d}$ and the soft masses ${\bf m}_{Q,U,D}^2$ used in this section correspond to the conventions of Martin and Vaughn \cite{Martin:1993zk} and are related to the trilinear couplings and to the soft masses in the convention of~\cite{Rosiek:1995kg} through the relations $h_{u,d}=-A_{u,d}^T$ and ${\bf m}_{Q}^2=m_Q^2$, ${\bf m}_{U,D}^2=(m_{U,D}^2)^T$. See also appendix~\ref{sec:appendixRosiek} for additional details.},
all mixing terms are suppressed by 1st or 2nd generation Yukawa couplings and
can therefore be safely neglected. In fact, neglecting 1st and 2nd generation Yukawa couplings, the RGEs
for the off-diagonal elements of ${\bf m}_{U,D}^2$ read at the one-loop level
\begin{equation}
 16\pi^2 \frac{d}{dt} ({\bf m}_U^2)_{ij} \stackrel{i \neq j}{=}
 2 \left(y_{t}^2\right) ({\bf m}_U^2)_{ij} (\delta_{i3}+\delta_{j3})
 + 4 (h_u h_u^\dagger)_{ij} \,,
\end{equation}
\begin{equation}
 16\pi^2 \frac{d}{dt} ({\bf m}_D^2)_{ij} \stackrel{i \neq j}{=}
 2 \left(y_{b}^2\right) ({\bf m}_D^2)_{ij} (\delta_{i3}+\delta_{j3})
+ 4 (h_d h_d^\dagger)_{ij} \,,
\end{equation}
where $t=\log(\mu/\mu_0)$. As can be easily seen, $({\bf m}_{U,D}^2)_{12}$ are RG invariant in this approximation;
we have checked that this holds numerically to an excellent approximation even if light generation Yukawas 
and two-loop effects are taken into account.

Concerning the remaining entries, we find that their values at low energies are well approximated by
\begin{equation}
\left( {\bf m}_U^2 \right)_{13} \simeq 0.87 \left( {\bf m}_U^2 \right)_{13}^0 \,,
\qquad
\left( {\bf m}_U^2 \right)_{23} \simeq 0.82 \left( {\bf m}_U^2 \right)_{23}^0 \,,
\label{eq:mU2ij}
\end{equation}
\begin{equation}
\left( {\bf m}_D^2 \right)_{13} \simeq (1-0.10 \,\tilde{t}^2-0.05 \,\tilde{t}^4) \left( {\bf m}_D^2 \right)_{13}^0 \,,
\label{eq:mD213}
\end{equation}
\begin{equation}
\left( {\bf m}_D^2 \right)_{23} \simeq (1-0.10 \,\tilde{t}^2-0.05 \,\tilde{t}^4) \left( {\bf m}_D^2 \right)_{23}^0 \,,
\label{eq:mD223}
\end{equation}
where we have defined $\tilde{t}=\tan\beta/50$, and where quantities with superscript 0 on the right-hand
side denote the values at the GUT scale predicted by the particular flavour model considered, while those
on the left-hand side are meant to be evaluated at the weak scale.

To summarize, the off-diagonal squark mass matrix elements in the RR sector are reduced by at most 15\%,
they do not mix among each other or with LL MIs, and they are not generated by the running once they 
vanish at the GUT scale.

The situation in the LL sector is different; there, also mixing takes place, and the elements can be 
generated by RG effects even if they vanish at the GUT scale. Of course, both these effects are suppressed
by combinations of CKM elements, since they would be absent if the CKM matrix were diagonal. The RG equation
for the off-diagonal elements of ${\bf m}_{Q}^2$ reads
\begin{equation}
\begin{split}
 16\pi^2 \frac{d}{dt} ({\bf m}_Q^2)_{ij}  \stackrel{i \neq j}{=} &\
 2 \left( y_{d,i} \, y_{d,j} \right) ({\bf m}_D^2)_{ij}
 + \left(y_{b}^2\right) ({\bf m}_Q^2)_{ij} (\delta_{i3}+\delta_{j3})
 + y_t^2( {\bf m}_Q^2)_{ik}  \lambda_{kj}
  + y_t^2 ({\bf m}_Q^2)_{kj}  \lambda_{ik}+ \\
& + y_t^2 2 m_{H_u}^2   \lambda_{ij} + 2 y_t^2 ({\bf m}_U^2)_{33} \lambda_{ij} + 2 (h_u^\dagger h_u)_{ij}
  + 2 (h_d^\dagger h_d)_{ij} \,,
\end{split}
\label{eq:RGEmQ2}
\end{equation}
where $\lambda_{ij} = V_{ti}^* V_{tj}$ and we have again neglected light generation Yukawas, except in the first term, which in the case of $(ij)=(23)$ is only suppressed by $y_s/y_b$, but unsuppressed by CKM angles and can therefore be comparable in size to the remaining terms.

Consequently we find that, numerically, the low-energy values of the  $({\bf m}_{Q}^2)_{ij}$ are well approximated
by the following formulae,
\begin{align}
\left( {\bf m}_Q^2 \right)_{13} &\simeq 
 (0.91 - 0.05 \,\tilde{t}^2)\left( {\bf m}_Q^2 \right)_{13}^0
- \Delta m^2_{13}
- 0.09 \left[ \lambda_{12} \left( {\bf m}_Q^2 \right)_{23}^0 + \lambda_{23} \left( {\bf m}_Q^2 \right)_{12}^0 \right] \,,
\label{eq:mQ213}
\\
\left( {\bf m}_Q^2 \right)_{23} &\simeq 
 (0.91 - 0.05 \,\tilde{t}^2)\left( {\bf m}_Q^2 \right)_{23}^0
- \Delta m^2_{23}
- 0.09 \left[ \lambda_{21} \left( {\bf m}_Q^2 \right)_{13}^0 + \lambda_{13} \left( {\bf m}_Q^2 \right)_{21}^0 
+  0.02 \tilde{t}^2  \left( {\bf m}_D^2 \right)_{23}^0
\right] \,,
\label{eq:mQ223}
\\
\left( {\bf m}_Q^2 \right)_{12} &\simeq 
\left( {\bf m}_Q^2 \right)_{12}^0
- \Delta m^2_{12}
- 0.09 \left[ \lambda_{13} \left( {\bf m}_Q^2 \right)_{32}^0+ \lambda_{32} \left( {\bf m}_Q^2 \right)_{13}^0 \right] \,,
\label{eq:mQ212}
\end{align}
where
\begin{equation}
\Delta m^2_{ij} =  \lambda_{ij} \left(0.33 \,m_0^2 + 0.89 \,M_{1/2}^2 + 0.03 \,A_0^2 - 0.14 \,M_{1/2} A_0 \right) \,,
\end{equation}
assuming CMSSM-like boundary conditions for the gaugino masses, trilinear couplings and the diagonal elements of sfermion mass matrices.

As in the RR sector, the Yukawa-induced reduction of the elements, cf. the first terms in~(\ref{eq:RGEmQ2}) 
and (\ref{eq:mQ213})--(\ref{eq:mQ212}), is only sizable in the $(13,23)$ sectors. The terms in square brackets 
describe the mixing among the LL elements, while $\Delta m^2_{ij}$ describes the CKM-induced generation of
LL MIs, which takes place even in a completely flavour blind situation at the GUT scale, such as in the CMSSM.

To summarize, the off-diagonal squark mass matrix elements in the LL sector mix among each other and they
can be generated even if vanishing at the GUT scale; however, these effects are suppressed by combinations
of CKM elements. Mixing of RR elements into LL elements only takes place in the 23-sector and is suppressed
by a factor $y_s y_b /y_t^2$.

Finally, let us also remind that the attained values for the MIs $\delta_{ij}$ are renormalization scale dependent as the diagonal elements are strongly affected by RGE effects.

\section{Strategy} \label{sec:strategy}

One of our main goals in the subsequent sections is to investigate the patterns of flavour violation
in flavour models, putting particular emphasis on $b\to s$ transitions. To this end, we select specific
flavour models showing representative flavour structures in the soft masses. As we will see in
sec.~\ref{sec:numerics}, in order to generate large NP effects in $S_{\psi\phi}$, sizable -- at least
CKM-like -- right-handed currents (driven by non-vanishing RR MIs) are unavoidable. The main reasons
for this are twofold: 1) right-handed currents are less constrained than left-handed currents by
low energy $b\to s$ transitions, especially $b\to s\gamma$ and
2) since LL MIs (and hence left-handed currents) even if not present at the GUT scale, are RGE generated
at the low scale via the CKM and the top Yukawa coupling (see sec.~\ref{sec:running}), non-vanishing
right-handed currents guarantee the presence of the large NP contributions provided by the left-right
$\Delta F = 2$ operators ($Q_4$ and $Q_5$ in~(\ref{eq:DF2_ops})) that are strongly enhanced by QCD
RGE effects and by a large loop function.

Therefore, we consider scenarios with
\begin{itemize}
\item [i)] large $\mathcal{O}(1)$ RR mass insertions,
\item [ii)] comparable LL and RR mass insertions that are CKM-like,
\item [iii)] only CKM-like LL mass insertions.
\end{itemize}
In sec. \ref{sec:numerics} we present their predictions for both CP violating and CP conserving effects 
occurring in $b\to s$ transitions requiring at the same time that the models we consider satisfy the FCNC
and CP violation bounds set by the experimental values of $\epsilon_{K}$, $\Delta{M}_{K}$, $\Delta{M}_{D}$, 
$d_{e}$, $d_{n}$, etc. It will also be useful to compare the results of such an analysis with the results
of the model-independent analysis discussed in sec. \ref{sec:model_independent}.

To the best of our knowledge, the current analysis represents the most complete analysis present
in the literature in this subject as for the number of processes considered and the inclusion of
all relevant SUSY contributions. In particular,
\begin{itemize}
\item We perform a full diagonalization of the sfermion mass matrices so that we do not make use of the MI 
approximation method. In fact, the latter method cannot be trusted when the flavour violating mixing angles
are $\mathcal O(1)$, as is predicted by many flavour models.
\item We systematically include the full set of one loop contributions to the FCNC processes, namely not
only the gluino contributions, but also the charged Higgs, the chargino and the neutralino contributions.
\item We systematically include the beyond-leading-order threshold corrections arising from
$\tan\beta$-enhanced non-holomorphic corrections in the presence of new sources of flavour
and CP violation hence, accounting also for FCNC effects driven by the neutral Higgs sector.
\item All the above contributions have been systematically included for a very complete set
of $\Delta F=2$, $\Delta F=1$ and $\Delta F=0$ processes.
\end{itemize}

We believe that the inclusion of all these contributions and observables is crucial in order to try to
understand the pattern of deviations from the SM prediction that may  be found in future measurements.

In order to increase the transparency of our presentation, we will now outline the strategy for our 
numerics that will proceed in five steps.

\subsubsection*{Step 1: Bounds on Mass Insertions}\label{subsec:bounds}

In order to get an idea of the size of departures from the SM expectations that are still allowed in the 
supersymmetric framework, we will perform a model-independent analysis by calculating the allowed ranges
for the mass insertions $(\delta_{d,u}^{AB})_{ij}$.
In this approach, as usual, only one MI of a given ``chirality'' ${\rm AB}$ and relative to a given
family transition $ij$ will be switched on at a time. Consequently, the bounds so obtained are valid 
barring accidental cancellations among different contributions.

\subsubsection*{Step 2: An Abelian Flavour Model}
%
In this step, we analyse an abelian flavour model by Agashe and Carone (AC)  with a $U(1)$
flavour symmetry \cite{Agashe:2003rj} which predicts large right-handed currents with order one CPV
phases in the $b\to s$ sector.

As discussed previously, flavour models with a single $U(1)$ are typically disfavoured by the $\epsilon_K$
and $\Delta M_K$ constraints~\cite{Nir:1993mx,Dudas:1995eq}. However, this is no longer true in the case
of the AC model, where a high degree of quark-squark alignment is realized by means of a particular localization of fermions in extra dimensions, suppressing unwanted FCNC effects.

In particular, the pattern of the relevant flavour off-diagonal MIs at the GUT scale, as function
of the Cabibbo angle $\lambda$, is given by
\begin{equation}
\delta_d^{LL}
\simeq
\begin{pmatrix}
\star & 0 & 0 \\
0 & \star & \lambda^2 \\
0 & \lambda^2 & \star \\
\end{pmatrix} ~,~~
\delta_d^{RR}
\simeq
\begin{pmatrix}
\star & 0 & 0 \\
0 & \star & e^{i\phi_R} \\
0 & e^{-i\phi_R} & \star \\
\end{pmatrix} ~,
\end{equation}
\begin{equation}
(\delta_u^{LL})_{12}\simeq \lambda,\hspace{0.2cm}
(\delta_u^{RR})_{12}\simeq \lambda^3 ~,
\end{equation}
where we have suppressed the $\mathcal O(1)$ coefficients which multiply the individual elements
of the matrices and $\phi_R$ is a free parameter.

In our analysis, the correlations between various observables will play the crucial role. We will show that
in the AC model the most interesting correlations involve the CP asymmetry $S_{\psi\phi}$ that can be much
larger than in the SM so that this model can accommodate the recent data from CDF and D0. Indeed, in this
context it is important to ask whether the desire to explain the latter data automatically implies other
departures from SM expectations. We will therefore calculate $\text{BR}(B_s\to\mu^+\mu^-)$, $S_{\phi K_S}$,
$\Delta a_\mu$ and $A_\text{SL}^s$ as functions of $S_{\psi\phi}$.

The first correlation in our list is of particular interest as the upper bound on $\text{BR}(B_s\to\mu^+\mu^-)$ should soon be improved by CDF, D0 and LHCb, possibly even finding
first events for this decay.
Next, the dependence of $S_{\phi K_S}$ on $S_{\psi\phi}$ will tell us whether large values of
the latter asymmetry are compatible with $S_{\phi K_S}$ significantly lower than $S_{\psi K_S}$
as signalled by BaBar~\cite{Barberio:2008fa}. Similar comments apply to the $(g-2)_\mu$ anomaly
$\Delta a_\mu$. In addition, we investigate the ratios ${\rm BR}(B_s\to\mu^+\mu^-)/\Delta M_s$
and ${\rm BR}(B_s\to\mu^+\mu^-)/{\rm BR}(B_d\to\mu^+\mu^-)$ emphasizing the powerful tool they
offer to unveil non-MFV structures of the model.
We will also analyze the CP asymmetries in $B\to K^*\mu^+\mu^-$ and the impact of the $D^0-\bar D^0$
mixing constraint on the correlations listed above. Finally, we will address the question how this
model faces the current UT tension discussed in sec.~\ref{sec:UT_tensions}.

\subsubsection*{\boldmath Step 3: Non-Abelian $SU(3)$ Models}
%
We next analyse a non-abelian model by Ross et al. (RVV) based on an $SU(3)$ flavour
symmetry~\cite{Ross:2004qn}. First, we recall that the observed structure in the Yukawa
couplings does not fix uniquely the K\"ahler potential hence, the soft sector is not
unambiguously determined. In the following, we analyse a particular case of the RVV
model to which we refer to as RVV2 model~\cite{Calibbi:2009ja}.
Similarly to the abelian AC model, also the RVV2 model predicts large right-handed currents.
More explicitly, at the GUT scale, again suppressing the $\mathcal{O} (1)$ coefficients,
the expressions for the flavour off-diagonal entries in the soft mass matrices in the SCKM
basis read~\cite{Calibbi:2009ja}~\footnote{In~(\ref{sckm2R}),~(\ref{sckm2L}), in order to
avoid accidental cancellations among different phases, we have set to zero an extra CPV phase, $\beta^{\prime}_2$~\cite{Calibbi:2009ja}, that is not constrained by the requirement of
reproducing a correct CKM matrix.}
\begin{eqnarray}
\delta^{RR}_{d}
& \simeq &
\left(\begin{array}{ccc}
\star & -\bar\varepsilon^3\,e^{i\omega_{us}} & -\bar\varepsilon^2\,y_b^{0.5} e^{i(\omega_{us}-\chi+\beta_3)}
\\
-\bar\varepsilon^3\,e^{-i\omega_{us}} & \star & \bar\varepsilon\,y_b^{0.5} e^{-i(\chi-\beta_3)}
\\
-\bar\varepsilon^2\,y_b^{0.5} e^{-i(\omega_{us}-\chi+\beta_3)} & \bar\varepsilon\,y_b^{0.5} e^{i(\chi-\beta_3)} & \star
\end{array}\right)\,,
\label{sckm2R}
\\
\delta^{LL}_{d}
& \simeq &
\left(\begin{array}{ccc}
\star & -\varepsilon^2\bar\varepsilon\,e^{i\omega_{us}} & \varepsilon\bar\varepsilon\,y_t^{0.5} e^{i(\omega_{us}-2\chi+\beta_3)}
\\
-\varepsilon^2\bar\varepsilon\,e^{-i\omega_{us}} & \star & \varepsilon\,y_t^{0.5} e^{-i(2\chi-\beta_3)}
\\
\varepsilon\bar\varepsilon\,y_t^{0.5} e^{-i(\omega_{us}-2\chi+\beta_3)} & \varepsilon\,y_t^{0.5} e^{i(2\chi-\beta_3)} & \star
\end{array}\right)\,,
\label{sckm2L}
\end{eqnarray}
where the parameters $\varepsilon \simeq 0.05$ and $\bar\varepsilon\simeq 0.15$ are defined, after
the flavour symmetry breaking, in order to reproduce the observed values for fermion masses and
mixing angles. Moreover, the phases $\omega_{us}$, $\chi$ and $\beta_3$ are set, to a large extent,
by the requirement of reproducing the CKM phase; in particular, it turns out that $\omega_{us}\approx-\lambda$~\cite{Calibbi:2009ja} and $(\chi,\beta_3)\approx (20^{\circ},-20^{\circ})$
(or any other values obtained by adding $180^{\circ}$ to each)~\cite{Calibbi:2009ja}. Additionally,
it is found that~\cite{Calibbi:2009ja}
\begin{equation}
(\delta_u^{LL})_{12}\simeq \lambda^4,\hspace{0.2cm}
(\delta_u^{RR})_{12}\simeq \lambda^6\,.
\end{equation}

The trilinear couplings follow the same symmetries as the Yukawas. In the SCKM basis, after rephasing
the fields, the trilinears lead to the following flavour off-diagonal LR MIs~\cite{Calibbi:2009ja}
\begin{eqnarray}
\delta^{LR}_{d}
\simeq
\left(\begin{array}{ccc}
\star & \bar\varepsilon^{3}\,e^{-i\omega_{us}} & \bar\varepsilon^{3}\,e^{-i\omega_{us}} \\
\bar\varepsilon^{3}\,e^{-i\omega_{us}} & \star & \bar\varepsilon^{2} \\
\bar\varepsilon^{3}\,e^{i(\omega_{us}+2\beta_3-2\chi)} & \bar\varepsilon^{2}\,e^{2i(\beta_3-\chi)} & \star
\end{array}\right)
\frac{A_0}{m^2_0}\,m_b\,.
\end{eqnarray}

Similarly to the flavour model considered in step 2, $S_{\psi\phi}$ can be large so that the
analysis can be done along the lines of the one done for the AC model making the comparison of
both models very transparent. In particular, as is already evident from the structure of the
MIs, NP enters $D^0-\bar D^0$ mixing and $\epsilon_K$ in a profoundly different manner in these
two models.

As the RVV model is embedded in a $SO(10)$ SUSY GUT model, correlations between flavour violating
processes in the lepton and quark sectors naturally occur making additional tests of this model
possible.

In particular, in the following, we list the flavour off-diagonal soft breaking terms of the
leptonic sector arising in the RVV2 model~\cite{Calibbi:2009ja}
\begin{eqnarray}
\delta^{RR}_{e}
& \simeq &
\left(\begin{array}{ccc}
\star & -\frac{1}{3}\bar\varepsilon^3 & -\frac{1}{3}\bar\varepsilon^2\,y_{b}^{0.5}e^{i(-\chi+\beta_3)}
\\
-\frac{1}{3}\bar\varepsilon^3 & \star & \bar\varepsilon\,y_{b}^{0.5} e^{-i(\chi-\beta_3)}
\\
-\frac{1}{3}\bar\varepsilon^2\,y_{b}^{0.5} e^{i(\chi-\beta_3)} & \bar\varepsilon\,y_{\tau}^{0.5} e^{i(\chi-\beta_3)} & \star
\end{array}\right)\,,
\\
\delta^{LL}_{\ell}
& \simeq &
\left(\begin{array}{ccc}
\star & -\frac{1}{3}\varepsilon^2\bar\varepsilon & \frac{1}{3}\varepsilon\bar\varepsilon\,y_t^{0.5} e^{i(-2\chi+\beta_3)}
\\
-\frac{1}{3}\varepsilon^2\bar\varepsilon & \star & \varepsilon\,y_t^{0.5} e^{-i(2\chi-\beta_3)}
\\
\frac{1}{3}\varepsilon\bar\varepsilon\,y_t^{0.5} e^{i(2\chi-\beta_3)} & \varepsilon\,y_t^{0.5}e^{i(2\chi-\beta_3)} & \star
\end{array}\right)\,,
\end{eqnarray}
while the leptonic off-diagonal LR MIs have the following structure~\cite{Calibbi:2009ja}
\begin{eqnarray}
\delta^{LR}_{e}
\label{aterms2}
\simeq
\left(\begin{array}{ccc}
\star & \bar\varepsilon^3 & \bar\varepsilon^3 \\
\bar\varepsilon^3 & \star & 3\bar\varepsilon^2 \\
\bar\varepsilon^3\,e^{i(2\beta_3-2\chi)} & 3\bar\varepsilon^2\,e^{2i(\beta_3-\chi)} & \star
\end{array}\right)
\frac{A_0}{m^2_0}\,m_{\tau}\,.
\end{eqnarray}

Finally, we consider a second example of {\it non-abelian} $SU(3)$ flavour model analyzed by
Antusch et al.~\cite{Antusch:2007re} to which we refer to as AKM model.
In the AKM model, in contrast to the RVV2 model, there is the freedom to suppress arbitrarily
the flavour changing soft terms $(\delta_{d}^{LL})_{ij}$, hence we take the limit where
$(\delta_{d}^{LL})_{ij}=0$ at the high scale. Therefore, our results have to be regarded 
as irreducible predictions of the AKM model, barring accidental cancellations among different
contributions to physical observables. In the following, we report the flavour structure for
the soft sector of the AKM model, in the SCKM basis, suppressing the $\mathcal{O} (1)$ coefficients~\cite{Antusch:2007re}
\begin{eqnarray}
\label{sckm2_king}
\delta^{RR}_{d}
\simeq
\left(\begin{array}{ccc}
\star & \bar\varepsilon^3 & \bar\varepsilon^3
\\
\bar\varepsilon^3 & \star & \bar\varepsilon^2 e^{i\Psi_f}
\\
\bar\varepsilon^{3} & \bar\varepsilon^{2} e^{-i\Psi_{f}} & \star
\end{array}\right)\,.
\end{eqnarray}

The trilinear couplings lead to the following flavour off-diagonal LR MIs~\cite{Antusch:2007re}
\begin{eqnarray}
\label{aterms_king}
\delta^{LR}_{d}
\simeq
\left(\begin{array}{ccc}
\star & \bar\varepsilon^3 & \bar\varepsilon^3 \\
\bar\varepsilon^3 & \star & \bar\varepsilon^2 \\
\bar\varepsilon^3 & \bar\varepsilon^2 & \star
\end{array}\right)
\frac{A_0 v}{m^2_0\tan\beta}\,,
\end{eqnarray}
where the above expressions are the same both for the down squark and the slepton sectors, modulo
the unknown $\mathcal O(1)$ coefficients. Concerning the up squark sector, the relevant flavour
mixing angle generating $D^0-\bar D^0$ mixing is $(\delta_u^{RR})_{12}\simeq\lambda^5$, hence,
we conclude that the $D^0-\bar D^0$ constraints are safely under control also in the AKM model.

An interesting feature of the AKM model is the presence of a leading $\mathcal O(1)$ CPV phase in
the 23 RR sector but not in the 12 and 13 sectors. This will turn out to be crucial to generate CPV
effects in the $B_s$ mixing amplitude. Moreover, we notice that, while the RVV2 model predicts
that $(\delta_{d}^{RR})_{23}\simeq\bar\varepsilon$, the AKM model predicts that $(\delta_{d}^{RR})_{23}\simeq\bar\varepsilon^2$ with $\bar\varepsilon\simeq 0.15$. Even if we expect
that the CPV effects in the AKM model are smaller than in the RVV2 model, the indirect constraints
(mainly coming from $\epsilon_K$, $b\to s\gamma$ and $\Delta M_s/\Delta M_d$), that are different
in the two models, can play a crucial role to establish the allowed NP room for CPV effects in $B_s$
systems. Hence, in order to distinguish between the RVV2 and the AKM model, by means of their
footprints in low energy processes, a careful and comparative analysis is necessary.

\subsubsection*{Step 4: Flavour Model with Purely Left-handed Currents}
%
We will next turn to the predictions for various low-energy processes induced by $b\to s$ transitions 
within flavour models predicting pure, CKM-like, left-handed currents, i.e. $\delta^{LL}_d\not=0$
and $\delta^{RR}_d=0$ (or better $\delta^{RR}_d\ll \delta^{LL}_d$), with a CPV phase in the $b\to s$ 
sector\footnote{We will refer to this model as $\delta$LL model in the following.}.

This study will give us general predictions for a broad class of abelian~\cite{Nir:2002ah}
and non-abelian~\cite{Hall:1995es} flavour models not containing RH currents.
Concerning the low energy predictions for the abelian case, we can draw clear cut conclusions
immediately: the marriage of the tight constraints from $D^0-\overline{D}^0$ mixing (typical
of abelian flavour models) with the absence of RH currents (preventing large CP violating effects
in $\Delta F=2$ observables in the down sector) allows non-standard and testable effects to occur 
only in CPV observables related to $D^0-\overline{D}^0$ mixing.

A non-abelian scenario has very different low energy predictions.
In sec. \ref{sec:numerics}, we will analyse a non-abelian model based on the following LL MIs
\begin{equation}
\delta_d^{LL}
\simeq
\begin{pmatrix}
\star & \lambda^5 & \lambda^3 \\
\lambda^5 & \star & \lambda^2 e^{i\phi_L}\\
\lambda^3 & \lambda^2 e^{-i\phi_L} & \star \\
\end{pmatrix}~,
\label{dLL_scenario}
\end{equation}
\noindent and all other MIs put to zero. Within that framework we will confirm our general statement
of sec. \ref{sec:DF2} that $S_{\psi\phi}$ does not receive large NP contributions in this context.
As the NP effects in $\Delta F=2$ processes in such models are relatively small and a large asymmetry $S_{\psi\phi}$ cannot be easily generated, this asymmetry is not an interesting variable in this scenario.

In contrast, large non-standard effects can be generated in $\Delta F=1$ observables like $S_{\phi K_S}$,
$A_\text{CP}(b\to s\gamma)$ and the CP asymmetries in $B\to K^*\mu^+\mu^-$. In particular, as we will see,
$S_{\phi K_S}$ can attain values low enough to explain the related anomaly.
We will therefore calculate this time $\Delta a_\mu$, $A_\text{CP}(b\to s\gamma)$, $S_{\eta^\prime K_S}$
and the CP asymmetries in $B\to K^*\mu^+\mu^-$ as functions of $S_{\phi K_S}$. The predictions of models
with purely left-handed currents will be systematically compared with those of somewhat similar models,
i.e. the FBMSSM and the MFV MSSM, considered in step~5.

\subsubsection*{Step 5: Comparison with the FBMSSM and the MFV MSSM}
%
Finally we comment on the predictions for CP violating effects arising within MFV MSSM models.
This issue was recently addressed in~\cite{Altmannshofer:2008hc} in the context of the so-called
{\it flavour blind} MSSM (FBMSSM)~\cite{Baek:1998yn,Baek:1999qy,Bartl:2001wc,Ellis:2007kb,Altmannshofer:2008hc},
where the CKM matrix is the only source of flavour violation and where additional CP violating phases
in the soft sector are allowed.

In the FBMSSM, universal soft masses for different squark generations are assumed. Such a strong assumption
gets somewhat relaxed in the framework of the general MFV ansatz~\cite{D'Ambrosio:2002ex} where the scalar
soft masses receive additional corrections. The most general expressions for the low-energy soft-breaking terms compatible with the MFV principle and relevant for our analysis read~\cite{D'Ambrosio:2002ex,Colangelo:2008qp}
\begin{equation} \label{eq:mQ_MFV}
m_Q^2 = \tilde m^2
\bigg[
\mathbf{1}+b_1 \mathbf{Y}_u^\dagger \mathbf{Y}_u + b_2 \mathbf{Y}_d^\dagger \mathbf{Y}_d + (b_3 \mathbf{Y}_d^\dagger \mathbf{Y}_d \mathbf{Y}_u^\dagger \mathbf{Y}_u + \mathrm{h.c.})
\bigg]~,
\end{equation}
where $\tilde m$ sets the mass scale of the left-left soft mass, while $b_i$ are unknown, order one,
numerical coefficients. The small departures from a complete flavour blindness of the soft terms
generate additional FCNC contributions by means of gluino and squark loops. 
The latter effects were neglected in the context of the FBMSSM~\cite{Altmannshofer:2008hc} since,
in principle, they can be very small if the parameters $b_i$ are small and/or if the gluino mass is
significantly larger than the chargino/up-squark masses. In this respect, the contributions to FCNC
processes discussed in~\cite{Altmannshofer:2008hc} can be regarded as irreducible effects arising in
MFV scenarios.

The natural question that arises is whether the findings of~\cite{Altmannshofer:2008hc}, remain valid 
in the general MFV framework. Actually, within a general MFV framework, there are even new CPV effects 
to both $\Delta F =2$ and $\Delta F =1$ transitions, hence, potential departures from the FBMSSM 
predictions could in principle be expected. In particular in (\ref{eq:mQ_MFV}) the SM Yukawa couplings
are not necessarily the only source of CPV. While $b_1$ and $b_2$ must be real, the parameter $b_3$ is 
generally allowed to be complex~\cite{Colangelo:2008qp}.

However, our analysis in sec. \ref{sec:step5_numerics} will confirm the general finding of~\cite{Altmannshofer:2008hc} that within SUSY MFV scenarios, large NP contributions can only
be expected in $\Delta F =1$ and $\Delta F =0$ transitions.

We will also see that the FBMSSM and the MFV MSSM bear similarities to the models with left-handed 
CKM-like currents discussed in step 4.

\section{Numerical Analysis} \label{sec:numerics}
\setcounter{equation}{0}

In this section, we present the numerical analysis of our study following the five steps described in
the previous section. To this end, we list a number of constraints we impose throughout the analysis:
\begin{itemize}
\item[i)]  the data on flavour physics observables (see tabs.~\ref{tab:DF2exp} and \ref{tab:observables}),
\item[ii)]  mass bounds from direct SUSY searches,
\item[iii)] requirement of a neutral lightest SUSY particle,
\item[iv)]  requirement of correct electroweak symmetry breaking and vacuum stability,
\item[v)]   constraints from electroweak precision observables (EWPO)\,.
\end{itemize}
We further assume a CMSSM spectrum where, in the case of the model-independent
analysis (Step 1) we take at the EW scale only one non-vanishing MI at a time.
Hence our bounds are valid barring accidental cancellations among amplitudes 
from different MIs. On the contrary, when we analyze  abelian and non-abelian 
flavour models, we impose the flavour structures of the soft terms at the GUT scale
where we assume the flavour models are defined. We then  run the SUSY spectrum
down to the EW scale by means of the MSSM RGEs at the 2-loop level~\cite{Martin:1993zk}, 
consistently taking into account all the effects discussed in sec.~\ref{sec:running}.

The usual CMSSM contains (assuming vanishing flavour blind phases) five parameters:
$M_{1/2}$, $m_0$, $A_0$, $\tan\beta$ and the sign of $\mu$. However, we take a
positive $\mu$ as is preferred by both the $b\to s\gamma$ and muon anomalous 
magnetic moment constraints. Consequently only four parameters are involved.

\subsection{Step 1: Bounds on Mass Insertions}
%
Starting with the model-independent analysis, we derive the allowed ranges for the MIs
$(\delta_{d,u}^{AB})_{ij}$ (with $A,B=L,R$ and $i\neq j=1,2,3$) under the constraints
listed above.
We scan the values of the CMSSM parameters in the following ranges: $M_{1/2}\le 200$~GeV,
$m_{0}\le 300$~GeV, $|A_{0}|\le 3 m_{0}$ and $5<\tan\beta<15$, where the bound on $A_{0}$
is set to avoid charge and/or colour breaking minima~\cite{Casas:1995pd}. This choice for
the input parameters would correspond to squark and gluino masses of order
$m_{\tilde Q},M_{\tilde g}\lesssim 600$~GeV, both possibly observable at the LHC.

\subsubsection{1-2 Sector}

\begin{figure}[t]\centering
\includegraphics[width=0.3\textwidth]{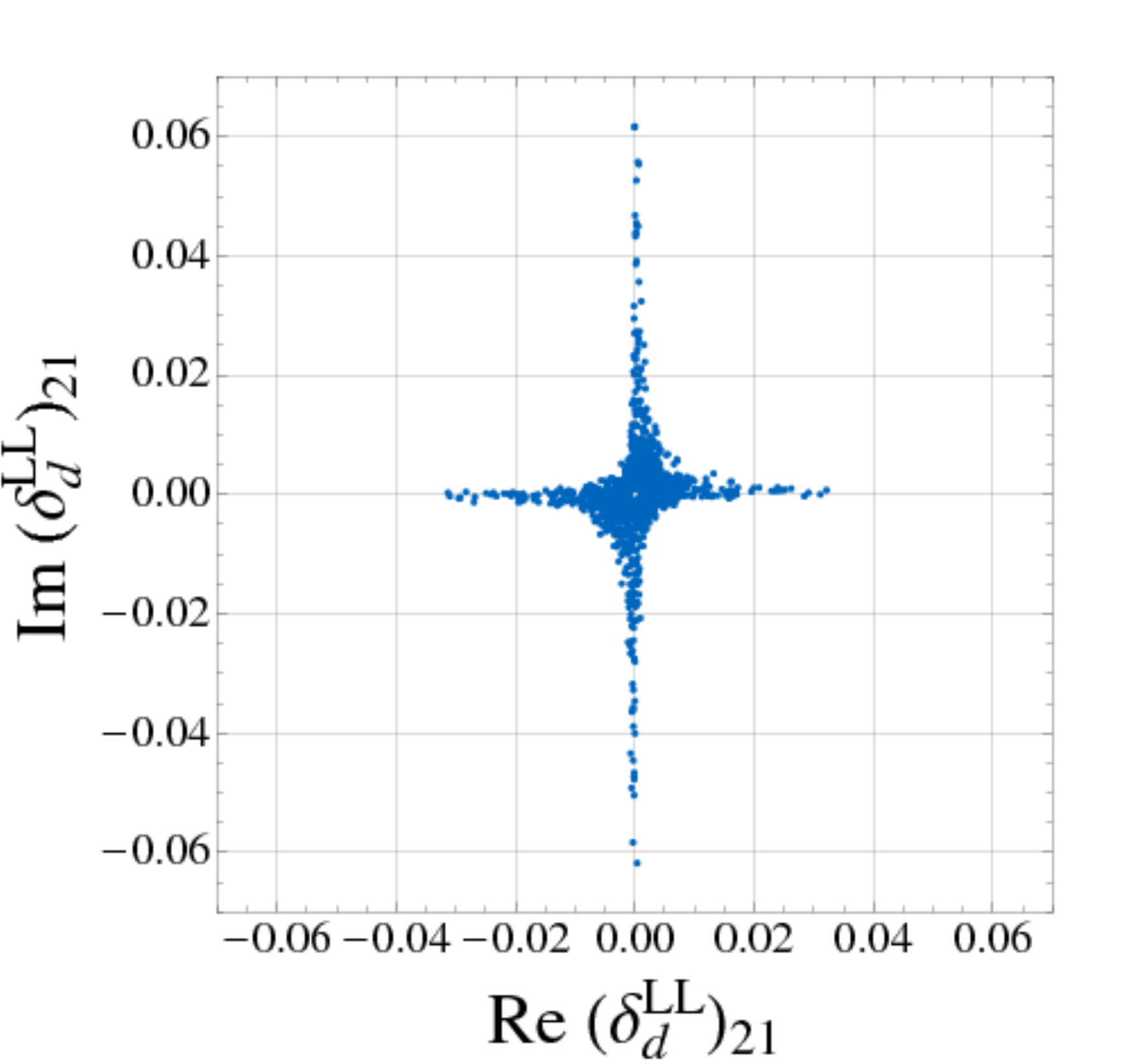}~~~
\includegraphics[width=0.3\textwidth]{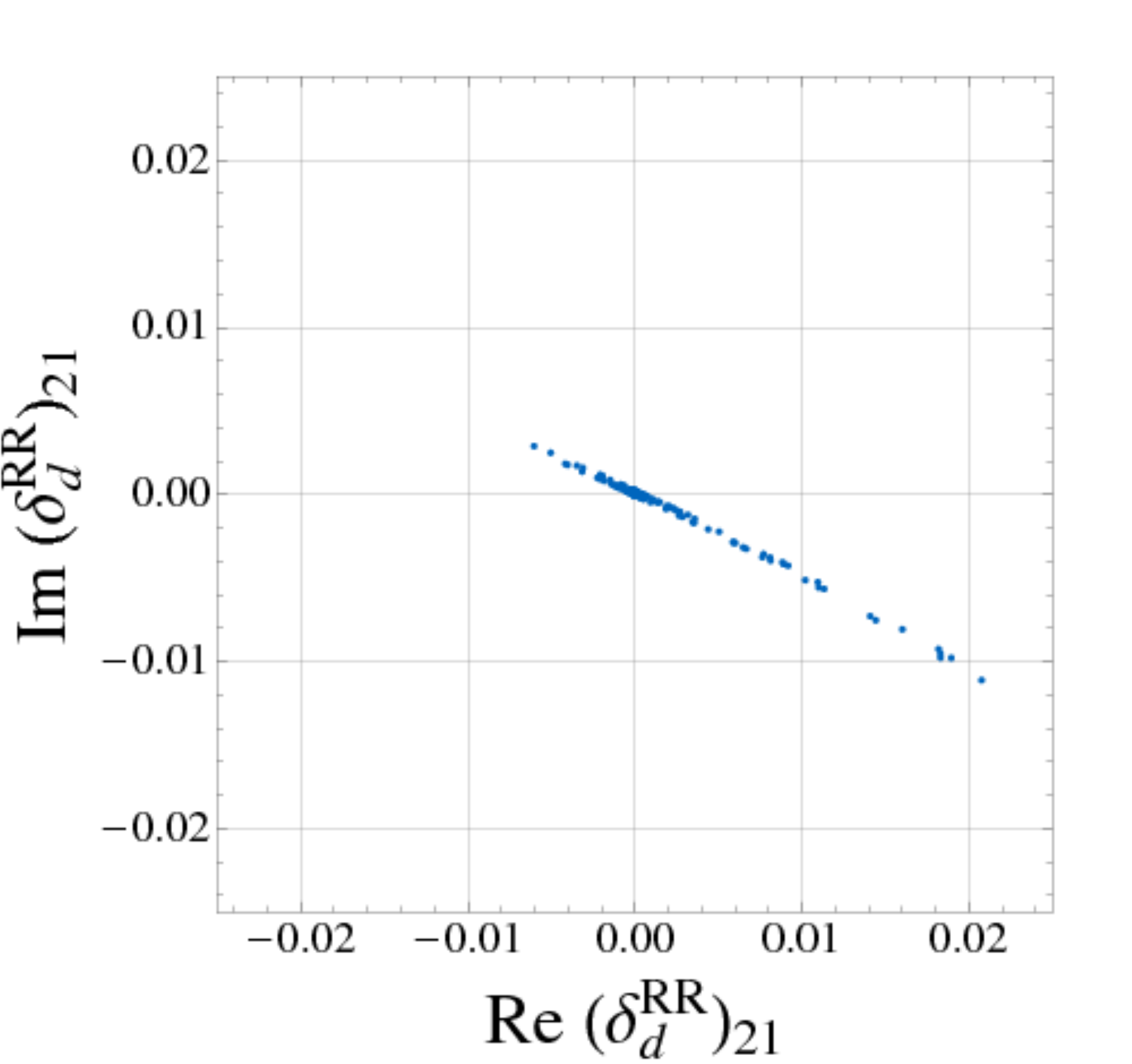}~~~
\includegraphics[width=0.31\textwidth]{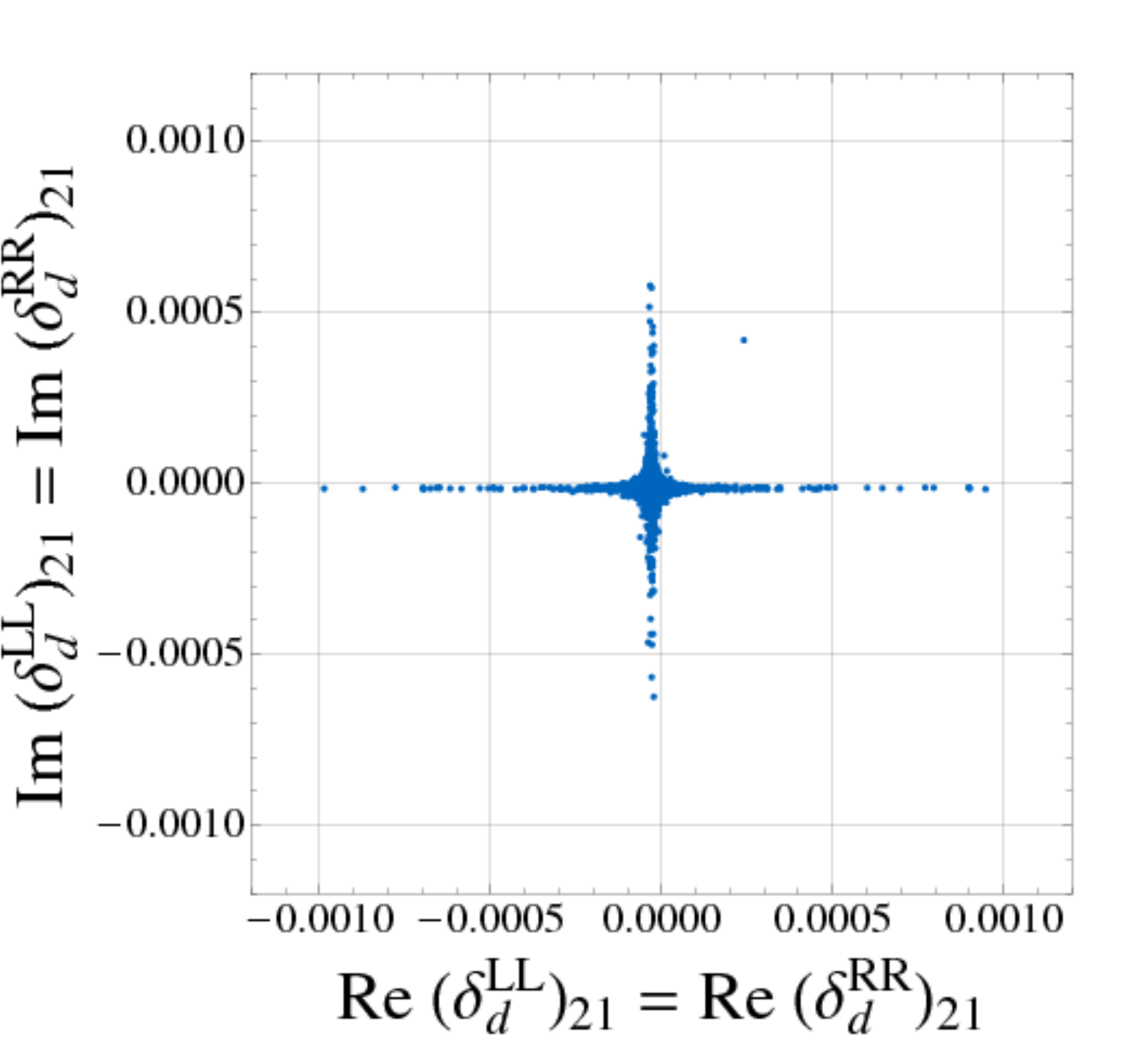}\\[20pt]
\includegraphics[width=0.305\textwidth]{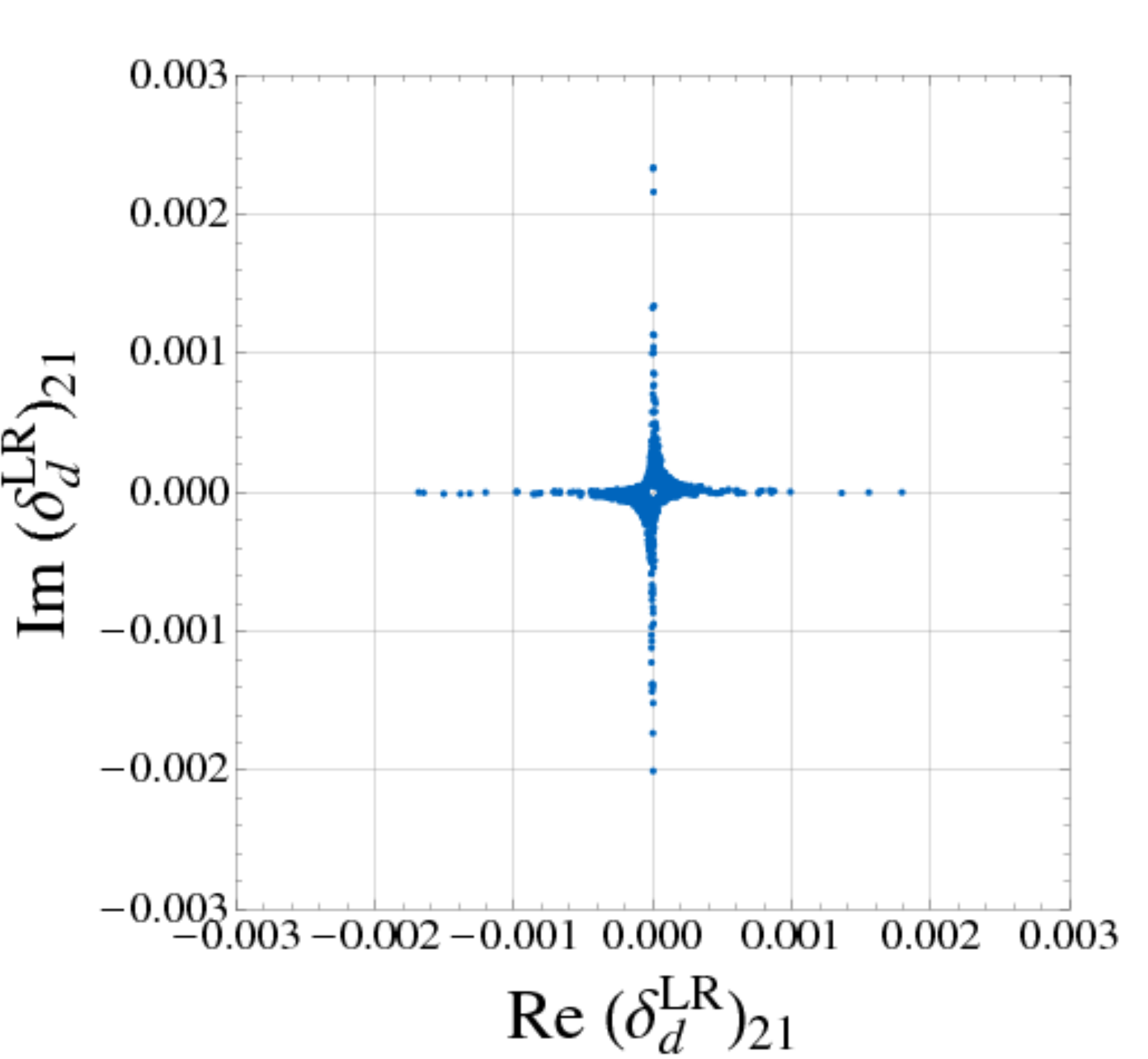}~~~
\includegraphics[width=0.305\textwidth]{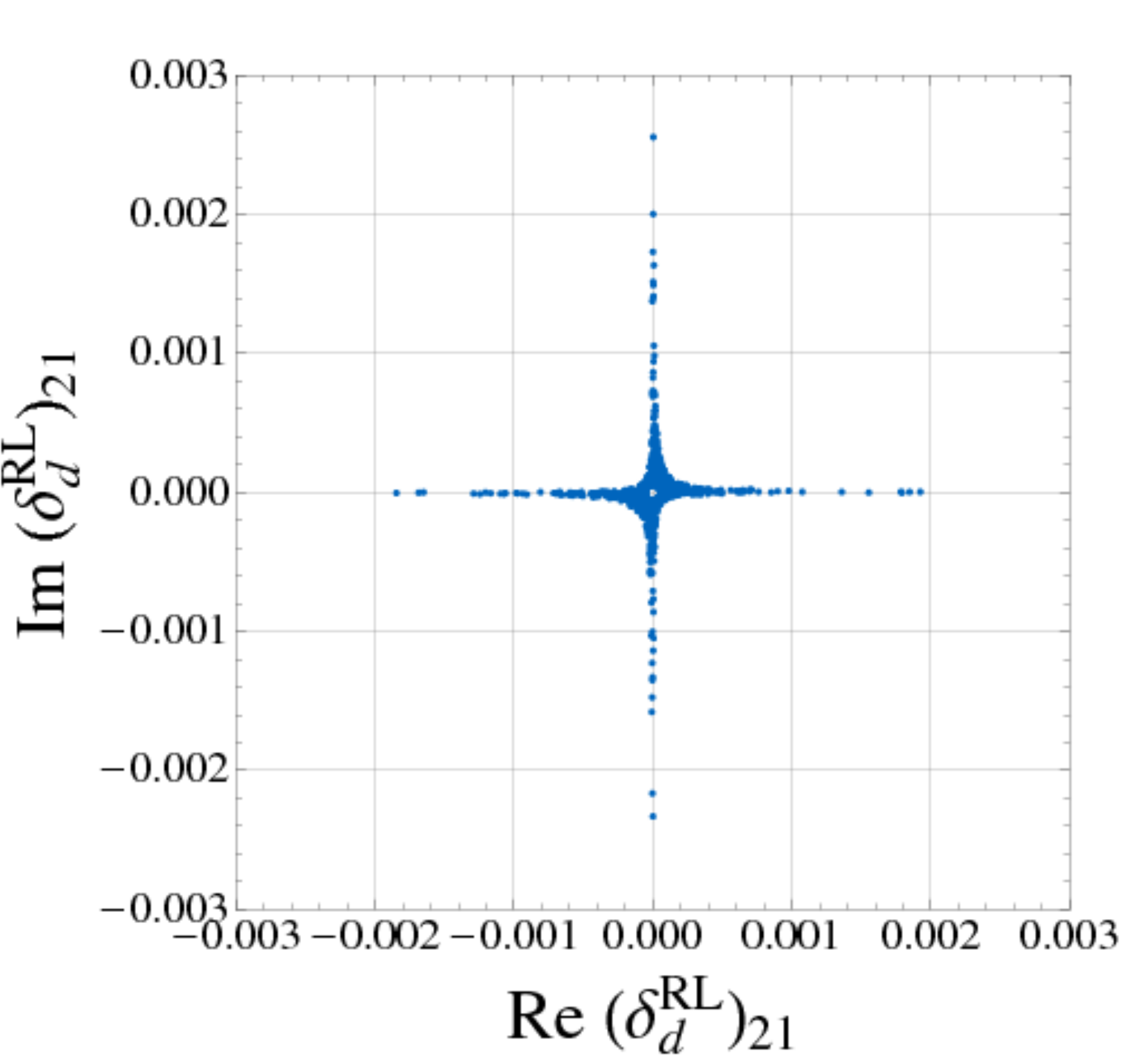}
\caption{\small
Bounds on the various MIs $(\delta^{AB}_d)_{21}$ (with $A,B=L,R$) as obtained by imposing
the experimental constraints from tabs.~\ref{tab:DF2exp} and \ref{tab:observables}, 
in particular $\epsilon_K$ and $\Delta M_K$.}
\label{fig:MI12}
\end{figure}

The measurements of $\Delta M_K$ and $\epsilon_K$ are used to constrain the $\left(\delta^{AB}_{d}\right)_{21}$, as shown in fig.~\ref{fig:MI12}. $\Delta M_K$ and $\epsilon_K$ 
constrain the real and imaginary parts of the product $\left(\delta^{AB}_d\right)_{21}\left(\delta^{AB}_d\right)_{21}$, respectively.
In the case of $\Delta M_K$, given the uncertainty coming from the long-distance SM contribution, 
we use the range $-\Delta M_K^{\rm exp}\le (\Delta M_K)_{\rm SD}^{\rm SM} + (\Delta M_K)^{\rm SUSY}\le\Delta M_K^{\rm exp}$ where $(\Delta M_K)_{\rm SD}^{\rm SM}$ refers to the short-distance SM contribution.
The measurement of $\epsilon^\prime/\epsilon$ could put an additional bound on $\text{Im}[(\delta^{AB}_d)_{21}]$ that is effective in the case of the LR MI only. However, given the 
large hadronic uncertainties in the SM calculation of $\epsilon^\prime/\epsilon$, to be conservative,
we do not use this bound. Notice that the bound on the RR MI is obtained in the presence of a radiatively 
induced LL MI $(\delta^{LL}_d)_{21} \propto V_{td}^*V_{ts}$, see~(\ref{eq:mQ212}).
In the Kaon sector, the product $\left(\delta^{LL}_{d}\right)_{21}\left(\delta^{RR}_{d}\right)_{21}$ 
generates left-right operators that are enhanced by the QCD evolution, a large loop function and by
the hadronic matrix element. Therefore, the bounds on RR MIs are more stringent than the ones on LL MIs.
For the same reason also the constraints in the other cases $\left(\delta^{LL}_{d}\right)_{21} = \left(\delta^{RR}_{d}\right)_{21}$, $\left(\delta^{LR}_{d}\right)_{21}$ and
$\left(\delta^{RL}_{d}\right)_{21}$ are particularly strong.

\begin{figure}[t]\centering
\includegraphics[width=0.3\textwidth]{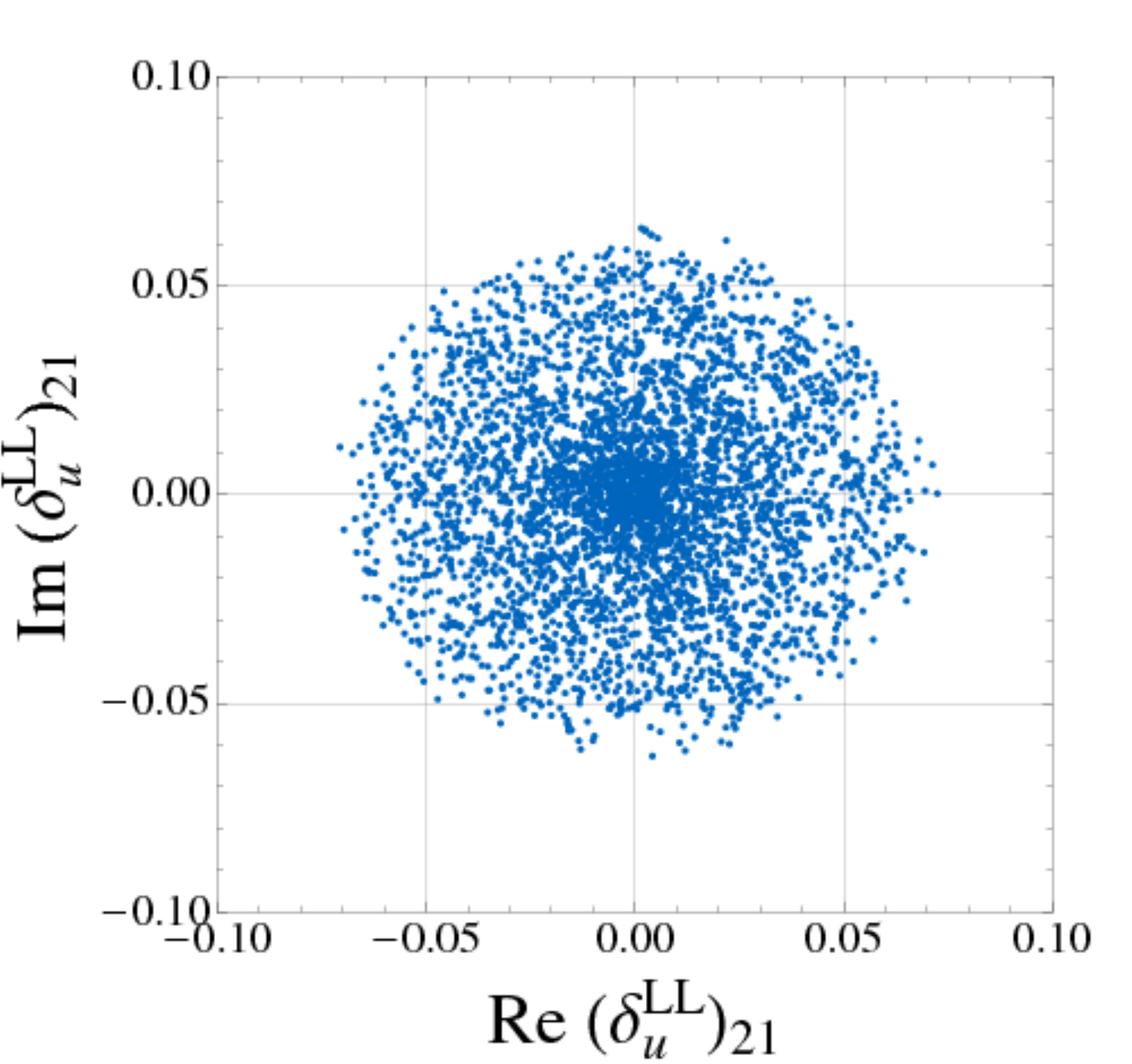}~~~
\includegraphics[width=0.3\textwidth]{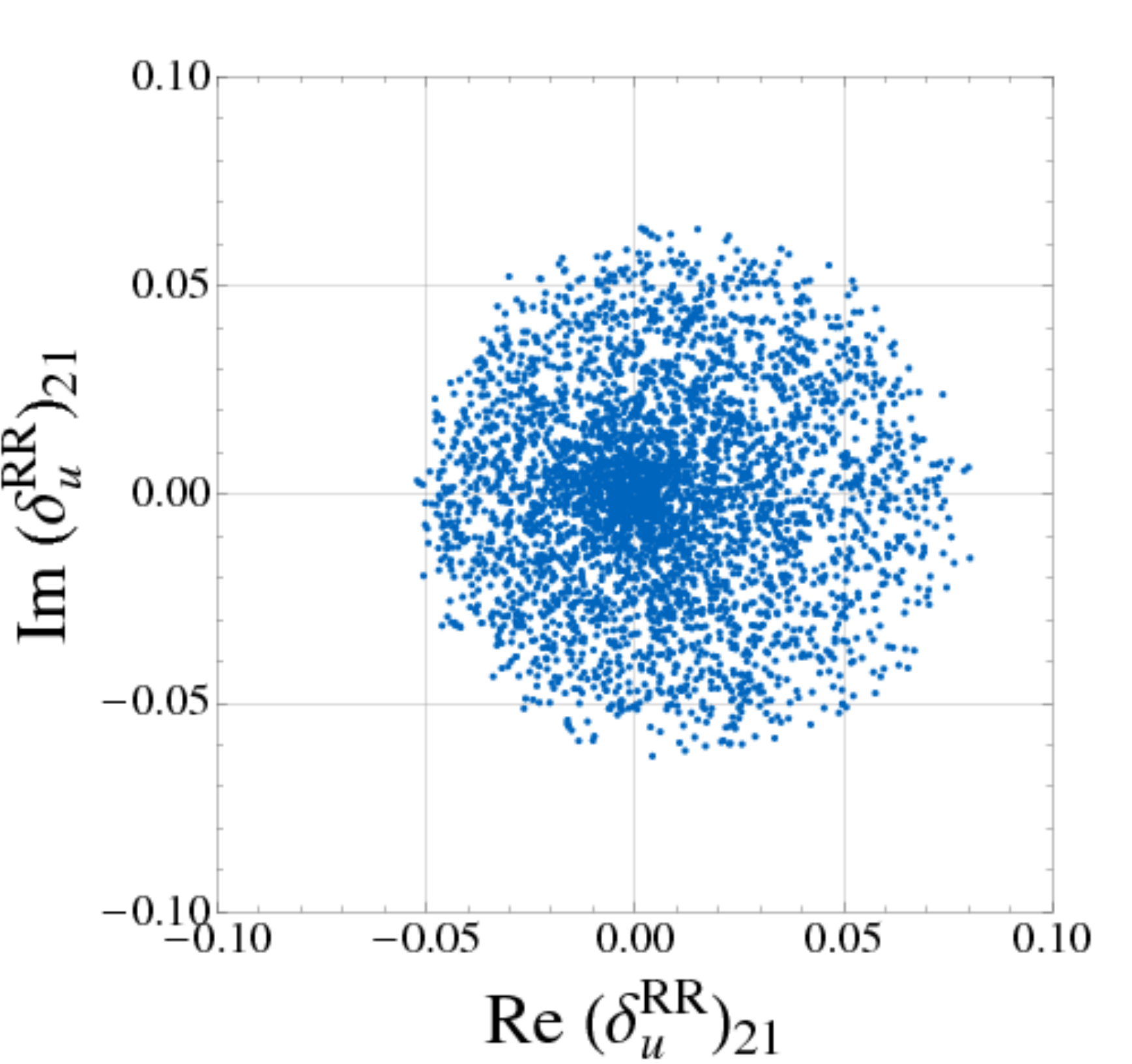}\\[20pt]
\includegraphics[width=0.3\textwidth]{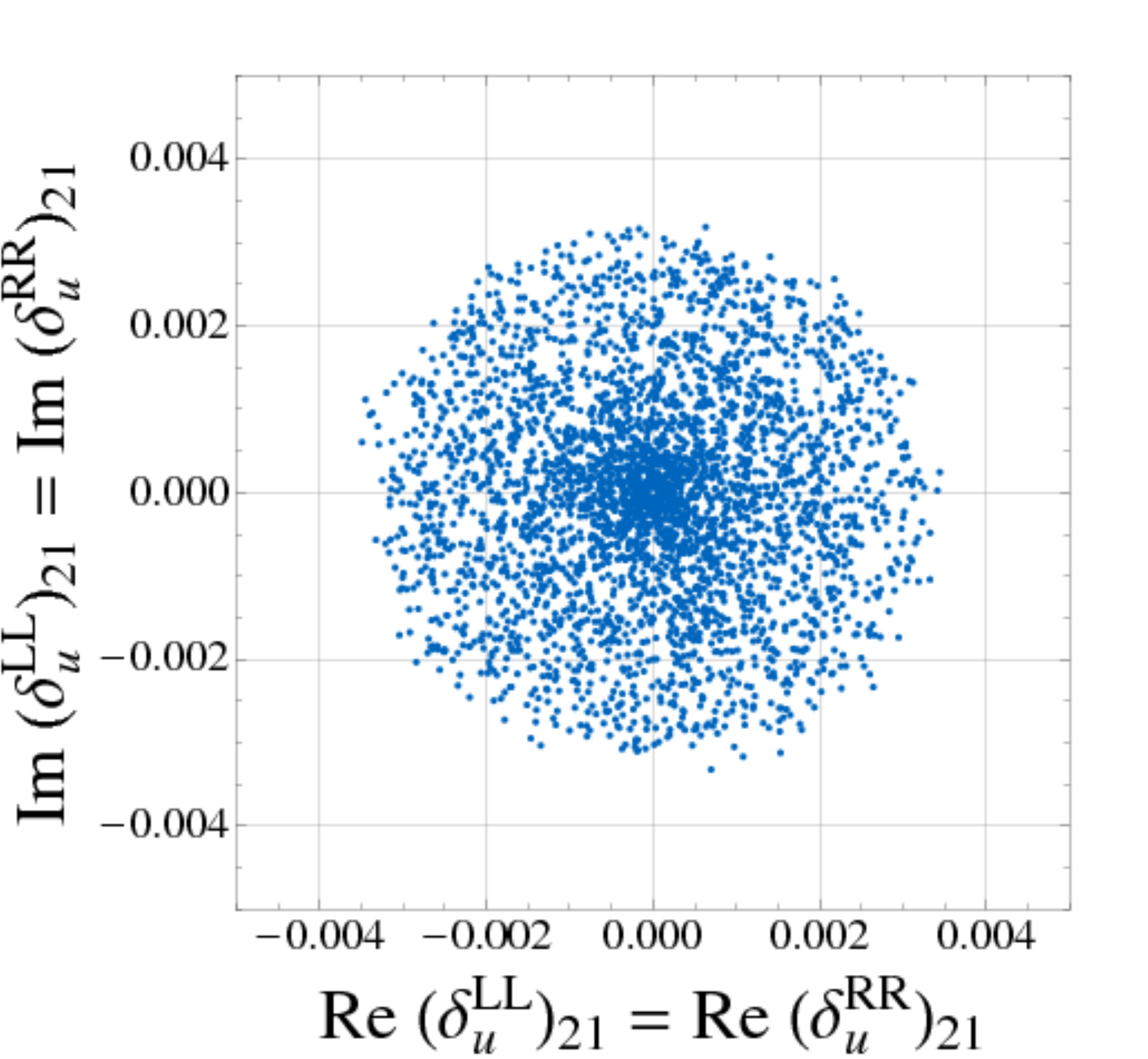}~~~
\includegraphics[width=0.3\textwidth]{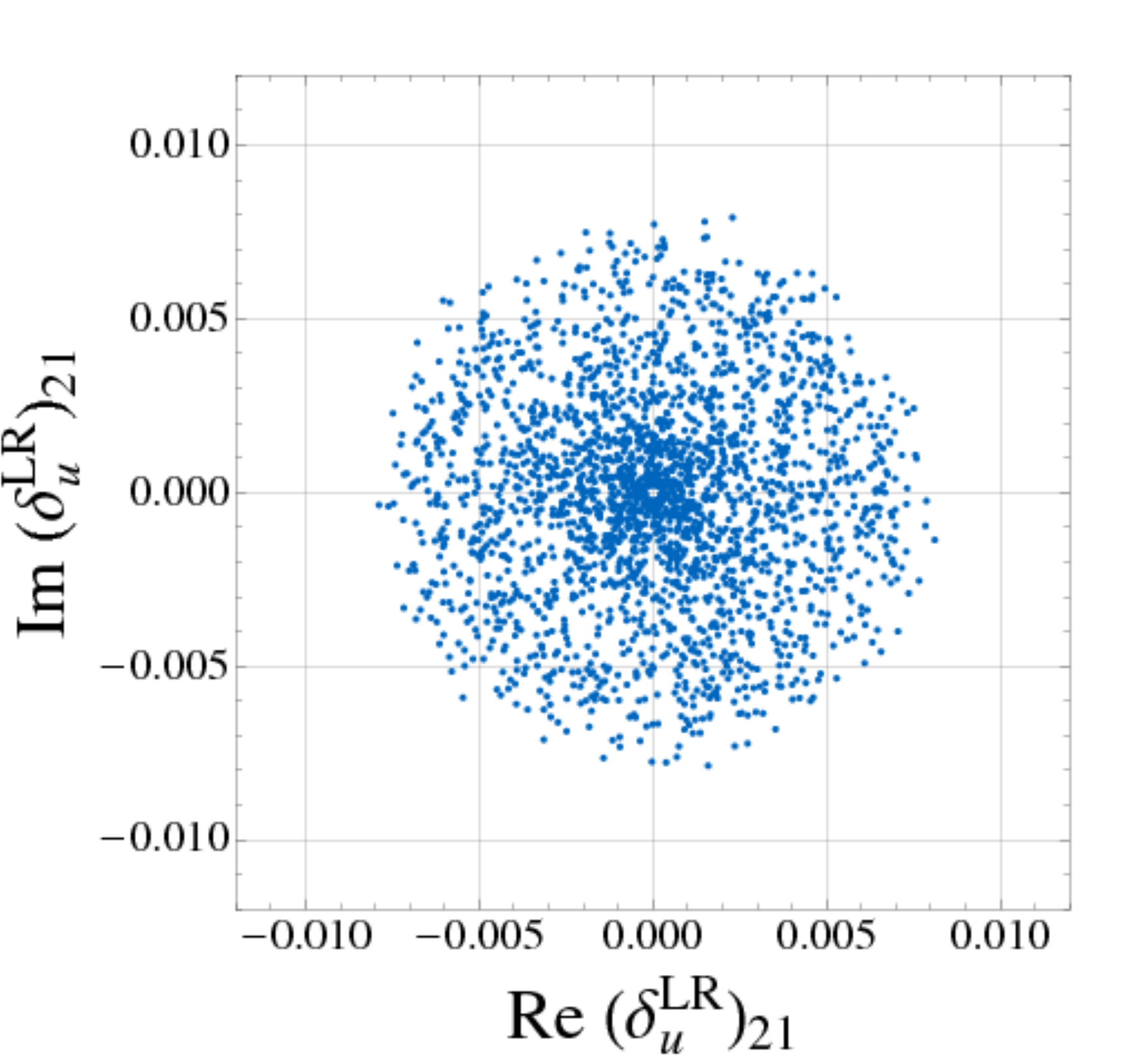}
\caption{\small
Bounds on the various MIs $(\delta^{AB}_u)_{21}$ (with $A,B=L,R$) as obtained 
by imposing the experimental constraint from $D^0-\bar D^0$ mixing and the 
constraints from tabs.~\ref{tab:DF2exp} and~\ref{tab:observables}.}
\label{fig:MI12up}
\end{figure}

Similar arguments apply to the $\left(\delta^{AB}_{u}\right)_{21}$ with $(A,B)=(L,R)$
where we make use of the recent experimental measurement of $D^0-\bar D^0$ mixing.
The corresponding bounds on $\left(\delta^{AB}_{u}\right)_{21}$ are based on
$|(M_{12}^D)_{\rm SUSY}| < 0.02 ~{\rm ps}^{-1}$~\cite{Ciuchini:2007cw} and shown
in fig.~\ref{fig:MI12up}.

Concerning the bounds on $\left(\delta^{LL}_{u}\right)_{21}$ (see the upper left plot
of fig.~\ref{fig:MI12up}), one would naively expect that they are almost the same as
the bounds found for $\left(\delta^{LL}_{d}\right)_{21}$, after the constraints from
$\epsilon_K$ and $\Delta M_K$ are imposed. In fact, the $SU(2)_{L}$ gauge symmetry
implies that $(M_u^{2})^{LL}_{21}=\left[V^*(M_d^{2})^{LL}V^{\rm T}\right]_{21}$,
hence, $(\delta_u^{LL})_{21}\simeq(\delta_d^{LL})_{21}$ for degenerate left-handed
squarks of the first two generations. However, as discussed in sec.~\ref{sec:SUSY_breaking},
within scenarios predicting non degenerate left-handed squarks the above argument is
no longer true, therefore $(\delta_u^{LL})_{21}\neq(\delta_d^{LL})_{21}$ and the bounds 
on $(\delta_u^{LL})_{21}$ and $(\delta_d^{LL})_{21}$ apply independently.
In fig.~\ref{fig:MI12up}, we consider such a general scenario. 
Obviously, all the bounds on the other MIs $\left(\delta^{AB}_{u}\right)_{21}$ with
$AB\neq LL$ are unrelated to the corresponding bounds for the down sector as no symmetry
principle is at work here.

\subsubsection{1-3 Sector}

\begin{figure}[t]\centering
\includegraphics[width=0.3\textwidth]{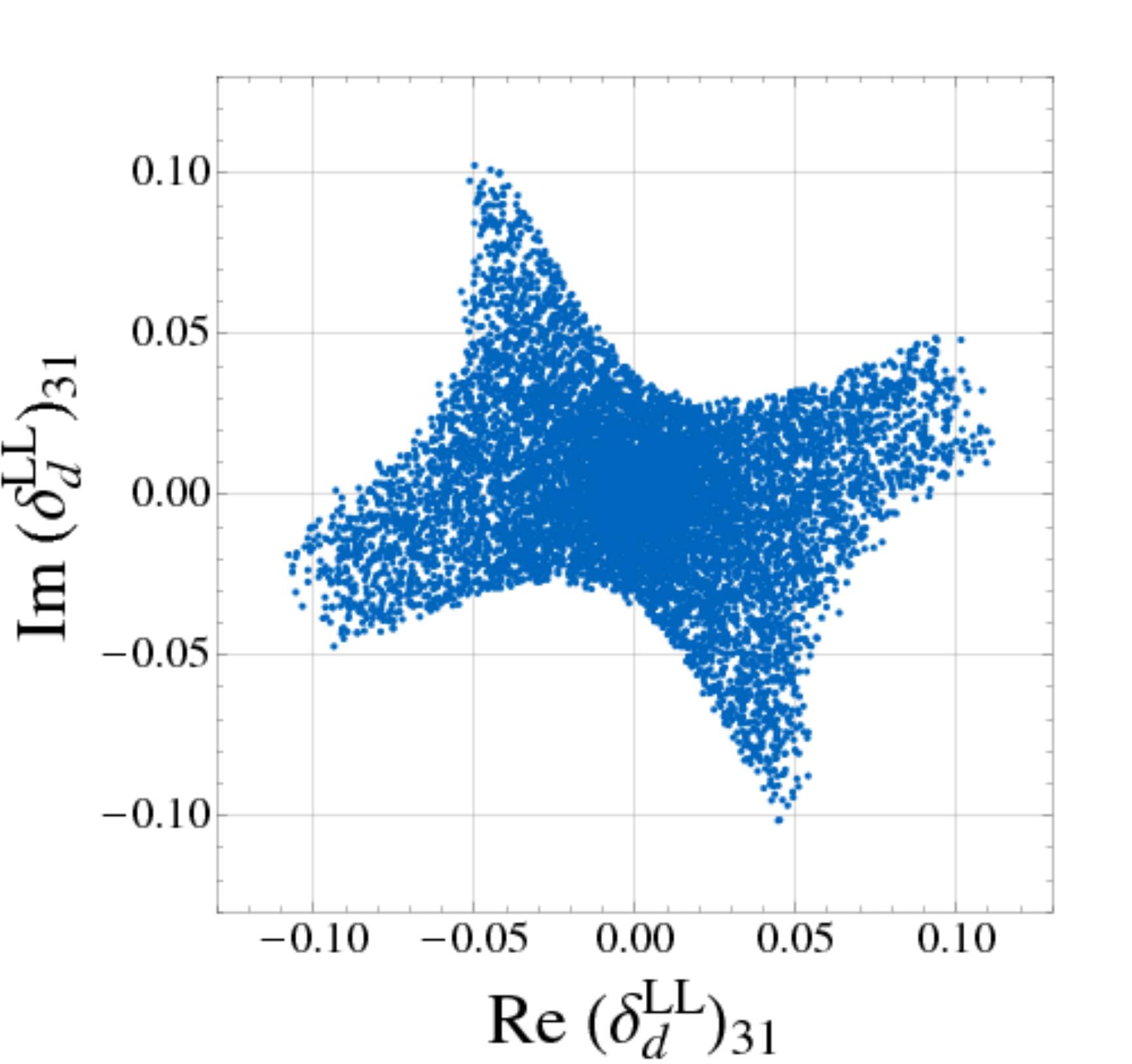}~~~
\includegraphics[width=0.295\textwidth]{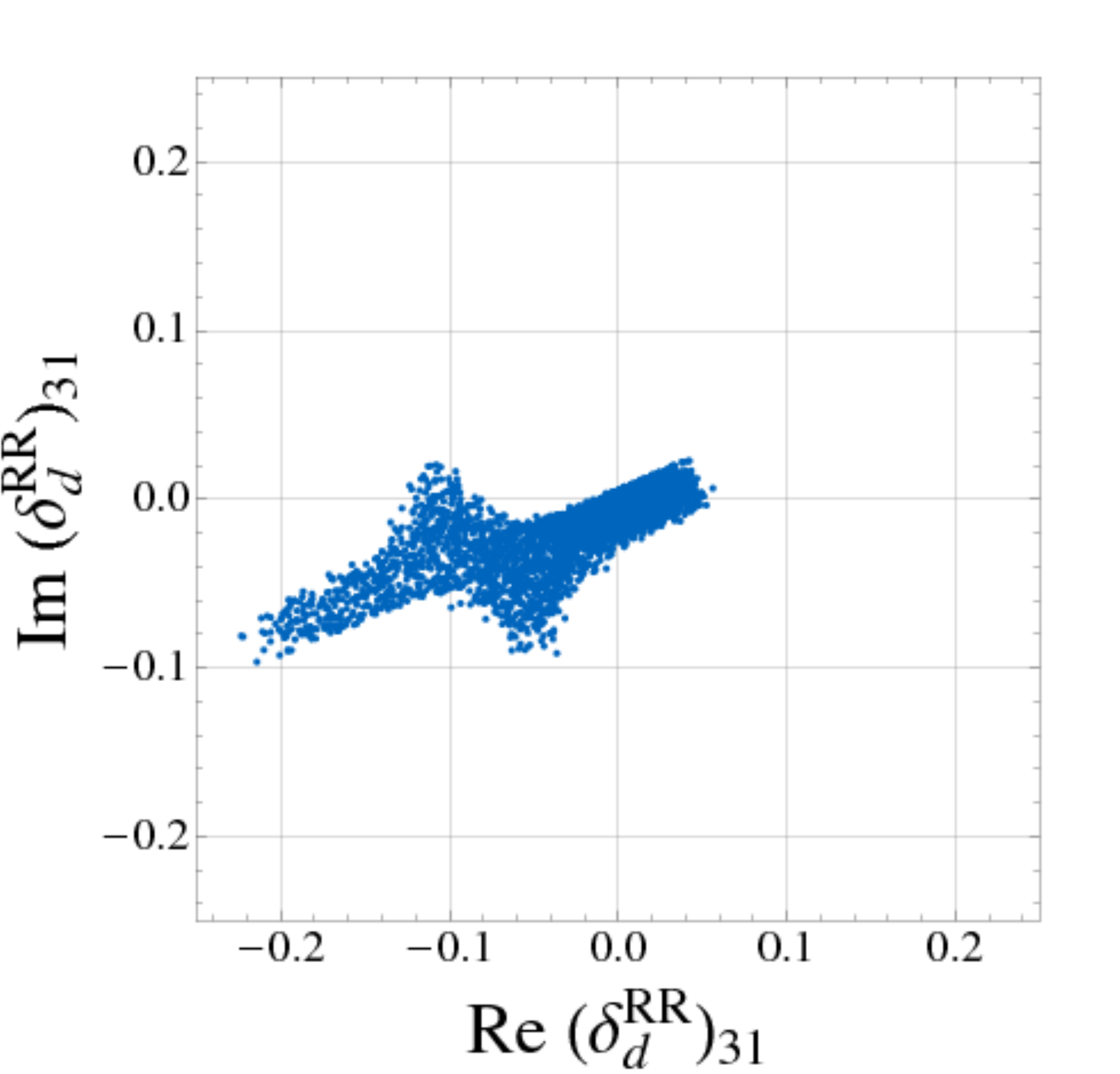}~~~
\includegraphics[width=0.305\textwidth]{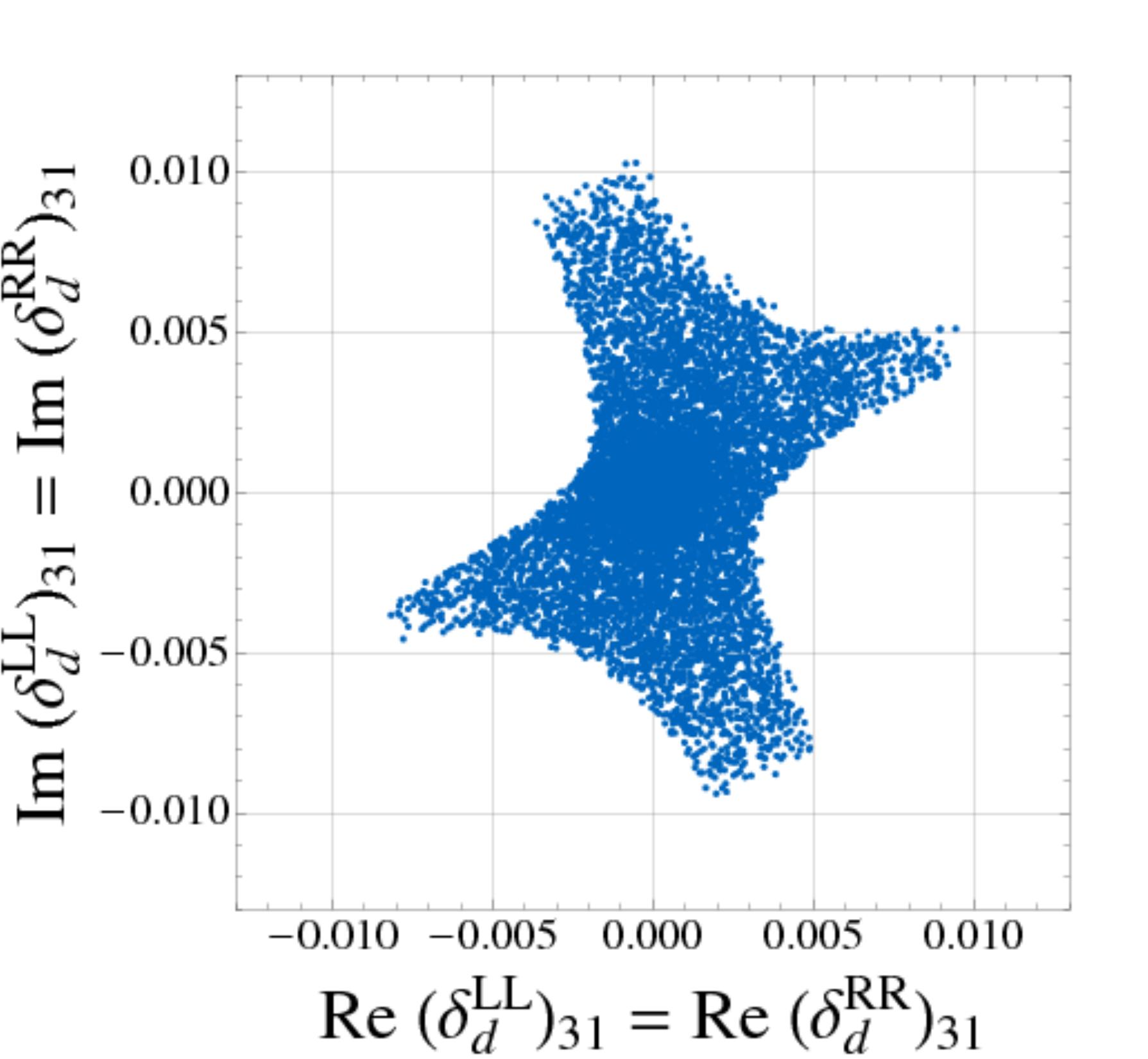}\\[20pt]
\includegraphics[width=0.3\textwidth]{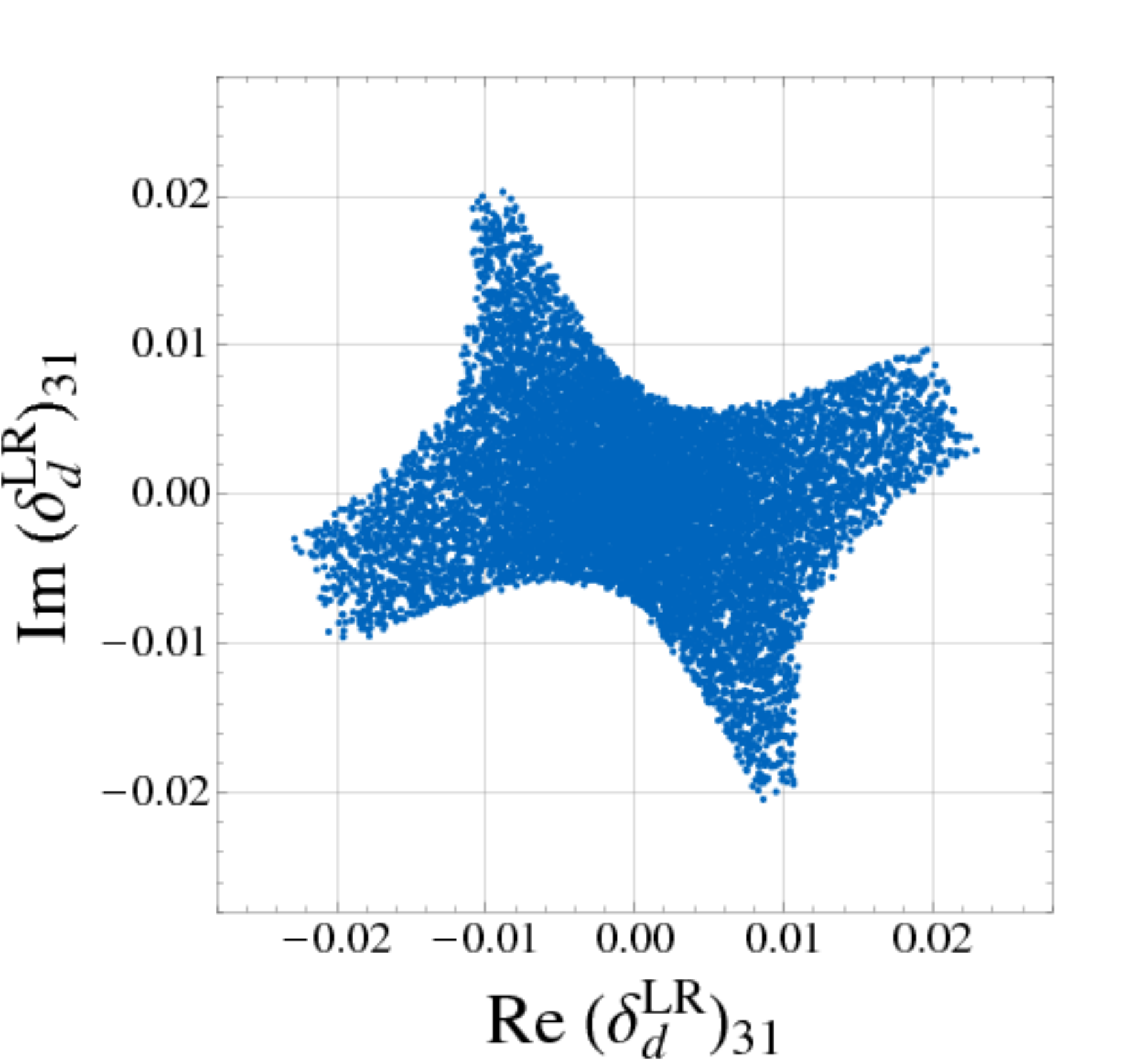}~~~
\includegraphics[width=0.3\textwidth]{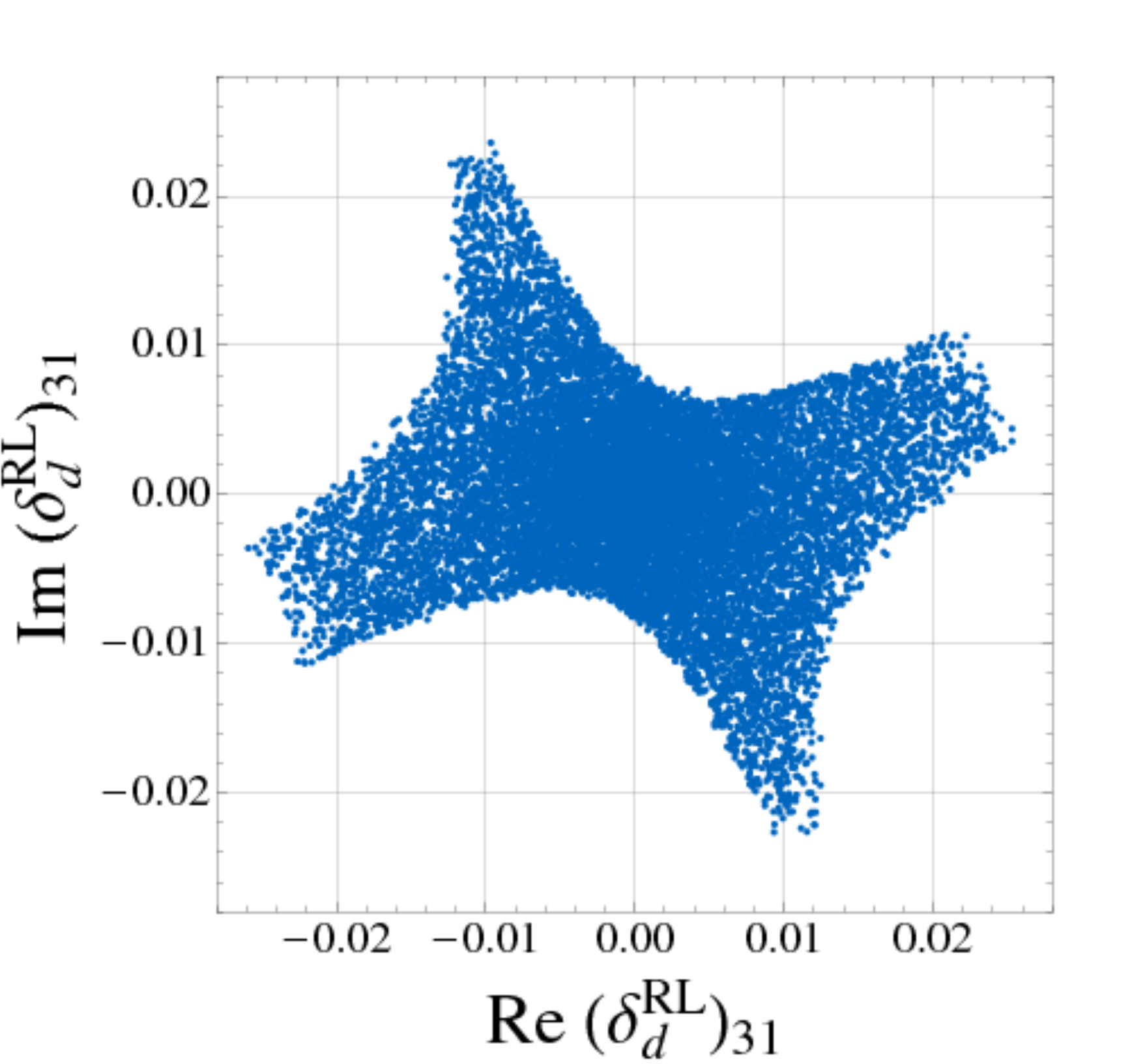}
\caption{\small
Bounds on the various MIs $(\delta^{AB}_d)_{31}$ (with $A,B=L,R$) as obtained by imposing the experimental
constraints from tabs.~\ref{tab:DF2exp} and \ref{tab:observables}, in particular $S_{\psi K_S}$, $\cos 2\beta$ and $\Delta M_d$.}
\label{fig:MI13}
\end{figure}

The measurements of $\Delta M_{d}$ and $S_{\psi K_S}$ constrain the modulus and the phase of the
$B_d$ mixing amplitude, respectively. The bounds on the various $\left(\delta^{AB}_{d}\right)_{31}$
are reported in fig.~\ref{fig:MI13}. The constraints on $\left(\delta^{RR}_{d}\right)_{31}$ and $\left(\delta^{LL}_{d}\right)_{31}$ are different because of the contributions of the left-right
operator (arising by means of the RGE induced $(\delta^{LL}_d)_{31} \propto V_{tb}V_{td}^*$) that 
is effective in the $\left(\delta^{RR}_{d}\right)_{31}$ case only. The constraints in the cases $\left(\delta^{LL}_{d}\right)_{31}=\left(\delta^{RR}_{d}\right)_{31}$, $\left(\delta^{LR}_{d}\right)_{31}$ 
and $\left(\delta^{RL}_{d}\right)_{31}$, are particularly strong due again to the large NP
contributions provided by the left-right operators that are strongly enhanced by renormalization
group effects and by a large loop function.

\subsubsection{2-3 Sector}

In this sector, we can exploit a large number of constraints. In particular, they arise from $\Delta M_{s}$
and $\Delta B=1$ branching ratios such as $b\to s\gamma$ and $b\to s\ell^+\ell^-$.

On the other hand, we do not impose the bounds from the time-dependent CP asymmetries in $B_d\to\phi K_S$
and $B_d\to\eta^{\prime}K_S$ and the direct CP asymmetry in $b\to s\gamma$, given the still rather large uncertainties. In fig.~\ref{fig:MI23} we show the allowed regions for the mass insertions and indicate in addition the resulting values for $S_{\psi\phi}$ with different colors. The following comments are in order:

\begin{figure}[t]\centering
\includegraphics[width=0.3\textwidth]{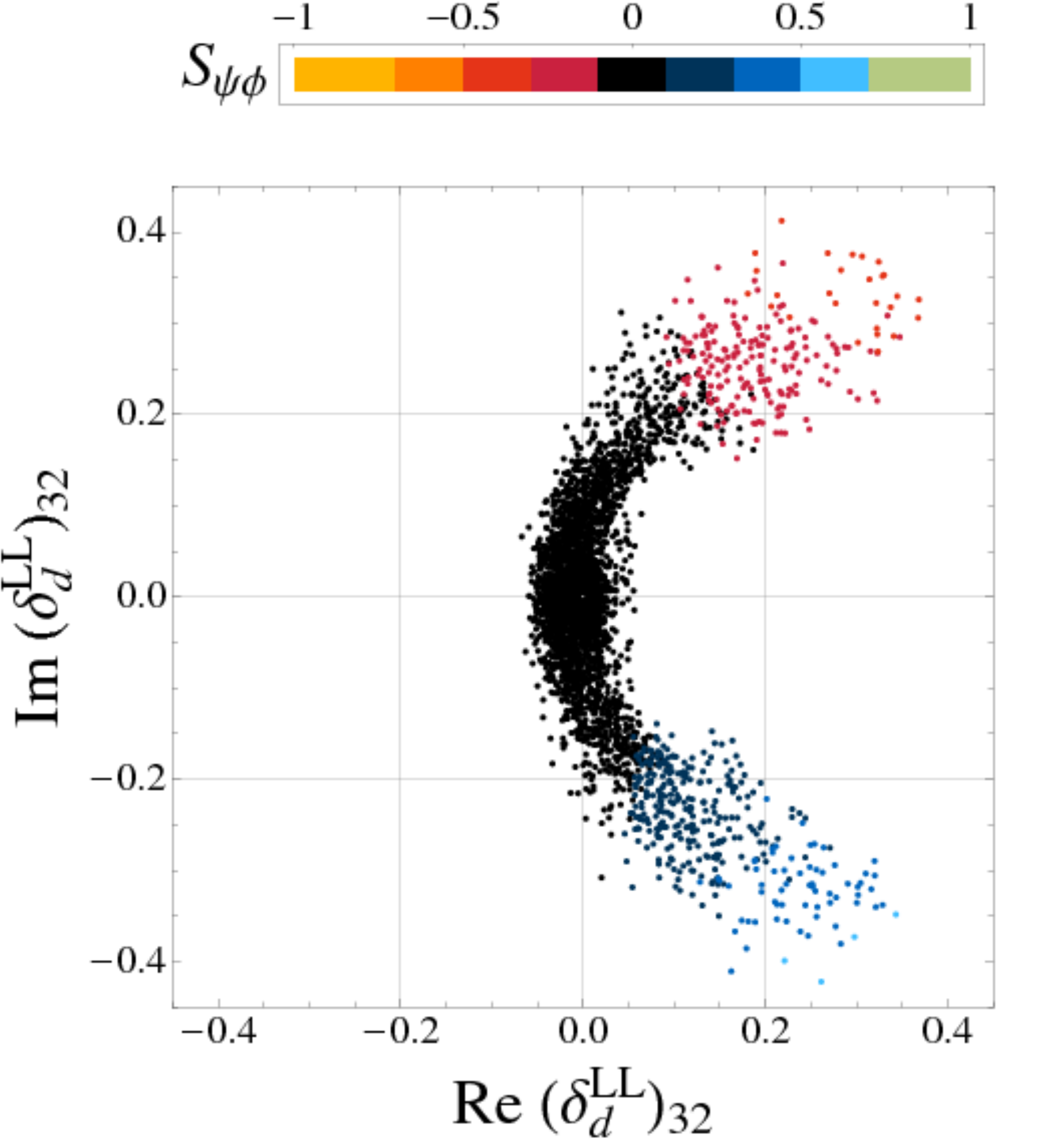}~~~
\includegraphics[width=0.3\textwidth]{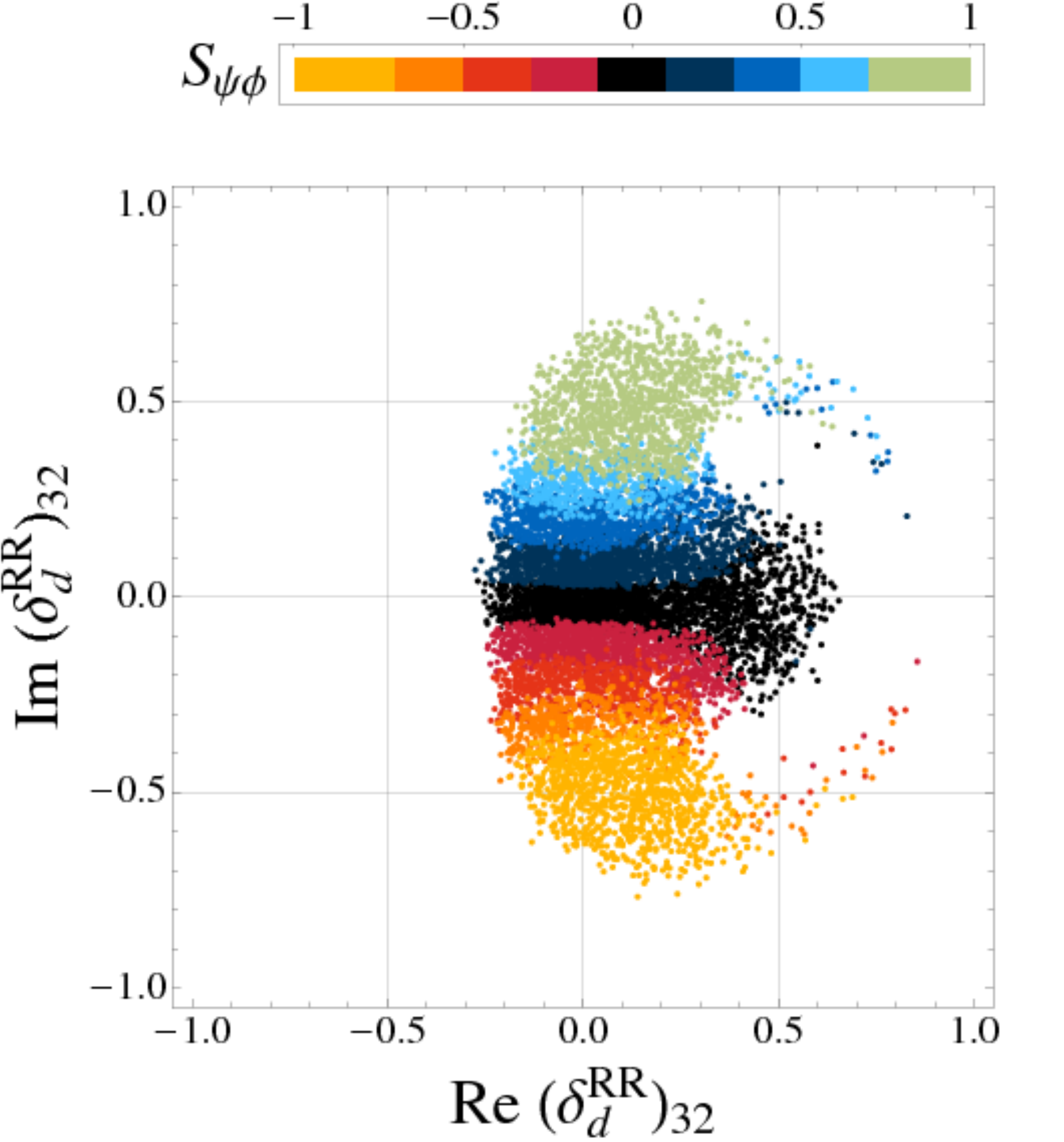}~~~
\includegraphics[width=0.3\textwidth]{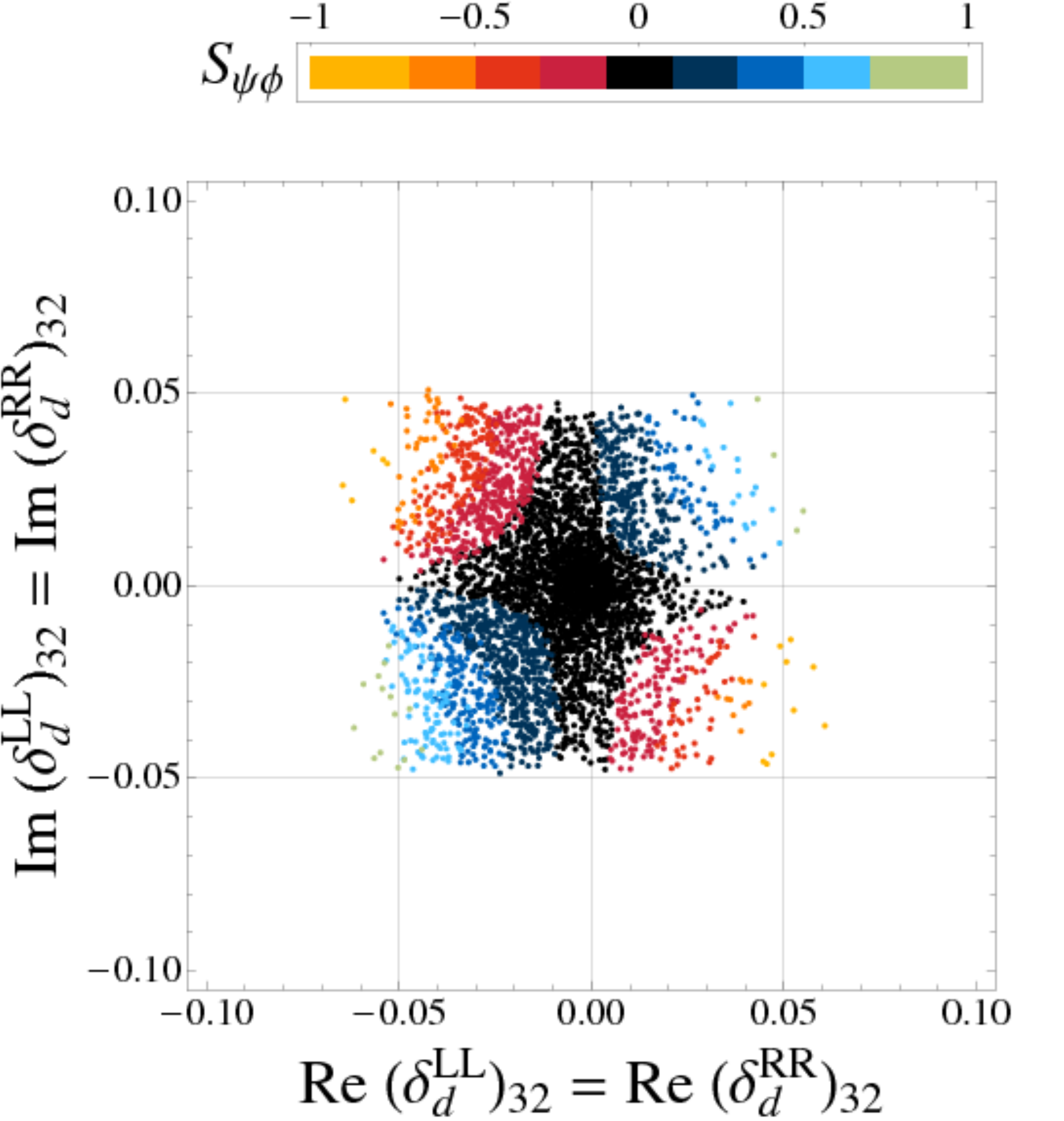} \\[20pt]
\includegraphics[width=0.3\textwidth]{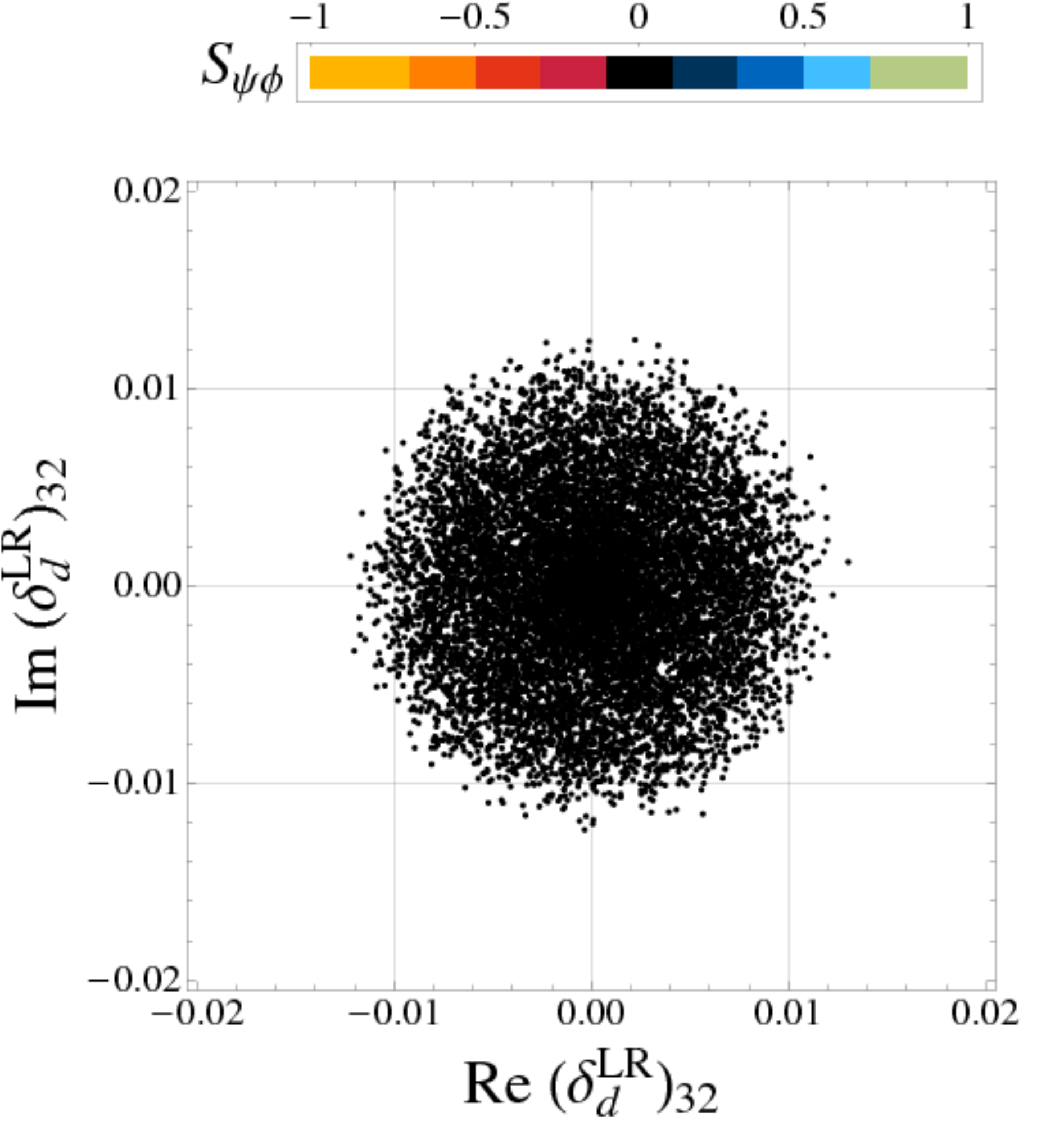}~~~
\includegraphics[width=0.3\textwidth]{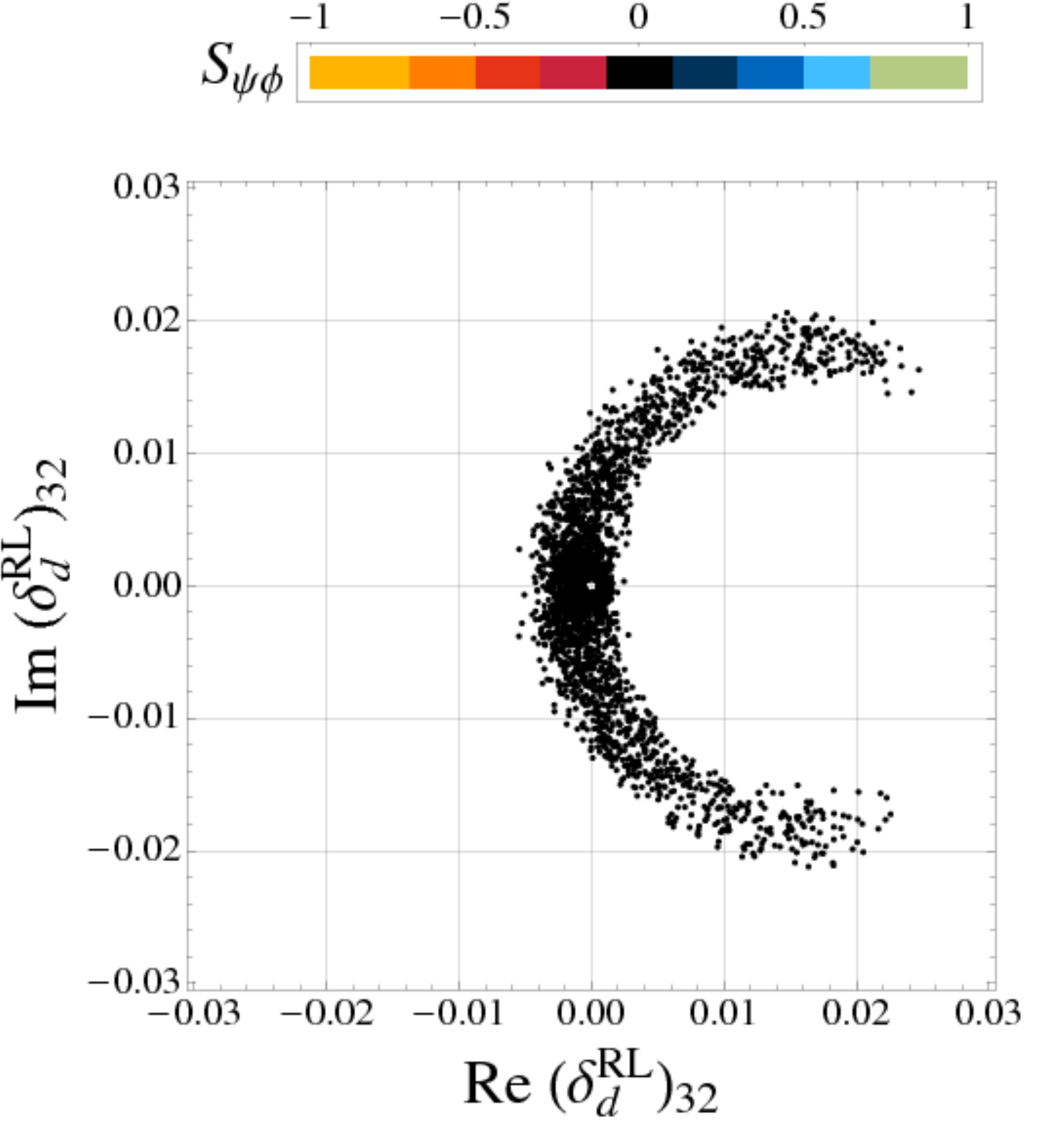}
\caption{\small
Bounds on the various MIs $(\delta^{AB}_d)_{32}$ (with $A,B=L,R$) as obtained by imposing the
experimental constraints from the $b\to s$ transitions listed in tabs.~\ref{tab:DF2exp} and \ref{tab:observables}. The different colors indicate the resulting values for $S_{\psi\phi}$.}
\label{fig:MI23}
\end{figure}

\begin{itemize}
\item In the $\delta^{LL}$ case, a strong constraint arises from $\text{BR}(b\to s\gamma)$ as the
related NP amplitude can interfere with the SM one.
Regions in parameter space with a large real part of $(\delta_{d}^{LL})_{32}$ that would be allowed by
$b\to s\gamma$ are excluded by the constraint from BR($B \to X_{s} \ell^+ \ell^-$), while $\Delta M_{s}$
does not provide any further constraint. We observe that in the considered framework the possible
values for $S_{\psi\phi}$ are rather moderate and lie in the range $-0.1< S_{\psi\phi}<0.1$
even for quite large values of $(\delta_{d}^{LL})_{32}\leq 0.1$.
\item In the $\delta^{RR}$ case, the situation is very different compared to the $\delta^{LL}$ case
as now $\text{BR}(b\to s\gamma)$ is not so much effective since the related NP amplitude (arising
from right-handed currents) does not interfere with the SM one. Also $\text{BR}(B\to X_s\ell^+\ell^-)$
is not effective for the same reason. Now, $\Delta M_{s}$ plays the main role in constraining $(\delta_d^{RR})_{32}$. As in the other cases, RGE induced effects generate at the low energy an
effective MI $(\delta^{LL}_d)_{32}\propto V_{tb}V_{ts}^*$.
The product $(\delta^{LL}_{d})_{32}(\delta^{RR}_{d})_{32}$ generates left-right operators that are
enhanced both by the QCD evolution and by a large loop function. Therefore, the bounds on RR MIs from
$\Delta M_{s}$ are more stringent than the ones on LL MIs. We find that huge effects in $S_{\psi\phi}$
in the entire range from $-1$ to $1$ are possible in this scenario.
\item In the case $(\delta^{LL}_{d})_{32}=(\delta^{RR}_{d})_{32}$, very strong constraints on the MIs
arise from $\Delta M_{s}$ due to the large contributions it receives from the left-right operator.
Furthermore also BR$(b\to s\gamma)$ provides additional constraints given the rather light SUSY spectrum
we consider here. Still in the remaining parameter space large effects in the $B_s$ mixing phase in the
range $-0.7\le S_{\psi\phi}\le 0.7$ are possible even for $(\delta^{LL}_{d})_{32}=(\delta^{RR}_{d})_{32}\simeq 0.05$.
\item Finally, in the last two scenarios with $(\delta^{LR}_{d})_{32}$ or $(\delta^{RL}_{d})_{32}$
switched on, BR$(b\to s\gamma)$ is the main constraint as the NP amplitude realizes the necessary
chirality flip for the dipole $b\to s\gamma$ transition without involving the bottom mass insertion.
In fact, the $b\to s\gamma$ constraint is so strong that $S_{\psi\phi}$ cannot depart significantly
from the SM prediction in these cases.
\end{itemize}

\begin{figure}[t]\centering
\includegraphics[width=0.3\textwidth]{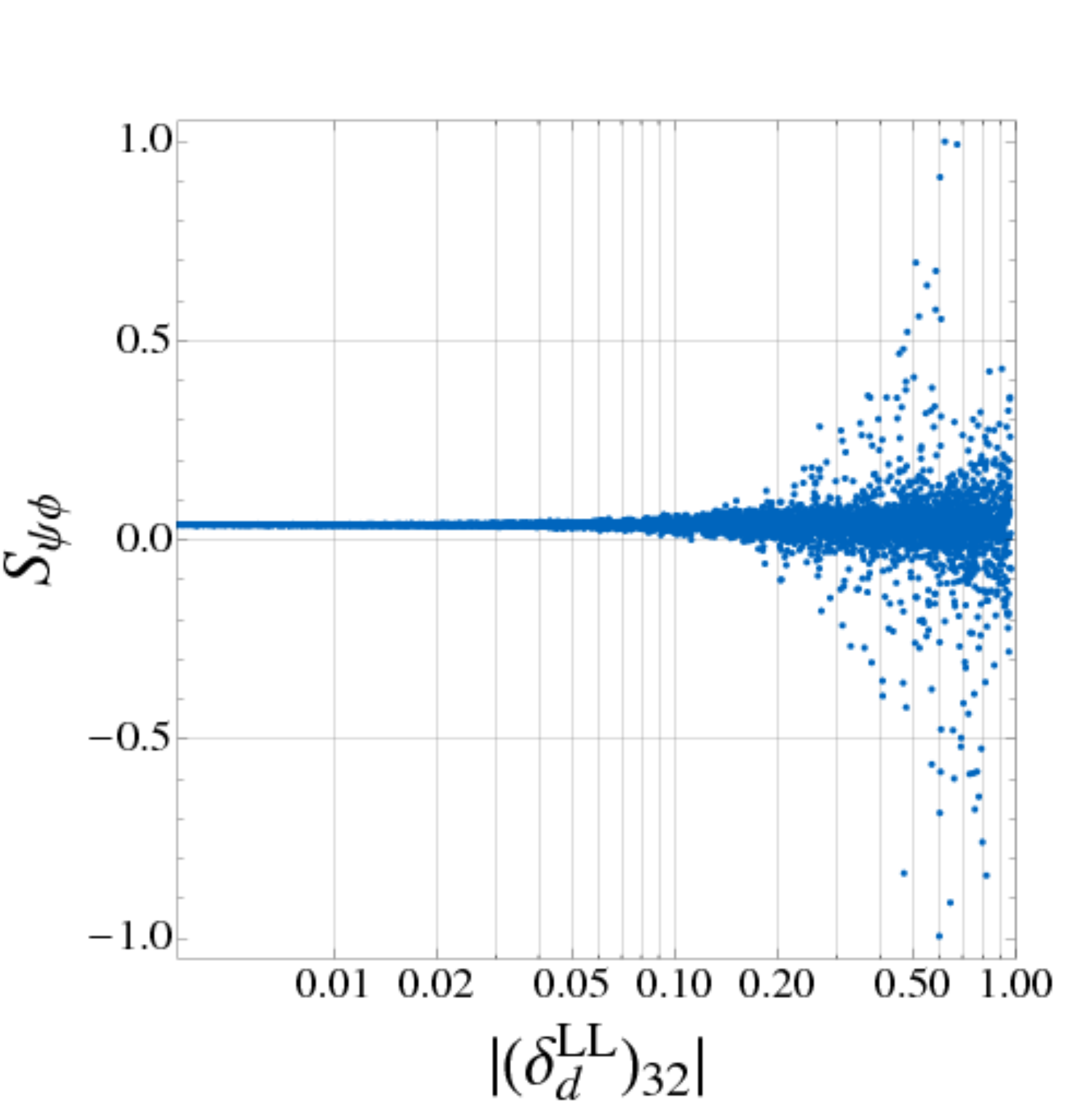}~~~
\includegraphics[width=0.3\textwidth]{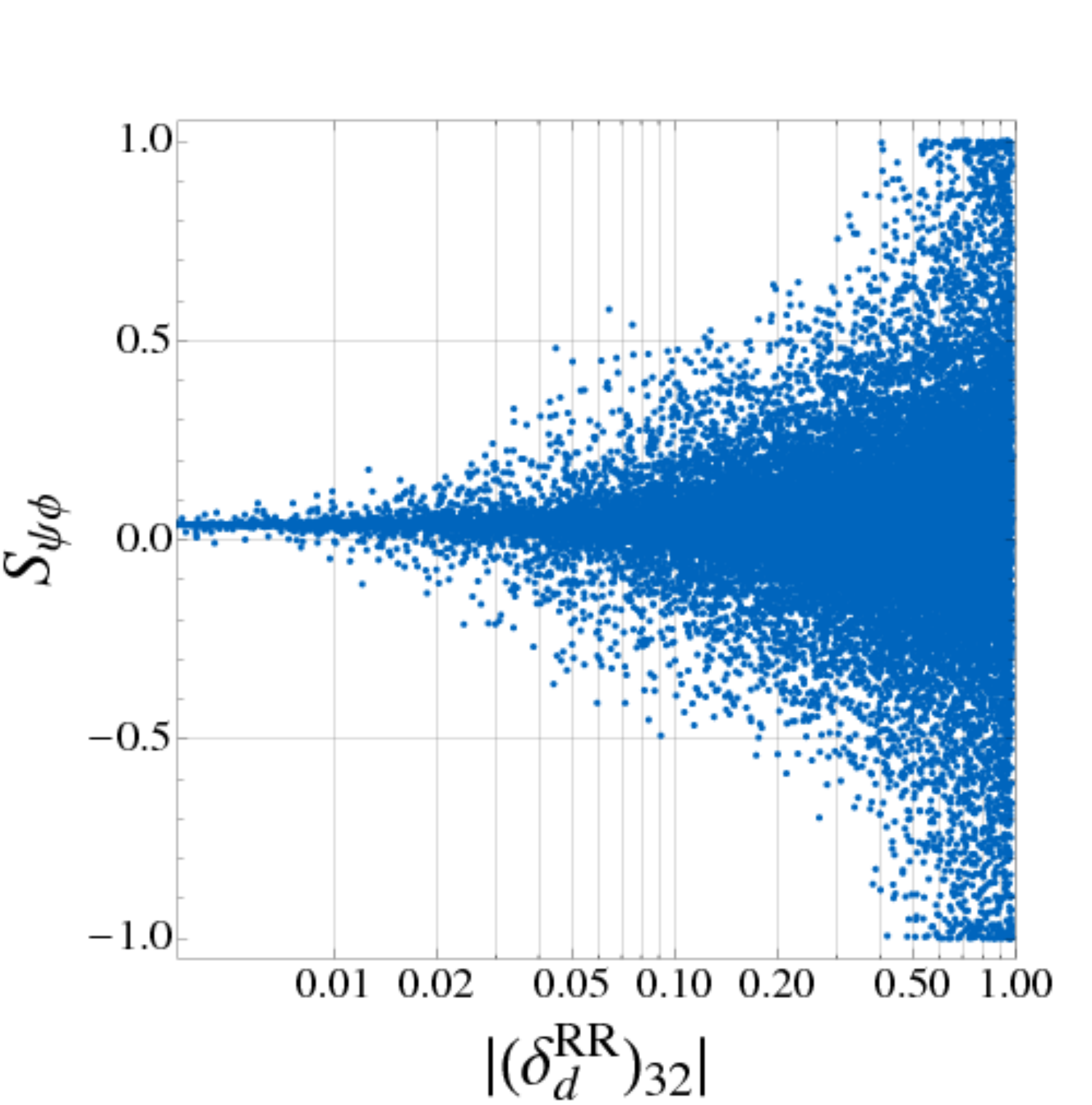}~~~
\includegraphics[width=0.3\textwidth]{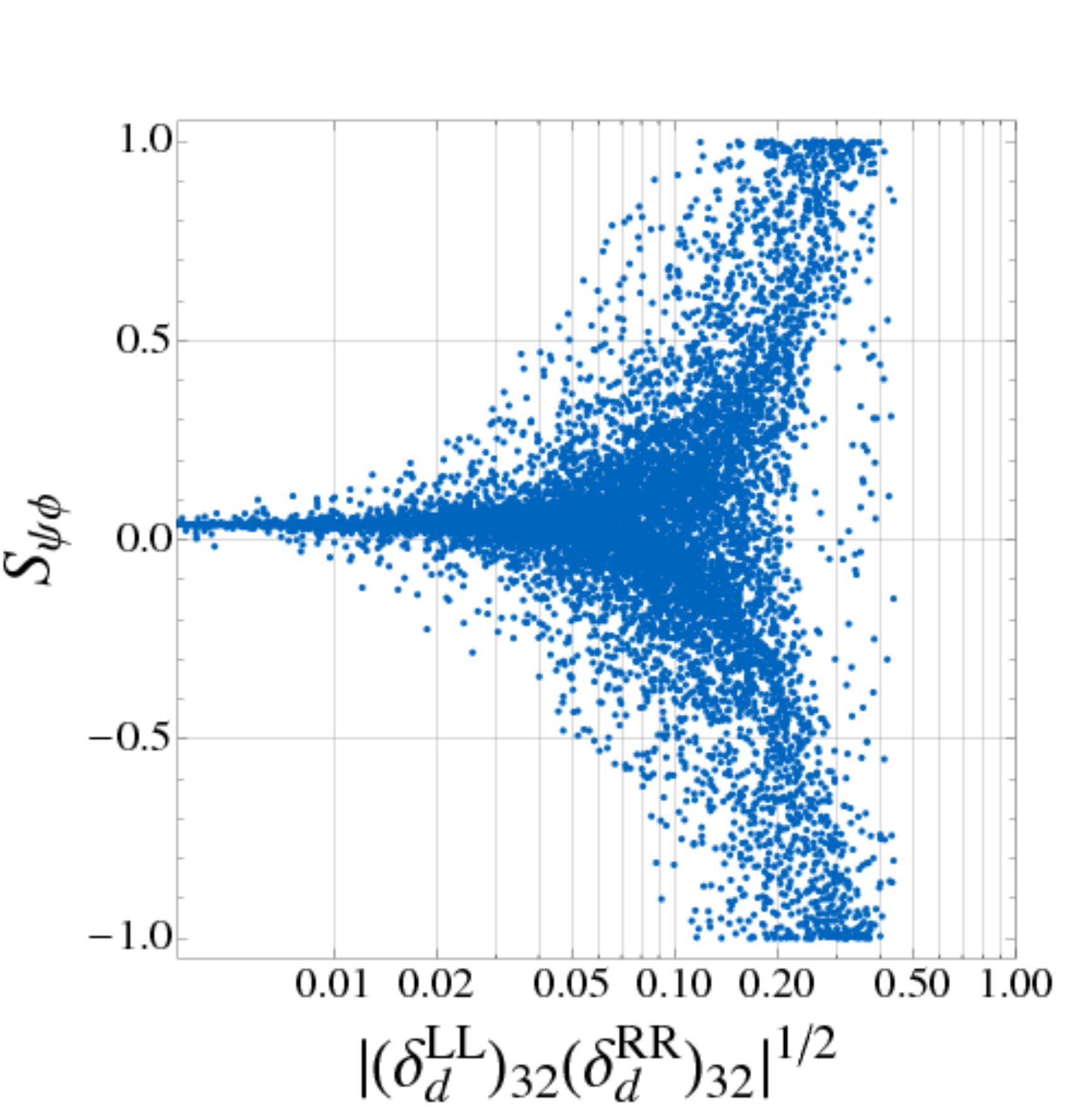}
\caption{\small
$S_{\psi\phi}$ vs. $|(\delta^{LL}_d)_{32}|$, $|(\delta^{RR}_d)_{32}|$
and $|(\delta^{LL}_d)_{32}(\delta^{RR}_d)_{32}|^{1/2}$, respectively.
All the points fulfill the indirect constraints from flavour physics.
}
\label{fig:plots_deltas}
\end{figure}

In fig.~\ref{fig:plots_deltas}, we show the values attained by $S_{\psi\phi}$ as a function of the
modulus of different MIs, i.e. $|(\delta^{LL}_d)_{32}|$ (plot on the left), $|(\delta^{RR}_d)_{32}|$
(plot in the middle) and $|(\delta^{LL}_d)_{32}(\delta^{RR}_d)_{32}|^{1/2}$ (plot on the right),
assuming a CMSSM spectrum with the following ranges for the input parameters: $m_0<1$~\rm{TeV},
$M_{1/2}< 1$~\rm{TeV}, $|A_0|<3m_0$, and $5<\tan\beta< 50$.

This choice of the parameter space is different from that employed before to get the bounds on the
various $\delta$s. In fact, in order to find the maximum allowed values for $S_{\psi\phi}$, it is
crucial to consider the large $\tan\beta$ regime ~--~in order to generate large Higgs mediated effects
to $S_{\psi\phi}$~--~ and to enlarge the ranges for the SUSY mass scale ~--~to relax the indirect
constraints from observables (especially $b\to s\gamma$) decoupling faster than $S_{\psi\phi}$ with
respect to the SUSY mass scale.

The plots of fig.~\ref{fig:plots_deltas} are complementary to those of fig.~\ref{fig:MI23} and they
show that the most natural scenario where it is possible to get large values for $S_{\psi\phi}$ even
for small, CKM-like mixing angles is the third scenario where simultaneously $\delta^{LL}\neq 0$ and
$\delta^{RR}\neq 0$.

\subsection{Step 2: Abelian Model}\label{subsec:abelian model}

\begin{figure}[t]\centering
\includegraphics[width=0.31\textwidth]{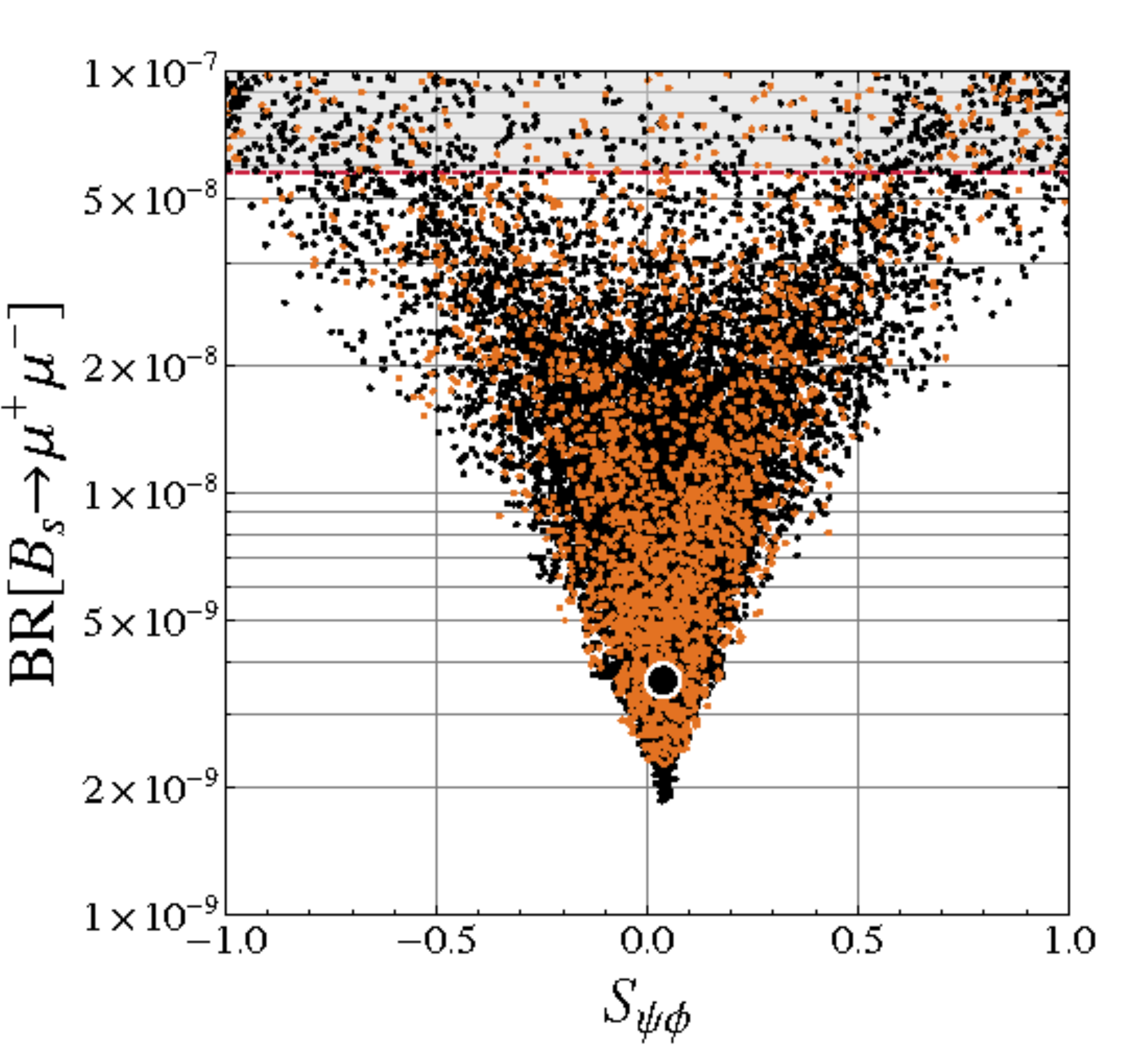}~~~~~
\includegraphics[width=0.29\textwidth]{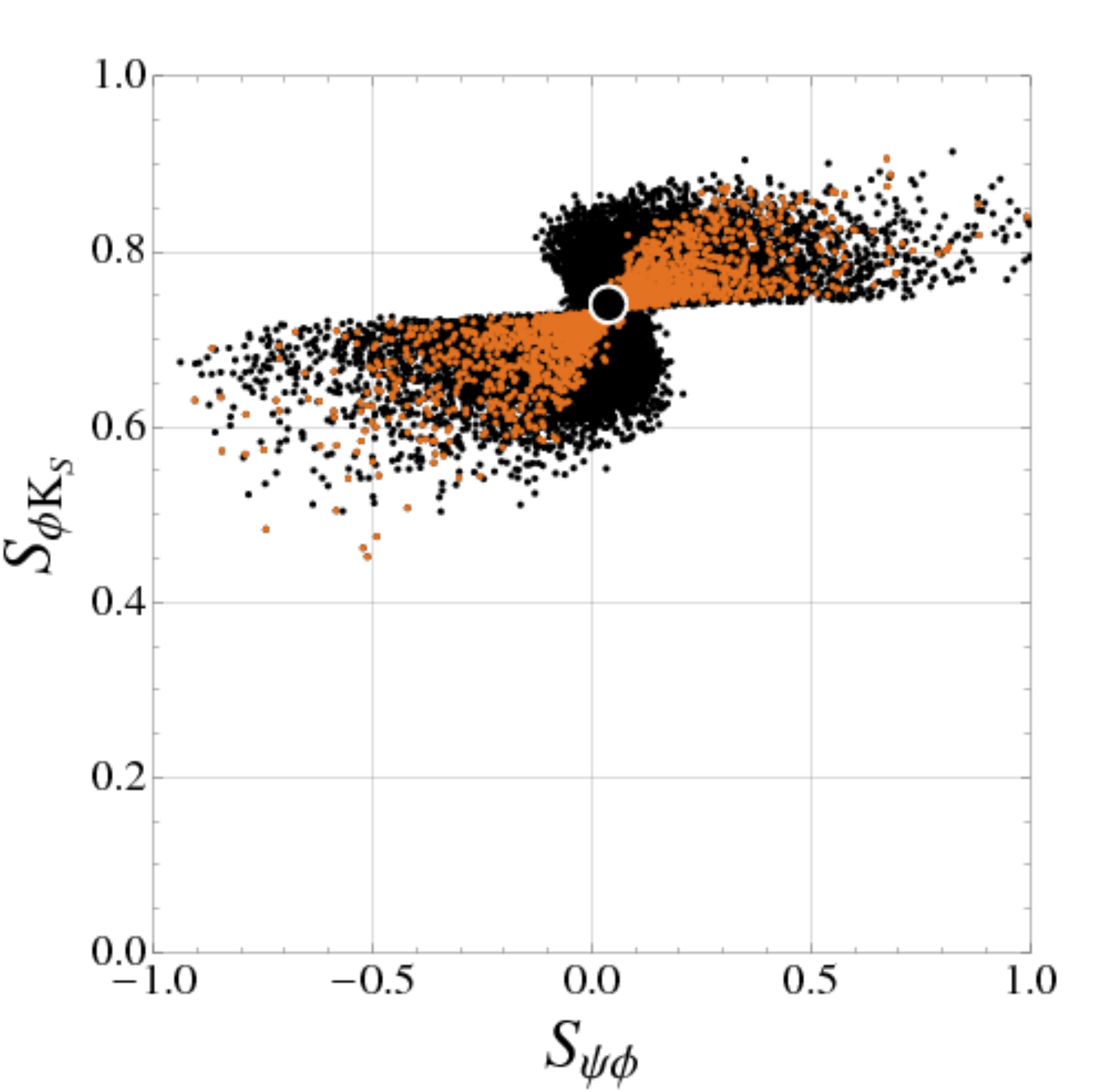}~~~~~
\includegraphics[width=0.295\textwidth]{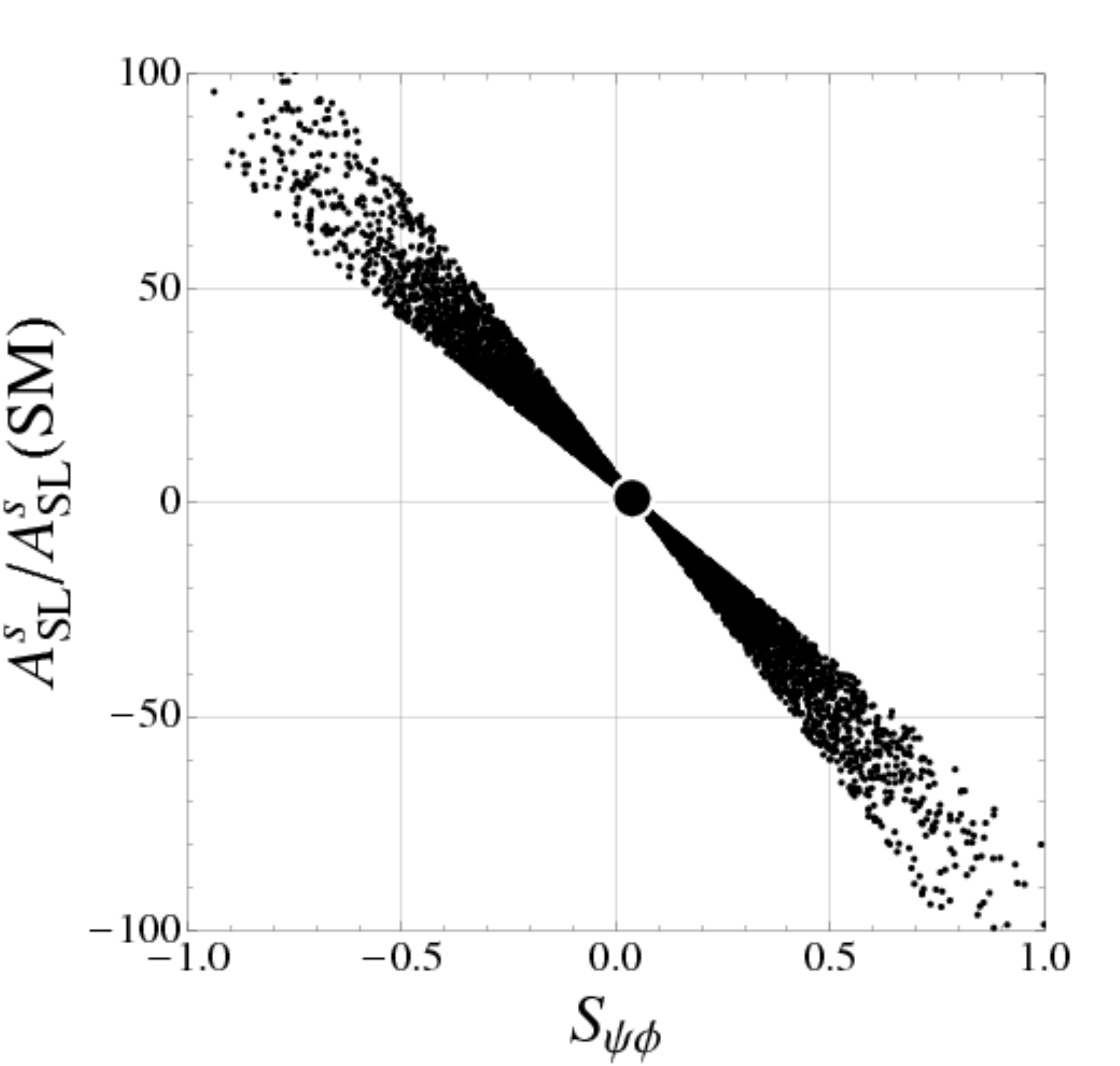}\\[20pt]
\includegraphics[width=0.315\textwidth]{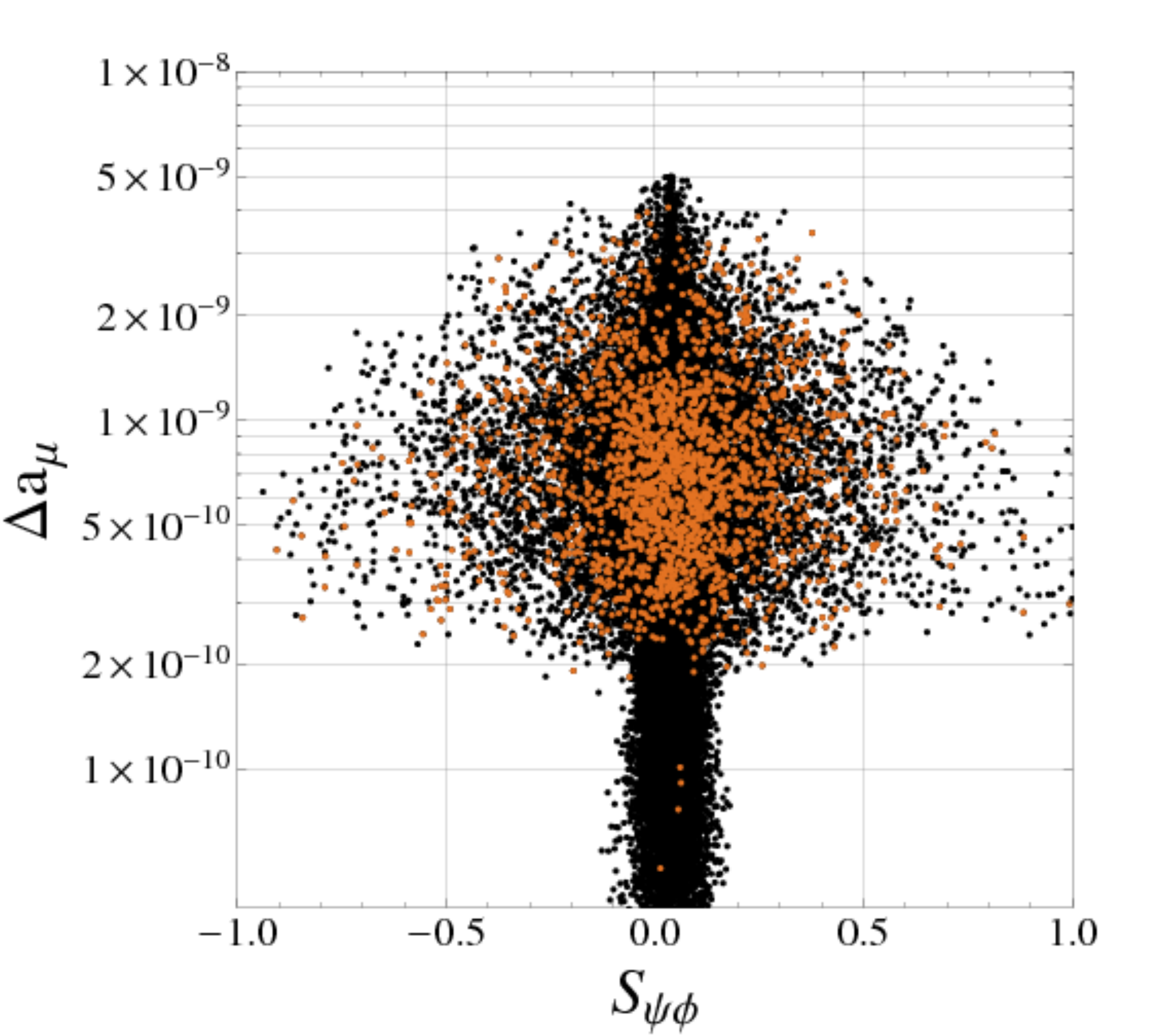}~~~
\includegraphics[width=0.315\textwidth]{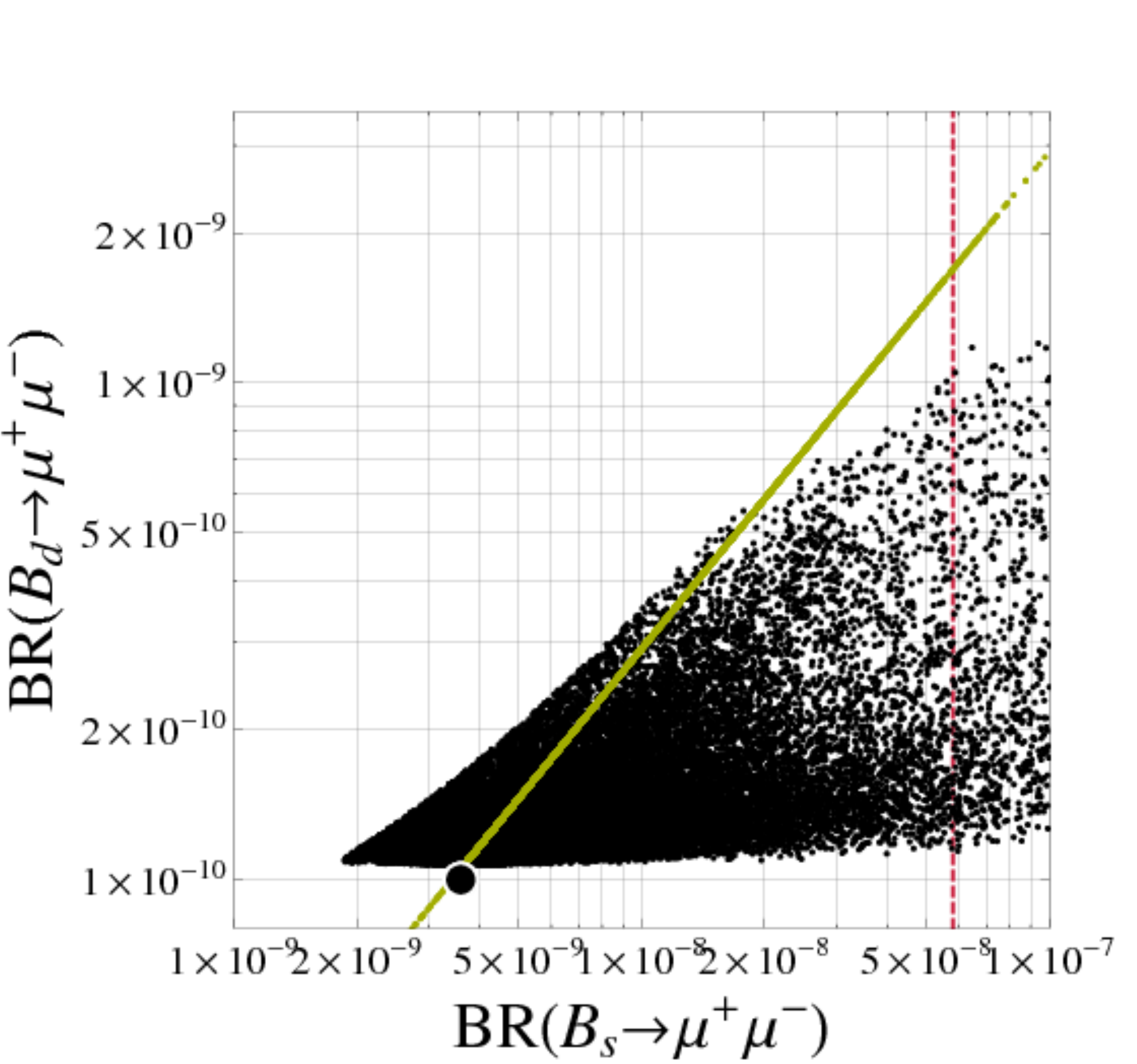}~~~
\includegraphics[width=0.31\textwidth]{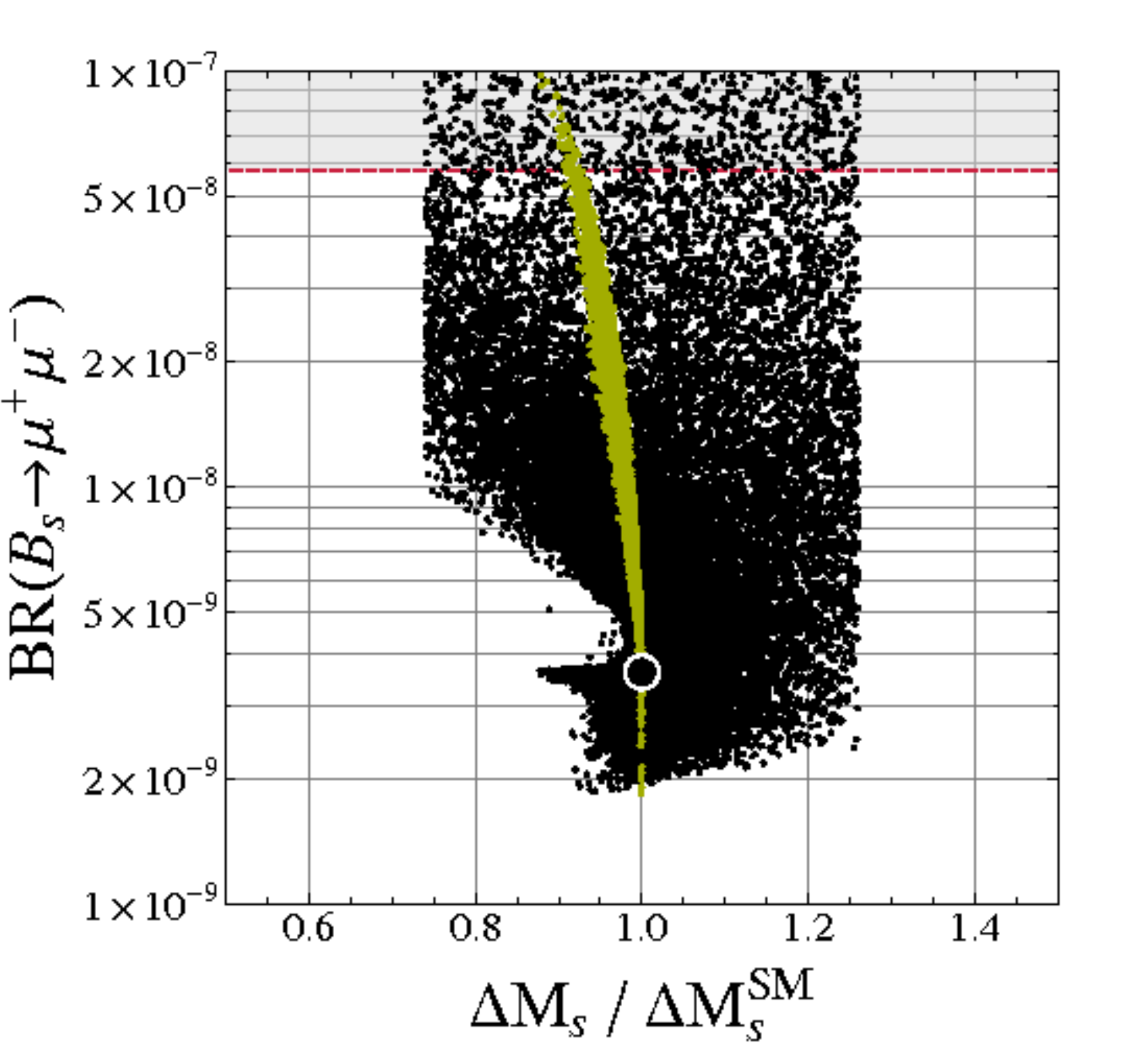}
\caption{\small
Predictions for various observables vs. $S_{\psi\phi}$ in the AC model.
Orange points correspond to negative NP contributions to $\Delta M_d/\Delta M_s$ at the level of $15\%-25\%$ times the SM one, able to solve the UT tension. The green points in the plot of BR$(B_s\to\mu^+\mu^-)$ vs. BR$(B_d\to\mu^+\mu^-)$ and $\Delta M_s/\Delta M_s^{\rm SM}$ vs. BR$(B_s\to\mu^+\mu^-)$ show the correlation
of these observables in the MFV MSSM.}
\label{fig:model_AC}
\end{figure}

In this section, we analyze the abelian flavour model by Agashe and Carone~\cite{Agashe:2003rj},
presented in sec. \ref{sec:strategy}, stressing, in particular, the correlations among various
observables. Here and in the analysis of the other flavour models, we assume a CMSSM spectrum with
the following ranges for the input parameters: $m_0<2$~\rm{TeV}, $M_{1/2}<1$~\rm{TeV}, $|A_0|<3m_0$,
and $5<\tan\beta<55$. Concerning the unknown $\mathcal{O}(1)$ coefficients multiplying the off-diagonal entries in the soft masses~(5.12), they have been varied in the range $\pm[0.5,2]$ in our numerical analysis.

As discussed in sec. \ref{sec:SUSY_breaking}, models with alignment naturally account for
the tight FCNC constraints from $\epsilon_K$ and $\Delta M_K$ by construction. Consequently
in this model the $\epsilon_K$ anomaly discussed in sec. 2 can only be solved by modifying
the value of $R_t$ as also $(\sin 2\beta)_{\psi K_S}$ is SM-like because of the negligibly
small NP contributions to the $b\to d$ transition in this model. Yet we should remark that the 
latter solution implies a value for the angle $\alpha$ significantly below its best determined 
value (see fig.~\ref{fig:UTfit}) and it is to be seen whether such a solution will remain viable.
On the other hand, in contrast to $\epsilon_K$, large effects in $D^0-\bar D^0$ mixing are
expected, provided the first two squark families are non-degenerate, as we expect from
naturalness principles.

In our numerical analysis, concerning the squark mass matrices at the GUT scale, we impose a large splitting between the 1st and 2nd squark generation masses such that $m_{\tilde{u}_L}=2 m_{\tilde{c}_L}=2m_0$,
then the GUT scale MI is effectively $(\delta^{LL}_{u})_{21}\sim \lambda$. However, at the low scale,
where we evaluate the SUSY contributions to the physical observables, $(\delta^{LL}_{u})_{21}$ is significantly smaller than $\lambda$. In fact, there is a degeneracy mechanism triggered mainly by the flavour blind $SU(3)$ interactions that restores a partial degeneracy between the 1st and 2nd generation squark masses~\cite{Dine:1990jd,Brignole:1993dj,Nir:2002ah}. In particular, while the off-diagonal entries
in the squark mass matrices stay almost unaffected during the running from the GUT to the low scale (as discussed in sec.~\ref{sec:running}), the diagonal masses, in contrast, get strongly renormalized by the $SU(3)$ interactions: their GUT scale values $m^2_{\tilde{u}_L}(M_{GUT})=4 m^2_0$ and $m^2_{\tilde{c}_L}(M_{GUT})=m^2_0$ become at the low scale $m^2_{\tilde{u}_L}(M_W)\simeq 4m^2_0+6M^2_{1/2}$ and $m_{\tilde{c}_L}(M_W) \simeq m^2_0+6 M^2_{1/2}$. As a result, the GUT MI
$(\delta^{LL}_u)_{21}\sim\lambda$ is typically reduced by one order of magnitude at the low scale
and the constraints from $D^0-\bar D^0$ mixing can be easily satisfied even for squark masses of
a few hundred $\rm{GeV}$. This is in contrast with the results of a low energy approach were $(\delta^{LL}_u)_{21}\sim \lambda$ holds at the low scale implying a lower bound on the squark
masses of around $2\,\rm{TeV}$~\cite{Ciuchini:2007cw,Nir:2007ac}.

In fig.~\ref{fig:model_AC}, we show the resulting predictions for several observables as functions of $S_{\psi\phi}$. In all the plots, the black points satisfy the constraints of tab.~\ref{tab:observables}
while the orange points correspond to negative NP effects in $\Delta M_d/\Delta M_s$ at the level of
$15\%-25\%$ times the SM one, able to solve the UT tension.

In the first row of fig.~\ref{fig:model_AC}, we show, from left to right, the allowed values for $B_s\to\,\mu^+\mu^-$ vs. $S_{\psi\phi}$, $S_{\phi K_S}$ vs. $S_{\psi\phi}$ and $A^{s}_\text{SL}$ vs. $S_{\psi\phi}$, respectively.
In particular, fig.~\ref{fig:model_AC} shows that large values for $S_{\psi\phi}$ in the range
$-1\lesssim S_{\psi\phi}\lesssim +1$ are still allowed while being compatible with all the constraints
such as $D^0-\bar D^0$ mixing, $\text{BR}(b\to s\gamma)$, $R_{\Delta M}$ and $\epsilon_K$. Interestingly,
the first plot shows that large effects in $S_{\psi\phi}$ predict a lower bound on BR$(B_s\to \mu^+\mu^-)$
at the level of BR$(B_s\to \mu^+\mu^-)> 10^{-8}$ for $|S_{\psi\phi}|\geq 0.3$ (the converse is obviously
not true). The correlation between $S_{\psi\phi}$ and BR$(B_s \to \mu^+\mu^-)$ signals that the double Higgs penguin contributions~(\ref{eq:C4HB}) are responsible for the large effects in $S_{\psi\phi}$ that
we find in this model.

Moreover, the correlation between $S_{\psi\phi}$ and $S_{\phi K_S}$ shown in the second plot indicates that
both asymmetries can simultaneously depart significantly from the SM expectations although an explanation
for the $B_d\to\phi K_S$ anomaly would unambiguously imply negative values for $S_{\psi\phi}$, in contrast
with the present data, in particular when the UT tensions are to be solved in this model. However, even in
this case $S_{\phi K_S}$ can be at most suppressed down to 0.55.

The shape of the correlation between $S_{\psi\phi}$ and $S_{\phi K_S}$, i.e. the fact that a positive
(negative) $S_{\psi\phi}$ implies an enhancement (suppression) of $S_{\phi K_S}$ with respect to its
SM prediction can also be understood analytically. In the considered model, the NP effects in
$S_{\phi K_S}$ are dominantly induced by the Wilson coefficient $\tilde C_8$ given in (\ref{eq:tildeC7_g}).
As we consider only the case of a real and positive $\mu$ parameter, the sign and phase of $\tilde C_8$
and therefore also of the NP contribution to $S_{\phi K_S}$ is fixed by $\left(\delta_{d}^{RR}\right)_{32}$.
Concerning $S_{\psi\phi}$, the dominant NP contribution to the mixing  amplitude $M_{12}^s$ is induced
by the double Higgs penguin contribution (\ref{eq:C4HB}).
In particular we find that in most parts of the parameter space the largest contribution comes from a
double penguin with one gluino and one Higgsino loop (see the diagram in the middle of
fig.~\ref{diagrams_DeltaFeq2}).
The corresponding analytical expression is stated in the second line of (\ref{eq:C4HB}).
As both the trilinear coupling $A_t$ and the loop functions $h_1$ and $h_3$ have a fixed
sign in almost the entire parameter space considered by us, the sign and phase of the NP
contribution to $M_{12}^s$ and hence to $S_{\psi\phi}$ is again determined by
$\left(\delta_d^{RR}\right)_{32}$ and the correlation in the second plot of fig.~\ref{fig:model_AC} emerges.\footnote{We note that if the dominant contribution to $M_{12}^s$ came from gluino
boxes (\ref{eq:MIA2_C4}) or from double penguins with two gluino loops (first line of
(\ref{eq:C4HB})), with $\left(\delta_d^{LL}\right)_{32}$ induced radiatively through RG effects,
the correlation between $S_{\psi\phi}$ and $S_{\phi K_S}$ would have the opposite sign.}

As the CDF and D0 data clearly favour $S_{\psi\phi}$ to be positive, we conclude that from the
point of view of the AC model
\begin{itemize}
\item
Either $0\le S_{\psi\phi}\le 0.15$, $S_{\phi K_S}\ge 0.60$ and the UT tensions will go away, or
\item
$S_{\psi\phi}$ can be as large as $1$, the UT tension is solved in this model
but $S_{\phi K_S}$  will eventually turn out to be SM-like.
\end{itemize}

The third plot confirms the model-independent correlation between $S_{\psi\phi}$ and $A_\text{SL}^s$ \cite{Ligeti:2006pm,Grossman:2009mn} discussed in sec.~\ref{sec:Bs_system}, and shows that 
the dileptonic asymmetry $A^{s}_\text{SL}$ normalized to its SM value  can lie in the range
$-100 \lesssim A^{s}_\text{SL}/(A^{s}_\text{SL})_\text{SM} \lesssim 100$ for
$-1\lesssim S_{\psi\phi}\lesssim 1$.

In the second row of the plot, from left to right, we show $\Delta a_{\mu}$ vs. $S_{\psi\phi}$, $B_d\to\,\mu^+\mu^-$ vs. $B_s\to\,\mu^+\mu^-$ and $B_s\to\,\mu^+\mu^-$ vs. $\Delta M_s/(\Delta M_s)_\text{SM}$, respectively. Interestingly, there are many points accounting for large (non-standard) values for $S_{\psi\phi}$ while providing a natural explanation of the $(g-2)_\mu$ anomaly. 
Moreover, the AC model can predict very striking deviations from the MFV SUSY expectations for $\text{BR}(B_d\to\,\mu^+\mu^-)/\text{BR}(B_s\to\,\mu^+\mu^-)$ and $\text{BR}(B_s\to\,\mu^+\mu^-)/\Delta M_s/(\Delta M_s)_\text{SM}$.
Hence, the above two ratios represent very clean and powerful observables to test the MFV hypothesis.
In particular, as the AC model predicts new flavour structures only in the $b\to s$ sector, it turns
out that only $B_s\to\,\mu^+\mu^-$ can depart from the MFV predictions.
More specifically, the ratio ${\rm BR}(B_d\to\mu^+\mu^-)/{\rm BR}(B_s\to\mu^+\mu^-)$ is
dominantly below its MFV prediction and can be much smaller than the latter.

Moreover, within a MFV framework, the current experimental constraints from $\text{BR}(B_s\to\,\mu^+\mu^-)$ already imply a very small NP room in $\Delta M_s/(\Delta M_s)_\text{SM}$, typically at a level smaller
than $10\%$.
In contrast, the AC model can predict quite sizable effects in $\Delta M_s/(\Delta M_s)_\text{SM}$,
as it is evident from fig.~\ref{fig:model_AC}, which might saturate the $R_{\Delta M}$ constraints.

Further, even if not shown in fig.~\ref{fig:model_AC}, the CP asymmetry in $b\to s\gamma$ is SM-like
as it is basically not sensitive to right-handed currents.

In this model, NP contributions to $K\to\pi\nu\bar\nu$ decays are very strongly suppressed and the
resulting branching ratios are SM-like. Consequently, if future experiments on these decays will
show large departures from the SM expectations, the AC model will be ruled out.

Concerning the CP asymmetries in $B\to K^*\mu^+\mu^-$, we have found only small effects, at most at the
percent level, after imposing all the indirect constraints, especially those from $D^0-\overline{D}^0$
mixing.

Finally, we point out that in the AC model, as well as in many other abelian flavour
models, large effects for the neutron EDM are also expected. In fact, the peculiar flavour
structure of the abelian flavour models ~--~predicting large mixing angles in the 12
up-squark sector~--~ leads to the following order of magnitude value for the up-quark EDM $d_u$
(see~(\ref{Eq:edm_u_gluino}))
\beq
\left|\left(\frac{d_{u}}{e}\right)_{\tilde g}\right|
\approx
10^{-26}\left(\frac{2 \rm{TeV}}{{\tilde m}}\right)^{2}
\frac{\left|{\rm Im}\left[(\delta^{LL}_{u})_{12}(\delta^{RR}_{u})_{21}\right]\right|}{\lambda^{4}}
\,,
\eeq
where we have assumed a degenerate SUSY spectrum, for simplicity. A similar expression holds
for the up-quark CEDM too. Since $d_n \sim d_u$~(\ref{Eq:dn_odd}), we conclude that
$\mathcal{O}(1)$ phases for $(\delta^{RR}_{u})_{12}$ ($\phi_{uR}$) of abelian flavour models 
lead to a $d_n$ close to its current experimental upper bound.

Analogously, CPV effects in $D^0-\overline{D}^0$ mixing receive the dominant contributions 
by ${\rm Im}\left[(\delta^{LL}_{u})_{12}(\delta^{RR}_{u})_{12}\right]$.
Hence, large CPV effects in $D^0-\overline{D}^0$ mixing would imply experimentally visible
values for the neutron EDM by means of the up-quark EDM.

Further, in the AC model, large CPV effects in $B_s$ systems unambiguously imply a very large
strange quark EDM. In particular, one can find that (see~(\ref{Eq:edm_d_gluino}))
\beq
\left|\left(\frac{d_{s}}{e}\right)_{\tilde g}\right|
\approx
10^{-23}\times\left(\frac{2 \rm{TeV}}{{\tilde m}}\right)^{2}
\frac{\left|{\rm Im}\left[(\delta^{LL}_{d})_{23}(\delta^{RR}_{d})_{32}\right]\right|}{\lambda^{2}}
\left(\frac{\tan\beta}{50}\right)\,,
\eeq
hence, the current experimental bounds on $d_n$ already tell us that either the strange quark
contributions to $d_n$ have to be very small, with a proportionality coefficient smaller than
$10^{-3}$, or that $\mathcal{O}(1)$ CPV phases for $(\delta^{LL}_{d})_{32}$ ($\phi_{dL}$) and/or
$(\delta^{RR}_{d})_{32}$ ($\phi_{dR}$) are not allowed (unless $\phi_{dL}=\phi_{dR}$).
In the former case, large CPV effects in $B_s$ systems are still allowed while they are excluded
in the latter case (unless $\phi_{dL}=\phi_{dR}$). In this respect, a reliable knowledge of the 
order of magnitude for the strange quark contributions to $d_n$, by means of lattice QCD techniques,
would be of the utmost importance to probe or to falsify flavour models with large RH currents in 
the 2-3 sector embedded in a SUSY framework.

In summary the striking predictions of the AC model in case the $S_{\psi\phi}$ anomaly will be
confirmed by more accurate data are:
\begin{itemize}
\item
  The enhancement of $\text{BR}(B_s\to\,\mu^+\mu^-)$ above $10^{-8}$,
\item
  $\Delta a_\mu\approx 10^{-9}$, thereby providing a natural explanation of the
  $(g-2)_\mu$ anomaly.
\item
  $\text{BR}(B_d\to\,\mu^+\mu^-)/\text{BR}(B_s\to\,\mu^+\mu^-)$ and
  $\text{BR}(B_s\to\,\mu^+\mu^-)/\Delta M_s$ possibly very different from the 
  MFV expectations with the first ratio dominantly smaller than its MFV value.
  Note however that $\text{BR}(B_d\to\,\mu^+\mu^-)$ can reach values by a factor 
  of 10 larger than in the SM.
\item
  $S_{\phi K_S}\approx S_{\psi K_S}$
\item
  Simultaneously UT tensions can be solved through the shift in $\Delta M_d/\Delta M_s$
  at the prize of a rather low $\alpha\approx 75^\circ$.
 \item
 Large effects in $D^0-\bar D^0$ mixing but very small effects in $K^0-\bar K^0$ mixing.
\item Large values for the neutron EDM ~--~very close to the current experimental bound~--~
generated either by the up-quark (C)EDM or by the strange-quark (C)EDM.
In the former case, visible CPV effects in $D^0-\bar D^0$ mixing are also expected while they
are not necessarily implied in the latter case. A correlated study of several hadronic EDMs,
with different sensitivity to the up-quark (C)EDM and the strange-quark (C)EDM, would provide
a precious tool to unveil the peculiar source of CPV that is generating the neutron EDM.
\end{itemize}


\subsection[Step 3: Non-abelian SU(3) Models]{\boldmath Step 3: Non-abelian $SU(3)$ Models}\label{sec:nonabeliannumerics}

We now turn to the analysis of another very interesting flavour model, the model by Ross and
collaborators (RVV)~\cite{Ross:2004qn}, or more specifically its particular version RVV2 recently
discussed in~\cite{Calibbi:2009ja}. It is  based on the $SU(3)$ flavour symmetry, as was presented
already in sec.~\ref{sec:strategy}. As already stressed in previous sections, the symmetry 
properties of non-abelian flavour models naturally account for degenerate squarks of the first two
generations (at least) solving thereby the flavour problem related to the experimental constraints 
from $\epsilon_K$ and $\Delta M_K$.
As we will see, quite large ~--~but still experimentally allowed effects in $\epsilon_K$~--~ can arise.
Thus, in contrast to the AC model, the first solution to the UT tensions, through NP contribution to 
$\epsilon_K$, becomes viable.
Concerning the phenomenology of the up sector, the RVV model~\cite{Ross:2004qn} and its RVV2
version predict very small (most likely untestable) effects in the $D^0-\bar D^0$ mixing observables.
These are probably the most peculiar features of non-abelian flavour models compared to the abelian
ones with alignment, where the NP effects in $\epsilon_K$ are completely negligible by construction
while large effects in $D^0-\bar D^0$ mixing are unavoidable.

On the other hand, similarly to the AC model, also the RVV2 model exhibits large RH currents.
Hence $S_{\psi\phi}$ can be large so that the analysis can be done along the lines of the one
done for the AC model making the comparison of both models very transparent.

A very interesting aspect of the RVV model (as well as its RVV2 version) is that it is
embedded in a SUSY GUT $\rm{SO}(10)$ model so that correlations among flavour violating processes
in the lepton and quark sectors naturally occur making additional tests of this model possible.

As a second non-abelian model, we also consider the $SU(3)$ model proposed by Antusch et al.~\cite{Antusch:2007re} (AKM model), already introduced in sec. \ref{sec:strategy}.
An interesting feature of the AKM model, that is in contrast to the RVV2 model, is the
presence of a leading $\mathcal O(1)$ CPV phase only in the 23 RR sector but not in the 
12 and 13 sectors.
This will turn out to be crucial to generate CPV effects in the $B_s$ mixing amplitude.
Moreover, in the RVV2 model it turns out that $(\delta_{d}^{RR})_{23}\simeq\bar\varepsilon$,
while in the AKM model we have $(\delta_{d}^{RR})_{23}\simeq\bar\varepsilon^2$ with
$\bar\varepsilon\simeq 0.15$.
Hence, we expect larger CPV effects in the RVV2 than in the AKM models. We stress that, even if
both models arise from an underlying $SU(3)$ flavour symmetry, their low energy predictions can
be quite different as their soft sectors, that cannot be set uniquely by the flavour symmetry,
are different.

\subsubsection{RVV2 Model: Results in the Hadronic and Leptonic Sectors}

\begin{figure}
\centering
\includegraphics[width=0.31\textwidth]{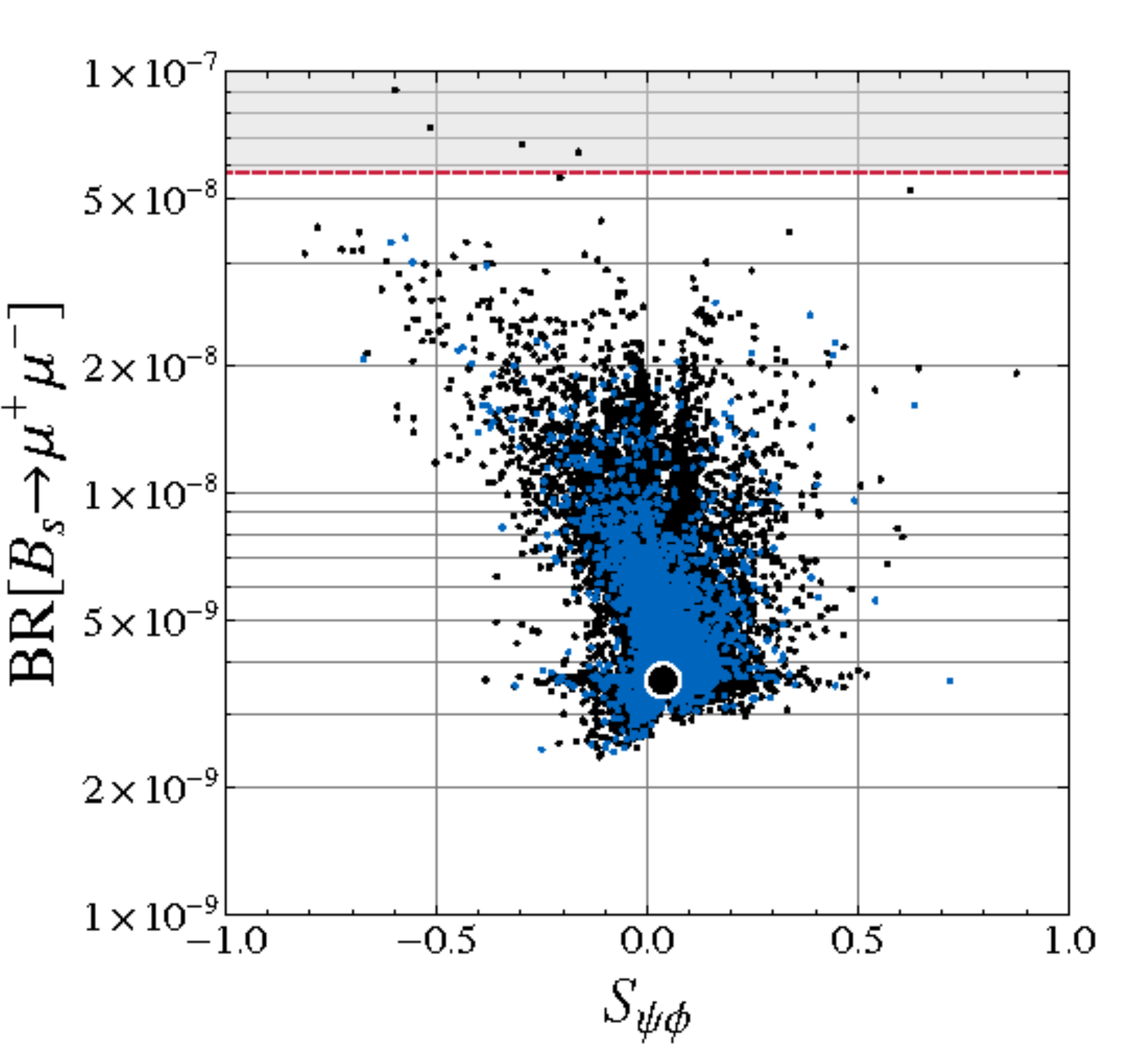}~~~~~
\includegraphics[width=0.29\textwidth]{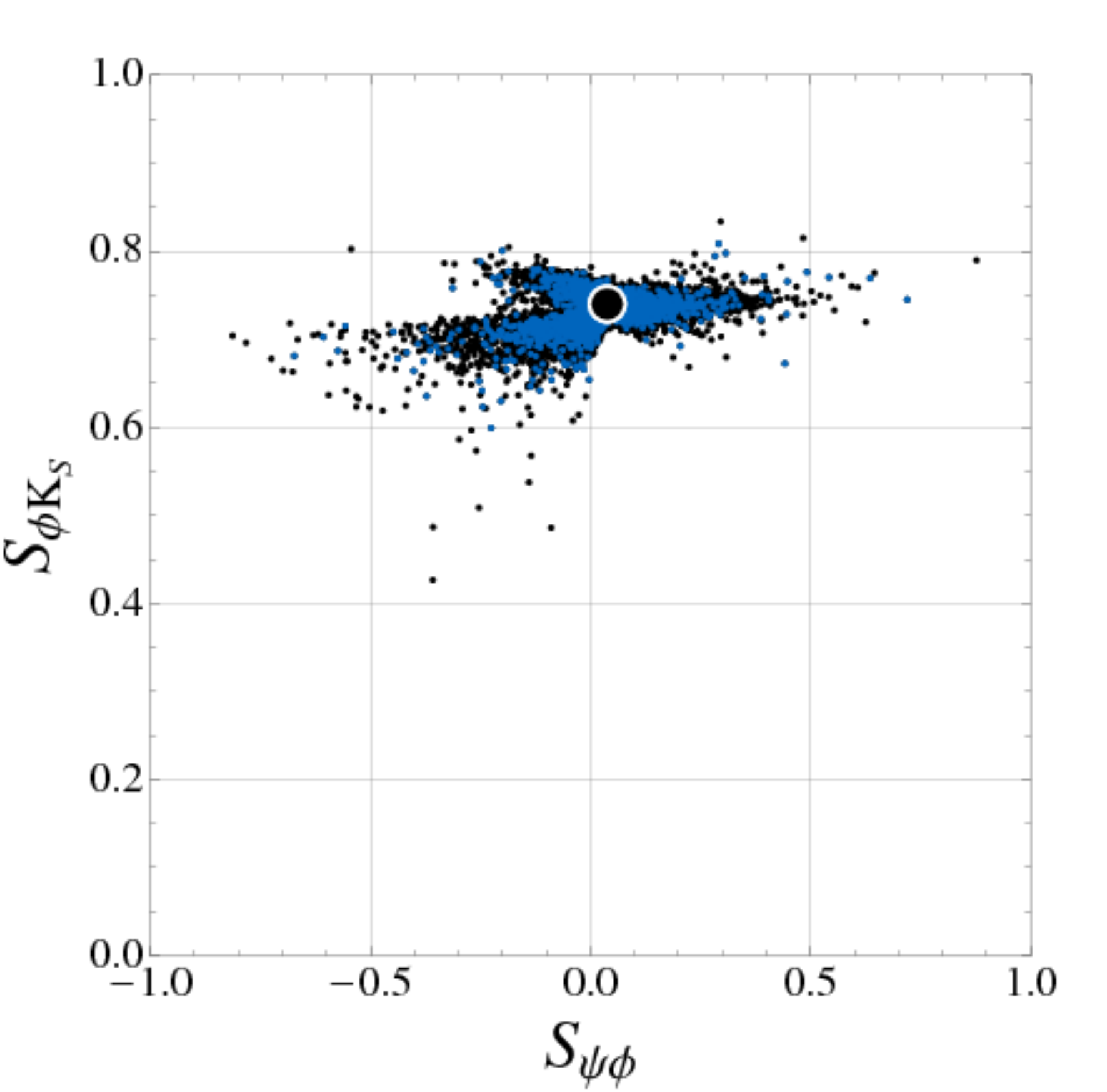}~~~~~
\includegraphics[width=0.295\textwidth]{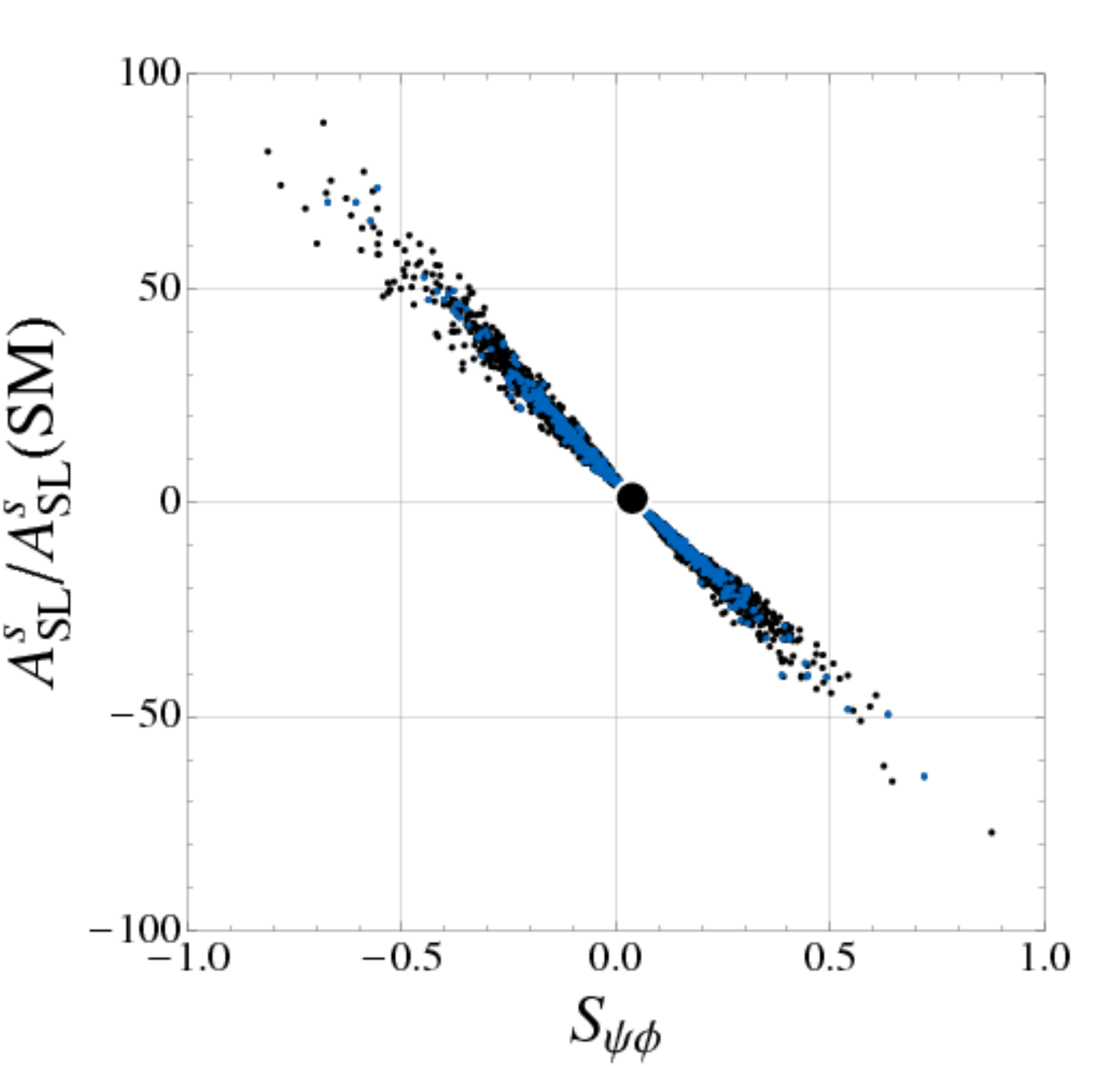}\\[20pt]
\includegraphics[width=0.315\textwidth]{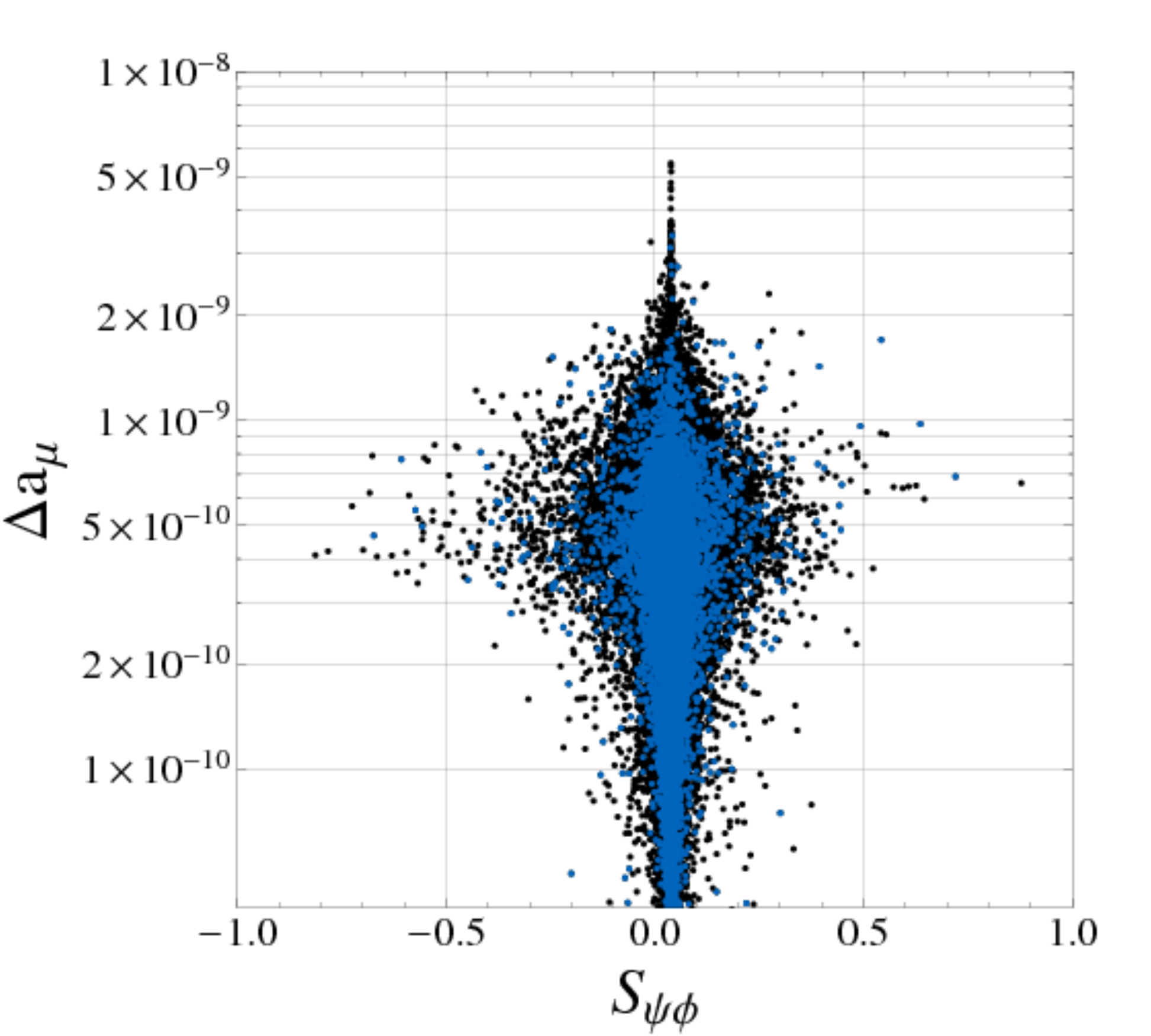}~~~
\includegraphics[width=0.315\textwidth]{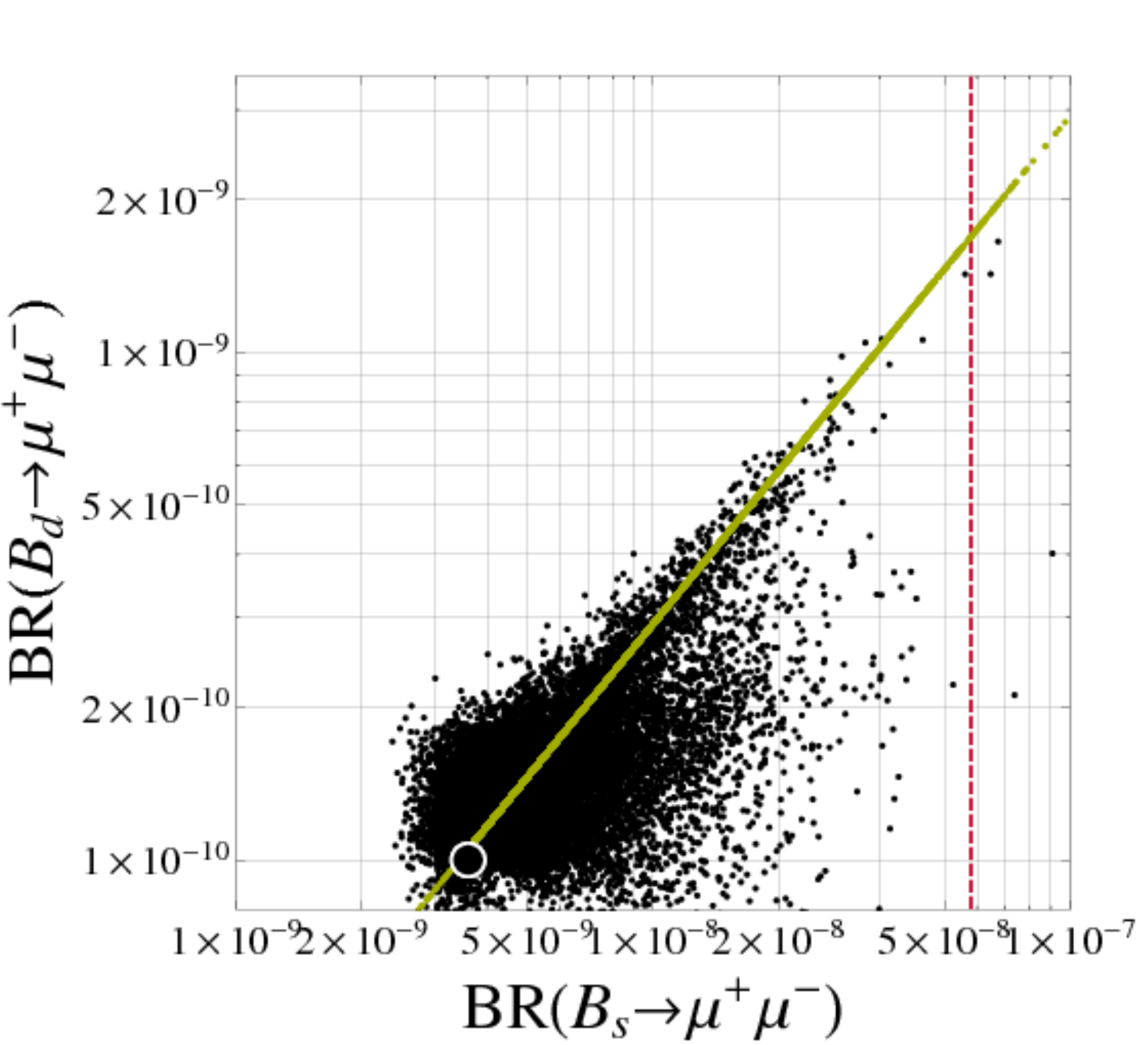}~~~
\includegraphics[width=0.31\textwidth]{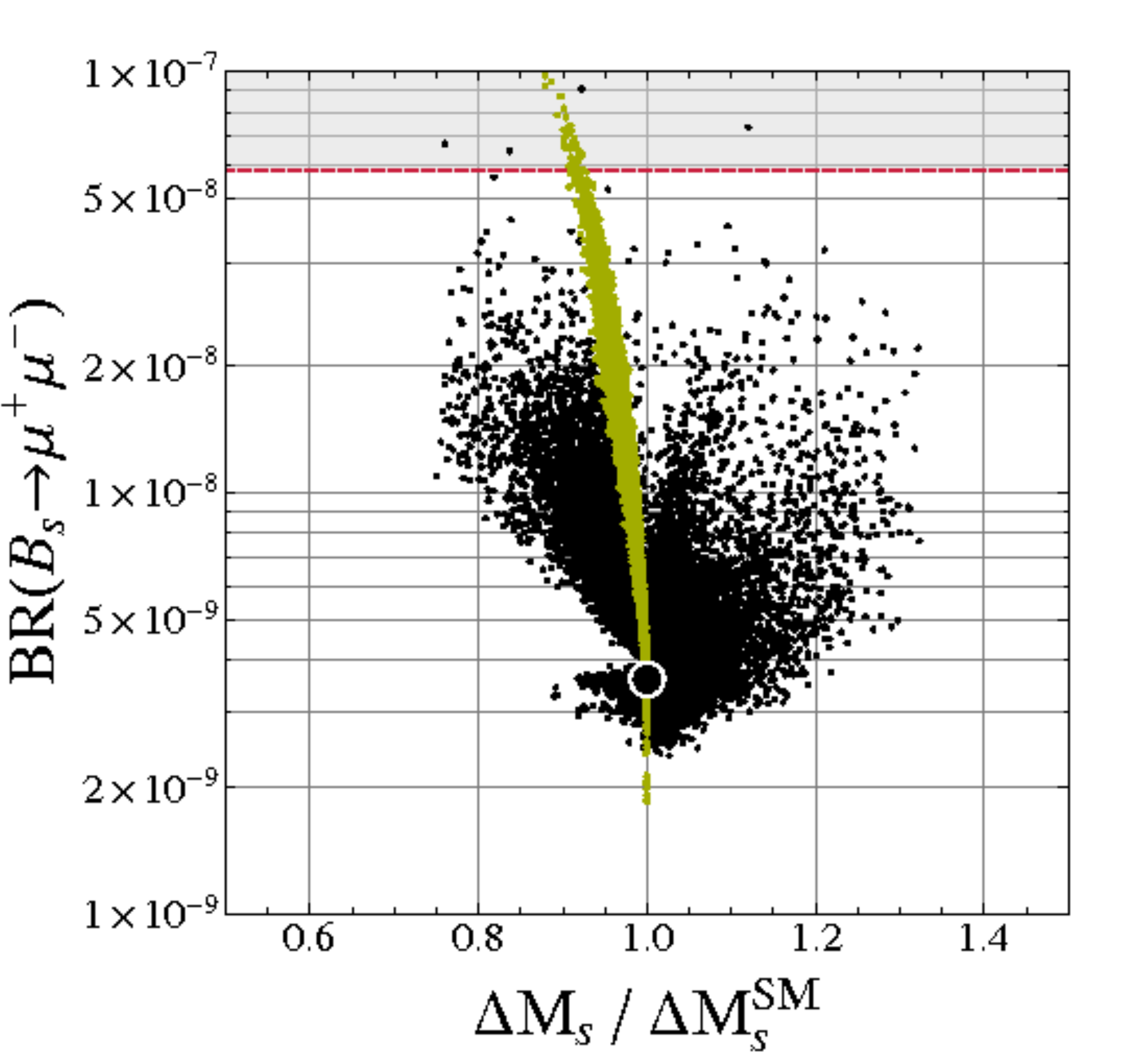}\\[20pt]
\includegraphics[width=0.305\textwidth]{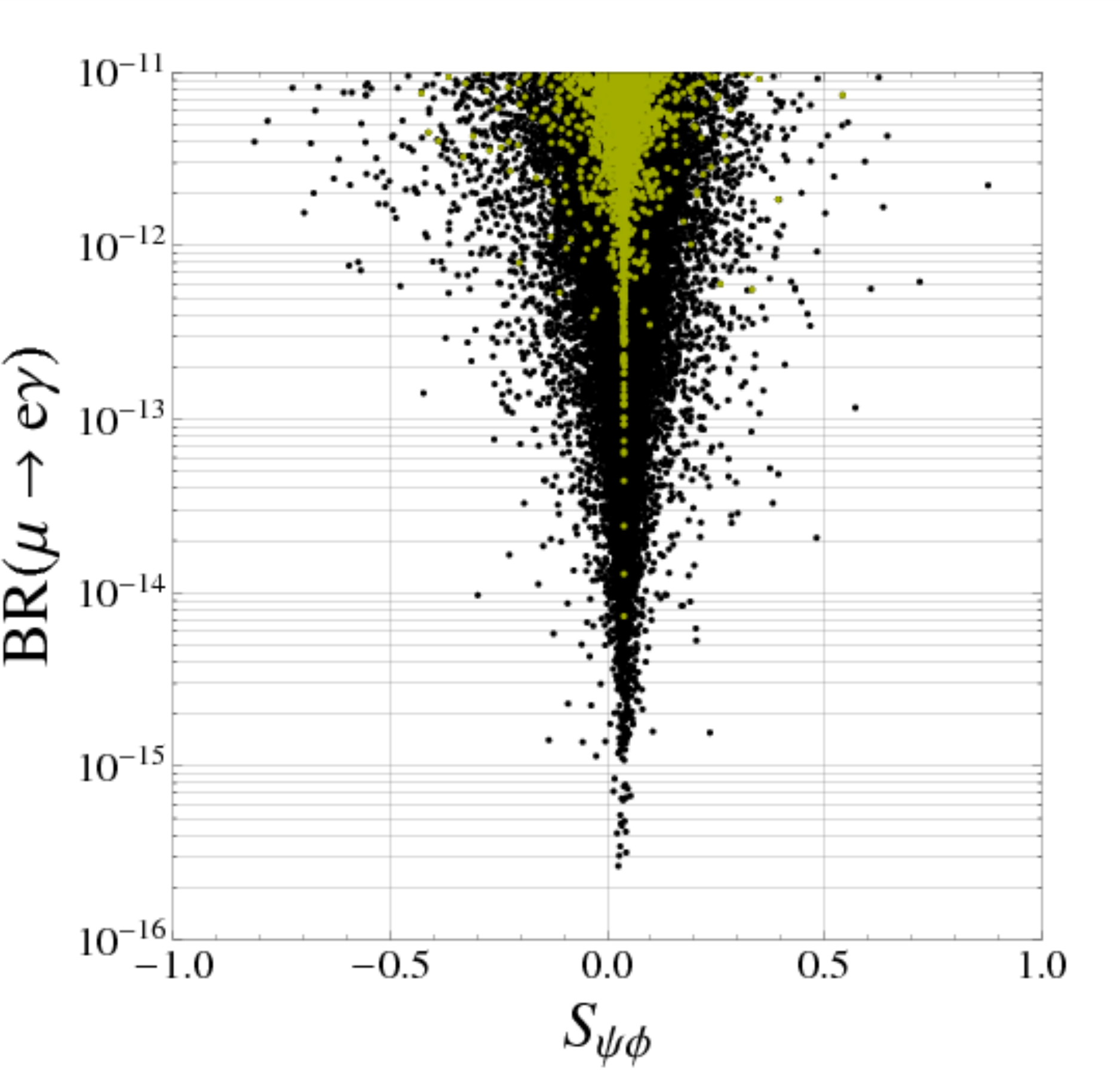}~~~~
\includegraphics[width=0.305\textwidth]{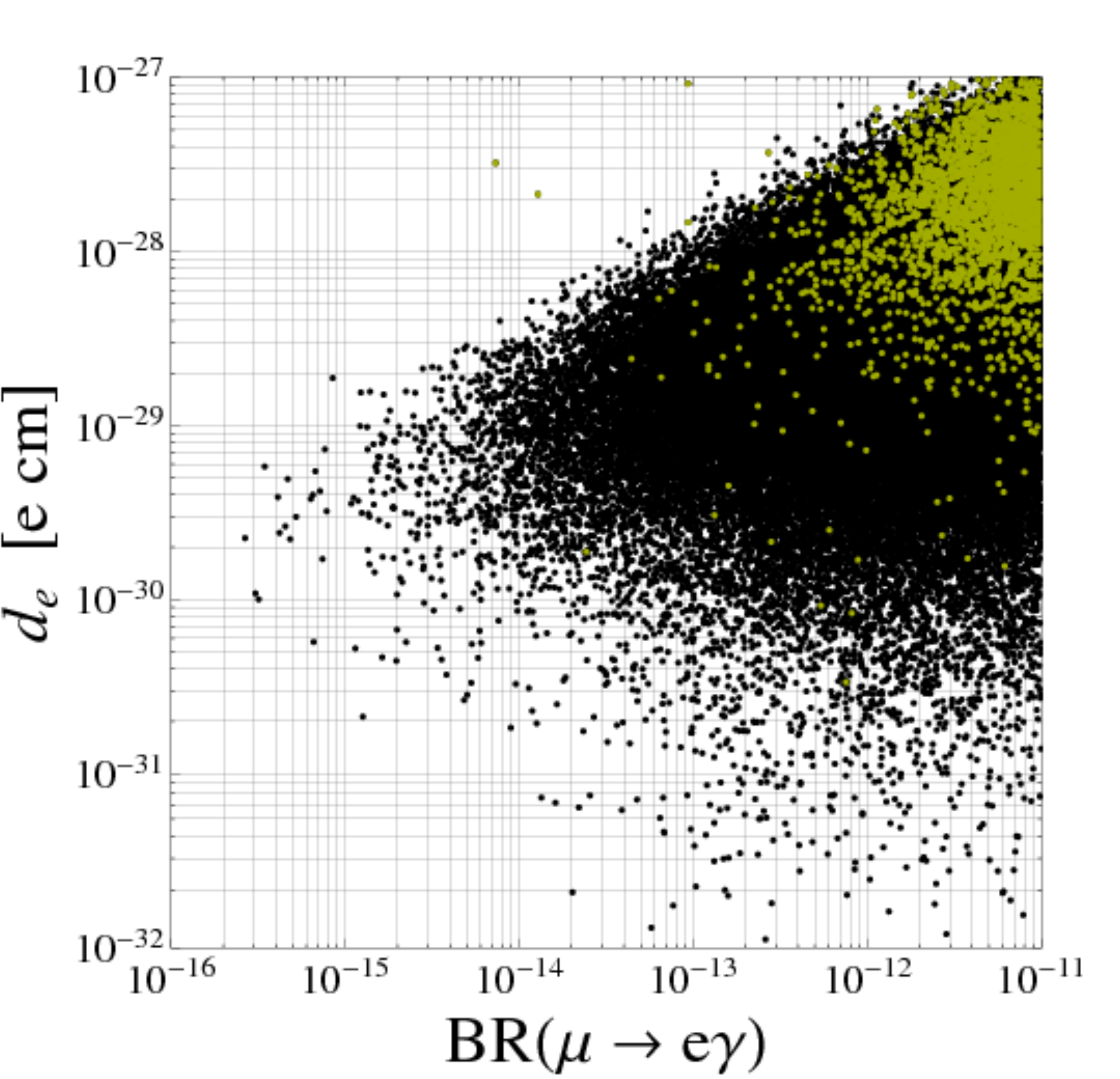}~~~~
\includegraphics[width=0.305\textwidth]{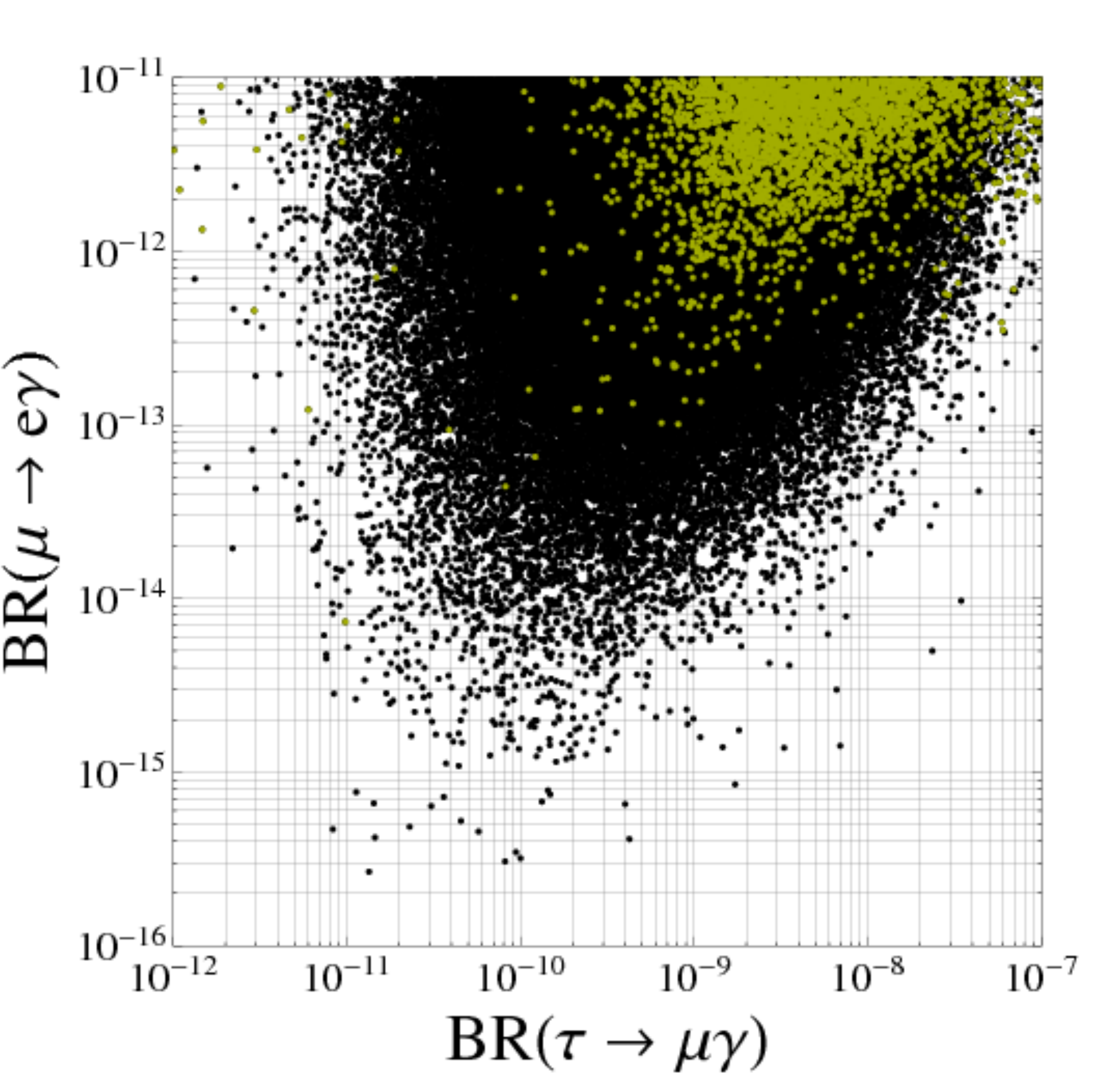}
\caption{\small
Predictions of the RVV2 model both in the hadronic and leptonic sectors.
In the first two rows we show the predictions for  various
observables vs. $S_{\psi\phi}$. The blue points correspond to positive NP effects in $\epsilon_K$ such
that $1.2<\epsilon_K/(\epsilon_K)_\text{SM}<1.3$ and $\Delta M_d/\Delta M_s$ is SM-like.
The green points in the plots of BR$(B_s\to\mu^+\mu^-)$ vs. BR$(B_d\to\mu^+\mu^-)$ and
$\Delta M_s/\Delta M_s^{\rm SM}$ vs. BR$(B_s\to\mu^+\mu^-)$ show the correlation of
these observables in the MFV MSSM. The last row refers to the predictions for leptonic
observables. The green points explain the $(g-2)_{\mu}$ anomaly at the $95\%$ C.L.,
i.e. $\Delta a_{\mu}> 1\times 10^{-9}$.}
\label{fig:model_ross}
\end{figure}

\begin{figure}
\centering
\includegraphics[width=0.31\textwidth]{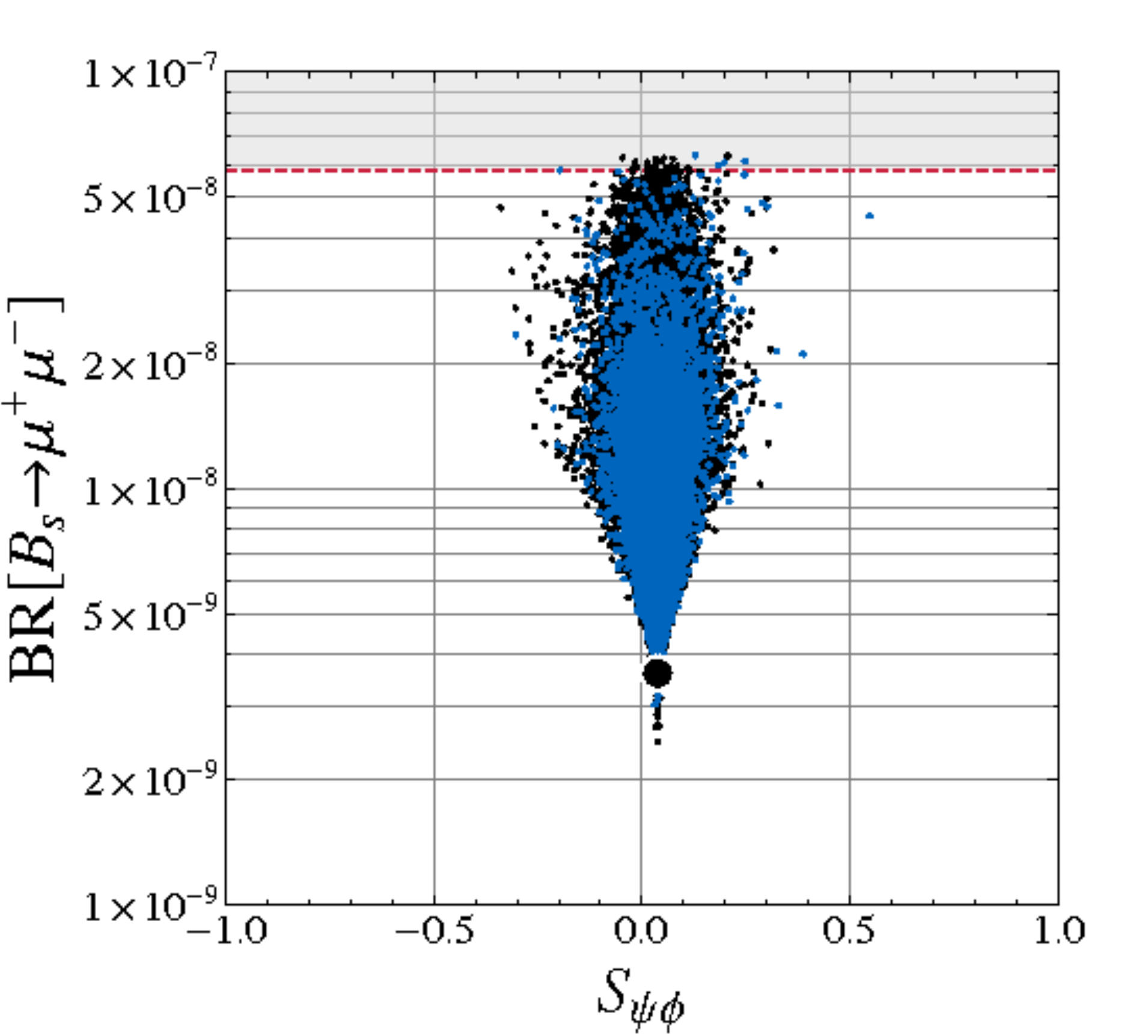}~~~~~
\includegraphics[width=0.29\textwidth]{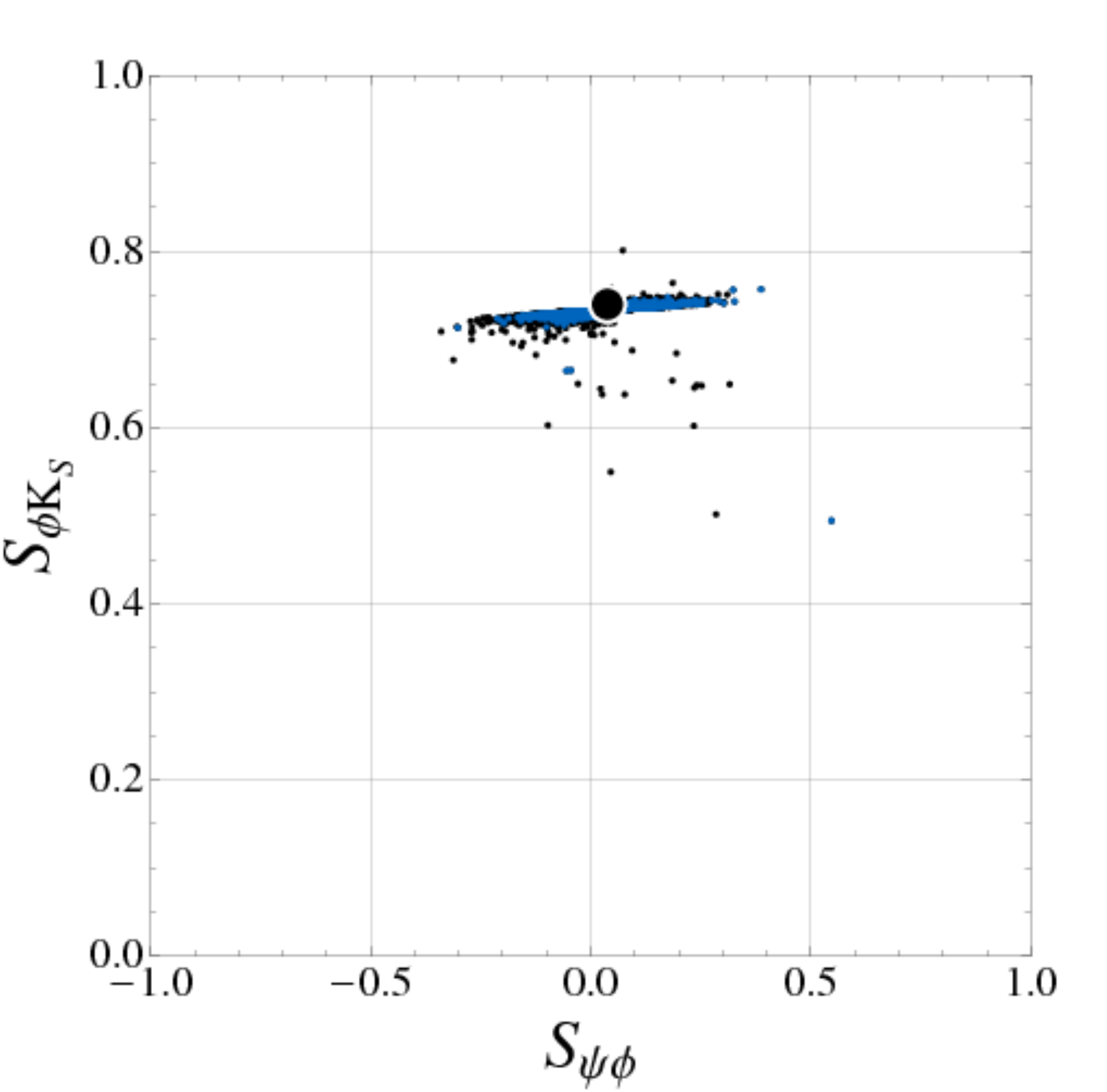}~~~~~
\includegraphics[width=0.295\textwidth]{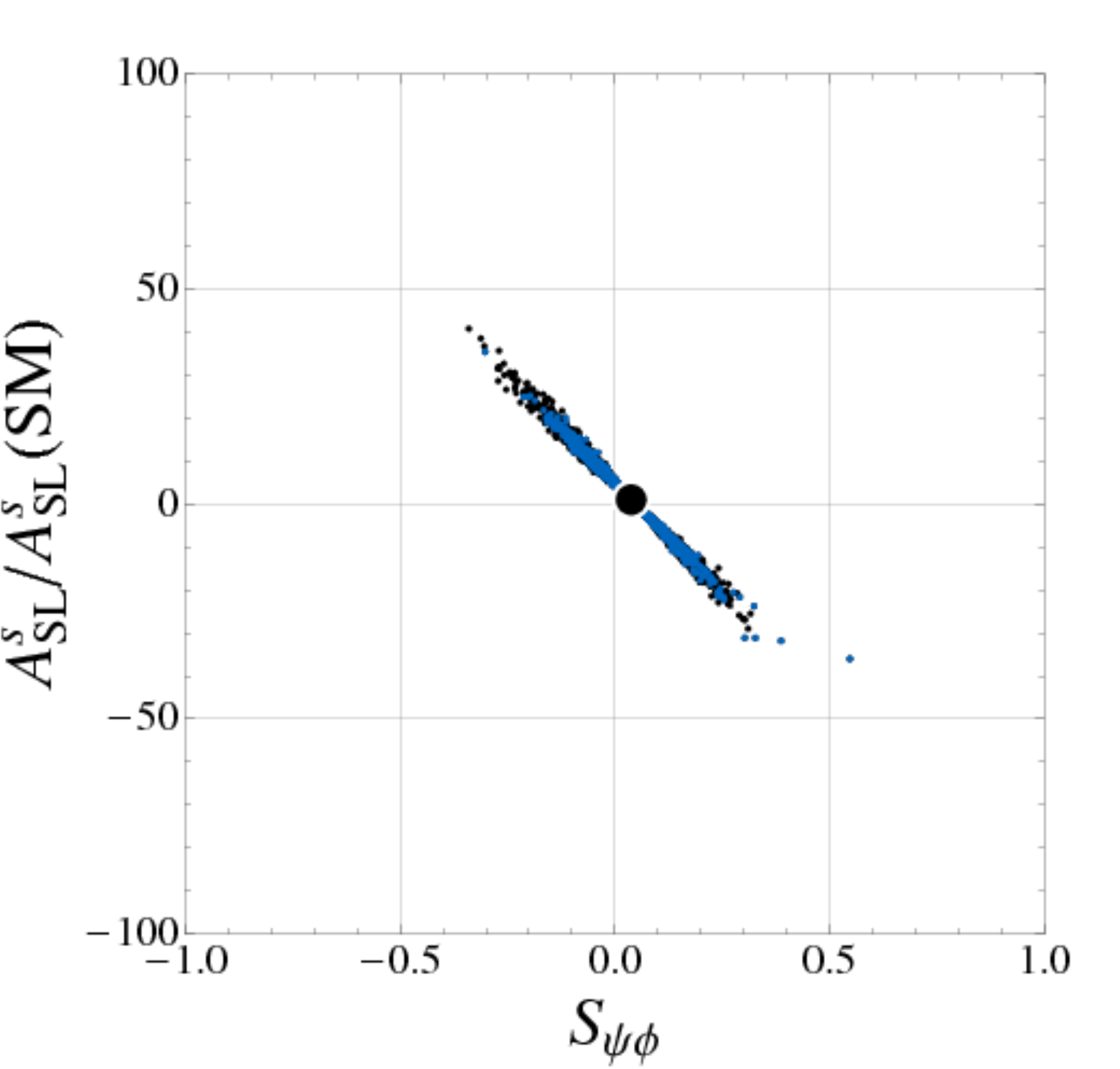}\\[20pt]
\includegraphics[width=0.315\textwidth]{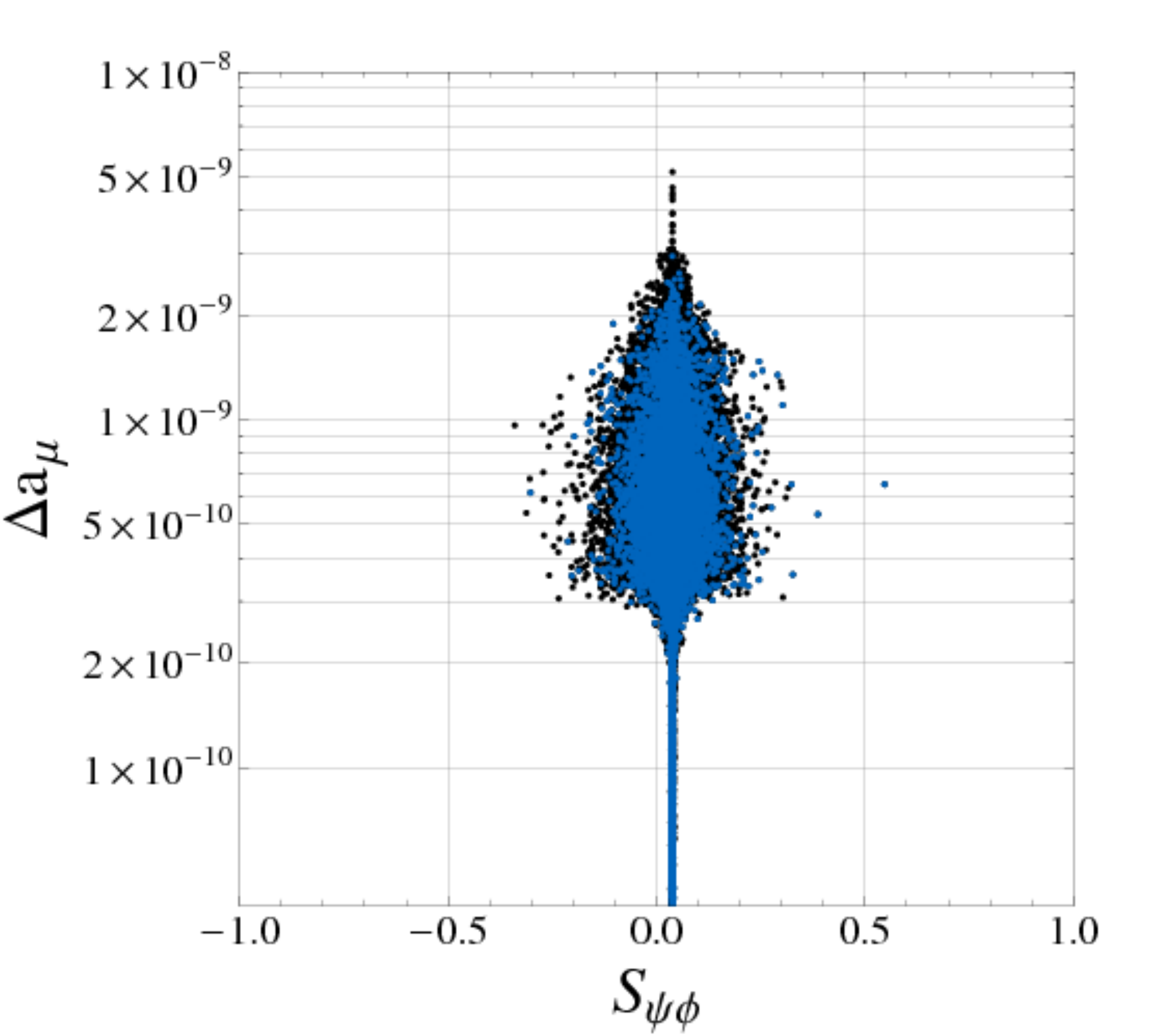}~~~
\includegraphics[width=0.315\textwidth]{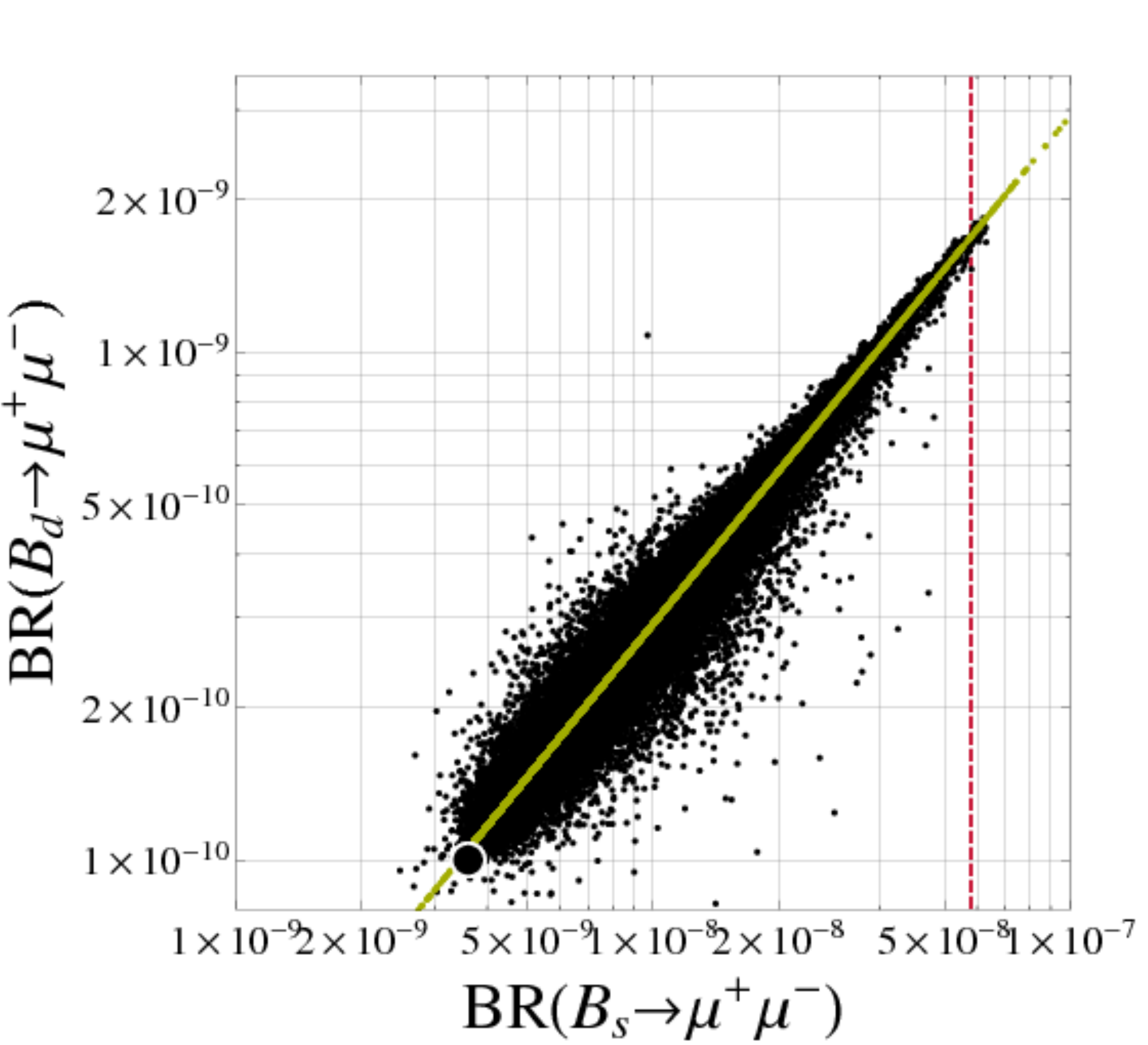}~~~
\includegraphics[width=0.31\textwidth]{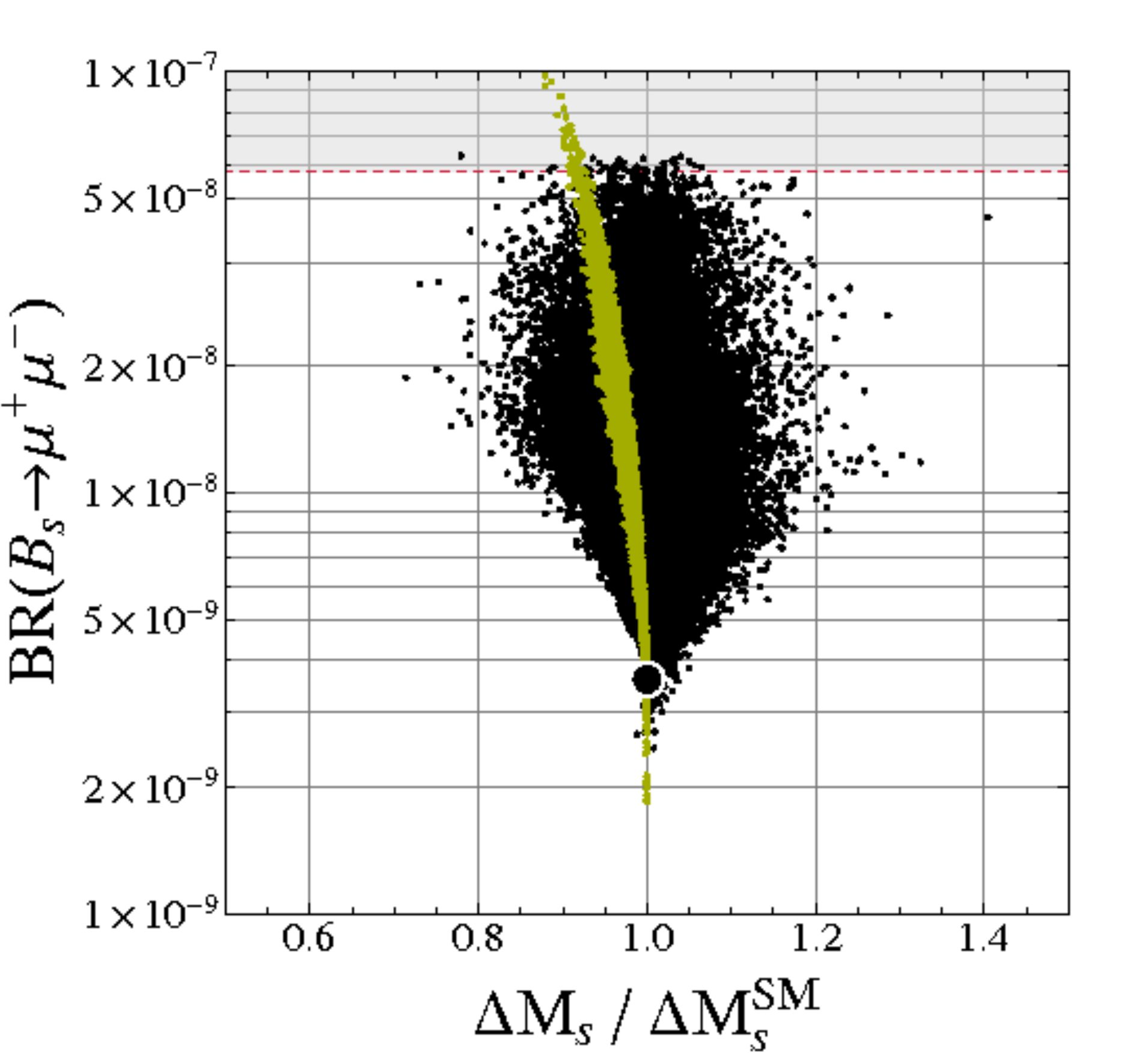}\\[20pt]
\includegraphics[width=0.305\textwidth]{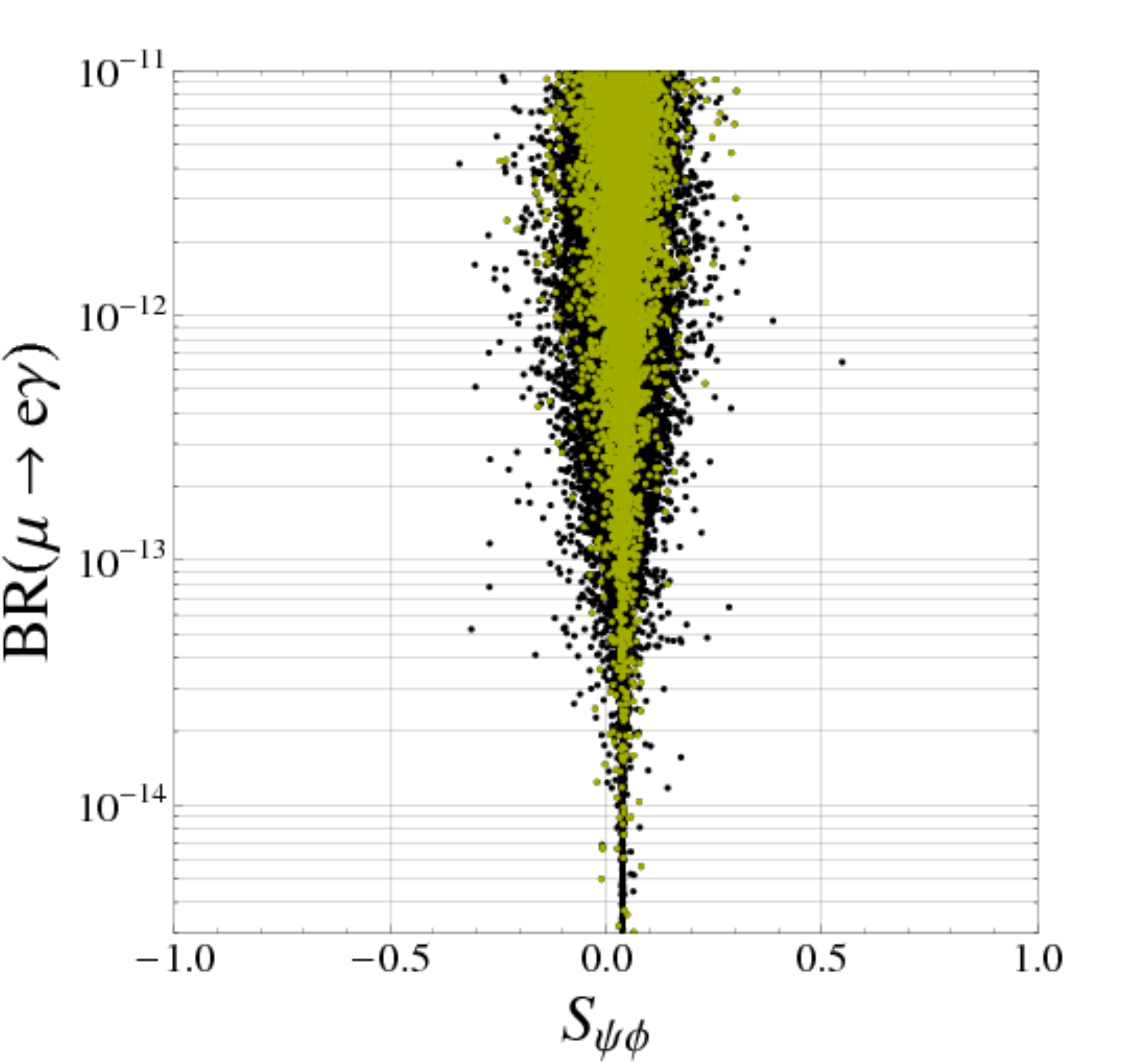}~~~~
\includegraphics[width=0.305\textwidth]{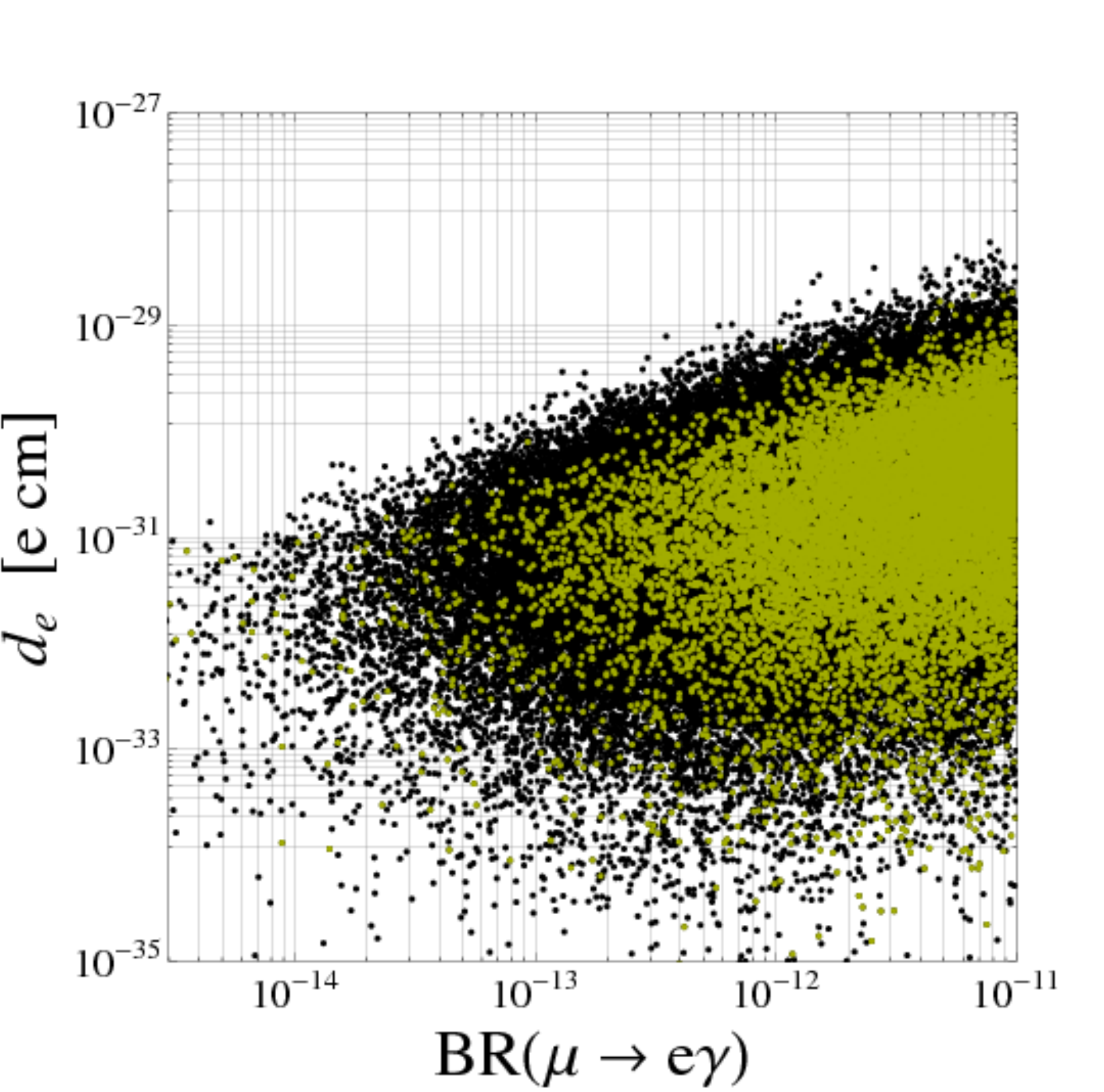}~~~~
\includegraphics[width=0.305\textwidth]{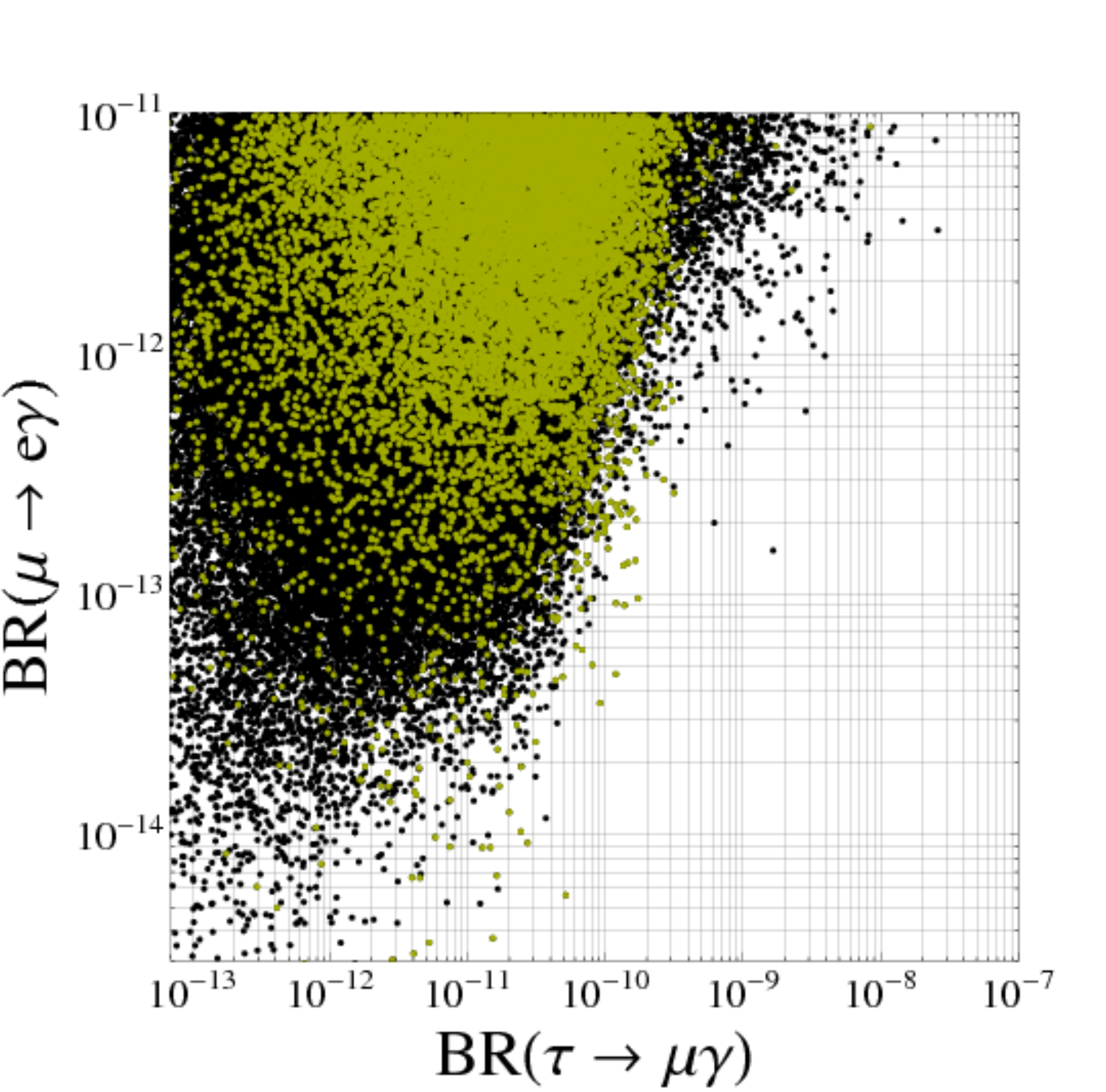}
\caption{\small
Predictions of the AKM model both in the hadronic and leptonic sectors.
In the first two rows we show the predictions for  various 
observables vs. $S_{\psi\phi}$. The blue points correspond to positive NP effects in $\epsilon_K$ such
that $1.2<\epsilon_K/(\epsilon_K)_\text{SM}<1.3$ and $\Delta M_d/\Delta M_s$ is SM-like.
The green points in the plots of BR$(B_s \to \mu^+\mu^-)$ vs. BR$(B_d\to\mu^+\mu^-)$
and $\Delta M_s/\Delta M_s^{\rm SM}$ vs. BR$(B_s\to\mu^+\mu^-)$ show the correlation
of these observables in the MFV MSSM. The last row refers to the predictions for
leptonic observables. The green points explain the $(g-2)_{\mu}$ anomaly at the
$95\%$ C.L., i.e. $\Delta a_{\mu}> 1\times 10^{-9}$.}
\label{fig:model_king}
\end{figure}

In fig.~\ref{fig:model_ross}, we show the predictions of the RVV2 model for several observables as
functions of $S_{\psi\phi}$ in analogy to fig.~\ref{fig:model_AC} for the abelian flavour model.

Our numerical results are obtained by implementing the flavour structures reported in
sec.~\ref{sec:strategy} assuming a CMSSM spectrum and scanning the unknown $\mathcal{O}(1)$
coefficients in the range $\pm [0.5,2]$.

Large values for $S_{\psi\phi}$ up to $-0.7\lesssim S_{\psi\phi}\lesssim 0.7$ are allowed
while being compatible with all the constraints, in particular from $\text{BR}(b\to s\gamma)$,
$R_{\Delta M}$ and $\epsilon_K$.

One of the most prominent differences between the RVV2 and the AC models is the loss of correlation
between $S_{\psi\phi}$ and $\text{BR}(B_s\to\,\mu^+\mu^-)$, although both observables can differ
spectacularly from their SM predictions. In particular, in the AC model, the Higgs mediated effects
were the only contributions able to generate large (non-standard) values for $S_{\psi\phi}$, after
imposing the indirect constraints, especially from $b\to s\gamma$ and $D^0-\overline{D}^0$ mixing.
As a result, $S_{\psi\phi}$ was correlated with $B_s\to\,\mu^+\mu^-$ as the latter can also receive
large SUSY effects only through the Higgs sector.

In the RVV2 model, the absence of the $D^0-\overline{D}^0$ constraints, as well as the more
complicated flavour structure of the model (as for instance the presence of left-handed currents
carrying new sources of CPV) destroy the above correlation as shown in the first plot of
fig.~\ref{fig:model_ross}.

The correlation between $S_{\psi\phi}$ and $S_{\phi K_S}$ reveals that $S_{\psi\phi}$, but probably
not $S_{\phi K_S}$, can significantly depart from its SM expectation. Still, similarly to the AC
model, negative NP values for $S_{\phi K_S}$, as suggested experimentally, imply negative values 
for $S_{\psi\phi}$, in contrast to the Tevatron results.

Moreover, even if the RVV2 model predicts sizable left-handed currents with order one CPV phases,
the CP asymmetry in $b\to s\gamma$, as well as the asymmetries in $B\to K^* \ell^+\ell^-$ turn out
to be almost SM-like after imposing all the indirect constraints.

In all the plots, all the points satisfy the constraints of tabs.~\ref{tab:DF2exp} and 
\ref{tab:observables} while the blue points correspond to a positive NP effect in $\epsilon_K$
such that $1.2<\epsilon_K/(\epsilon_K)_\text{SM}<1.3$ and $\Delta M_d/\Delta M_s$ SM-like,
allowing to solve the UT tension discussed in sec.~\ref{sec:model_independent}. However,
it is interesting to observe that a simultaneous solution to the $S_{\psi\phi}$, $(g-2)_\mu$
and $\epsilon_K$ anomalies is not very likely in this model, which constitutes an important
test for this model. Similarly to the AC model the pattern of deviations from the SM predictions
for $B_{s,d}\to\mu^+\mu^-$ and $\Delta M_s$ can differ spectacularly from MFV expectations.
In particular, as in the AC model, the ratio ${\rm BR}(B_d\to\mu^+\mu^-)/{\rm BR}(B_s\to\mu^+\mu^-)$
is dominantly below its MFV prediction and can be much smaller than the latter.

On general grounds, we can conclude that the predictions of the AC and RVV2 models in the hadronic
sector are quite similar. The NP effects in the RVV2 model are a bit smaller than in the AC model
as the size of the right-handed couplings in the latter model is larger. However, one of the
most important differences in predictions of these two models regards $\varepsilon_K$ and
$D^0-\bar D^0$ mixing. We stress again that the RVV2 model naturally predicts sizable effects
in $\varepsilon_K$ (able to solve the current UT tension) while predicting negligibly small
effects in $D^0-\bar D^0$ mixing; in the AC model exactly the opposite situation occurs.

Concerning the predictions for the hadronic EDMs, we have explicitly checked that
the down-quark (C)EDM reaches interesting values ~--~up to $\approx 10^{-26}e$~cm~--~
while being compatible with all the constraints. Moreover, large CPV effects in $B_s$
systems typically imply predictions for the neutron EDM within the expected future
experimental resolutions $\approx 10^{-28}e$~cm. The strange-quark (C)EDM is enhanced
compared to the down-quark (C)EDM by a factor of $\approx \epsb^{-2}\approx 50$ hence,
it would be very interesting to know precisely how it enters the neutron EDM.
This would be also of great interest to establish the NP room left to CPV in $B_s$
systems under the constraints from the hadronic EDMs.

In the last row of fig.~\ref{fig:model_ross}, we show the predictions of the RVV2 model for
observables in the leptonic sector. From left to right of fig.~\ref{fig:model_ross}, we report the
correlations between $S_{\psi\phi}$ vs. ${\rm BR}(\mu\to e\gamma)$, ${\rm BR}(\mu\to e\gamma)$ vs.
the electron EDM $d_e$ and ${\rm BR}(\tau\to\mu\gamma)$ vs. ${\rm BR}(\mu\to e\gamma)$, respectively.
The green points are such that $\Delta a_{\mu}> 1\times 10^{-9}$, thus they explain the $(g-2)_{\mu}$
anomaly at the $95\%$ C.L.

We observe that, both $\mu\to e\gamma$ and $\tau\to\mu\gamma$ are very sensitive probes of LFV in
the RVV2 model and they both can turn out to be the best probes of LFV in this model. In particular,
it is interesting to observe that the desire to solve the $(g-2)_{\mu}$ anomaly in the RVV2 model
implies values of ${\rm BR}(\tau\to \mu\gamma)$ in the reach of LHCb and future SuperB facilities
as well as ${\rm BR}(\mu\to e\gamma)\ge 10^{-13}$, within the MEG resolution~\cite{Maki:2008zz}.

As we can see, both ${\rm BR}(\ell_i\to\ell_j\gamma)$ and $d_e$ span over many orders of magnitude.
Their behavior can be understood looking at~(\ref{MIamplL}),~(\ref{MIamplR}) and~(\ref{Eq:lEDM_LO}),
respectively. In fact, given that
${\rm BR}(\ell_i\to\ell_j\gamma)\sim(t^{2}_{\beta}/\tilde{m}^4)\times|\delta_{ij}|^2$
and since $t_{\beta}$ and $\tilde{m}$ can vary roughly by one order of magnitude in our setup
while the $|\delta_{ij}|$s are defined modulo unknown coefficients in the range $\pm[0.5,2]$,
we expect ${\rm BR}(\ell_i\to\ell_j\gamma)$ to vary by roughly seven orders of magnitude.

Moreover, the loss of the correlation between ${\rm BR}(\tau\to\mu\gamma)$ and ${\rm BR}(\mu\to e\gamma)$
can be traced back noting that ${\rm BR}(\mu\to e\gamma)$ receives several contributions (of comparable
size) by both $\delta_{21}$ and $(\delta_{21})_{\rm eff.}\sim\delta_{23}\delta_{31}$ while the only
relevant MIs for ${\rm BR}(\tau\to\mu\gamma)$ are the $\delta_{32}$ ones.
Further, the impact of the MIs $\delta^{LR}$ and $\delta^{RL}$ is much more relevant for
${\rm BR}(\mu\to e\gamma)$ ~--~that receives an enhancement factor $(m_{\tau}/m_{\mu})^2$ from the
amplitude generated by $\delta^{LR}$ and $\delta^{RL}$~--~ than for ${\rm BR}(\tau\to\mu\gamma)$
and this also contributes to destroy their correlation.

Concerning $d_e$, we note that it is bounded from above by ${\rm BR}(\mu\to e\gamma)$ even if their
correlation is loose given their very different sensitivity to the flavor structures of the model.

In summary, the distinct patterns of flavour violation in the RVV2 model are
\begin{itemize}
\item
Large enhancements of $S_{\psi\phi}$ and $\text{BR}(B_{s,d}\to\,\mu^+\mu^-)$ but not in a
correlated manner as in the abelian AC model,
\item
Similarly to the AC model, $\text{BR}(B_d\to\mu^+\mu^-)/\text{BR}(B_s\to\mu^+\mu^-)$ and $\text{BR}(B_s\to\mu^+\mu^-)/\Delta M_s$ might be very different from the MFV expectations with the first
ratio dominantly smaller than its MFV value, especially for $\text{BR}(B_s\to\mu^+\mu^-)\gtrsim 10^{-8}$.
Also in this model, $\text{BR}(B_d\to\mu^+\mu^-)$ can reach values by a factor of 10 larger than in the SM,
\item
Removal of the UT tension through NP contributions to $\epsilon_K$, requiring then typically
$S_{\psi\phi}\le 0.3$ and $S_{\phi K_S}\approx S_{\psi K_S}$,
 \item
Small effects in $D^0-\bar D^0$ mixing,
\item Large CPV effects in $B_s$ systems typically imply predictions for the neutron EDM
within the expected future experimental resolutions $d_n\approx 10^{-28}e$~cm,
\item Large values for $-0.25<S_{\psi\phi}<0.25$ are still allowed even for
${\rm BR}(\mu\to e\gamma)\lesssim 10^{-13}$. However, the desire of an explanation
for the $(g-2)_{\mu}$ anomaly implies that ${\rm BR}(\mu\to e\gamma)\gtrsim 10^{-13}$
and $|S_{\psi\phi}|\lesssim 0.25$,
\item
${\rm BR}(\mu\to e\gamma)\ge 10^{-13}$, $d_e > 10^{-29}~e$cm and ${\rm BR}(\tau\to\mu\gamma)\ge 10^{-9}$
required by the solution of the $(g-2)_\mu$ anomaly.
\end{itemize}
%

\subsubsection{AKM Model: Results in the Hadronic and Leptonic Sectors}

In fig.~\ref{fig:model_king}, we show the predictions for the AKM model. As done for the other
flavour models, we obtain the numerical results by implementing the flavour structures reported
in sec.~\ref{sec:strategy} assuming a CMSSM spectrum and scanning the unknown $\mathcal{O}(1)$
coefficients in the range $\pm [0.5,2]$.

The main differences between the RVV2 and the AKM models can be traced back remembering the peculiar
flavour structures in the soft sector of the two models. In particular, the AKM model predicts a
CKM-like RH current while the corresponding mixing angle in the RVV2 model for the $b\to s$ transition
is larger. This implies that in the AKM model, the effects in CPV observables are typically smaller,
but still very interesting, compared to the RVV2 model.
In particular, it is found that, in the AKM model $S_{\psi\phi}$ lies most likely in the range
$-0.3<S_{\psi\phi}<0.3$ while being compatible with all the constraints. Interestingly enough,
similarly to the abelian case discussed before, large non-standard values for $S_{\psi\phi}$
would unambiguously point towards non-standard values for $\text{BR}(B_s\to\mu^+\mu^-)$ as shown
in the first plot of fig.~\ref{fig:model_king}. However, fig.~\ref{fig:model_king} also shows 
that within this model the $S_{\phi K_S}$ anomaly cannot be accounted for. Large values for $S_{\psi\phi}$
can also be compatible with an explanation of the $(g-2)_{\mu}$ anomaly. Moreover, we also observe that
departures from the MFV SUSY expectations for $\text{BR}(B_d\to\,\mu^+\mu^-)/\text{BR}(B_s\to\,\mu^+\mu^-)$
and $\text{BR}(B_s\to\,\mu^+\mu^-)/\Delta M_s/(\Delta M_s)_\text{SM}$(in both directions with respect
to the SM predictions) are also expected in the AKM model, even if the ratio 
${\rm BR}(B_d\to\mu^+\mu^-)/{\rm BR}(B_s\to\mu^+\mu^-)$ stays much closer to the MFV value of roughly
$1/33$~\cite{Buras:2003td,Hurth:2008jc}.

Passing to the hadronic EDMs, we observe that, even if the AKM does not contain leading
CPV phases in the 13 sector, the down-quark (C)EDM is always generated by means of the CKM phase,
in the presence of RH MIs (see sec.~\ref{sec:EDMs}). We have explicitly checked that the down-quark
(C)EDM might reach values up to $\approx 10^{-28}e$~cm after imposing all the constraints.
As a result, the constraints from the hadronic EDMs are well under control.
However, large CPV effects in $B_s$ systems would likely imply predictions for the hadronic 
EDMs within the expected future experimental resolutions $\approx 10^{-28}e$~cm.
The strange-quark (C)EDM is enhanced compared to the down-quark (C)EDM by a factor of
$\approx \epsb^{-2}\approx 50$, as in the RVV2 model. Once again, we stress that it would
be of crucial importance to know how the strange-quark (C)EDM enters the hadronic EDMs
in order to establish which are the CPV signals we can still expect in $B_s$ systems.

In the last row of fig.~\ref{fig:model_king}, we show the predictions of the AKM model 
for observables in the leptonic sector as done for the RVV2 model in fig.~\ref{fig:model_ross}.
In particular, both $\mu\to e\gamma$ and $\tau\to\mu\gamma$ are sensitive probes of LFV in the
AKM model. However, in contrast to the RVV2 model, $\mu\to e\gamma$ will represent the best
probe of LFV in this model, especially after the MEG sensitivity will be fully exploited.
In fact, it turns out that when ${\rm BR}(\mu\to e\gamma)\approx 10^{-13}$ then
${\rm BR}(\tau\to \mu\gamma)\leq 10^{-10}$, far from the SuperB reach.

Moreover, the predictions for the electron EDM in the AKM model are well below those of the RVV2 model.
In fact, in the RVV2 model, the two largest contributions to $d_e$ are proportional to either
${\rm Im}(\delta_{\ell}^{LR})_{13}(\delta_{\ell}^{RR})_{31}$ or
${\rm Im}(\delta_{\ell}^{LL})_{13}(\delta_{\ell}^{RR})_{31}$ with $(\delta_{\ell}^{RR})_{31}$
carrying a leading $\mathcal{O}(1)$ CPV phase.
In contrast, in the AKM model, the first non vanishing contribution to $d_e$ is generated at
the third order in the MI expansion by means of the combination
${\rm Im}(\delta_{\ell}^{LR})_{13}(\delta_{\ell}^{RR})_{32}(\delta_{\ell}^{RR})_{21}$
as $(\delta_{\ell}^{RR})_{32}$ is the only MI containing an $\mathcal{O}(1)$ CPV phase.
Hence, there is a higher order suppression in terms of small mixing angles in the AKM model
compared to the RVV2 model.

An order of magnitude for the upper bound on $d_e$, compatible with the constraints from
${\rm BR}(\mu\to e\gamma)$, can be obtained considering a degenerate SUSY spectrum
and assuming that the dominant contributions to ${\rm BR}(\mu\to e\gamma)$ come from
$(\delta^{RL}_{\ell})_{21}$ and $(\delta^{LR}_{\ell})_{21}$.

Taking the specific expressions for the MIs arising in the AKM model~(\ref{sckm2_king})
and~(\ref{aterms_king}), one can easily find that
\beq
\frac{d_{e}}{e}\lesssim
10^{-29}\sqrt{\frac{{\rm BR}(\mu\to e\gamma)}{10^{-11}}}\,\sin\Psi\,\,e\,{\rm cm}\,,
\label{edm_AKM_bound}
\eeq
as is fully confirmed numerically by fig.~\ref{fig:model_king}.

Still, visible values for $d_e$ can be reached in the AKM model; in particular, it turns out that
$d_e \lesssim 10^{-29} (10^{-30})~e$cm for ${\rm BR}(\mu\to e\gamma)\approx 10^{-11} (10^{-13})$.

Finally, we note that, for ${\rm BR}(\mu\to e\gamma)\lesssim 10^{-13}$, only values for $S_{\psi\phi}$ at the level of $|S_{\psi\phi}|< 0.15$ are still possible if one wants to be fully compatible with an explanation for the $(g-2)_{\mu}$ anomaly.

Therefore, the disentangling of the RVV2 and the AKM models might be problematic from
a low-energy point of view, even if not impossible. Clearly the knowledge of some
SUSY parameter, as the value for $\tan\beta$ and the charged Higgs mass, for instance,
would be of outmost importance to make access to the flavour structure of the flavour 
model at work.

In summary, several of the predictions in the AKM model are similar to the ones found in the
RVV2 model, but the following most significant differences should be noted:
\begin{itemize}
\item
$|S_{\psi\phi}|$ can reach values up to $|S_{\psi\phi}|\lesssim 0.3$, values that are fully
consistent with an explanation of the UT tension through NP contributions to $\epsilon_K$
\item
$S_{\psi\phi}\ge 0.2$ uniquely implies $\text{BR}(B_s\to\,\mu^+\mu^-)\ge 10^{-8}$.
\item
$\text{BR}(B_d\to\,\mu^+\mu^-)/\text{BR}(B_s\to\,\mu^+\mu^-)$ departs less from MFV expectation
than in the AC and RVV2 models and can be both larger and smaller relative to its MFV value.
\item A non standard $S_{\psi\phi}$ typically implies $d_n\approx 10^{-28}e$~cm that is
within the expected future experimental resolutions.
\item
$\text{BR}(\tau\to\,\mu\gamma)$ is likely out of the reach of SuperB machines, in particular,
${\rm BR}(\tau\to \mu\gamma)\lesssim (10^{-8},10^{-9},10^{-10})$ when
${\rm BR}(\mu\to e\gamma)\lesssim (10^{-11},10^{-12},10^{-13})$, respectively.
\item
$d_e\lesssim 10^{-29}(10^{-30})~e$cm for ${\rm BR}(\mu\to e\gamma)\lesssim 10^{-11}(10^{-13})$ where
${\rm BR}(\mu\to e\gamma)\gtrsim 10^{-13}$ is required by the solution of the $(g-2)_\mu$ anomaly.
\end{itemize}
%
\subsection{Step 4: Flavour Model with Purely Left-handed Currents}


\begin{figure}
\centering
\includegraphics[width=0.295\textwidth]{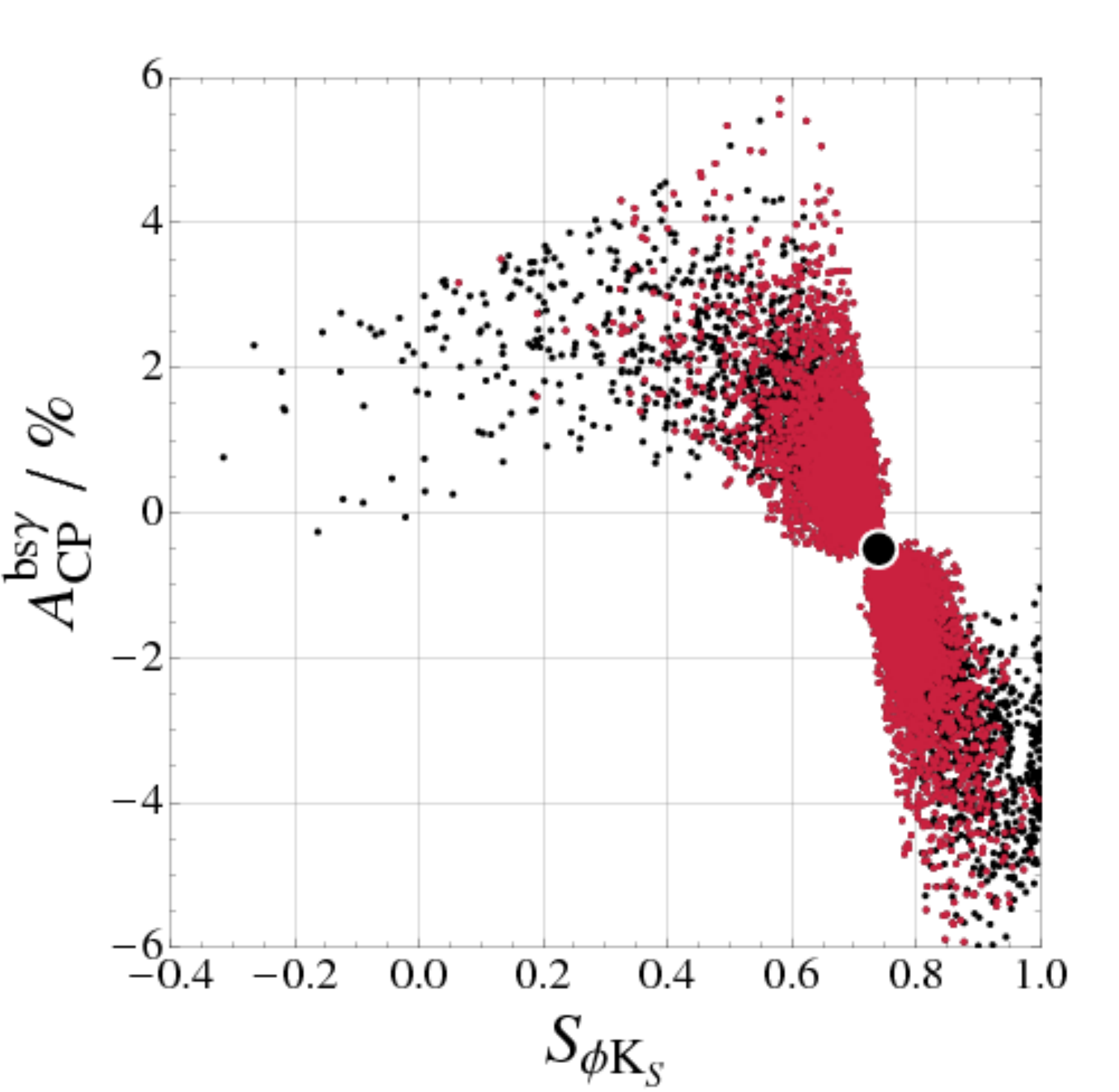}~~~~~
\includegraphics[width=0.295\textwidth]{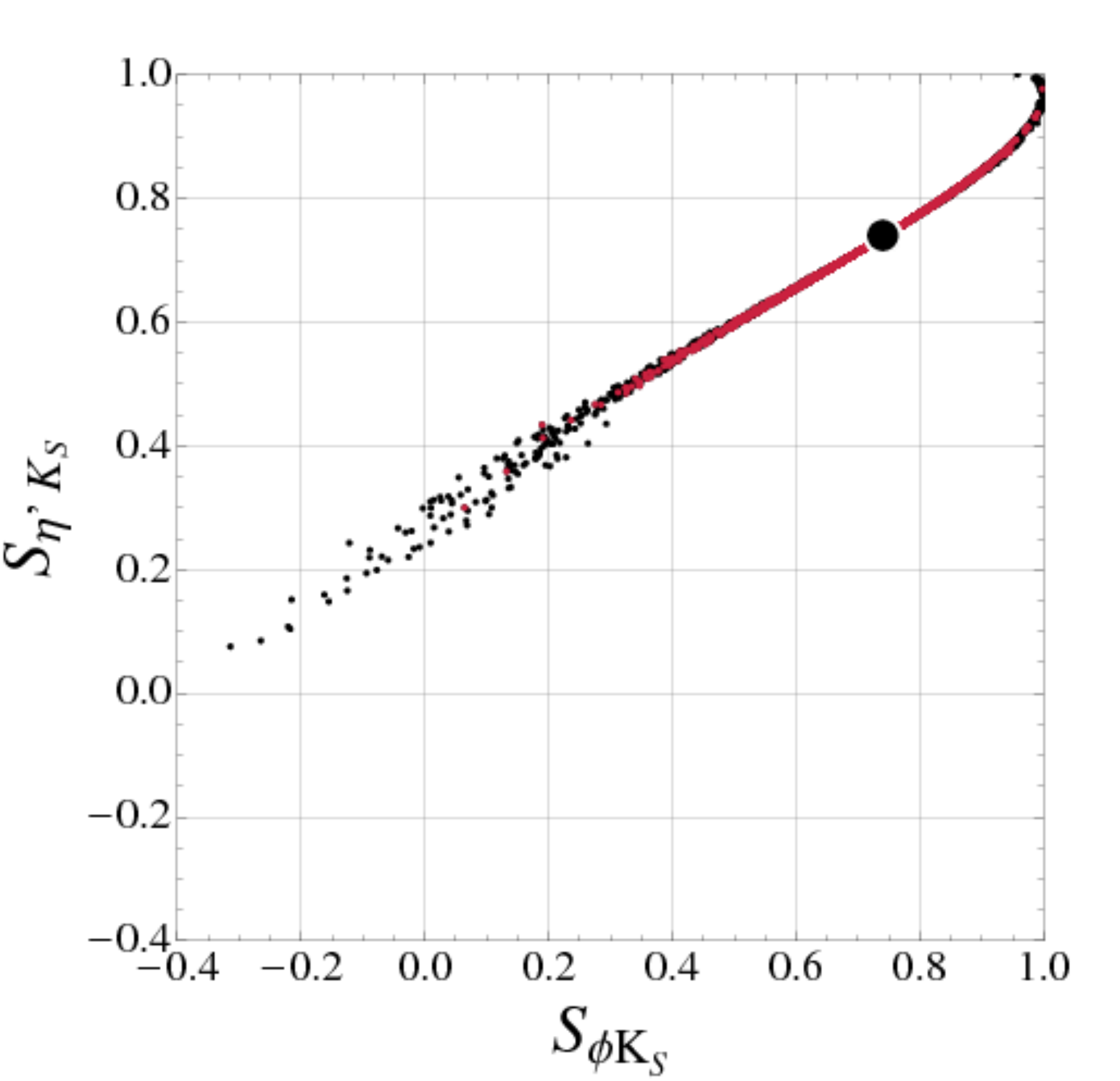}~~~~~
\includegraphics[width=0.29\textwidth]{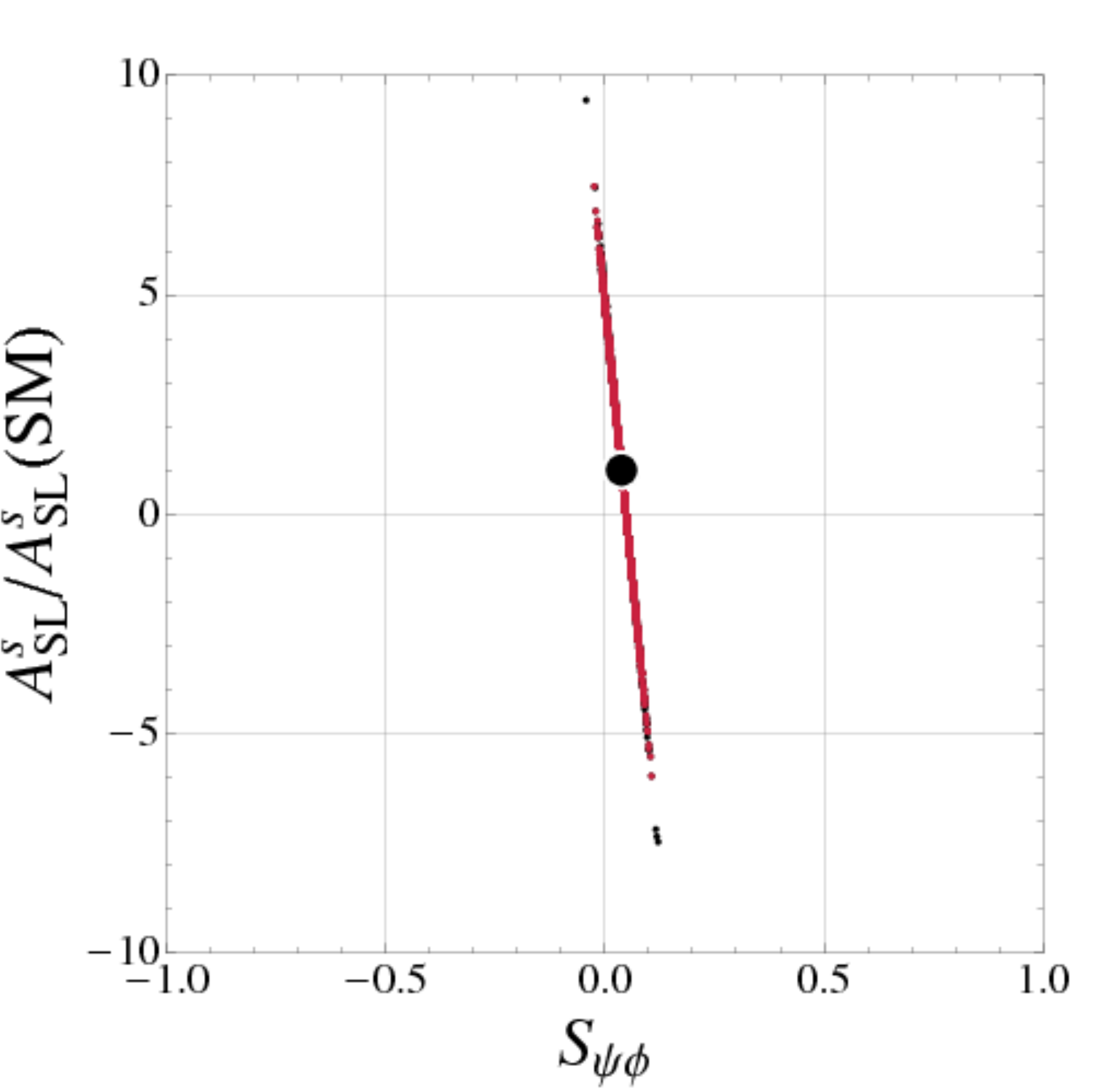}\\[20pt]
\includegraphics[width=0.315\textwidth]{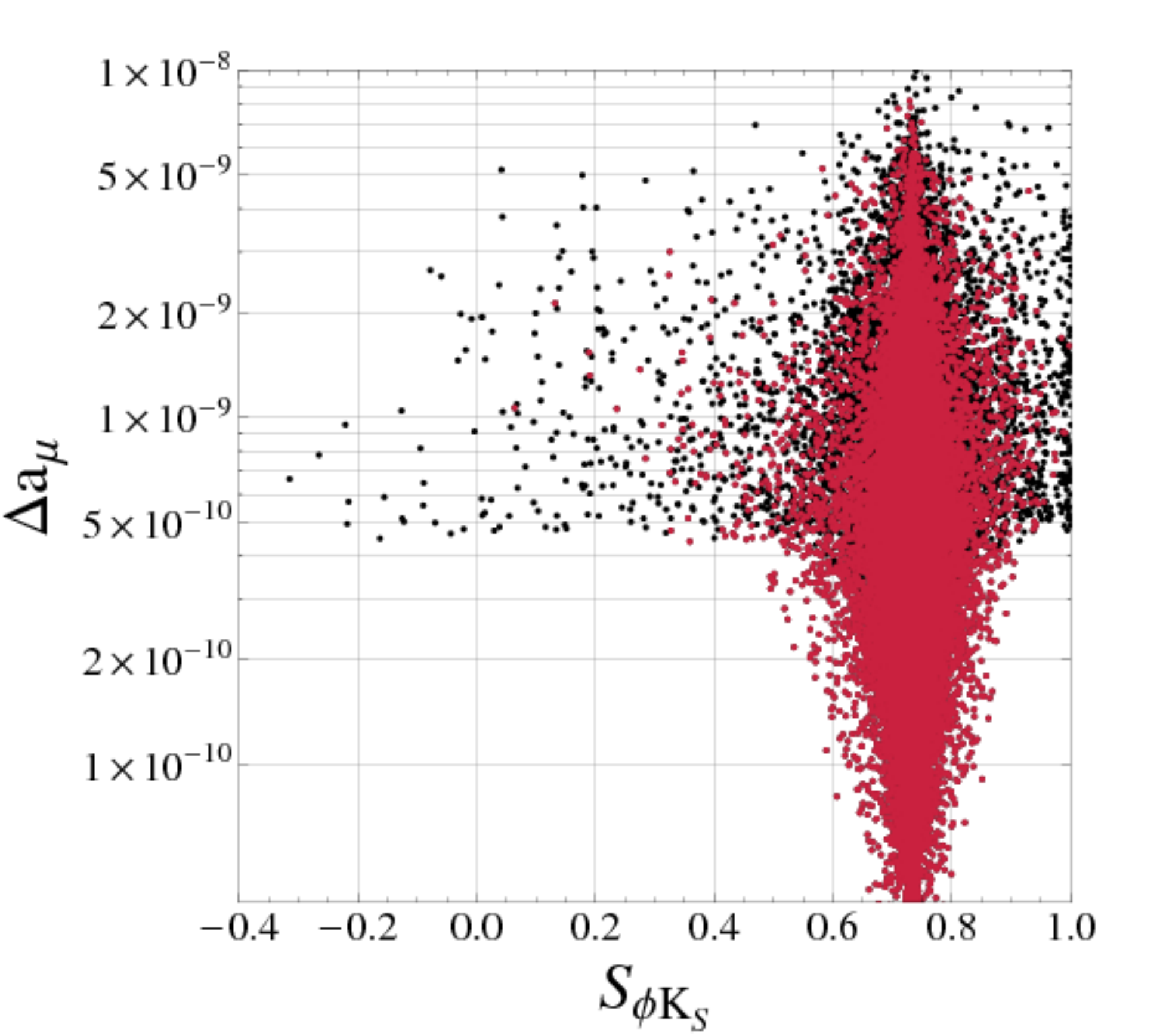}~~~
\includegraphics[width=0.315\textwidth]{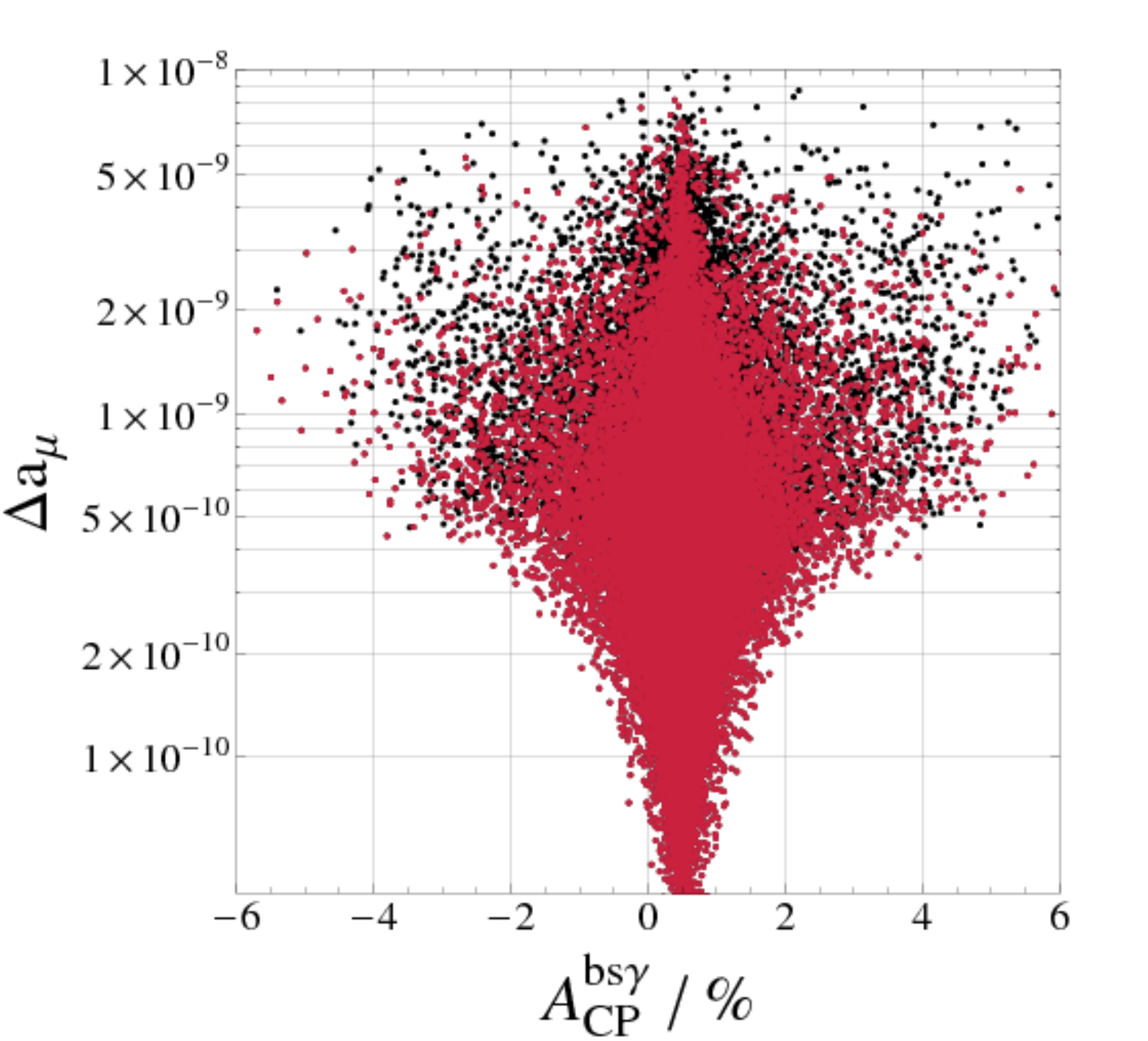}~~~
\includegraphics[width=0.31\textwidth]{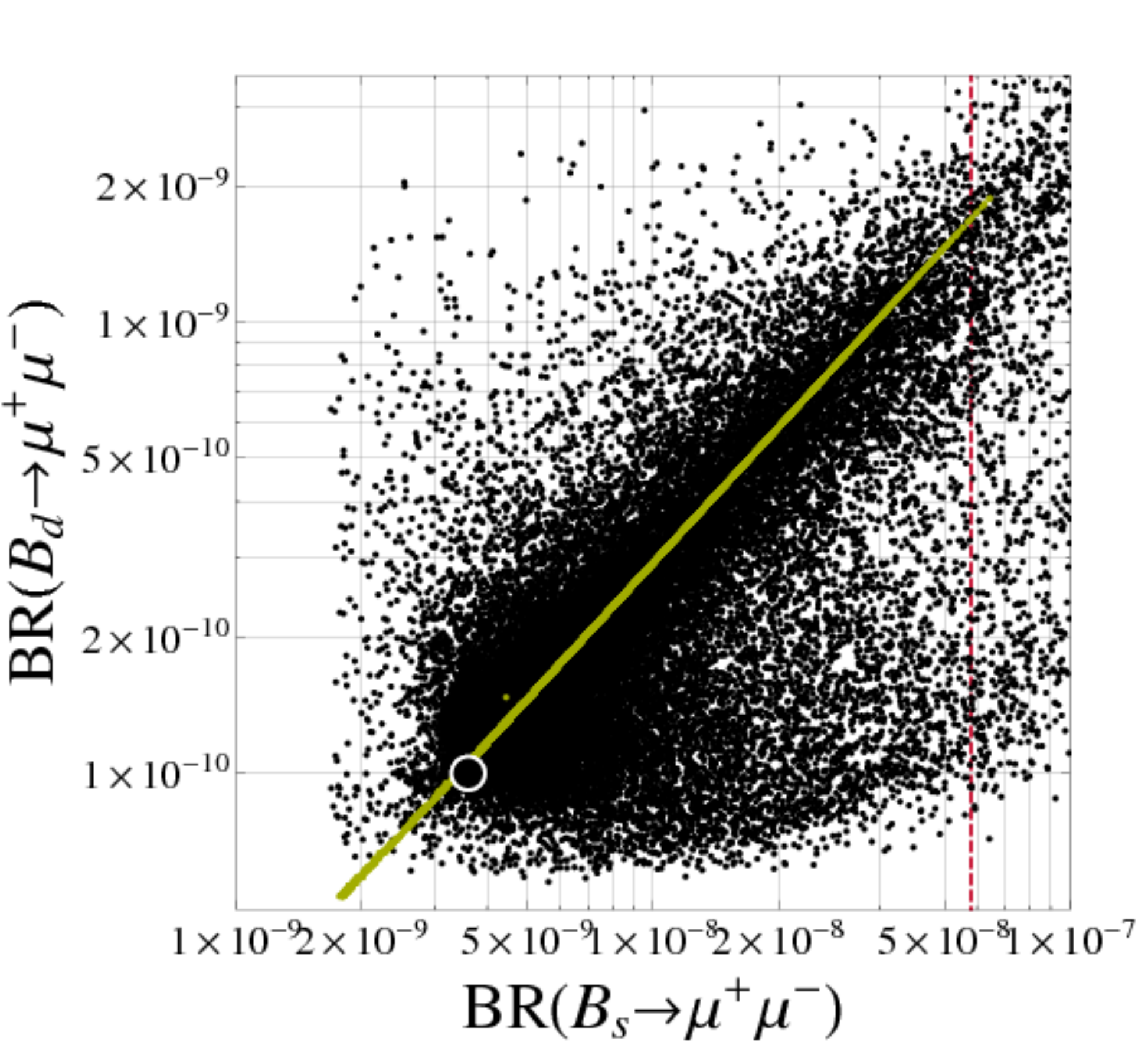}\\[20pt]
\includegraphics[width=0.3\textwidth]{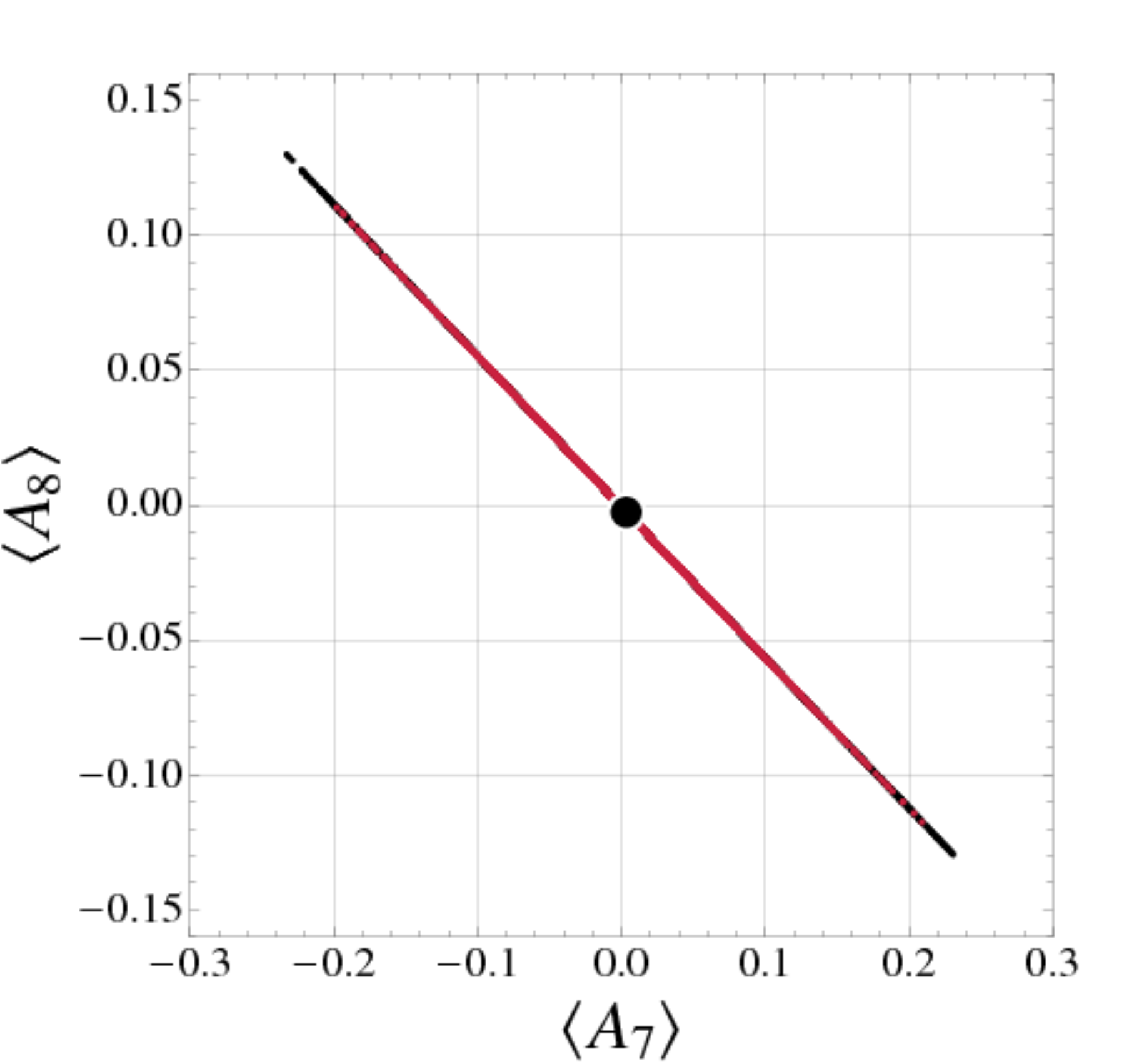}~~~~
\includegraphics[width=0.3\textwidth]{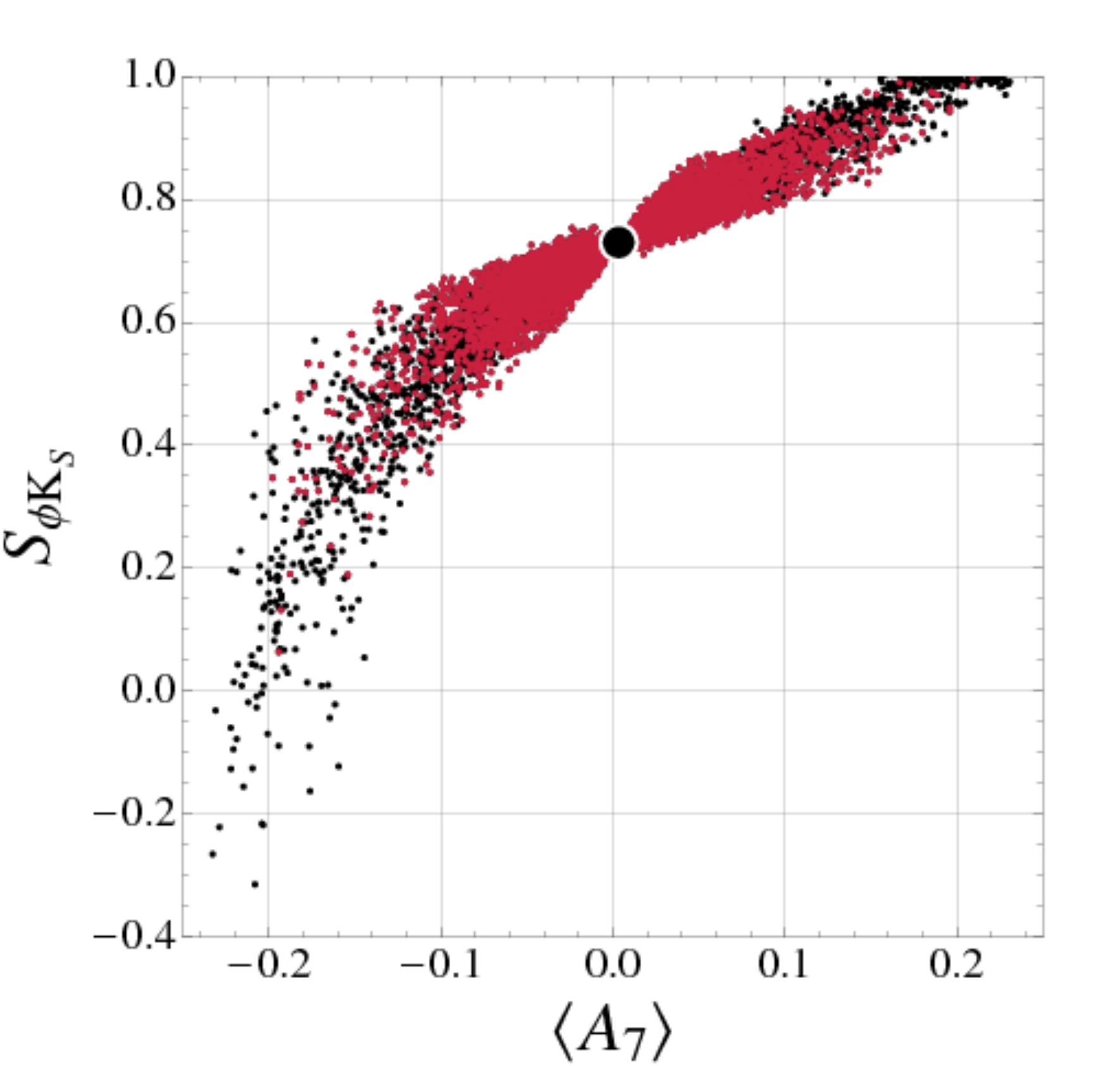}
\caption{\small
Predictions for various low energy processes induced by $b\to s$ transitions in the non-abelian flavour
model predicting pure, CKM-like, left-handed flavour mixing angles in the soft sector~\cite{Hall:1995es}.
In all the plots, all the black points satisfy the constraints of tab.~\ref{tab:observables} while
the red ones additionally satisfy $\text{BR}(B_s\to\mu^+\mu^-)< 6\times 10^{-9}$.
The green points in the plots of BR$(B_s \to \mu^+\mu^-)$ vs. BR$(B_d \to \mu^+\mu^-)$ show the correlation
of these observables in the MFV MSSM.}
\label{fig:model_CKM}
\end{figure}

Another typical flavour structure for the soft sector emerging in many abelian
and non-abelian flavour models are purely left-handed currents with CKM-like mixing
angles~\cite{Hall:1995es,Nir:2002ah}.

Based on the flavour structures given in~(\ref{dLL_scenario}) and scanning the unknown
$\mathcal{O}(1)$ coefficients in the range $\pm [0.5, 2]$, we present in fig.~\ref{fig:model_CKM}
the numerical results of our study of the non-abelian model of~\cite{Hall:1995es}.

As already discussed in sec.~\ref{sec:strategy}, in the abelian case, non-standard and testable
NP effects can be generated only in observables related to $D^0-\overline{D}^0$ mixing hence,
in the following, we focus on the non-abelian case.

The first upper plot refers to the correlation between $A_\text{CP}(b\to s\gamma)$ vs. $S_{\phi K_S}$.
We find that $-0.2\lesssim S_{\phi K_S}\lesssim 1$ and $-6\% \lesssim A_\text{CP}(b\to s\gamma)\lesssim 6\%$.
Interestingly, it turns out unambiguously that positive (negative) NP contributions for $S_{\phi K_S}$
are associated with negative (positive) NP effects in $A_\text{CP}(b\to s\gamma)$.

This correlation can also be understood analytically. While the NP contributions to $S_{\phi K_S}$ 
arise dominantly from the Wilson coefficient $C_8$, the CP asymmetry $A_{\rm CP}(b\to s\gamma)$ 
crucially depends on the relative size of the imaginary parts of $C_7$ and $C_8$, see (\ref{eq:acp_bsg}).
In the considered model, gluino loops typically give $C_7^{\tilde g} < C_8^{\tilde g}$ (\ref{eq:C7_g}),
while Wino loops lead to the opposite situation, i.e. $C_7^{\tilde W} > C_8^{\tilde W}$ (\ref{eq:C7_cha}).
In fact the correlation that we find corresponds to the latter case, implying that Wino contributions
dominate over gluino ones in large parts of the parameter space, since Winos are typically a factor 3
lighter than gluinos.

In the second upper plot we show that $S_{\eta^{\prime} K_S}$ turns out to be highly correlated with
$S_{\phi K_S}$, as they both receive the dominant NP contribution (that is larger in the $S_{\phi K_S}$
case) from the chromomagnetic Wilson coefficient $C_8$. Moreover, a solution of the $S_{\phi K_S}$ 
anomaly can also lead to the solution of the $(g-2)_{\mu}$ anomaly and large values for the direct CP 
asymmetry $A_\text{CP}(b\to s\gamma)$ are typically associated with large values for $(g-2)_{\mu}$.
The departures of $S_{\psi\phi}$ from the SM expectations $(S_{\psi\phi})_\text{SM}\approx 0.036$ are
quite small, in the range $0\lesssim S_{\psi\phi}\lesssim 0.1$, as expected from our model-independent 
analysis. The corresponding attained values for the semileptonic asymmetry $A^{s}_\text{SL}$ normalized
to the SM value lie in the range $-6\lesssim A^{s}_\text{SL}/(A^{s}_\text{SL})_\text{SM}\lesssim 6$,
hence, NP effects could be still experimentally visible in $A^{s}_\text{SL}$. Yet the confirmation of
the large values of $S_{\psi\phi}$ observed by CDF and D0 will be a serious problem for this kind of
models with purely left-handed currents. From fig.~\ref{fig:model_CKM} we also show that models with 
purely left-handed currents can predict very large deviations (in both directions) from the MFV SUSY
expectations for $\text{BR}(B_d\to\,\mu^+\mu^-)/\text{BR}(B_s\to\,\mu^+\mu^-)$. In particular, in
contrast to the models considered so far, the ratio in question can also be significantly larger than
in the MFV models and $\text{BR}(B_d\to\,\mu^+\mu^-)$ can reach values as high as $2\times 10^{-9}$,
while staying consistent with the bound on ${\rm BR}(B_s\rightarrow\mu^+\mu^-)$. 

Finally, in the last row of fig.~\ref{fig:model_CKM}, we show the correlations involving the CP asymmetries 
in $B\to K^*\mu^+\mu^-$, $\langle A_7 \rangle$ vs. $\langle A_8 \rangle$ and $\langle A_7 \rangle$ vs 
$S_{\phi K_S}$. We find a very stringent correlation between $\langle A_7 \rangle$ and $\langle A_8 \rangle$ 
with the NP effect in $\langle A_7 \rangle$ being a factor of two larger and with opposite sign with respect
to $\langle A_8\rangle$. This strong correlation is due to the fact  that, in the considered framework, 
effects in both $\langle A_7 \rangle$ and $\langle A_8 \rangle$ are almost exclusively induced by the NP contributions to $C_7$. Moreover, we also find a clean correlation between $\langle A_7\rangle$ and 
$S_{\phi K_S}$ with $-0.2\lesssim\langle A_7 \rangle \lesssim -0.1$ in the region accounting for the 
$S_{\phi K_S}$ anomaly.

Concerning the hadronic EDMs, the main difference between models with purely LH currents with respect 
to models containing RH currents is that in the former case the quark (C)EDMs turn out to be proportional
to the external light quark masses. As a result, the down- and up-quark (C)EDMs generated by flavour effects
are suppressed by small flavour mixing angles while not being enhanced by the heaviest quark masses, hence, 
the hadronic EDMs are safely under control. Assuming CKM-like MIs, ${\tilde m}=500\,{\rm GeV}$ and $\tan\beta=10$, $d_d/e$ and $d^c_d$ are of order $\sim 10^{-28}~{\rm cm}$.
The strange quark (C)EDM might reach interesting values as it is enhanced, compared to $d_d/e$ and $d^c_d$,
by a factor of $(m_s/m_d)\times\lambda^{-2}\approx 400$. However, it is still not possible to draw any 
clear-cut conclusion due to the uncertainties relating $d_s$ to physical hadronic EDMs.

In summary, the most striking predictions of the models with purely LH currents as opposed
to models with large RH currents are
\begin{itemize}
\item
the ability to explain the $S_{\phi K_S}$ anomaly with simultaneous possible explanation
of the $(g-2)_\mu$ anomaly and significantly enhanced direct CP asymmetry in $b\to s\gamma$,
\item
SM-like $S_{\psi\phi}$,
\item
in the considered $\delta$LL model, the possibility of very large deviations for $\text{BR}(B_d\to\,\mu^+\mu^-)/\text{BR}(B_s\to\,\mu^+\mu^-)$ in both directions compared
to its MFV value, with $\text{BR}(B_d\to\,\mu^+\mu^-)$ and $\text{BR}(B_s\to\,\mu^+\mu^-)$
reaching values as high as $2\times 10^{-9}$ and $6\times 10^{-8}$, respectively,

\item Small NP effects in $\Delta F=2$ processes, like $\epsilon_K$, $S_{\psi K_s}$ and 
$\Delta M_d/\Delta M_s$. Therefore difficulty in addressing the UT tension.
\end{itemize}

\subsection{Step 5: Comparison with the FBMSSM and the MFV MSSM} \label{sec:step5_numerics}

Finally, we want to recall the results for the FBMSSM~\cite{Altmannshofer:2008hc}, to discuss the general MFV framework and to make a comparison with the models with left-handed CKM-like currents discussed in step 4.

In~\cite{Altmannshofer:2008hc}, it was found that the best probes of the FBMSSM are the EDMs of the 
electron ($d_e$) and the neutron ($d_n$), as well as flavour changing and CP violating processes in 
$B$ systems, like the CP asymmetries in $b\to s\gamma$ and $B\to\phi(\eta^{\prime})K_{S}$, i.e. 
$A_\text{CP}(b\to s\gamma)$ and $S_{\phi(\eta^{\prime})K_{S}}$, respectively. 
The non-standard values for $S_{\phi(\eta^{\prime})K_{S}}$, measured at the $B$~factories, can find a 
natural explanation within the FBMSSM with the effect being typically by a factor of $1.5$  larger in 
$S_{\phi K_{S}}$ in agreement with the pattern observed in the data.

Interestingly, it was found that the desire of reproducing the observed low values of $S_{\phi K_{S}}$
and $S_{\eta^{\prime} K_{S}}$ implies:

\begin{itemize}
\item a lower bounds on the electron and neutron EDMs $d_{e,n} \gtrsim 10^{-28}\,e\,$cm,
\item a positive and sizable (non-standard) $A_\text{CP}(b\to s\gamma)$ asymmetry in the
ballpark of $1\% - 5\%$,
\item small NP effects in $S_{\psi K_S}$ and $\Delta M_d/\Delta M_s$ so that in the FBMSSM
these observables can be used to extract $\beta$ and $|V_{td}|$,
\item $|\epsilon_K|$ is enhanced over its SM value at most at a level of $\lesssim 15\%$,
\item very small effects in $S_{\psi\phi}$ which could, however, be still visible through
the semileptonic asymmetry $A^s_\text{SL}$,
\item a natural explanation of the $\Delta a_\mu$ anomaly under the mild assumption that the
slepton masses are not much heavier than the squark masses (this is true in almost all the known
models of SUSY breaking).
\end{itemize}

The question we want to address now is, to which extent this pattern of effects gets modified within the
framework of the MFV MSSM. In particular we will analyse if the richer flavour and CP violating structure
of the MFV MSSM allows to generate sizeable effects in $S_{\psi\phi}$.

As discussed in sec.~\ref{sec:DF2}, within a MFV MSSM scenario, the charginos and the charged Higgses 
can induce the FCNC amplitude $ M^{(M)}_{12}$ by means of $C_{1}$ and $\tilde{C}_{3}$, with the chargino
box contributions to $\tilde{C}_{3}$ given in (\ref{eq:C3_cha_MFV}) being the only non-negligible
contributions sensitive to the new phases.

Moreover, as mentioned in sec.~\ref{sec:strategy}, the additional terms in the LL soft masses in the
MFV MSSM lead to potentially complex MIs
\begin{equation} \label{eq:MI_MFV}
(\delta_d^{LL})_{3i} \simeq (b_1 + b_3 y^2_b)  V_{ti}~,
\end{equation}
that will lead to gluino contributions to $\Delta F=2$ processes by means of the Wilson coefficient $C_1$
given in (\ref{eq:MIA2_C1}).

\begin{figure}[t]
\centering
\includegraphics[width=0.5\textwidth]{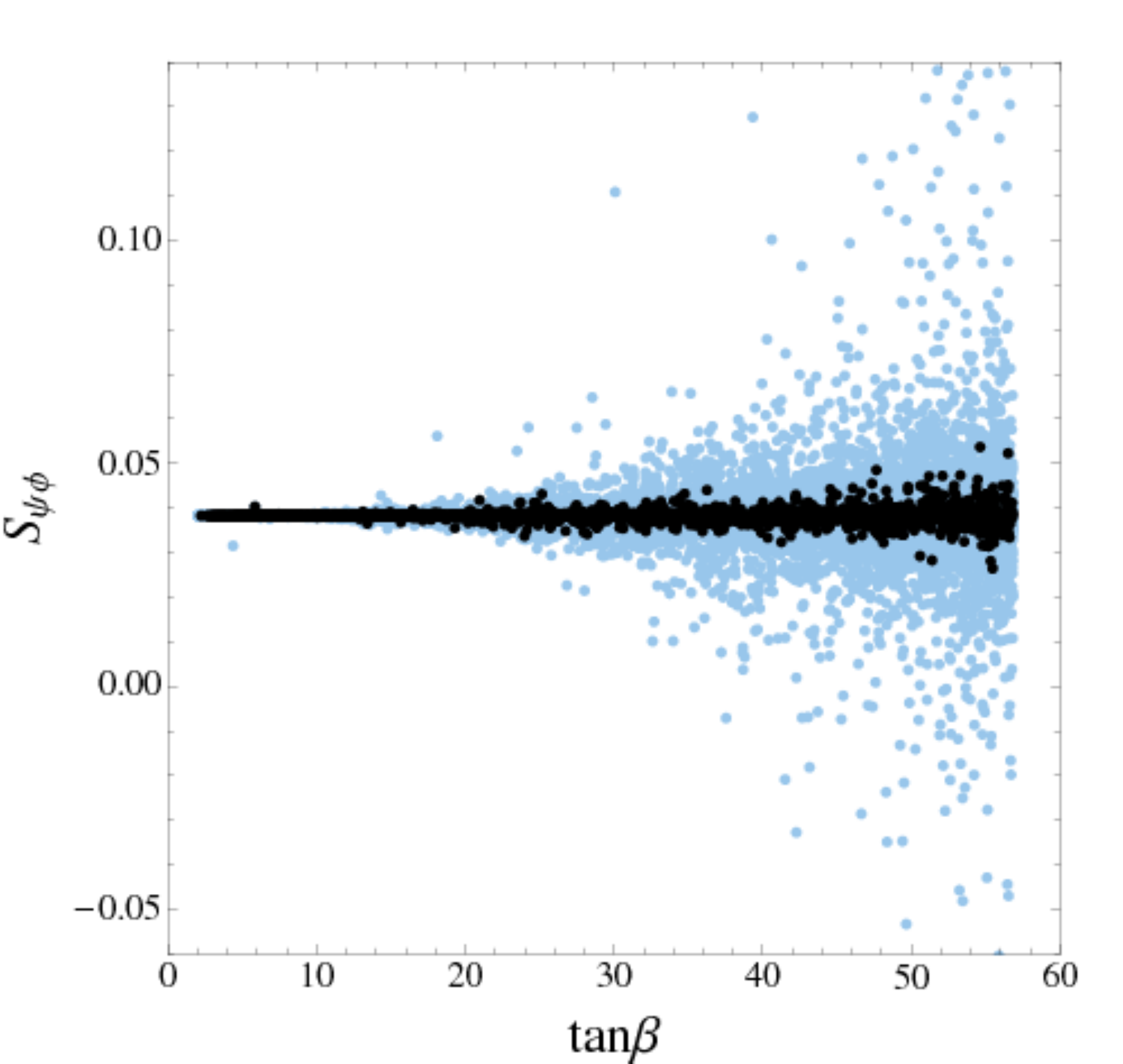}
\caption{\small
$S_{\psi\phi}$ vs. $\tan\beta$ in the MFV scenario. Light blue points do not satisfy the $B$-physics
constraints while the black ones do.}
\label{fig:Spsiphi_MFV}
\end{figure}

In fig.~\ref{fig:Spsiphi_MFV}, we show the predictions for $S_{\psi\phi}$ vs. $\tan\beta$ in a
MFV MSSM scenario taking into account the SUSY contributions listed above.
As we can see, potentially sizable effects to $S_{\psi\phi}$ would be possible only in the very
large $\tan\beta$ regime (light blue points); however, in this case the constraints from both
${\rm BR}(B_s\to\mu^+\mu^-)$ and ${\rm BR}(b\to s\gamma)$ become very powerful and they prevent
any visible effect in $S_{\psi\phi}$ (black points).
This confirms and strengthens the general finding of~\cite{Altmannshofer:2008hc} that within SUSY
MFV scenarios, large NP contributions can only be expected in $\Delta B =1$  transitions.

\bigskip

The models with left-handed CKM-like currents discussed in step 4 share many similarities with
the FBMSSM and the MFV MSSM. In particular, all the correlations among $B$-physics asymmetries
are almost the same in these scenarios and also the size of the effects are comparable.
As argued in~\cite{Altmannshofer:2008hc}, the $(g-2)_\mu$ anomaly is naturally solved once we
account for the non-standard value for $S_{\phi K_S}$; this claim has been fully confirmed in
the present work where we deal with a concrete SUSY breaking scenario, that is a SUGRA scenario.

The major difference discriminating these scenarios regards their predictions for the
leptonic and hadronic EDMs.
Within the FBMSSM, we predicted $d_{e,n} \gtrsim 10^{-28}\,e\,$cm while solving the
$S_{\phi K_{S}}$ anomaly. On the other hand, within the general class of models with
pure left-handed CKM-like currents, such a lower bound is significantly relaxed.
In fact, in this latter case, the source of CP violation is assumed to come from
flavour effects and the resulting EDMs are suppressed by small flavour mixing angles.
This suppression cannot be compensated by enhancement factors, as the heavy-light
Yukawa ratios, as it happens in presence of RH currents.

Hence, a potential discovery of some of the above CP asymmetries with those peculiar correlations
without any NP signal in $d_{e,n}$ at the level of $d_{e,n}<10^{-28}\,e\,$cm would most likely rule
out the FBMSSM and favour non-MFV models with purely left-handed currents.

\section{Comparison with Other Models}\label{sec:comparison}
\setcounter{equation}{0}

\subsection{Comparison with the LHT and RS Models}\label{sec:lht-rs}

In the present paper we have discussed several specific supersymmetric models that exhibit
different patterns of flavour and CP violation. In this section we would like to compare
briefly these patterns with the ones found in the extensive analyses of flavour violation
in the Littlest Higgs model with T-parity (LHT)
\cite{Blanke:2006sb,Blanke:2006eb,Blanke:2007db,Goto:2008fj,delAguila:2008zu,Blanke:2009am}
and in a Randall-Sundrum (RS) Model with custodial protection and with a Kaluza-Klein (KK)
scale in the reach of the LHC \cite{Blanke:2008zb,Blanke:2008yr}.

Let us first recall that the LHT and RS models contain new sources of flavour and CP violation but
while in the LHT model the operator structure in the effective weak Hamiltonians is the same as in
the SM  \cite{Blanke:2006sb,Blanke:2006eb,Blanke:2007db}, in the RS model new operators are present \cite{Blanke:2008zb,Blanke:2008yr}. In particular the left-right operators contributing to 
$\epsilon_K$ imply some fine tuning of the parameters in order to keep $\epsilon_K$ under control
if one wants to have the KK scale in the reach of the LHC \cite{Blanke:2008zb}. Analogous tunings,
albeit not as serious, are necessary in the RS model, in order to be in agreement with the experimental
bounds on $\text{BR}(B\rightarrow X_s\gamma)$ \cite{Agashe:2008uz}, $\text{BR}(\mu\rightarrow e\gamma)$ \cite{Agashe:2006iy,Davidson:2007si,Agashe:2009tu} and EDMs \cite{Agashe:2004cp,Iltan:2007sc},
that are all dominated by dipole operators. Also contributions to $\epsilon^{\prime}/{\epsilon}$
can be large \cite{Gedalia:2009ws}.

On the other hand in the LHT model $\text{BR}(B\to X_s\gamma)$ and EDMs are fully under control \cite{Blanke:2006sb} as NP contributions to the Wilson coefficients of dipole operators are small.
Still the branching ratio for the decay $\mu\to e\gamma$ can reach the present upper bound
\cite{Blanke:2007db,delAguila:2008zu} and sizable contributions to $\epsilon^{\prime}/{\epsilon}$
are possible~\cite{Blanke:2007wr}

We now compare the predictions of the LHT and RS models with the ones of the supersymmetric models
analyzed in the present paper. To this end we confine our discussion to four very interesting channels: $S_{\psi\phi}$, $B_s\to\mu^+\mu^-$, $K^+\to\pi^+\nu\bar\nu$ and $\mu\to e\gamma$ that are best suited
for the distinction of supersymmetric models from LHT and RS models.

\begin{itemize}
\item First $S_{\psi\phi}$ can be large in LHT, RS, AC, RVV2 and AKM models, however with the following
hierarchy for the maximal possible values
\begin{equation}\label{eq:hierarchy}
(S_{\psi\phi})_\text{LHT}^\text{max}\approx (S_{\psi\phi})_\text{AKM}^\text{max} < (S_{\psi\phi})_\text{RVV2}^\text{max} < (S_{\psi\phi})_\text{RS}^\text{max} \approx (S_{\psi\phi})_\text{AC}^\text{max}
\end{equation}
with $(S_{\psi\phi})_\text{LHT}^\text{max}\approx 0.3$ and $(S_{\psi\phi})_\text{AC}^\text{max}\simeq 1$~\footnote{In the LHT model, $S_{\psi\phi}$ can be significant in spite of the purely left-handed 
structure of the flavour violating currents because the $b\to s\gamma$ and $B_s\rightarrow\mu^+\mu^-$ 
constraints are easier satisfied than in the $\delta$LL model.}.

Instead the values of $S_{\psi\phi}$ in the $\delta$LL model, in the FBMSSM and the MFV MSSM are SM-like.
\item While $\text{BR}(B_s\rightarrow\mu^+\mu^-)$ in all supersymmetric models discussed by us can be as
large as the present experimental upper bound, the enhancements of $\text{BR}(B_s\to\mu^+\mu^-)$ in
the RS and LHT models do not exceed $10\%$ and $30\%$, respectively.
\item The opposite pattern is found for $K^+\to\pi^+\nu\bar\nu$ decays. In all the supersymmetric flavour
models discussed by us, $\text{BR}(K^+\to \pi^+\nu\bar\nu)$ is basically SM-like. On the other hand in
the RS and LHT models $\text{BR}(K^+\to\pi^+\nu\bar\nu)$ can be enhanced by as much as a factor 1.5 and 2.5, respectively.
\item Of particular interest will be the impact of a measurement of $S_{\psi\phi}$ much larger
than the SM value:
\begin{enumerate}
\item On one hand this would exclude the $\delta$LL, FBMSSM and MFV MSSM models.
\item On the other hand in the AC and AKM models it would imply a lower bound on
$\text{BR}(B_s\to\mu^+\mu^-)$ significantly higher than possible values in the LHT and RS models
and consequently the measurement of $\text{BR}(B_s\to\mu^+\mu^-)$ could distinguish these two
classes of models. In the RVV2 model large values of the branching ratio in question are possible
but not necessarily implied by the anomalously large value of $S_{\psi\phi}$.
\item In the case of $K^+\to\pi^+\nu\bar\nu$, a measurement of a sizable $S_{\psi\phi}$ precludes
significant enhancements of $\text{BR}(K^+\to\pi^+\nu\bar\nu)$ in RS and LHT models. Consequently
in such a scenario it will be difficult to distinguish these two models from the supersymmetric ones,
just taking into account the decay $K^+\rightarrow\pi^+\nu\bar\nu$. On the other hand, the measurement
of $\text{BR}(K^+\to\pi^+\nu\bar\nu)$ much larger than the SM value would favour RS and LHT models to
all the supersymmetric flavour models discussed by us if $S_{\psi\phi}$ is SM-like but would
rule out also the RS and LHT models if $S_{\psi\phi}$ is large.
\end{enumerate}
\item Finally let us recall that while the supersymmetric models and LHT models can reach the
experimental bound for $\text{BR}(\mu\to e\gamma)$ \cite{Blanke:2007db,delAguila:2008zu,Blanke:2009am},
the ratios ${\text{BR}(\mu\to 3e)}/{\text{BR}(\mu\to e\gamma)}$ and
${\text{BR}(\tau\to 3\mu)}/{\text{BR}(\tau\to e\gamma)}$ are in supersymmetric models roughly
by one order of magnitude smaller than in the LHT model.
\end{itemize}

In summary, we observe that already the combination of $S_{\psi\phi}$,
$\text{BR}(B_s\to\mu^+\mu^-)$ and $\text{BR}(K^+\to\pi^+\nu\bar\nu)$ should 
be able to distinguish between the AC, RVV2, AKM,  LHT and RS models
unless all these three observables are found SM-like. The distinction between
(AC,RVV2,AKM) and (LHT,RS) models can easily be made with the help of
$\text{BR}(B_s\to\mu^+\mu^-)$ and $\text{BR}(K^+\to \pi^+\nu\bar\nu)$ only 
but the inclusion in this test of $S_{\psi\phi}$ will be very helpful.
The distinction between the AC and RVV2 models has been discussed in the previous
sections while the one between LHT and RS in \cite{Blanke:2008yr,Blanke:2009am}.
Here the correlation between $K_L\to\pi^0\nu\bar\nu$ and $K^+\to\pi^+\nu\bar\nu$
markedly different in both models and the hierarchy in (\ref{eq:hierarchy})
could play important roles.

\definecolor{green1}{rgb}{0.06,0.66,0.06}
\definecolor{orange1}{rgb}{0.98,0.60,0.07}
\newcommand{\green}{{\color{green1}$\bigstar$}}
\newcommand{\orange}{{\color{orange1}\LARGE \protect\raisebox{-0.1em}{$\bullet$}}}
\newcommand{\red}{{\color{red}\small \protect\raisebox{-0.05em}{$\blacksquare$}}}
\newcommand{\three}{{\color{red}$\bigstar\bigstar\bigstar$}}
\newcommand{\two}{{\color{blue}$\bigstar\bigstar$}}
\newcommand{\one}{{\color{black}$\bigstar$}}
%
\begin{table}[t]
\addtolength{\arraycolsep}{4pt}
\renewcommand{\arraystretch}{1.5}
\centering
\begin{tabular}{|l|c|c|c|c|c|c|c|}
\hline
&  AC & RVV2 & AKM  & $\delta$LL & FBMSSM & LHT & RS
\\
\hline\hline
$D^0-\bar D^0$& \three & \one & \one & \one & \one & \three & ?
\\
\hline
$\epsilon_K$& \one & \three & \three & \one & \one & \two & \three 
\\
\hline
$ S_{\psi\phi}$ & \three & \three & \three & \one & \one & \three & \three 
\\
\hline\hline
$S_{\phi K_S}$ & \three & \two & \one & \three & \three & \one & ? \\
\hline
$A_{\rm CP}\left(B\rightarrow X_s\gamma\right)$ & \one & \one & \one & \three & \three & \one & ?
\\
\hline
$A_{7,8}(B\to K^*\mu^+\mu^-)$ & \one & \one & \one & \three & \three & \two & ?
\\
\hline
$A_{9}(B\to K^*\mu^+\mu^-)$ & \one & \one & \one & \one & \one & \one & ? 
\\
\hline
$B\to K^{(*)}\nu\bar\nu$  & \one & \one & \one & \one & \one & \one & \one 
\\
\hline
$B_s\rightarrow\mu^+\mu^-$ & \three & \three & \three & \three & \three & \one & \one
\\
\hline
$K^+\rightarrow\pi^+\nu\bar\nu$ & \one & \one & \one & \one & \one & \three & \three 
\\
\hline
$K_L\rightarrow\pi^0\nu\bar\nu$ & \one & \one & \one & \one & \one & \three & \three
\\
\hline
$\mu\rightarrow e\gamma$& \three & \three & \three & \three & \three & \three & \three \\
\hline
$\tau\rightarrow \mu\gamma$ & \three & \three & \one & \three & \three  & \three & \three \\
\hline
$\mu + N\rightarrow e + N$& \three & \three & \three & \three & \three & \three & \three \\
\hline\hline
$d_n$& \three & \three & \three & \two & \three & \one & \three
\\
\hline
$d_e$& \three & \three & \two & \one & \three & \one & \three
\\
\hline
$\left(g-2\right)_\mu$& \three & \three & \two & \three & \three & \one & ?
\\
\hline

\end{tabular}
\renewcommand{\arraystretch}{1}
\caption{\small
``DNA'' of flavour physics effects for the most interesting observables in a selection of SUSY
and non-SUSY models \three\ signals large effects, \two\ visible but small effects and \one\
implies that the given model does not predict sizable effects in that observable.}
\label{tab:DNA}
\end{table}

\subsection{DNA-Flavour Test of New Physics Models}\label{sec:dna}

We have seen in the previous sections and in sec. \ref{sec:lht-rs} that the patterns of flavour
violation found in various extensions of the SM differed from model to model, thereby allowing
in the future to find out which of the models considered by us, if any, can survive the future
measurements. Undoubtedly, the correlations between various observables that are often
characteristic for a given model will be of the utmost importance in these tests.

In tab.~\ref{tab:DNA}, we show a summary of the potential size of deviations from the SM results
allowed for a large number of observables considered in the text, when all existing constraints
from other observables not listed there are taken into account. We distinguish among:
\begin{itemize}
\item large effects (three {\it red} stars),
\item moderate but still visible effects (two {\it blue} stars),
\item vanishingly small effects (one {\it black} star).
\end{itemize}
This table can be considered as the collection of the DNA's for various models.
These DNA's will be modified as new experimental data will be availabe and in certain
cases we will be able to declare certain models to be disfavoured or even ruled out.

In constructing the table we did not take into account possible correlations among
the observables listed there. We have seen that in some models, it is not possible to
obtain large effects simultaneously for certain pairs or sets of observables and
consequently future measurements of a few observables considered in tab.~\ref{tab:DNA}
will have an impact on the patterns shown in this DNA table. It will be interesting to
monitor the changes in this table when the future experiments will provide new results.

\section{Flavour vs. LHC Data}\label{sec:flavour_vs_lhc}

In sec. \ref{sec:numerics}, we have performed a correlated analysis of low-energy predictions,
arising in SUSY flavour models, for a complete set of flavour observables with the aim of
outlining clear patterns of deviation from the SM predictions.

In the following, we will try to address the question of the synergy and interplay existing
between direct and indirect NP searches.

For instance, one relevant question could be: which is the expected SUSY spectrum at the LHC
in case sizable indirect NP effects will be detected in some flavour channels at the upcoming
experimental facilities?
Vice versa, if the LHC will measure the masses of some SUSY particles (within a given accuracy),
which additional information could we obtain from some non-standard effects in flavour observables?

To answer the above questions, we analyze the flavour physics predictions of the flavour models
discussed in sec.~\ref{sec:numerics} with respect to the expected spectrum at the LHC.

\begin{figure}
\centering
\includegraphics[width=0.3\textwidth]{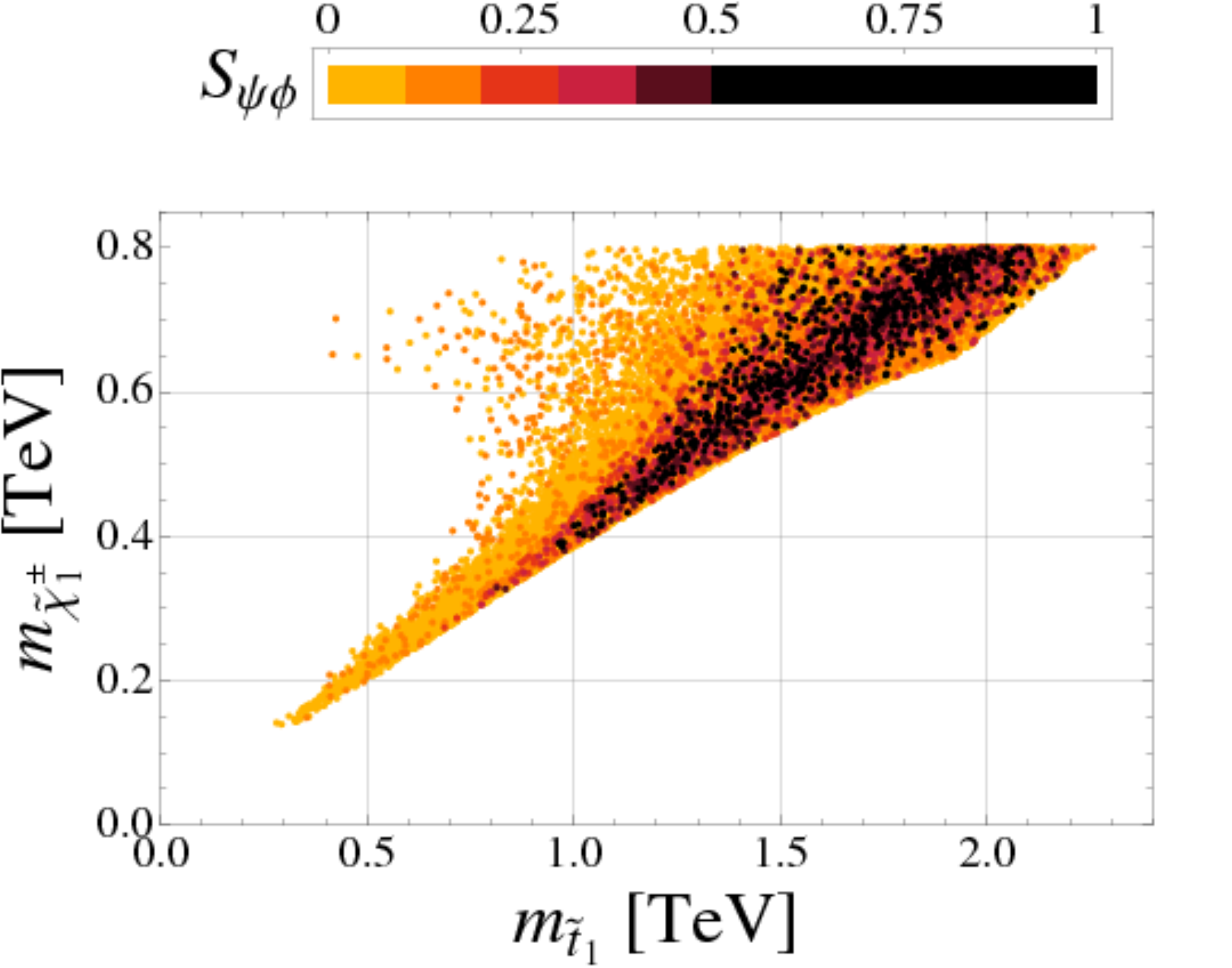}~~~~
\includegraphics[width=0.3\textwidth]{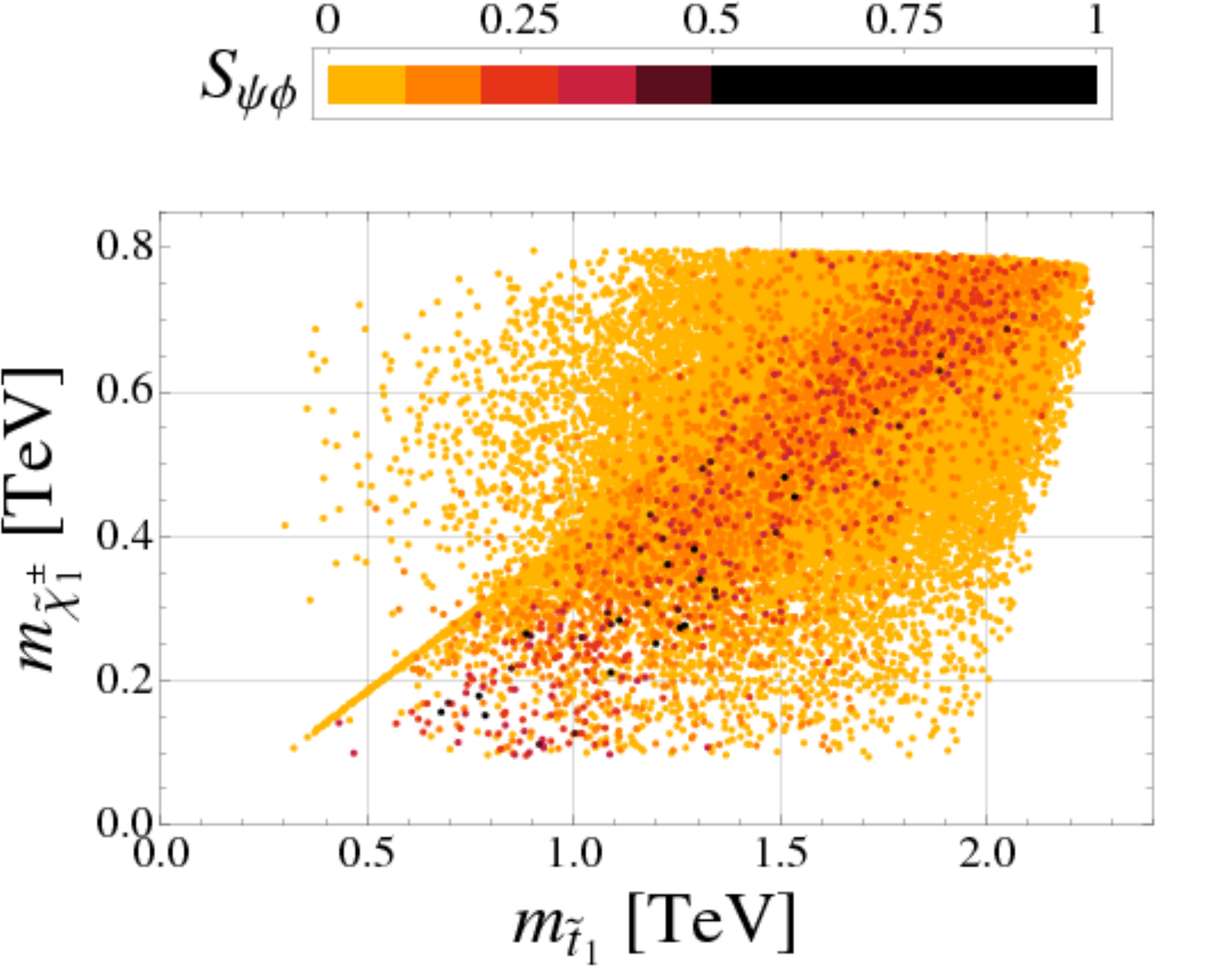}~~~~
\includegraphics[width=0.3\textwidth]{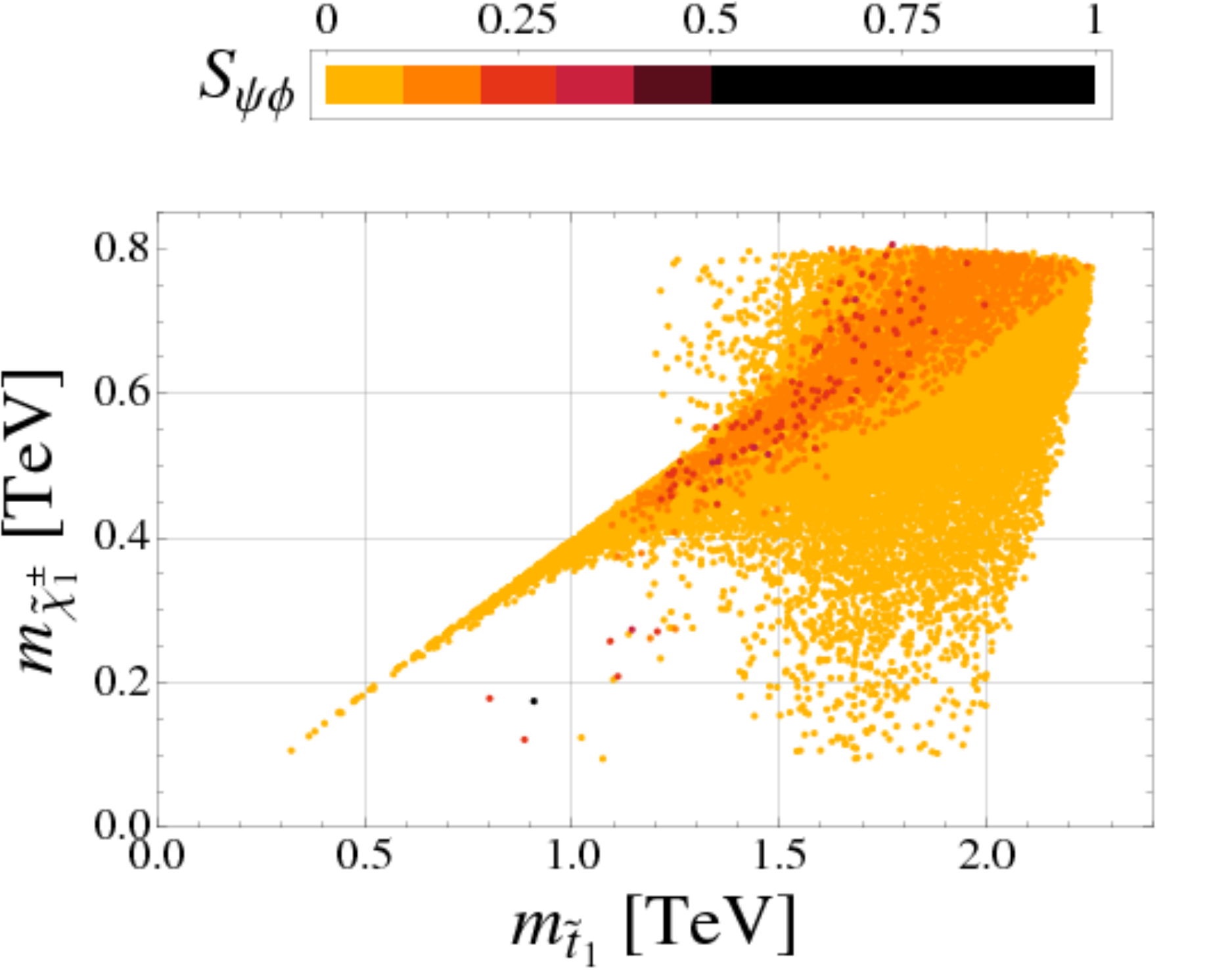}\\[20pt]
\includegraphics[width=0.3\textwidth]{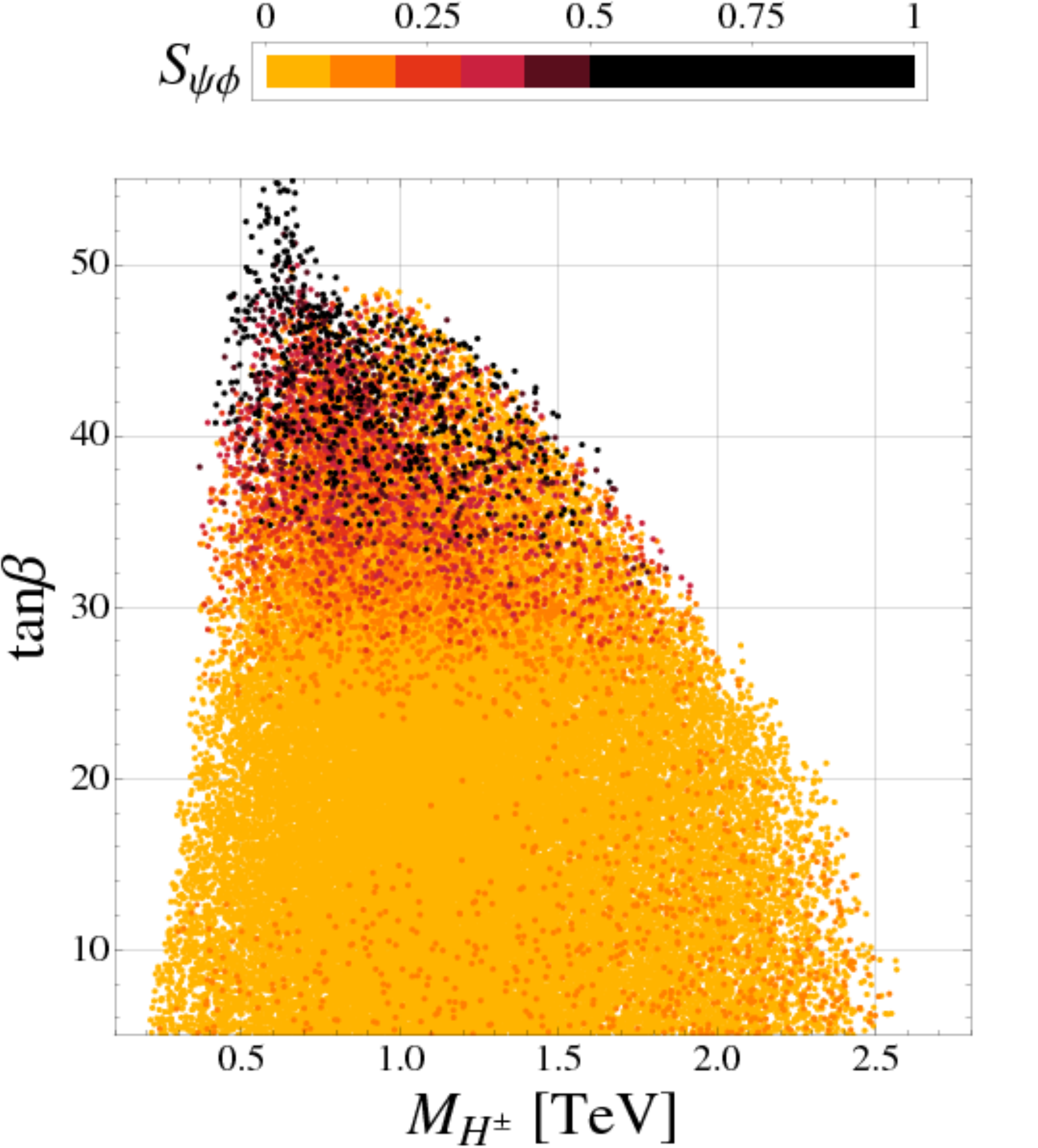}~~~
\includegraphics[width=0.3\textwidth]{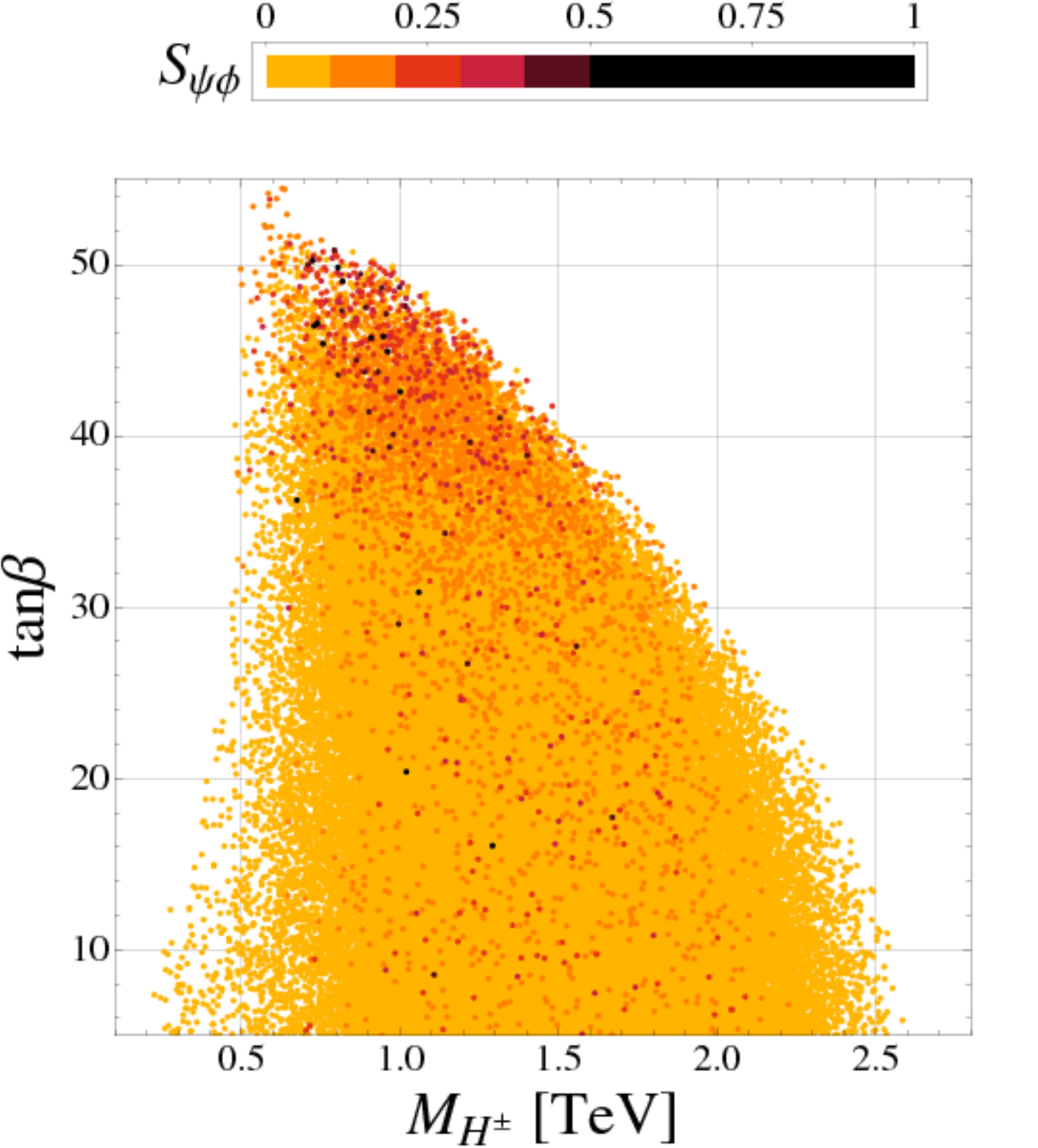}~~~
\includegraphics[width=0.3\textwidth]{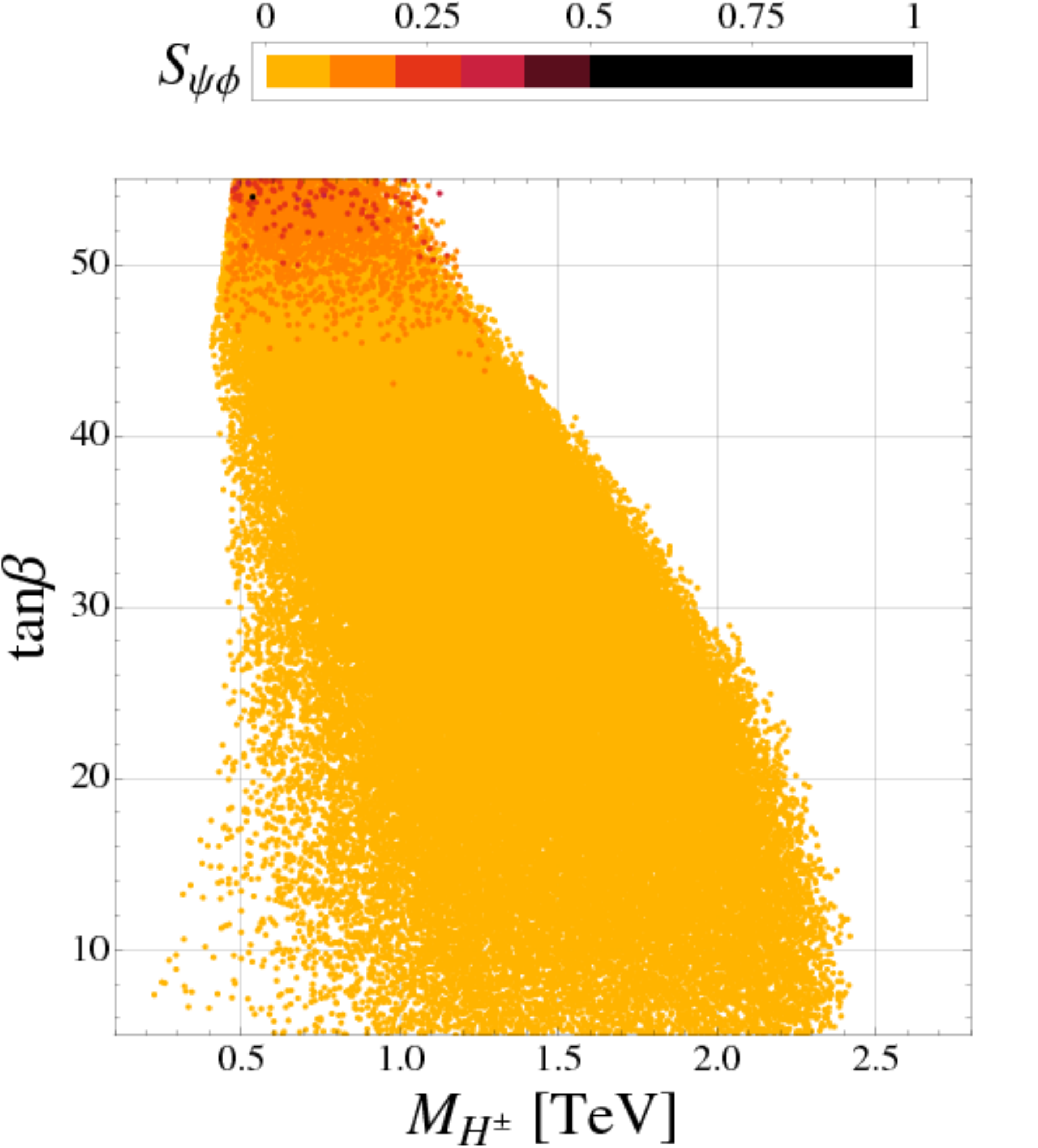}\\[20pt]
\includegraphics[width=0.3\textwidth]{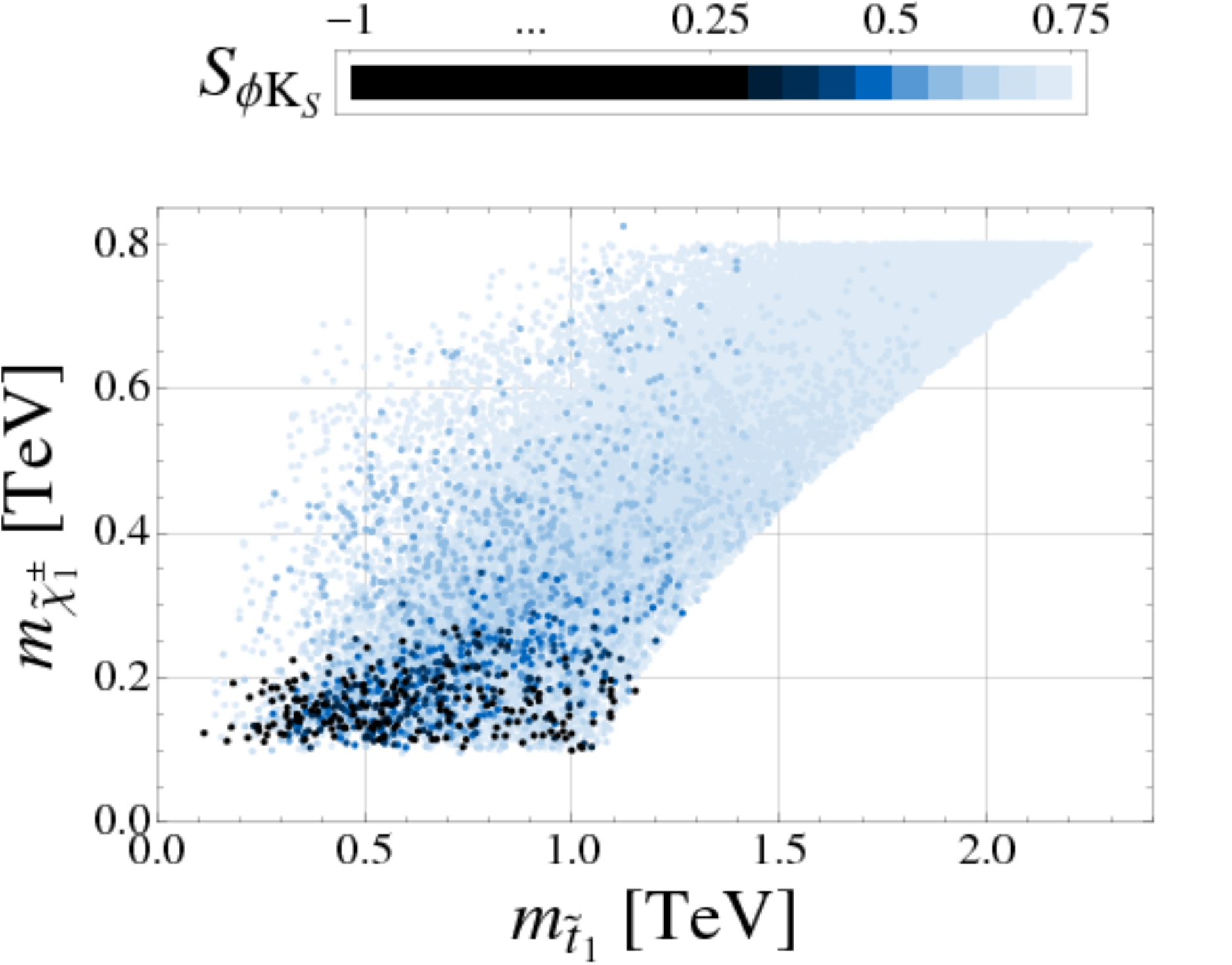}~~~~
\includegraphics[width=0.3\textwidth]{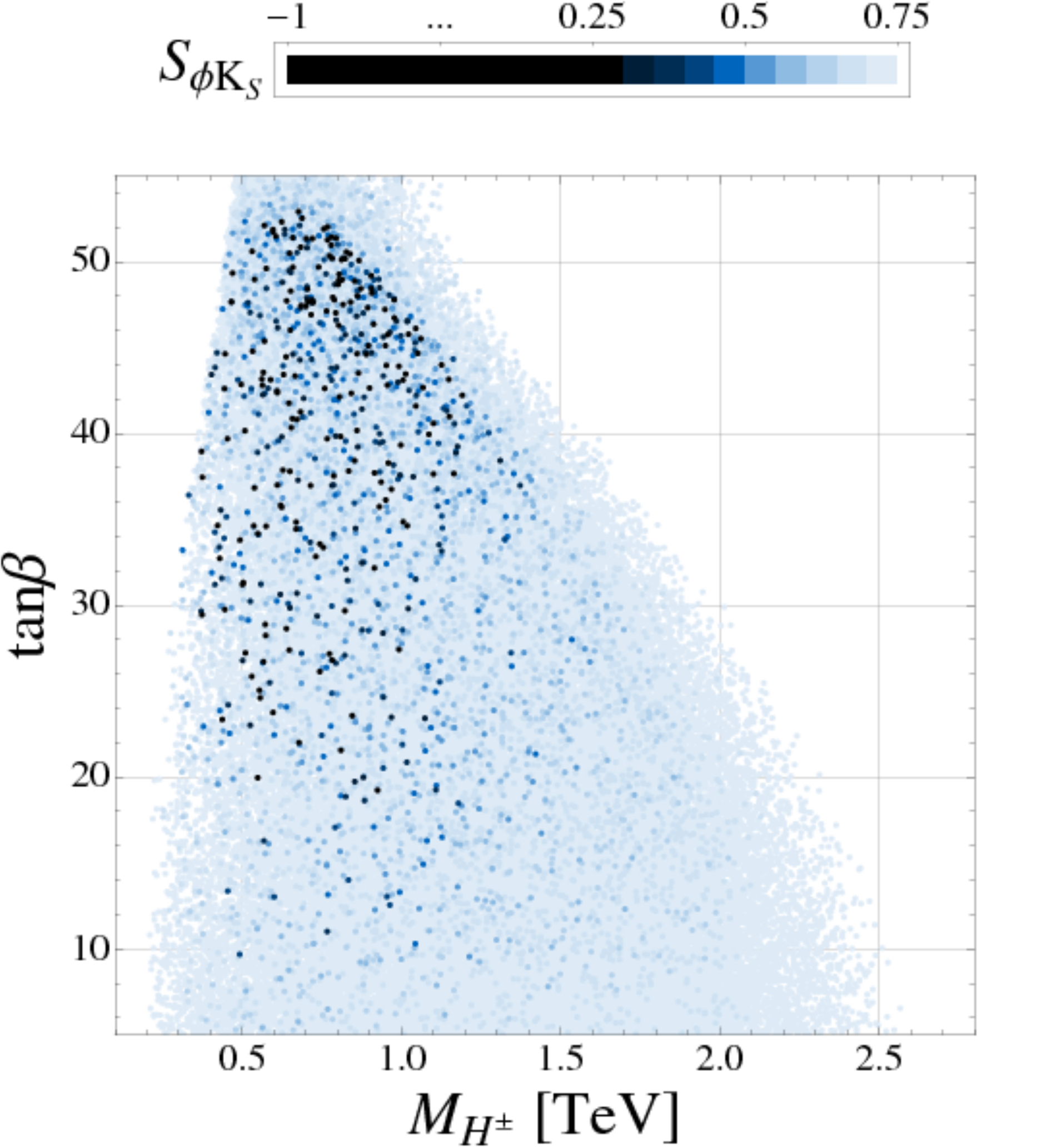}
\caption{\small
The plane of the lightest stop mass ($m_{\tilde{t}_{1}}$)
vs. the lightest chargino mass ($m_{\tilde{\chi}_{1}}$) as well as the $M_H-\tan\beta$ planes.
The first two rows of fig.~\ref{fig:LHC} correspond to the AC model~\cite{Agashe:2003rj}
(plots on the left), the RVV2 model~\cite{Ross:2004qn} (plots in the middle) and the AKM model~\cite{Antusch:2007re} (plots on the right), respectively. The different colors show
the possible values for $S_{\psi\phi}$ in these models as indicated in the overall bar.
The last lower plots correspond to models with pure left-handed currents with CKM-like
mixing angles~\cite{Hall:1995es} and the different colors indicate the attained values
for $S_{\phi K_S}$.}
\label{fig:LHC}
\end{figure}

In fig.~\ref{fig:LHC}, we show the planes for the lightest stop mass ($m_{\tilde{t}_{1}}$)
vs. the lightest chargino mass ($m_{\tilde{\chi}_{1}}$) as well as the $M_H-\tan\beta$ planes.
The first two rows of fig.~\ref{fig:LHC} correspond to the AC model~\cite{Agashe:2003rj}
(plots on the left), the RVV2 model~\cite{Ross:2004qn} (plots in the middle) and the AKM model~\cite{Antusch:2007re} (plots on the right), respectively.
The different colors show the possible values for $S_{\psi\phi}$ in these models as indicated
in the overall bar. The last lower plots correspond to models with pure left-handed currents
with CKM-like mixing angles~\cite{Hall:1995es} and the different colors indicate the attained
values for $S_{\phi K_S}$.

An interesting feature emerging from fig.~\ref{fig:LHC} is that, in the case of the AC, RVV2 and AKM
models, large effects in $S_{\psi\phi}$ are possible ~--~and indeed even favoured~--~ for a quite
heavy soft SUSY spectrum, even beyond the LHC reach. However, this is not an evasion of the decoupling
theorem but, instead, a confirmation of the dominance of the Higgs mediated effects in $S_{\psi\phi}$.
In fact, since the effective FCNC couplings of the Higges with the SM fermions arise from dimension-4
operators, they do not decouple with the soft SUSY scale. Clearly, the double Higgs penguin contributions generating $S_{\psi\phi}$ decouple with the mass of the heavy Higgs sector, as confirmed by the second
row of fig.~\ref{fig:LHC}, in agreement with the decoupling theorem. Moreover, $S_{\psi\phi}$ attains
the maximum values for heavy soft masses since the most stringent indirect constraints arising from
$\rm{BR}(b\to s\gamma)$ and $\epsilon_K$, decoupling with the soft SUSY sector, are relaxed in this case.

Hence, there exist regions of the SUSY parameter space at the border or even beyond the LHC reach where
we can expect clear non-standard signals in flavour processes. In these regions, flavour phenomena, and
thus the indirect search, represent the most powerful tool to shed light on NP in case of SUSY.
In such a case, where we will obtain only an improved upper bound on the scale of SUSY masses, the study
of the correlations among flavour observables, departing from their SM expectations, would be the only
tool at our disposal to unveil the nature of the NP theory that is operating.

An opposite behaviour, compared to the AC, RVV2 and AKM models, is shown by the $\delta$LL model (last
row of fig.~\ref{fig:LHC}) where the double Higgs penguins do not play a significant role and the flavour observables decouple with the soft SUSY masses.

Large ~--~non-standard~--~ effects in flavour observables necessarily imply, in this case, a SUSY
spectrum within the LHC reach. Hence, combining the information on the spectrum from the LHC with
the information from the low-energy flavour processes, we would be able, to some extent, to measure
the mixing angle regulating the flavour transition. Such an achievement would represent a crucial
step forward towards the reconstruction of the underlying flavour symmetry at work. The described
situation represents one clear example of the synergy and interplay of direct and indirect NP searches.

\section{Summary}\label{sec:summary}
\setcounter{equation}{0}

The coming years will witness tremendous progress at the high energy frontier accomplished primarily
at the LHC, where the available energy will be increased by one order of magnitude. Equivalently, for
the first time we will be able to resolve directly distances well below $10^{-18}$~m, that have been
explored so far. Parallel to these developments, important advances are expected at the high precision
frontier through the improved $B_s$-physics experiments at the Tevatron and in particular LHCb at CERN.
At later stages in the coming decade these efforts will be strengthened by new rare $K$ experiments at
J-PARC, the NA62 collaboration at CERN and possibly Project~X at Fermilab as well as Belle-II at KEK
and the planned Super$B$ facility in Rome. The latter two will also provide new insights into the FCNC
processes in the $D$ meson system and in charged lepton decays.

While the main goal at the high energy frontier is the discovery of new particles and the determination
of their masses, the main goal of flavour physics is the search for the footprints of these new
particles in rare processes and the determination of their couplings. As the latter exploration of
very short distance scales is indirect, only measurements of a large number of observables and the
study of correlations between them in a given extension of the SM and in particular of patterns of
flavour violation characteristic for a given model can allow us to identify the correct NP scenario.

In the present paper, we have performed an extensive study of processes governed by $b\to s$ transitions
and of their correlations with processes governed by $b\to d$ transitions, $s\to d$ transitions,
$D^0-\bar D^0$ oscillations, lepton flavour violating decays, electric dipole moments and $(g-2)_{\mu}$.
To this end, we have considered a number of supersymmetric flavour models that on one hand aim at the 
explanation of the observed hierarchical fermion masses and mixings and on the other hand provide natural suppression of FCNC transitions.
In particular we have analyzed the following representative scenarios:
\begin{itemize}
\item [i)] dominance of right-handed currents (abelian model by Agashe and Carone~\cite{Agashe:2003rj}),
\item [ii)] comparable left- and right-handed currents with CKM-like mixing angles represented by the
special version (RVV2) of the non abelian $SU(3)$ model by Ross, Velasco and Vives~\cite{Ross:2004qn}
  as discussed recently in~\cite{Calibbi:2009ja} and the model by Antusch, King and Malinsky (AKM)
  \cite{Antusch:2007re},
\item [iii)] dominance of left-handed currents in non-abelian models~\cite{Hall:1995es}.
\end{itemize}

In the choice of these three classes of flavour models, we were guided by our model independent
analysis, as these three scenarios predicting quite different patterns of flavour violation should
give a good representation of most SUSY flavour models discussed in the literature.

The distinct patterns of flavour violation found in each scenario have been listed in detail in
sec. 6 and the corresponding plots can be found in figures \ref{fig:model_AC}--\ref{fig:model_CKM}.

The main messages from our analysis of the models in question are as follows:
\begin{itemize}
\item
Supersymmetric models with right-handed currents (AC, RVV2, AKM) and those with left-handed currents
can be globally distinguished by the values of the CP-asymmetries $S_{\psi\phi}$ and $S_{\phi K_S}$
with the following important result: none of the models considered by us can simultaneously solve the $S_{\psi\phi}$ and $S_{\phi K_S}$ anomalies observed in the data. In the models with RH currents,
$S_{\psi\phi}$ can naturally be much larger than its SM value, while $S_{\phi K_S}$ remains either
SM-like or its correlation with $S_{\psi\phi}$ is inconsistent with the data. On the contrary, in the
models with LH currents, $S_{\psi\phi}$ remains SM-like, while the $S_{\phi K_S}$ anomaly can easily
be solved. Thus, already future precise measurements of $S_{\psi\phi}$ and $S_{\phi K_S}$ will select
one of these two classes of models, if any.
\item
The desire to explain the $S_{\psi\phi}$ anomaly within the models with RH currents unambiguously
implies, in the case of the AC and the AKM models, values of $\text{BR}(B_s\to\mu^+\mu^-)$ as high
as several $10^{-8}$, while in the RVV2 model, such values are also possible but not necessarily
implied by the large value of $S_{\psi\phi}$.
Interestingly enough, in all these models large values of $S_{\psi\phi}$ can also explain the
$(g-2)_\mu$ anomaly. Moreover, in the AC and RVV2 models, the ratios $\text{BR}(B_d\to\mu^+\mu^-)/\text{BR}(B_s\to\mu^+\mu^-)$ and $\text{BR}(B_d\to\mu^+\mu^-)/\Delta M_s$ 
can significantly differ from the MFV prediction, providing a splendid opportunity to shed light 
on new sources of flavour violation beyond the CKM ones.
Further, $\text{BR}(B_d\to\mu^+\mu^-)/\text{BR}(B_s\to\mu^+\mu^-)$ is predicted to be dominantly
smaller than its MFV value especially for $\text{BR}(B_s\to\mu^+\mu^-)\gtrsim 10^{-8}$.
Values of $\text{BR}(B_d\to\mu^+\mu^-)$ as high as $1\times 10^{-9}$ are still possible in these models.
\item The hadronic EDMs represent very sensitive probes of SUSY flavour models with right-handed
currents. In the AC model, large values for the neutron EDM might be easily generated by both the
up- and strange-quark (C)EDM. In the former case, visible CPV effects in $D^0-\bar D^0$ mixing
are also expected while in the latter case large CPV effects in $B_s$ systems are unavoidable.
The RVV2 and AKM models predict values for the down-quark (C)EDM and, hence for the neutron
EDM, above the $\approx 10^{-28}\,e$cm level when a large ~--~non-standard~--~ $S_{\psi\phi}$
is generated. All the above models predict a large strange-quark (C)EDM, hence, a reliable
knowledge of its contribution to the hadronic EDMs, by means of lattice QCD techniques, would
be of the utmost importance to probe or to falsify flavour models embedded in a SUSY framework.
\item
In the RVV2 and AKM models, the desire to explain the $(g-2)_\mu$ anomaly implies that
$\text{BR}(\mu\to e\gamma)\gtrsim 10^{-13}$, in the reach of the MEG experiment, and
$d_e>10^{-29}\,e$cm ($d_e\lesssim 10^{-30}\,e$cm) in the RVV2 (AKM) model is predicted.
Moreover, in the case of the RVV2 model, $\text{BR}(\tau\to\mu\gamma)\gtrsim 10^{-9}$ is
then in the reach of Super-B machines, while $\tau\to\mu\gamma$ remains most likely beyond
the Super-B reach in the case of the AKM model.
The explanation of the $(g-2)_\mu$ anomaly, combined with non-standard values for $S_{\psi\phi}$,
would imply $\text{BR}(\mu\to e\gamma)\gtrsim 10^{-12}$ in both the RVV2 and AKM models.
\item
Finally, while the abelian AC model resolves the present UT tension through the modification
of the ratio $\Delta M_d/\Delta M_s$, the non-abelian flavour models RVV2 and AKM provide the
solution through NP contributions to $\epsilon_K$. Moreover, while the AC model predicts
sizable NP contributions to $D^0-\bar D^0$ mixing, such contributions are negligibly small
in the RVV2 and AKM models.
\item
In the supersymmetric models with LH currents only, the desire to explain the $S_{\phi K_S}$
anomaly unambiguously implies that the direct CP asymmetry in $b\to s\gamma$ is much
larger than its SM value. This is in contrast to the models with RH currents where this
asymmetry remains SM-like. Also the solution to the $(g-2)_\mu$ anomaly can be easily
accounted for.
\item
Interestingly, also in the LH-current-models --that are in many aspects similar to MFV models--,
the ratio $\text{BR}(B_d\to\mu^+\mu^-)/\text{BR}(B_s\to\mu^+\mu^-)$ can not only deviate significantly
from its MFV value, but in contrast to the models with RH currents considered by us can also be
much larger than the latter value. Consequently, values for $\text{BR}(B_d\to\mu^+\mu^-)$ as high
as $2\times 10^{-9}$ are still possible while being consistent with the bounds on all other
observables, in particular the one on $\text{BR}(B_s\to\mu^+\mu^-)$. Also interesting correlations
between $S_{\phi K_S}$ and CP asymmetries in $B\to K^*\ell^+\ell^-$ are found.
\item The FBMSSM as well as the MFV MSSM bear several similarities to the models with
LH currents only. In particular, in all these models, $S_{\psi\phi}$ is basically SM like.
The FBMSSM and the MFV MSSM differ however from the $\delta$LL model especially in their
predictions of the EDMs. In the FBMSSM and the MFV MSSM, large effects in CP violating
$\Delta F=1$ observables naturally imply values for the EDMs close to the current experimental
limits. In the $\delta$LL model, on the other hand, the EDMs can be easily well under control.
\end{itemize}

This summary of our most important results already shows that the simultaneous study of various 
flavour violating processes can allow us to distinguish various NP scenarios. More results can be
found in the numerous figures presented by us.

We have also presented a comprehensive model-independent analysis of the MSSM with general
sources of flavour and CP violation that goes in several aspects, listed at the beginning of our
paper, beyond many analyses found in the literature.
The bounds on the mass insertions $(\delta_{u,d}^{AB})_{i,j}$ can be found in figures
\ref{fig:MI12}--\ref{fig:MI23}. Any version of the MSSM  based on a SUGRA spectrum
has to satisfy them. Beyond our MSSM analysis, we have also emphasized the usefulness of
the $R_b-\gamma$ plane in exhibiting transparently various tensions in present UT analyses.

It will be exciting to monitor upcoming results from Tevatron and the LHC on $S_{\psi\phi}$
and $B_s\to\mu^+\mu^-$. Already these two measurements will be capable of excluding some of
the models considered by us and distinguish them from the LHT and RS model with custodial
protection in which $S_{\psi\phi}$ can be large but $B_{s,d}\to\mu^+\mu^-$ remain SM-like.
Further observables analyzed by us will help to identify more precisely the correct extension
of the SM. In particular, while the branching ratios for $K\to\pi \nu\bar\nu$ decays in the
supersymmetric models considered by us remain SM-like, they can be significantly enhanced in
the LHT and RS models.

A DNA-Flavour Test proposed by us should give still a deeper insight into the patterns of flavour
violation in various scenarios, in particular when it is considered simultaneously  with various
correlations present in concrete models. The interplay of these efforts with the direct searches 
for NP will be most exciting.

In conclusion, the above physics cases are representative of the richness which is present in flavour
physics once we assume flavour models embedded within a gravity mediated SUSY breaking framework.
It could be that, at the end, flavour physics is one of the very few tools we have to understand from
low-energy physics whether nature possesses supersymmetry at the root of its symmetries, or whether
other scenarios like LHT or RS are realized in nature.

\section*{Acknowledgments}

We would like to thank M.~Blanke for discussions and especially J.~Jones-Perez and O.~Vives
for many informative correspondences about the RVV2 model~\cite{Calibbi:2009ja}.

This work has been supported in part by the Cluster of Excellence ``Origin and
Structure of the Universe'', the Deutsche Forschungsgemeinschaft
(DFG) under contract BU 706/2-1 and the German Bundesministerium f\"ur Bildung
und Forschung under contract 05HT6WOA.
WA acknowledges support by the Graduiertenkolleg GRK 1054 of DFG.
SG acknowledges support by the European Community's Marie Curie Research Training Network
under contract MRTN-CT-2006-035505 [``HEP-TOOLS''].

\appendix

\setcounter{equation}{0}
\renewcommand{\theequation}{A.\arabic{equation}}

\section{Explicit Expressions for the Loop Functions} \label{sec:appendix}

\subsection*{Loop Functions for the \boldmath $\Delta F=2$ Mixing Amplitudes}

\begin{equation}
g_1^{(1)}(x) = - \frac{11+144x+27x^2-2x^3}{108(1-x)^4} - \frac{x(13+17x)}{18(1-x)^5}\log{x}~,
\end{equation}
\begin{equation}
g_1^{(2)}(x) = \frac{33+665x+237x^2-39x^3+4x^4}{216(1-x)^5} + \frac{x(26+49x)}{18(1-x)^6}\log{x}~,
\end{equation}
\begin{equation}
g_1^{(3)}(x) = - \frac{66+1835x+1005x^2-255x^3+55x^4-6x^5}{1080(1-x)^6} - \frac{x(13+32x)}{18(1-x)^7}\log{x}~,
\end{equation}

\begin{equation}
g_4^{(1)}(x) = \frac{2-99x-54x^2+7x^3}{18(1-x)^4} - \frac{x(5+19x)}{3(1-x)^5}\log{x}~,
\end{equation}
\begin{equation}
g_4^{(2)}(x) = - \frac{3-212x-192x^2+48x^3-7x^4}{18(1-x)^5} + \frac{10x(1+5x)}{3(1-x)^6}\log{x}~,
\end{equation}
\begin{equation}
g_4^{(3)}(x) = \frac{12-1117x-1452x^2+528x^3-152x^4+21x^5}{180(1-x)^6} - \frac{x(5+31x)}{3(1-x)^7}\log{x}~,
\end{equation}

\begin{equation}
g_5^{(1)}(x) = - \frac{10+117x+18x^2-x^3}{54(1-x)^4} - \frac{x(11+13x)}{9(1-x)^5}\log{x}~,
\end{equation}
\begin{equation}
g_5^{(2)}(x) = \frac{15+272x+84x^2-12x^3+x^4}{54(1-x)^5} + \frac{2x(11+19x)}{9(1-x)^6}\log{x}~,
\end{equation}
\begin{equation}
g_5^{(3)}(x) = - \frac{60+1507x+732x^2-168x^3+32x^4-3x^5}{540(1-x)^6} - \frac{x(11+25x)}{9(1-x)^7}\log{x}~,
\end{equation}


\begin{equation}
h_1(x) = \frac{4(1+x)}{3(1-x)^2} + \frac{8x}{3(1-x)^3}\log{x}~,
\end{equation}
\begin{equation}
h_2(x) = -\frac{4(2+5x-x^2)}{9(1-x)^3} - \frac{8x}{3(1-x)^4}\log{x}~,
\end{equation}
\begin{equation}
h_3(x) = -\frac{1}{2(1-x)} - \frac{x}{2(1-x)^2}\log{x}~,
\end{equation}
\begin{equation}
h_4(x,y) = -\frac{1}{(1-x)(1-y)} + \frac{x}{(1-x)^2(y-x)}\log{x} + \frac{y}{(1-y)^2(x-y)}\log{y}~,
\end{equation}


\begin{equation}
f_1(x) = - \frac{x + 1}{4(1-x)^2} - \frac{x }{2(1-x)^3}\log x~,
\end{equation}
\begin{equation}
f_3(x) =  \frac{x^2 - 6x - 17}{6(1-x)^4} - \frac{3x + 1}{(1-x)^5}\log x~,
\end{equation}

\begin{equation}
f(x) =  \frac{1}{1-x} + \frac{x}{(1-x)^2}\log x~.
\end{equation}

\subsection*{Loop Functions for \boldmath $b \to s \gamma$}

\begin{equation}
h_7(x) = - \frac{5x^2-3x}{12(1-x)^2} - \frac{3x^2-2x}{6(1-x)^3}\log{x}~,
\end{equation}
\begin{equation}
h_8(x) = -\frac{x^2-3x}{4(1-x)^2} + \frac{x}{2(1-x)^3}\log{x}~,
\end{equation}
\begin{equation}
f_7^{(2)}(x) = - \frac{13-7x}{24(1-x)^3} - \frac{3+2x-2x^2}{12(1-x)^4}\log{x}~,
\end{equation}
\begin{equation}
f_8^{(2)}(x) = \frac{1+5x}{8(1-x)^3} + \frac{x(2+x)}{4(1-x)^4}\log{x}~,
\end{equation}
\begin{equation}
f_{7,8}^{(1)}(x,y) = \frac{2}{x-y} \left( f_{7,8}^2(x) - f_{7,8}^2(y) \right)~,
\end{equation}
\begin{equation}
g_7^{(1)}(x) = - \frac{2(1+5x)}{9(1-x)^3} - \frac{4x(2+x)}{9(1-x)^4}\log{x}~,
\end{equation}
\begin{equation}
g_7^{(2)}(x) = - \frac{2(1+10x+x^2)}{9(1-x)^4} - \frac{4x(1+x)}{3(1-x)^5}\log{x}~,
\end{equation}
\begin{equation}
g_8^{(1)}(x) = \frac{11+x}{3(1-x)^3} + \frac{9+16x-x^2}{6(1-x)^4}\log{x}~,
\end{equation}
\begin{equation}
g_8^{(2)}(x) = \frac{53+44x-x^2}{12(1-x)^4} + \frac{3+11x+2x^2}{2(1-x)^5}\log{x}~.
\end{equation}

\subsection*{Loop Functions for \boldmath $\ell_i \to \ell_j\gamma$}

\begin{equation}
f_{2n}(x) = (-5x^2+4x+1+2x(x+2)\log x)/(4(1-x)^4)~, 
\end{equation}
\begin{equation}
f_{3n}(x) = (1+9x-9x^2-x^3+6x(x+1)\log x)/(3(1-x)^5)~,
\end{equation}
\begin{equation}
f_{2c}(x) = (-x^2-4x+5+2(2x+1)\log x)/(2(1-x)^4)~.
\end{equation}

\subsection*{Loop Functions for the EDMs}

\beq
 f^{d}_{\tilde g}(x)
 = \left\{
  - \frac{4}{9} f_0^{(3)}(x),
  - \frac{1}{6} f_0^{(3)}(x)
  + \frac{3}{2} f_1^{(3)}(x)
   \right\}~,
\eeq
\beq
 f^{u}_{\tilde g}(x)
 = \left\{
    \frac{8}{9} f_0^{(3)}(x),
  - \frac{1}{6} f_0^{(3)}(x)
  + \frac{3}{2} f_1^{(3)}(x)
   \right\}~,
\eeq
\beq
 f_{H^\pm}(z) =
 \left\{- f_0^{(0)}(z) + \frac{2}{3} f_1^{(0)}(z),~f_1^{(0)}(z)\right\}~,
\eeq
\beq
 f_{{\tilde H}^\pm}(y) = \left\{\frac{2}{3}f_0^{(1)}(y)-f_1^{(1)}(y),~
 f_0^{(1)}(y)\right\}~,
\eeq
\beq
g_{{\tilde H}^\pm}(x)= \left\{\frac{2}{3}f^{(2)}_{0}(x) - f^{(2)}_{1}(x),~ 
f_{0}^{(2)}(x)\right\}\,,
\eeq
\begin{eqnarray}
 f_0^{(0)}(x)
 \!\!\! &=& \!\!\!
  \frac{1-x^2+2x\log x}{(1-x)^3}~,
 \nonumber\\
 f_0^{(1)}(x)
 \!\!\! &=& \!\!\!
  \frac{-1-4x+5x^2-2x(x+2)\log x}{(1-x)^4}~,
 \nonumber\\
 f_0^{(2)}(x)
 \!\!\! &=& \!\!\!
  \frac{1+9x-9x^2-x^3+6x(x+1)\log x}{(1-x)^5}\,,
 \nonumber\\
 f_0^{(3)}(x)
 \!\!\! &=& \!\!\!
  \frac{-3-44x+36x^2+12x^3-x^4-12x(3x+2)\log x}{3(1-x)^6}~,
 \nonumber\\
 f_1^{(0)}(x)
 \!\!\! &=& \!\!\!
  \frac{3-4x+x^2+2\log x}{(1-x)^3}~,
 \nonumber\\
 f_1^{(1)}(x)
 \!\!\! &=& \!\!\!
  \frac{-5+4x+x^2-2(1+2x)\log x}{(1-x)^4}~,
 \nonumber\\
 f_1^{(2)}(x)
 \!\!\! &=& \!\!\!
  \frac{2(3-3x^2+(1+4x+x^2)\log x)}{(1-x)^5}\,,
 \nonumber\\
 f_1^{(3)}(x)
 \!\!\! &=& \!\!\!
  2~\frac{-10-9x+18x^2+x^3-3(1+6x+3x^2)\log x}{3(1-x)^6}\,.
\end{eqnarray}

\subsection*{SM Loop Functions}

\begin{eqnarray}
S_0(x)&=&\frac{4x-11x^2+x^3}{4(1-x)^2}-\frac{3x^3\log x}{2(1-x)^3}\,,\\
Y_0(x)&=&\frac{x}{8}\left(\frac{x-4}{x-1}+\frac{3x}{(x-1)^2}\log x\right)\,,\\
S_0(x_c,x_t)&=&x_c \left(\log\frac{x_t}{x_c}-\frac{3x_c x_t}{4(1-x_t)}-\frac{3x_t^2\log x_t}{4(1-x_t)^2}\right)\,.
\end{eqnarray}

\newpage
\section{MSSM parameter convention} \label{sec:appendixRosiek}

Throughout sec.~\ref{sec:DF012} we adopted the convention of~\cite{Rosiek:1995kg} for squark matrices and trilinear terms. In tab.~9 we show a dictionary between this convention and the ``SUSY Les Houches Accord 2'' (SLHA 2) convention in~\cite{Allanach:2008qq}.

\begin{table}[h]\label{tab:dictionary}
\addtolength{\arraycolsep}{3pt}
\renewcommand{\arraystretch}{1.4}
\begin{center}
\begin{tabular}{|c|c|}
\hline
SLHA 2~\cite{Allanach:2008qq} & ~\cite{Rosiek:1995kg} \\
\hline\hline
$\hat T_U, \hat T_D, \hat T_E$  &  $-A_u^T, +A_d^T, +A_l^T$ \\
$\hat m_{\tilde Q}^2, \hat m_{\tilde L}^2$  &  $m_{Q}^2, m_{L}^2$ \\
$\hat m_{\tilde u}^2, \hat m_{\tilde d}^2, \hat m_{\tilde e}^2$  &  $(m_{U}^2)^T, (m_{D}^2)^T, (m_{R}^2)^T$ \\
${\mathcal M}_{\tilde u}^2, {\mathcal M}_{\tilde d}^2$  &  $({\mathcal M}_{U}^2)^T, ({\mathcal M}_{D}^2)^T$ \\
\hline
\end{tabular}
\end{center}
\caption{Dictionary between the SLHA 2 convention of~\cite{Allanach:2008qq} and the convention of~\cite{Rosiek:1995kg}. Note that $A_t$ and $A_b$ used in sec.~\ref{sec:DF012} are defined as $m_t A_t=\frac{v_2}{\sqrt 2}\left(A_u\right)_{33}$ and $m_b A_b=\frac{v_1}{\sqrt 2}\left(A_d\right)_{33}$.}
\end{table}


\bibliographystyle{utphys}
\bibliography{ABGPS}


\end{document}